\documentclass[final,1p,times,authoryear]{elsarticle}
\usepackage{longtable}
\usepackage{graphicx}
\usepackage{amsmath,amsfonts,amssymb}
\usepackage{wrapfig}
\usepackage{multirow}
\usepackage{epsfig}
\usepackage{psfrag}
\usepackage{subfigure}
\usepackage{xcolor}
\usepackage{mathrsfs}
\usepackage{stackengine}
\usepackage{float}
\usepackage{longtable}
\usepackage{tabularx}
\journal{New Astronomy}

\begin{document}

\begin{frontmatter}

\title{A study of dynamical effects in the observed second time-derivative of the spin or orbital frequencies of pulsars}

\author[a,b]{Dhruv Pathak\corref{cor1}}
\cortext[cor1]{Corresponding Author}
\ead{dhruvpathak@imsc.res.in}
\author[a,b]{Manjari Bagchi}

\address[a]{The Institute of Mathematical Sciences, C. I. T. campus, Taramani, Chennai, 600113, India}
\address[b]{Homi Bhabha National Institute, Training School Complex, Anushakti Nagar, Mumbai 400094, India}

\begin{abstract}
The observed values of the time-derivatives of the spin or orbital frequency of pulsars are affected by their dynamical properties. We derive thorough analytical expressions for such dynamical contributions in terms of the Galactic coordinates, the proper motion, the pulsar distance, and the radial velocity. We find that the effects of the dynamical terms in the second-derivative of frequencies or parameters based on such second derivatives, e.g., braking index, are usually negligible. However, unique pulsars for which the effects of the dynamical terms are significant can exist. In particular, dynamical effects can make the magnitude of the observed value of the braking index to be in the order of thousand while the true value of it is close to the theoretically expected value three, especially if the pulsars lie close to the Galactic centre. Dynamics can also affect the value of the second derivative of the orbital frequency of a binary pulsar at the first decimal place. We also emphasize the fact that our expressions provide more accurate results than pre-existing approximate ones that exclude some of the terms. Comparison with a set of pulsars showed that the median value of the difference between the results obtained by our method and a pre-existing method is about 50 percent.

\end{abstract}


\end{frontmatter}

\section{Introduction}
\label{sec:intro}

Pulsar timing, which tracks the rotation of a pulsar, is a powerful tool to study a wide range of physics, e.g., dense matter equations of state \citep{bagchi18}, properties of interstellar medium \citep{kcs13}, alternative theories of gravity \citep{fwe12,bt14}, and even low-frequency gravitational waves \citep{det79}. The set of parameters that leads to a good fit between the expected pulse time of arrivals and observed time of arrivals is called a timing solution. The timing solution of radio pulsars contains various measured parameters such as the coordinates of the pulsar, the proper motion, the spin frequency and its derivatives, the dispersion measure, the parallax (if available), etc. In case of binary pulsars, timing solutions give Keplerian parameters like the orbital period (or the orbital frequency), the orbital eccentricity, the longitude of periastron, the projected semi-major axis of the orbit, and the epoch of periastron passage, as well as post-Keplerian parameters like the time-derivative of the orbital period (or the time derivative of the orbital frequency), the periastron advance, the Einstein delay, and the range and shape of the Shapiro delay. Sometimes, even the higher-order time-derivatives of the spin and/or the orbital frequency are fitted. Among all these timing parameters, the measured values of the frequency derivatives are affected by the velocity and acceleration of the pulsars. These effects are known as `dynamical contributions'. A thorough discussion on these dynamical contributions can be found in our earlier work \citep{pb18}, where we restricted ourselves in the investigation of the dynamical effects in the first derivatives of the spin and orbital periods.

Higher-order time derivatives of the spin and the orbital frequencies are also affected by dynamics. In one of the seminal works in this topic, \cite{jr97} discussed how the higher derivatives of the spin frequency bear the imprints of unmodelled orbital motion. More specifically, they expressed the spin frequency derivative as the line of sight acceleration, the spin frequency second derivative as the line of sight jerk, and so on. These line of sight acceleration, jerk, jounce, etc., might arise due to the motion of the pulsar in the gravitational potential of the Galaxy as well as due to the orbital motion (if any). The expressions for the line of sight acceleration, jerk, jounce, etc. due to the orbital motion are given in \cite{blw13} and \cite{bas16}. The main limitation of the work by  \cite{jr97} is the fact that they ignored a few effects, e.g., the intrinsic values of the frequency derivatives, the acceleration and jerk of the pulsar due to the gravitational potential of the Galaxy and other nearby stars, the acceleration of the local inertial frame (the Solar system barycentre), and the change in the direction of the line-of-sight due to the motion of the pulsar.

Some interesting results by modeling the higher-order time derivatives of spin frequencies came in the past few years. One example is the revelation of millisecond pulsars PSRs J1024$-$0719 and J1823-3021A being members of very wide orbit binaries \citep{gl16,kap16,bas16, psl17}. Although for the second pulsar, which is located in the globular cluster NGC 6624, there is some controversy on the conclusion of the mass modeling of the cluster \citep{gby18}, the binary nature of the pulsar is still unchallenged.

Lately, \cite{liu18} provided an analytical derivation for the dynamical effects on the first and second derivative of the pulsar spin frequency. However, the expressions that they provided were in the frame where the motion of the Sun is taken to be negligible. Since the radial velocity ($v_r$) of only a few pulsars is known (five as per \cite{liu18}), courtesy optical spectroscopy of their binary companions, they vary $v_r$ from -200 ${\rm km~s^{-1}}$ to 200 ${\rm km~s^{-1}}$ to get an estimate the radial velocity dependent terms. They assumed braking index ($n$) as three in order to estimate the order of magnitude of the intrinsic spin frequency second derivative from the expression $\ddot{f}_{\rm s, int}=n{\dot{f}^{2}_{\rm s, int}}/{f_{\rm s}}$, where $\ddot{f}_{\rm s, int}$ is the intrinsic value of second time derivative of the spin frequency, $\dot{f}_{\rm s,int}$ is the intrinsic value of the first time derivative of the spin frequency, and $f_{\rm s}$ is the spin frequency. Instead of deriving and using analytical expressions for the terms involving acceleration, jerk etc., they used a model of the gravitational potential of the Galaxy and an orbit integrator to integrate the motion of the pulsars and numerically obtained the acceleration and jerk terms by a polynomial fitting. 

In the present work, we derive complete analytical expressions for all dynamical terms and provide a way to calculate their individual contribution to the measured second derivative of the frequency without any numerical fitting. These expressions are valid for both of the second derivative of the spin frequency, as well as of the orbital frequency. Note that, unlike our earlier work, here we derive expressions in terms of derivatives of frequencies (instead of periods) as it is a common practice in timing analysis to fit for derivatives of frequencies.

One might ask the question that how useful would these expressions be as measuring the second derivative of the spin frequency is not easy. In the ATNF catalogue (version 1.63), there are only 434 pulsars in the Galactic filed for which measured values of the second derivative of the spin frequency have been reported out of total 2634 of such pulsars. Moreover, this parameter has not been fitted in the timing solutions of any of the pulsars in the IPTA second data release \citep{pdd19}. 

These facts do not make the study of dynamical contributions in the second derivativee of the frequency irrelevant. \citet{liu19} suggested that the timing solutions would be more accurate if this parameter is fitted for at least some of the Pulsar Timing Array pulsars (e.g., PSRs J1024$-$0719, J1939+2134, J0621+1002, J1022+1001, and B1821$-$24A). Moreover, in the future, the square kilometer array (SKA) will detect  many millisecond pulsars, the simulations by \citet{kbk14} showed that that SKA1-MID would detect about 9000 normal pulsars and about 1400 millisecond pulsars while the numbers for SKA1-LOW were about 7000 normal pulsars and about 900 millisecond pulsars. Out of this large number of new pulsars, some will have measurably large values of the second derivative of the spin frequency. Additionally, there are ongoing efforts to understand the causes of uncertainty in measurement of this parameter, as an example, Indian Pulsar Timing Array can measure the variation of DM accurate up to the fourth decimal place (\citet{kk20}; see also the talk by Bhal Chandra Joshi in the IPTA catch-up meeting). There are many efforts to model the red noise in pulsar timing data too. The implementation of these results and models in the timing analysis will reduce the uncertainty of measurement of the second derivative of the spin frequency. With a larger sample of measured values of this parameter, a good model to the decouple dynamical terms from the intrinsic term will help better understanding of pulsar properties like braking index etc. The additional advantage is that this same formalism can be applied to decouple dynamical terms from the intrinsic term in the second derivative of the orbital frequency. In short, the motivation of this paper is not to explore the detectability of the second derivative of the frequency, it is to provide a formalism based on which one can decouple the dynamical terms from the intrinsic terms.

This paper is organized as follows. In section \ref{se:der1}, we present the analytical expressions for dynamical contributions of the second derivatives of the frequency with details reported in appendices A and B. In section \ref{sec:MCMC}, we present the numerical exploration to understand relative importance of these dynamical terms in case of both synthetic, as well as, real pulsars. In section \ref{se:App}, we present some applications of correcting the dynamical effects from the second derivatives of the frequencies. In particular, first we discuss the properties of PSR J1024$-$0719, then we investigate how dynamical terms can affect the measured values of braking index and finally we explore the contributions from the dynamical terms in the second derivatives of orbital frequencies. Then in section \ref{se:con}, we present the conclusions of our work. A comparison between the results using our method and an existing approximate method \citep{liu18} can be found in \ref{appendix:compareLiu}.

\section{Analytical expression for the intrinsic second derivative of the frequency}
\label{se:der1}

The Doppler shift of the frequency (either the spin or the orbital) of the pulsar can be written as:

\begin{equation}
f_{\rm int} = (c + \vec{v}_{\rm p} . \widehat{n}_{\rm sp}) (c + \vec{v}_{\rm s} . \widehat{n}_{\rm sp})^{-1} f_{\rm obs} 
\label{eq:doppler1}
\end{equation} where $c$ is the speed of light in vacuum, $\vec{v}_{\rm p}$ is the velocity of the pulsar, $\vec{v}_{\rm s}$ is the velocity of the Sun, $\widehat{n}_{\rm sp}$ is the unit vector from the Sun to the pulsar and is taken to be the radial direction, $f_{\rm obs}$ is the observed (measured) frequency, and $f_{\rm int}$ is the intrinsic frequency. This frequency can be either the spin or the orbital. As the pulse arrival times on the earth are first translated to the Solar system barycentre before doing any timing analysis, the Solar system barycentre plays the role of the receiver in the Doppler shift eq. (\ref{eq:doppler1}), and due to the proximity of the barycentre to the Sun, we simply write it as `the Sun'. Differentiating eq. (\ref{eq:doppler1}) with respect to time, we get:

\begin{equation}
\dot{f}_{\rm int} 
= \frac{\left(\vec{a}_{\rm p}\cdot \widehat{n}_{\rm sp} + \vec{v}_{\rm p}\cdot\dot{\widehat{n}}_{\rm sp}\right)}{(c + \vec{v}_{\rm s}\cdot\widehat{n}_{\rm sp})}\,{f_{\rm obs}} - \frac{(c + \vec{v}_{\rm p}\cdot\widehat{n}_{\rm sp})}{(c + \vec{v}_{\rm s}\cdot\widehat{n}_{\rm sp})^2}\left(\vec{a}_{\rm s}\cdot \widehat{n}_{\rm sp} + \vec{v}_{\rm s}\cdot\dot{\widehat{n}}_{\rm sp} \right)\,{f_{\rm obs}}+\frac{(c + \vec{v}_{\rm p}\cdot\widehat{n}_{\rm sp})}{(c + \vec{v}_{\rm s}\cdot\widehat{n}_{\rm sp})}\,\dot{f}_{\rm obs} ~.
\label{eq:dplr2}
\end{equation}

Here, $\vec{a}_{\rm p}$ is the acceleration of the pulsar, $\vec{a}_{\rm s}$ is the acceleration of the Sun, and the dot over any parameter corresponds to the time derivative of that parameter, double dot representing the second time derivative and so on. Dividing of eq. (\ref{eq:dplr2}) by eq. (\ref{eq:doppler1}) and assuming $1 + \frac{\vec{v}_{\rm s} . \widehat{n}_{\rm sp}}{c} \simeq 1$ and $1 + \frac{\vec{v}_{\rm p} . \widehat{n}_{\rm sp}}{c} \simeq 1$, we get:

\begin{equation} \frac{\dot{f}_{\rm int} }{ {f}_{\rm int}} = \frac{ (\vec{a}_{\rm p} - \vec{a}_{\rm s}) \cdot \widehat{n}_{\rm sp}}{c} + \frac{1}{c} (\vec{v}_{\rm p} - \vec{v}_{\rm s}) \cdot \frac{d}{dt} (\widehat{n}_{\rm sp})+ \frac{\dot{f}_{\rm obs} }{ {f}_{\rm obs}}  ~.
\label{eq:doppler2b}
\end{equation}

The above assumptions just before eq. (\ref{eq:doppler2b}) also enable us to write ${f}_{\rm int} \simeq {f}_{\rm obs} = {f}$ and eq. (\ref{eq:doppler2b}) is written as:

\begin{equation}
\left(\frac{\dot{f}}{f}\right)_{\rm ex} = \left(\frac{\dot{f}}{f}\right)_{\rm obs} - \left( \frac{\dot{f}}{f}\right)_{\rm int} = - \left[ \frac{ (\vec{a}_{\rm p} - \vec{a}_{\rm s}) \cdot \widehat{n}_{\rm sp}}{c} + \frac{1}{c} (\vec{v}_{\rm p} - \vec{v}_{\rm s}) \cdot \frac{d}{dt} (\widehat{n}_{\rm sp}) \right] ~.
\label{eq:doppler2bn}
\end{equation}

The second term in the square brackets of eq. (\ref{eq:doppler2bn}) is the well known Shklovskii term. Both of the terms have been discussed in details in \cite{pb18} and a python package `GalDynPsr' has been created to estimate the value of $\left( \frac{\dot{f}}{f}\right)_{\rm ex}$ when all of the relevant parameters are known. After computing $\left( \frac{\dot{f}}{f}\right)_{\rm ex}$, one can estimate the value of $\dot{f}_{\rm int}$ using the relation:

\begin{equation}
\dot{f}_{\rm int} = f\,\left[ \left(\frac{\dot{f}}{f}\right)_{\rm obs} - \left( \frac{\dot{f}}{f}\right)_{\rm ex} \right] ~, 
\label{eq:fdotint}
\end{equation} if $f$ and $\dot{f}_{\rm obs}$ are known. Now, differentiating of eq. (\ref{eq:dplr2}) with respect to time, we get:

\begin{align}
\ddot{f}_{\rm int} = &\frac{\left(\dot{\vec{a}}_{\rm p}\cdot \widehat{n}_{\rm sp} + 2\vec{a}_{\rm p}\cdot\dot{\widehat{n}}_{\rm sp} + \vec{v}_{\rm p}\cdot\ddot{\widehat{n}}_{\rm sp}\right)}{(c + \vec{v}_{\rm s}\cdot\widehat{n}_{\rm sp})}\,f_{\rm obs}-\frac{(c + \vec{v}_{\rm p}\cdot\widehat{n}_{\rm sp})}{(c + \vec{v}_{\rm s}\cdot\widehat{n}_{\rm sp})^2}\left(\dot{\vec{a}}_{\rm s}\cdot \widehat{n}_{\rm sp} + 2\vec{a}_{\rm s}\cdot\dot{\widehat{n}}_{\rm sp} + \vec{v}_{\rm s}\cdot\ddot{\widehat{n}}_{\rm sp}\right)\,f_{\rm obs} \nonumber \\
 &-2\frac{\left(\vec{a}_{\rm p}\cdot \widehat{n}_{\rm sp} + \vec{v}_{\rm p}\cdot\dot{\widehat{n}}_{\rm sp}\right)\left(\vec{a}_{\rm s}\cdot \widehat{n}_{\rm sp} + \vec{v}_{\rm s}\cdot\dot{\widehat{n}}_{\rm sp} \right)}{(c + \vec{v}_{\rm s}\cdot\widehat{n}_{\rm sp})^2}\,f_{\rm obs}+2\frac{(c + \vec{v}_{\rm p}\cdot\widehat{n}_{\rm sp})}{(c + \vec{v}_{\rm s}\cdot\widehat{n}_{\rm sp})^3}\left(\vec{a}_{\rm s}\cdot \widehat{n}_{\rm sp} + \vec{v}_{\rm s}\cdot\dot{\widehat{n}}_{\rm sp} \right)^2\,f_{\rm obs} \nonumber \\
 &+2\frac{\left(\vec{a}_{\rm p}\cdot \widehat{n}_{\rm sp} + \vec{v}_{\rm p}\cdot\dot{\widehat{n}}_{\rm sp}\right)}{(c + \vec{v}_{\rm s}\cdot\widehat{n}_{\rm sp})}\,\dot{f}_{\rm obs}-2\frac{(c + \vec{v}_{\rm p}\cdot\widehat{n}_{\rm sp})}{(c + \vec{v}_{\rm s}\cdot\widehat{n}_{\rm sp})^2}\left(\vec{a}_{\rm s}\cdot \widehat{n}_{\rm sp} + \vec{v}_{\rm s}\cdot\dot{\widehat{n}}_{\rm sp} \right)\,\dot{f}_{\rm obs}+\frac{(c + \vec{v}_{\rm p}\cdot\widehat{n}_{\rm sp})}{(c + \vec{v}_{\rm s}\cdot\widehat{n}_{\rm sp})}\,\ddot{f}_{\rm obs}
\label{eq:pddot1}
\end{align}

\vskip 0.5cm

Dividing eq. (\ref{eq:pddot1}) by eq. (\ref{eq:doppler1}), we get:

\begin{align}
\frac{\ddot{f}_{\rm int} }{{f}_{\rm int}} = &\frac{\left(\dot{\vec{a}}_{\rm p}\cdot \widehat{n}_{\rm sp} + 2\vec{a}_{\rm p}\cdot\dot{\widehat{n}}_{\rm sp} + \vec{v}_{\rm p}\cdot\ddot{\widehat{n}}_{\rm sp}\right)}{(c + \vec{v}_{\rm p}\cdot\widehat{n}_{\rm sp})}-\frac{\left(\dot{\vec{a}}_{\rm s}\cdot \widehat{n}_{\rm sp} + 2\vec{a}_{\rm s}\cdot\dot{\widehat{n}}_{\rm sp} + \vec{v}_{\rm s}\cdot\ddot{\widehat{n}}_{\rm sp}\right)}{(c + \vec{v}_{\rm s}\cdot\widehat{n}_{\rm sp})} \nonumber \\
&-2\frac{\left(\vec{a}_{\rm p}\cdot \widehat{n}_{\rm sp} + \vec{v}_{\rm p}\cdot\dot{\widehat{n}}_{\rm sp}\right)\left(\vec{a}_{\rm s}\cdot \widehat{n}_{\rm sp} + \vec{v}_{\rm s}\cdot\dot{\widehat{n}}_{\rm sp} \right)}{(c + \vec{v}_{\rm p}\cdot\widehat{n}_{\rm sp})(c + \vec{v}_{\rm s}\cdot\widehat{n}_{\rm sp})} +2\frac{\left(\vec{a}_{\rm s}\cdot \widehat{n}_{\rm sp} + \vec{v}_{\rm s}\cdot\dot{\widehat{n}}_{\rm sp} \right)^2}{(c + \vec{v}_{\rm s}\cdot\widehat{n}_{\rm sp})^2} + 2\frac{\left(\vec{a}_{\rm p}\cdot \widehat{n}_{\rm sp} + \vec{v}_{\rm p}\cdot\dot{\widehat{n}}_{\rm sp}\right)}{(c + \vec{v}_{\rm p}\cdot\widehat{n}_{\rm sp})}\frac{\dot{f}_{\rm obs}}{{f}_{\rm obs}} \nonumber \\
& -2\frac{\left(\vec{a}_{\rm s}\cdot \widehat{n}_{\rm sp} + \vec{v}_{\rm s}\cdot\dot{\widehat{n}}_{\rm sp} \right)}{(c + \vec{v}_{\rm s}\cdot\widehat{n}_{\rm sp})}\frac{\dot{f}_{\rm obs}}{{f}_{\rm obs}} + \frac{\ddot{f}_{\rm obs} }{{f}_{\rm obs}} \nonumber \\ 
= &\frac{\left(\dot{\vec{a}}_{\rm p}-\dot{\vec{a}}_{\rm s}\right)\cdot \widehat{n}_{\rm sp}}{c} + 2\frac{\left(\vec{a}_{\rm p} - \vec{a}_{\rm s}\right)\cdot\dot{\widehat{n}}_{\rm sp}}{c} + \frac{(\vec{v}_{\rm p} - \vec{v}_{\rm s})\cdot\ddot{\widehat{n}}_{\rm sp}}{c} \nonumber \\ 
&- 2\left(\frac{(\vec{a}_{\rm p} - \vec{a}_{\rm s}) \cdot \widehat{n}_{\rm sp}}{c} + \frac{1}{c}(\vec{v}_{\rm p} - \vec{v}_{\rm s})\cdot\dot{\widehat{n}}_{\rm sp})\right)\left(\frac{\vec{a}_{\rm s}\cdot \widehat{n}_{\rm sp}}{c} + \frac{\vec{v}_{\rm s}\cdot\dot{\widehat{n}}_{\rm sp}}{c} \right) \nonumber \\
&+ 2\left(\frac{(\vec{a}_{\rm p} - \vec{a}_{\rm s}) \cdot \widehat{n}_{\rm sp}}{c} + \frac{1}{c}(\vec{v}_{\rm p} - \vec{v}_{\rm s})\cdot\dot{\widehat{n}}_{\rm sp})\right)\frac{\dot{f}_{\rm obs}}{{f}_{\rm obs}} + \frac{\ddot{f}_{\rm obs} }{{f}_{\rm obs}}
\label{eq:pddot2}
\end{align} where in the second step we have again assumed $1 + \frac{\vec{v}_{\rm s} . \widehat{n}_{\rm sp}}{c} \approx 1$ and $1 + \frac{\vec{v}_{\rm p} . \widehat{n}_{\rm sp}}{c} \approx 1$. The above equation can be written in a neater way as:

\begin{align}
\left(\frac{\ddot{f}}{f}\right)_{\rm int} = &\frac{\left(\dot{\vec{a}}_{\rm p}-\dot{\vec{a}}_{\rm s}\right)\cdot \widehat{n}_{\rm sp}}{c} + 2\frac{\left(\vec{a}_{\rm p} - \vec{a}_{\rm s}\right)\cdot\dot{\widehat{n}}_{\rm sp}}{c} + \frac{(\vec{v}_{\rm p} - \vec{v}_{\rm s})\cdot\ddot{\widehat{n}}_{\rm sp}}{c} +2\left(\frac{\dot{f}}{f} \right)_{\rm ex}\left(\frac{\vec{a}_{\rm s}\cdot \widehat{n}_{\rm sp}}{c} + \frac{\vec{v}_{\rm s}\cdot\dot{\widehat{n}}_{\rm sp}}{c} \right)\nonumber \\ 
&-2\left(\frac{\dot{f}}{f}\right)_{\rm ex}\left(\frac{\dot{f}}{f}\right)_{\rm obs} + \left(\frac{\ddot{f}}{f}\right)_{\rm obs}
\label{eq:pddot4}
\end{align}

Defining

\begin{equation}
\left( \frac{\ddot{f}}{f}\right)_{\rm ex} = \left(\frac{\ddot{f}}{f}\right)_{\rm obs} - \left( \frac{\ddot{f}}{f}\right)_{\rm int} ~,
\label{eq:fddotex}
\end{equation} eq. (\ref{eq:pddot4}) can be written as

\begin{align}
\left( \frac{\ddot{f}}{f}\right)_{\rm ex} = &-\left[ \frac{\left(\dot{\vec{a}}_{\rm p}-\dot{\vec{a}}_{\rm s}\right)\cdot \widehat{n}_{\rm sp}}{c} + 2\frac{\left(\vec{a}_{\rm p} - \vec{a}_{\rm s}\right)\cdot\dot{\widehat{n}}_{\rm sp}}{c} + \frac{(\vec{v}_{\rm p} - \vec{v}_{\rm s})\cdot\ddot{\widehat{n}}_{\rm sp}}{c} +2\left(\frac{\dot{f}}{f} \right)_{\rm ex}\left(\frac{\vec{a}_{\rm s}\cdot \widehat{n}_{\rm sp}}{c} + \frac{\vec{v}_{\rm s}\cdot\dot{\widehat{n}}_{\rm sp}}{c} \right)\right. \nonumber \\ 
&\left.-2\left(\frac{\dot{f}}{f}\right)_{\rm ex}\left(\frac{\dot{f}}{f}\right)_{\rm obs}\right]
\label{eq:fddotex1}
\end{align}

Here, $\dot{\vec{a}}_{\rm p}$ is the jerk of the pulsar and $\dot{\vec{a}}_{\rm s}$ is the jerk of the Sun. As we have already mentioned, the value of $\left( \frac{\dot{f}}{f}\right)_{\rm ex}$ that appears in the last two terms in eq. (\ref{eq:fddotex1}) can be estimated using GalDynPsr. The unit vector $\widehat{n}_{\rm sp}$ and its derivatives can be expressed in terms of other measurable parameters (\ref{appendix:unitvector}). Those expressions are used in simplifying various terms of eq. (\ref{eq:fddotex1}) in subsequent appendices. In particular, in \ref{appendix:B1}, we derive an expression for $\frac{\left(\vec{a}_{\rm p} - \vec{a}_{\rm s}\right)\cdot\dot{\widehat{n}}_{\rm sp}}{c}$ in terms of measurable parameters. We perform the same task for $\frac{(\vec{v}_{\rm p} - \vec{v}_{\rm s})\cdot\ddot{\widehat{n}}_{\rm sp}}{c}$ in \ref{appendix:B2}, for $\left(\frac{\vec{a}_{\rm s}\cdot \widehat{n}_{\rm sp}}{c} + \frac{\vec{v}_{\rm s}\cdot\dot{\widehat{n}}_{\rm sp}}{c} \right)$ in \ref{appendix:B3}, and for $\frac{\left(\dot{\vec{a}}_{\rm p}-\dot{\vec{a}}_{\rm s}\right)\cdot \widehat{n}_{\rm sp}}{c}$ in \ref{appendix:B4}. Using the expressions derived in those appendices, eq. (\ref{eq:fddotex1}) can be written as:

\begin{align}
\left( \frac{\ddot{f}}{f}\right)_{\rm ex} = &-\left[ \frac{1}{c}\,\left( \dot{a}_{r}- a_{\rm T}\mu_T\,\cos \alpha \right)\right] - \left[2\,\left(\mu_{b}\,\frac{\sin b}{c}\,\left\{{a}_{\rm p,pl}\,\sqrt{\left(1-\frac{R_{\rm s}^{2}\,\sin^{2} l}{{R_{\rm p'}}^{2}}\right)} + {a}_{\rm s,pl}\,\cos l\right\} - \mu_{b}\,\frac{\cos b}{c}\,a_{\rm p,z} \right.\right. \nonumber \\  
&\left.\left. - \mu_{l}\,\frac{\sin l}{c}\,\left\{{a}_{\rm p,pl}\,\frac{R_{\rm s}}{R_{\rm p'}} - {a}_{\rm s,pl} \right\}\right) + 2\,\left(\frac{\dot{f}}{f} \right)_{\rm ex}\left(\cos b\,\cos l\,\frac{{a}_{\rm s,pl}}{c} + \mu_{b}\,\frac{{v}_{\rm s,pl}}{c}\,\sin b\,\sin l - \mu_{l}\,\frac{{v}_{\rm s,pl}}{c}\,\cos l \right)\right]  \nonumber \\ 
 & - \left[ \frac{1}{c}\,\left(\mu_{\rm T}\,a_{\rm T}\,\cos \alpha - 3\,v_{r}\,{\mu_{\rm T}}^{2}\right)\right] + \left[2\,\left(\frac{\dot{f}}{f}\right)_{\rm ex}\left(\frac{\dot{f}}{f}\right)_{\rm obs}\right] ~. 
\label{eq:fddotex4}
\end{align}

Here, $l$ is Galactic longitude, $b$ is the Galactic latitude, $\mu_{l}$ is the proper motion in $l$, $\mu_{b}$ is the proper motion in $b$, $\mu_{\rm T}$ is the total transverse proper motion, $v_{r}$ is the radial component of the relative velocity of the pulsar with respect to the Sun, $R_{\rm s}$ is the Galactocentric distance of the Sun, $R_{\rm p'}$ is Galactocentric distance of the pulsar projection on the Galactic plane, ${v}_{\rm s,pl}$ is the component of the velocity of the Sun parallel to the Galactic plane, $a_{\rm T}$ is the transverse component of the relative acceleration of the pulsar with respect to the Sun, ${a}_{\rm p,pl}$ is the component of the acceleration of the pulsar parallel to the Galactic plane, ${a}_{\rm s,pl}$ is the component of the acceleration of the Sun parallel to the Galactic plane, ${a}_{\rm p,z}$ is the component of the acceleration of the pulsar perpendicular to the Galactic plane, $\alpha$ is the angle between the transverse velocity and transverse acceleration, and $\dot{a}_{r}$ is the rate of change of the radial component of the relative acceleration of the pulsar with respect to the Sun. Expressions of $\dot{a}_{r}$ and $a_{\rm T}$ in terms of observables like $l$, $b$, $\mu_{l}$, $\mu_{b}$, $d$ (distance of the pulsar from the Sun), and $v_{r}$ are derived in \ref{appendix:B4}. Expression of $\alpha$ in terms of these observables is derived in \ref{appendix:B3}.

We can now use the values of $\left( \frac{\ddot{f}}{f}\right)_{\rm ex}$, $f$, and $\ddot{f}_{\rm obs}$ to estimate intrinsic frequency second derivative ($\ddot{f}_{\rm int}$) by the following relation:

\begin{equation}
\ddot{f}_{\rm int} = f\,\left[ \left(\frac{\ddot{f}}{f}\right)_{\rm obs} - \left( \frac{\ddot{f}}{f}\right)_{\rm ex} \right]
\label{eq:fddotint}
\end{equation}

Note that, $\dot{f}_{\rm int}$ and $\ddot{f}_{\rm int}$ are the values of the first and second time derivatives of the frequency after eliminating contributions from the velocity of the pulsar and its acceleration and jerk due to the gravitational potential of the Galaxy. There might be additional case specific dynamical/kinematic contributions. So, although we call these `intrinsic' they might not be the true intrinsic values. If the presence of additional dynamical terms are confirmed, then the value of the second derivative of the frequency obtained after eliminating the effects due to the velocity, acceleration and jerk of the pulsar due to the Galactic potential, as given in eq. (\ref{eq:fddotint}) should be rather called the `residual' value or $\ddot{f}_{\rm res}$. Same is true for the first derivative.

\section{Numerical exploration to understand relative importance of various terms in the expression of $\left( \frac{\ddot{f}}{f}\right)_{\rm ex}$:}
\label{sec:MCMC}

The last square bracket term in eq. (\ref{eq:fddotex4}) can be computed using the observed values of the frequency and the first time-derivative of the frequency with the help of GalDynPsr if all the relevant parameters like the coordinates, the distance and the proper motion of the pulsar are known, which is the case for most of the well-timed pulsars as these are the parameters one fits for while doing a timing analysis. In the present work, we use the best available model in GalDynPsr, i.e., model-Lb to estimate $\left( \frac{\dot{f}}{f}\right)_{\rm ex}$ \citep[for more details about various models available in GalDynPsr]{pb18}.

\subsection{Pulsar population in the Galactic field}
\label{subsec:MCMCField}

However, the other terms in eq. (\ref{eq:fddotex4}) contain parameters that might not be easily measurable, especially the first square bracket term. So, it is worth investigating the relative significance of various terms within square brackets in eq. (\ref{eq:fddotex4}) to decide whether any one of those can be ignored. As there are not many pulsars with all relevant parameters (appearing in the first three square bracket terms) known, we decided to perform simulation. In particular, we used `PsrPopPy' \citep{bat14}, a python based population synthesis package to generate a population model. From the version 1.63 of the ATNF\footnote{http://www.atnf.csiro.au/research/pulsar/psrcat/} pulsar catalogue \citep{mhth05}, we find that there are 1085 normal pulsars (spin period $>$ 30 ms) that are detected by the Parkes Multibeam Pulsar Survey \citep{mlc01}, PMPS henceforth, and which had well defined distance values, obtained from their Dispersion Measure (DM) values using the NE2001 model of electron density \citep{cl02,cl03}. We did not use independent distance measurements (e.g. using parallax etc) as PsrPopPy uses DM values to model detectability of pulsars. This set excludes pulsars in globular cluster and in both Large and Small Magellanic Clouds as in these cases, there will be extra dynamical effects due to the local gravitational potential. We used only the PMPS as this is the survey best modeled in PsrPopPy. We used this number as the target number for detection in PsrPopPy with default settings, i.e. the best model parameters, and survey set to PMPS. We found that a population of 124310 pulsars is generated in order to detect 1085 cases by that survey. However, 180 synthetic pulsars had to be discarded as these had DM values which exceeded the Galactic contribution and hence only a lower bound was available on their distances calculated using NE2001 model. From the remaining ones, we excluded 15 cases with spin period less than 30 ms. In this way, we were left with a set of 124115 synthetic normal pulsars.

Similarly, from ATNF catalogue, we find that there are 29 millisecond pulsars (spin period $<$ 30 ms) that are detected by the PMPS and which had well defined NE2001 model based distance. Using this number as the target number for detection in PsrPopPy, with survey set to PMPS and spin period set to follow a log-normal distribution with mean 1.45 ms and standard deviation 0.36 ms, and all other distributions as the best model of PsrPopPy, we generated a population model of 5338 pulsars. We chose these values for the mean and the standard deviation of the log-normal distribution to ensure that the minimum period value (1.518 ms) in the distribution does not go below the minimum period value (1.396 ms) seen in the ATNF catalogue. Out of these 5338 synthetic pulsars, seven had DM values that exceeded Galactic contribution and hence only a lower bound was available on their distances calculated using NE2001 model. We excluded these cases and got 5331 simulated pulsars. Out of these, we further excluded the cases with spin period greater than 30 ms and worked with the remaining 2791 synthetic millisecond pulsars.

From the population models generated by PsrPopPy, we extracted the values of the Galactic longitude ($l$), Galactic latitude ($b$), spin period ($P_{\rm s, int}$), and distance ($d$) based on NE2001 model in order to use in our calculations. We convert the intrinsic value of the spin period to that of the spin frequency ($f_{\rm s, int}$). For the parameters that are not available in PsrPopPy, e.g., the proper motion in the Galactic longitude ($\mu_l$), the proper motion in the Galactic latitude ($\mu_b$), the first time-derivative of the spin frequency ($\dot{f}_{\rm s, int}$), and the second time-derivative of the spin frequency ($\ddot{f}_{\rm s, int}$), we generated synthetic values based on the the distribution followed by the values of the parameters given in the ATNF catalogue. We performed this task separately for the normal and millisecond pulsars, i.e., generated 124115 synthetic values based on the the distributions followed by the values of the parameters for the normal pulsars and generated 2791 synthetic values based on the the distributions followed by the values of the parameters for the millisecond pulsars. For millisecond pulsars, we additionally generated 2791 synthetic values of the orbital frequency ($f_{\rm b, int}$), the first time-derivative of the orbital frequency ($\dot{f}_{\rm b, int}$), and the second time-derivative of the orbital frequency ($\ddot{f}_{\rm b, int}$). The underlying assumption here is that these parameters by themselves are not affected much by selection effects. For radial velocity, we took a uniform distribution between $-200$ and 200 km/s, same as the range used by \cite{liu18}.

We calculated all square bracket terms of eq. (\ref{eq:fddotex4}) separately as well as the sum of first three square bracket terms, which we call the `combined' term. Each term can have positive or negative values depending on the values of the parameters. However, the absolute value or the magnitude of these terms are more useful for the purpose of comparison.

\begin{figure*}
\begin{center}
\hskip-0.8cm\subfigure[]{\label{subfig:nogc123SBmsp}\includegraphics[width=0.52\textwidth, angle=0]{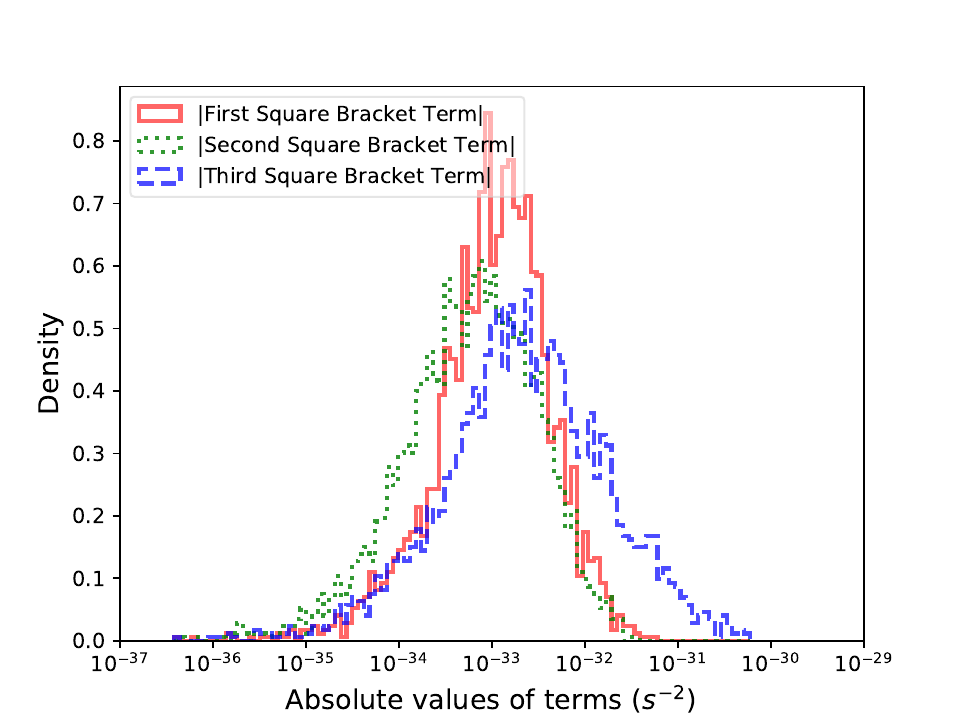}}
 \subfigure[]{\label{subfig:nogc4SBspincombmsp}\includegraphics[width=0.52\textwidth, angle=0]{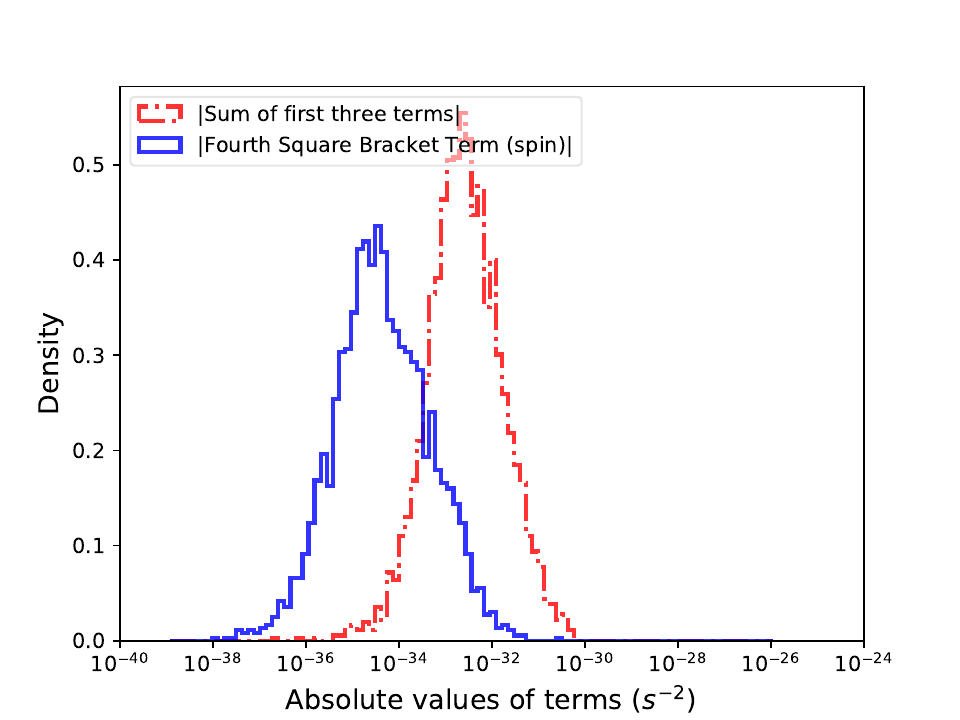}}\\
  \subfigure[]{\label{subfig:nogc4SBorbcombmsp}\includegraphics[width=0.52\textwidth, angle=0]{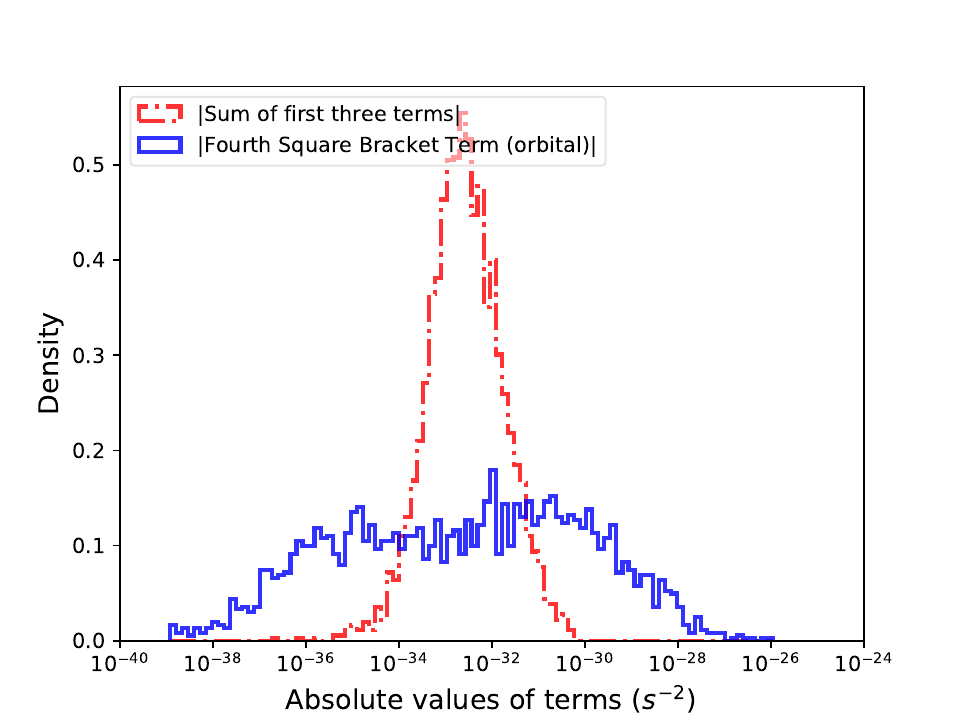}}\\
\end{center}
\caption{Density distribution of the absolute values of various square bracket terms in eq. (\ref{eq:fddotex4}) for simulated millisecond pulsars. The subplots are as follow: a) comparison of line histograms for the first square bracket term, the second square bracket term, and the third square bracket term, b) comparison of line histograms for the sum of the first three terms and the fourth square bracket term using the spin frequency and its derivatives, and c) comparison of line histograms for the sum of the first three terms and the fourth square bracket term using the orbital frequency and its derivatives.}
\label{fig:histMSP}
\end{figure*}

\begin{figure*}
\begin{center}
\hskip-0.8cm\subfigure[]{\label{subfig:nogc123SBnorm}\includegraphics[width=0.52\textwidth, angle=0]{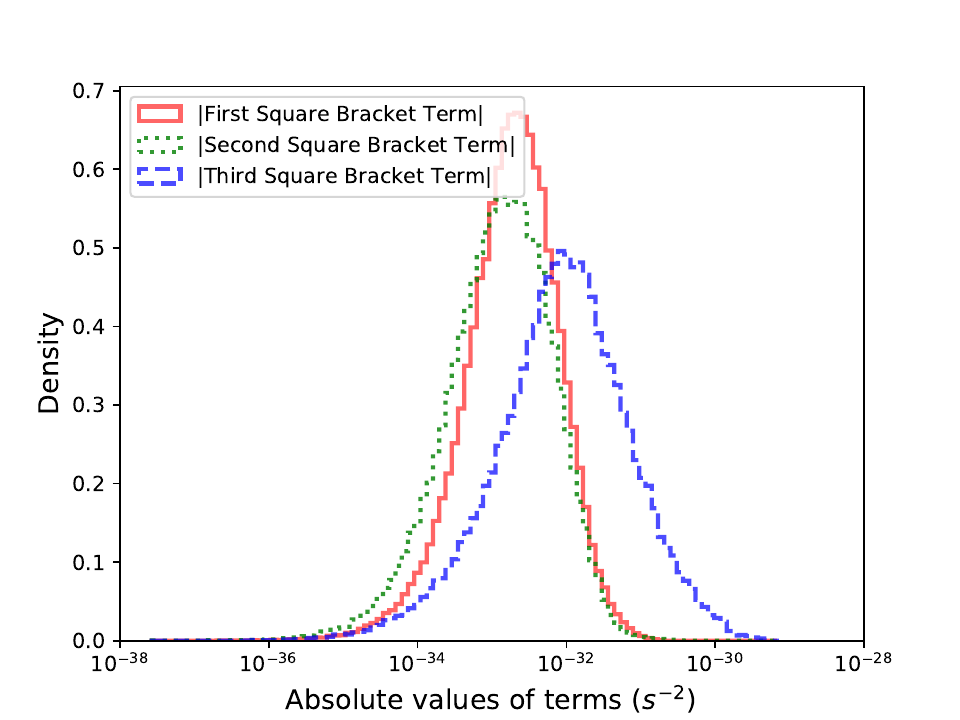}}
 \subfigure[]{\label{subfig:nogc4SBspincombnorm}\includegraphics[width=0.52\textwidth, angle=0]{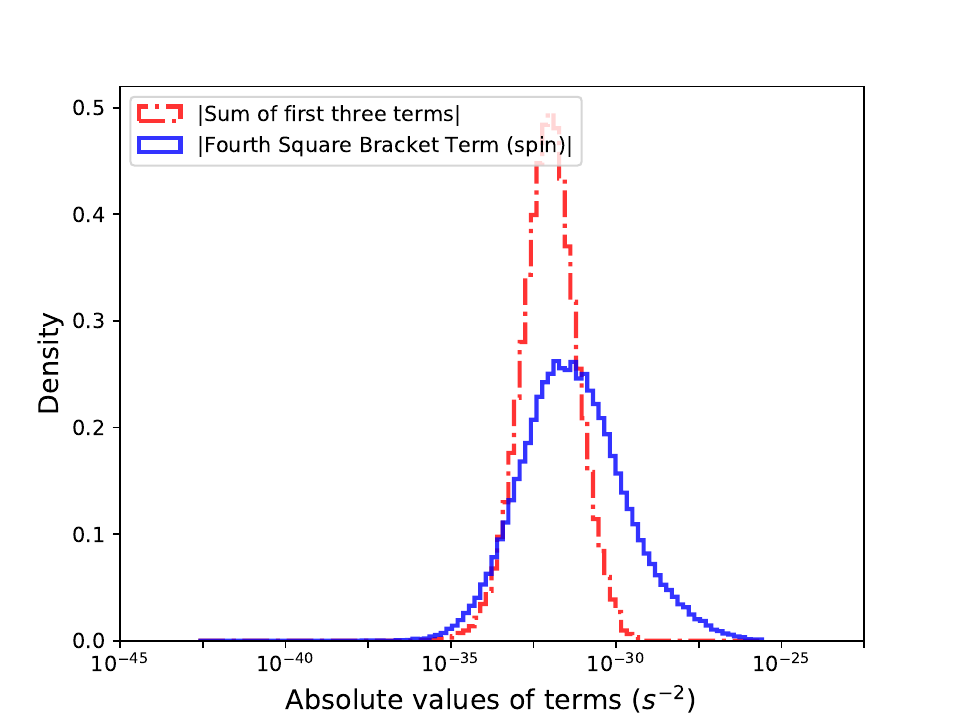}}\\
\end{center}
\caption{Density distribution of the absolute values of various square bracket terms in eq. (\ref{eq:fddotex4}) for the simulated normal pulsars. The subplots are as follow: a) comparison of line histograms for the first square bracket term, the second square bracket term, and the third square bracket term, and b) comparison of line histograms for the sum of the first three terms and the fourth square bracket term using the spin frequency and its derivatives.}
\label{fig:histNorm}
\end{figure*}

We summarize these results in the histograms of figure \ref{fig:histMSP} for synthetic millisecond pulsars and figure \ref{fig:histNorm} for synthetic normal pulsars. Additionally, table \ref{tb:simulrealvalsMSP} contains the statistical summary for the synthetic millisecond pulsars while table \ref{tb:simulrealvalsNorm} contains the statistical summary for the synthetic normal pulsars. From these, it is clear that all of the first three square bracket terms contribute `almost' equally to the observed value of the second derivative of the frequency.

We next aimed to compare the combined term with the remaining one, i.e., the fourth square bracket term of eq. (\ref{eq:fddotex4}). We can see from eq. (\ref{eq:fddotex4}) that the fourth square bracket term contains a factor $\left(\frac{\dot{f}}{f}\right)_{\rm obs}$, i.e., the ratio of observed frequency derivative and the frequency. The spin frequency and its derivative and the orbital frequency and its derivative are physically different entities, we can measure values of the spin frequency and its derivative even for isolated pulsars. Moreover, the magnitudes of these parameters and the sign of the derivatives are usually different. The spin frequency derivative is intrinsically a result of the slow down of the pulsar spin as it loses its rotational kinetic energy in the form of electromagnetic energy and hence expected to be negative while the orbital frequency derivative is intrinsically a result of the shrinking of the orbit due to the emission of the gravitational waves and hence expected to be positive. Although the dynamics affects the value of both the spin frequency derivative and the orbital frequency derivatives, it alters the sign very rarely. Hence the ratio of the observed frequency derivative and the frequency, and consequently, the entire fourth square bracket term will be different for the case when we consider spin frequency and its derivative as compared to when we consider orbital frequency and its derivative.

We first calculated the values of this term using the spin frequency and its derivative for both of the millisecond pulsar and the normal pulsar populations. Then, for the millisecond pulsar population, we calculated this term using the orbital frequency and its derivative.  Figures \ref{fig:histMSP} and \ref{fig:histNorm} as well as tables \ref{tb:simulrealvalsMSP} and \ref{tb:simulrealvalsNorm} contain results for the fourth square bracket term too. We did not calculate this term using the orbital frequency and its derivative for the normal pulsar population, as most of the real normal pulsars are isolated, and the orbital frequency and its first  derivative for the handful of normal pulsars does not bear the characteristic of the population. To be more specific, the ATNF catalogue (version 1.63) reports total 255 millisecond pulsars in the Galactic field; out of which 189 have orbital frequency reported and 30 have orbital frequency derivative measured. On the other hand, out of 2353 normal pulsars in the Galactic field, only 48 have orbital frequency reported and only 9 have orbital frequency derivative measured. Also, remember that this term is actually a part of the the excess (due to dynamics) in the second derivative of the orbital frequency as expressed in eq. (\ref{eq:fddotex4}), and so far no second derivative of orbital frequency has been ever measured for any normal pulsars, partly because millisecond pulsars are more stable and precise timing is possible.

The median of the absolute value of the combined term of millisecond pulsars is within about an order of magnitude of the absolute value of their fourth square bracket term for the orbital frequency. The median of the absolute value of the combined term of normal pulsars is within about an order of magnitude of the absolute value of their fourth square bracket term for the spin frequency while the median of the absolute value of the combined term of millisecond pulsars is within about two orders of magnitude of the absolute value of their fourth square bracket term for the spin frequency. On the other hand, due to their extreme stability, millisecond pulsars are timed better, so the second derivative of the spin frequency of millisecond pulsars are likely to be measured more accurately. Thus, we conclude that it is wise to retain all of the terms in eq. (\ref{eq:fddotex4}) when accurate values are aimed for.

\begin{table*}
\begin{center}
\caption{The top section of the table shows the statistical summary of all the square bracket terms appearing in eq. (\ref{eq:fddotex4}) for the simulated millisecond pulsars. We compare the absolute values of all the terms here. The middle section of the table shows the comparison of absolute values of the combined and the fourth square bracket (spin) terms for the 140 real millisecond pulsars in ATNF catalogue for which all relevant parameters to calculate $\left(\frac{\dot{f}}{f}\right)_{\rm ex}$, for the spin frequency and its derivative, are available. The bottom section of the table shows the comparison of absolute values of the combined and the fourth square bracket (orbital) terms for the 31 millisecond pulsars in ATNF catalogue for which all relevant parameters to calculate $\left(\frac{\dot{f}}{f}\right)_{\rm ex}$, for the orbital frequency and its derivative, are available.}
\begin{tabular}{|l|l|l|l|l|}
\hline
 \multicolumn{5}{|c|}{\addstackgap{Simulated Millisecond Pulsars}}  \\ 
 \hline
Term & Minimum & Mean & Median & Maximum \\ 
  & (${\rm s}^{-2}$) & (${\rm s}^{-2}$) & (${\rm s}^{-2}$)  & (${\rm s}^{-2}$) \\ \hline
  First Square Bracket Term & $3.49\times10^{-37}$ & $2.34\times10^{-33}$ & $1.15\times10^{-33}$ & $5.80\times10^{-32}$ \\ \hline
 Second Square Bracket Term & $4.51\times10^{-37}$ & $1.71\times10^{-33}$ & $6.54\times10^{-34}$ & $3.48\times10^{-32}$ \\ \hline
 Third Square Bracket Term & $3.78\times10^{-37}$ & $1.34\times10^{-32}$ & $2.20\times10^{-33}$ & $5.90\times10^{-31}$ \\ \hline
 Combined Term & $2.38\times10^{-37}$ & $1.38\times10^{-32}$ & $2.53\times10^{-33}$ & $5.77\times10^{-31}$ \\ \hline
 Fourth Square Bracket Term (spin) & $1.23\times10^{-38}$ & $6.32\times10^{-34}$ & $3.89\times10^{-35}$ & $2.60\times10^{-31}$ \\ \hline
 Fourth Square Bracket Term (orbital) & $1.17\times10^{-39}$ & $1.84\times10^{-29}$ & $3.40\times10^{-33}$ & $1.12\times10^{-26}$ \\ \hline  
  \multicolumn{5}{|c|}{\addstackgap{Real Millisecond Pulsars (spin)}}  \\ 
  \hline
    Combined Term & $1.08\times10^{-35}$ & $8.81\times10^{-33}$ & $1.06\times10^{-33}$ & $2.39\times10^{-31}$ \\ \hline
 Fourth Square Bracket Term (spin) & $7.01\times10^{-39}$ & $1.34\times10^{-35}$ & $2.23\times10^{-36}$ & $5.98\times10^{-34}$ \\ 
 \hline  
  \multicolumn{5}{|c|}{\addstackgap{Real Millisecond Pulsars (orbital)}}  \\ \hline
   Combined Term & $4.90\times10^{-35}$ & $1.18\times10^{-32}$ & $1.26\times10^{-33}$ & $2.39\times10^{-31}$ \\ \hline
 Fourth Square Bracket Term (orbital) & $1.62\times10^{-38}$ & $1.39\times10^{-32}$ & $1.21\times10^{-35}$ & $1.85\times10^{-31}$ \\
 \hline   
\end{tabular}
\label{tb:simulrealvalsMSP}
\end{center}
\end{table*}

\begin{table*}
\begin{center}
\caption{The top section of the table shows the statistical summary of all the square bracket terms appearing in eq. (\ref{eq:fddotex4}) for the simulated normal pulsars. We compare the absolute values of all the terms here. The bottom section of the table shows the comparison of absolute values of the combined and the fourth square bracket (spin) terms for the 238 real normal pulsars in ATNF catalogue for which all relevant parameters to calculate $\left(\frac{\dot{f}}{f}\right)_{\rm ex}$, for the spin frequency and its derivative, are available. }
\begin{tabular}{|l|l|l|l|l|}
\hline
 \multicolumn{5}{|c|}{\addstackgap{Simulated Normal Pulsars}}  \\ 
 \hline
Term & Minimum & Mean & Median & Maximum \\ 
  & (${\rm s}^{-2}$) & (${\rm s}^{-2}$) & (${\rm s}^{-2}$)  & (${\rm s}^{-2}$) \\ \hline
  First Square Bracket Term & $6.22\times10^{-39}$ & $4.54\times10^{-33}$ & $1.98\times10^{-33}$ & $3.57\times10^{-31}$ \\ \hline
 Second Square Bracket Term & $2.29\times10^{-39}$ & $3.83\times10^{-33}$ & $1.36\times10^{-33}$ & $4.36\times10^{-31}$ \\ \hline
 Third Square Bracket Term & $2.68\times10^{-38}$ & $5.57\times10^{-32}$ & $9.38\times10^{-33}$ & $6.73\times10^{-30}$ \\ \hline
 Combined Term & $1.01\times10^{-37}$ & $5.59\times10^{-32}$ & $9.62\times10^{-33}$ & $7.07\times10^{-30}$ \\ \hline
 Fourth Square Bracket Term (spin) & $1.57\times10^{-39}$ & $5.16\times10^{-29}$ & $4.33\times10^{-32}$ & $1.97\times10^{-25}$ \\ \hline 
 \multicolumn{5}{|c|}{\addstackgap{Real Normal Pulsars}}  \\ 
 \hline
   Combined Term & $2.63\times10^{-36}$ & $3.59\times10^{-32}$ & $4.25\times10^{-33}$ & $1.65\times10^{-30}$ \\ \hline
 Fourth Square Bracket Term (spin) & $1.12\times10^{-37}$ & $1.85\times10^{-30}$ & $1.47\times10^{-32}$ & $2.32\times10^{-28}$ \\ \hline   
\end{tabular}
\label{tb:simulrealvalsNorm}
\end{center}
\end{table*}

Here, one might be curious whether our result is affected by our choice of the Galactic electron density model, i.e., the use of NE2001 model instead of YMW16 \citep{ymw17} model. We actually explored both of the models although report only the results based on NE2001. To perform simulation based on YMW16 model, we used PsrPopPy2\footnote{\tt https://github.com/devanshkv/PsrPopPy2}. The use of different electron density models did not alter our conclusions, as over such a large number of samples, they provide almost the same statistics. Particularly, for MSPs (table \ref{tb:simulrealvalsMSP}), using NE2001 model the median of the combined term, the 4th square bracket spin term, and the 4th square bracket orbital term are $2.53 \times 10^{-33}$, $3.89 \times 10^{-35}$, and $3.40 \times 10^{-33}$ respectively while for YMW16 model, these terms are  $2.72 \times 10^{-33}$, $3.79 \times 10^{-35}$, and $3.48 \times 10^{-33}$. For normal pulsars (table \ref{tb:simulrealvalsNorm}) the median of the combined term and the 4th square bracket spin term are $9.62 \times 10^{-33}$ and $4.33 \times 10^{-32}$ respectively while for YMW16 model, these terms are  $9.51 \times 10^{-33}$ and $4.22 \times 10^{-32}$. Other statistics are also not affected much. 

We also need to remember that YMW16 is newer but not necessarily better than NE2001. In some directions, YMW16 is better, but in some other directions, NE2001 is better. The following examples will substantiate this claim. (i) For PSR J0034$-$0721, YMW16 agrees better with the independent distance measurement. For this pulsar, YMW16 distance, NE2001 distance and the independent distance values are 1.00, 0.39, and 1.03 kpc respectively. (ii) For PSR J0751$+$1807 NE2001 agrees better with independent distance measurement. For this pulsar YMW16 distance, NE2001 distance and the independent distance values are  0.43, 1.15, and 1.11 kpc respectively. (iii) Both of the models are bad (disagree with the independent distance measurement) for PSR J2337$+$6151. For this pulsar, YMW16 distance, NE2001 distance and independent distance values are 2.08, 3.15, and 0.70 kpc respectively. (iv) Both of the models are good (agree with independent distance measurement) for PSR J1833$-$0827. For this pulsar, YMW16 distance, NE2001 distance and independent distance values are  4.38, 4.66, 4.50 kpc respectively. There are more examples in each of these categories. We have decided to keep only NE2001 results in this paper as it will ease the comparison with older literature.

For the sake of comparison, we also computed the values of these terms for real pulsars too. The ATNF catalogue (version 1.63) has 378 pulsars for which all relevant parameters to calculate the value of $\left(\frac{\dot{f}}{f}\right)_{\rm ex}$, appearing in the fourth square bracket of eq. (\ref{eq:fddotex4}) for the spin frequency and its derivative, are available. Out of these 378 pulsars, 140 are millisecond pulsars and rest 238 are normal pulsars\footnote{Note that earlier we mentioned that ATNF catalogue reports 255 millisecond pulsars, but for all of those, all the parameters needed to compute the dynamical terms have not been reported, hence, here we are working with less number of millisecond pulsars.}. For these pulsars, all parameters required to calculate the first three square bracket terms are also known, except $v_r$ that appears in the first and the third square bracket terms. So, we needed to choose reasonable values for $v_r$. We performed our calculations (separately for both the sets- normal and millisecond pulsars) for some chosen values of $v_r$, i.e., -200, -100, -50, 0, 50, 100, and 200 km/s, except for two millisecond pulsars, PSRs J1024$-$0719 and J1903$+$0327, for which the values of $v_{r}$ are known to be 185 km/s and 42 km/s respectively \citep{liu18}. 

For the millisecond pulsars, when we took a nominal value of $v_{r}=$ 50 km/s (the same assumed value used by \cite{liu19}), we found the maximum difference between the absolute value of the combined and the fourth term for PSR J0437$-$4715, values being $2.39\times10^{-31}~{\rm s}^{-2}$ and $1.55\times10^{-34}~{\rm s}^{-2}$ respectively. Whereas, we found the minimum difference between the absolute value of the combined and the fourth term for PSR J0931$-$1902, values being $1.08\times10^{-37}~{\rm s}^{-2}$ and $1.80\times10^{-37}~{\rm s}^{-2}$ respectively. The change in the value of $v_r$ does not change the results in terms of order of magnitude, e.g., PSR J0437$-$4715 has the absolute value of the combined term as $9.54\times10^{-31}~{\rm s}^{-2}$ when $v_r$ is chosen to be 200 km/s and $9.51\times10^{-31}~{\rm s}^{-2}$ when $v_r$ is chosen to be $-200$ km/s. Note that the fourth term is independent of $v_r$.

Similarly, for the normal pulsars, when we took a nominal value of $v_{r}=$ 50 km/s, we found the maximum difference between the absolute value of the combined and the fourth term for PSR J1808$-$2024, values being $2.66\times10^{-34}~{\rm s}^{-2}$ and $2.32\times10^{-28}~{\rm s}^{-2}$ respectively. Whereas, we found the minimum difference between the absolute value of the combined and the fourth term for PSR J2018$+$2839, values being $2.70\times10^{-35}~{\rm s}^{-2}$ and $2.82\times10^{-35}~{\rm s}^{-2}$ respectively. With change in the value of $v_r$, the results varies within one order of magnitude, e.g., PSR J1808$-$2024 has the absolute value of the combined term as $1.63\times10^{-33}~{\rm s}^{-2}$ when $v_r$ is chosen to be 200 km/s and $2.00\times10^{-33}~{\rm s}^{-2}$ when $v_r$ is chosen to be $-200$ km/s. Note that the fourth term is independent of $v_r$.

Additionally, we report the statistical summary of the absolute values of the combined and the fourth square bracket term for the real millisecond pulsars in table \ref{tb:simulrealvalsMSP} and the real normal pulsars in table \ref{tb:simulrealvalsNorm} for $v_r=$ 50 km/s. We see that the median of the absolute values of the `combined term' is within three orders of magnitude of that of the values of the fourth square bracket term for both millisecond, as well as, normal pulsars. 

We have also found that, for 65.55\% of the normal pulsars, the absolute value of the fourth square bracket term is larger than the absolute value of the combined term. Further, 32.05\% of this particular set of pulsars (or 21.01\% of the total normal pulsar population) have the absolute value of the fourth square bracket term larger than the absolute value of the combined term by more than one order of magnitude. In case of the 140 millisecond pulsars, all of them had the absolute value of the fourth square bracket term smaller than the absolute value of the combined term. Furthermore, for 97.14\% of the millisecond pulsars,  the absolute value of the combined term was found to be larger than the absolute value of the fourth square bracket term by more than one order of magnitude. These facts again support our conclusion that the first three terms should not be ignored.

To complete our exploration with real pulsars, we find that the ATNF catalogue (version 1.63) has 31 millisecond pulsars for which all relevant parameters to calculate $\left(\frac{\dot{f}}{f}\right)_{\rm ex}$ for the orbital frequency and its derivative are available. For these pulsars too, all parameters required to calculate the first three square bracket terms are also known, except $v_r$, and we choose same values of $v_r$ as discussed above. When we took a nominal value of $v_{r}=$ 50 km/s, we found the maximum difference between the absolute value of the combined and the fourth term for PSR J0437$-$4715, values being $2.39\times10^{-31}~{\rm s}^{-2}$ and $1.17\times10^{-34}~{\rm s}^{-2}$ respectively. Whereas, we found the minimum difference between the absolute value of the combined and the fourth term for PSR J1603$-$7202, values being $4.90\times10^{-35}~{\rm s}^{-2}$ and $1.62\times10^{-38}~{\rm s}^{-2}$ respectively. The change in the value of $v_r$ does not change the results in the order of magnitude, e.g., PSR J0437$-$4715 has the absolute value of the combined term as $9.54\times10^{-31}~{\rm s}^{-2}$ when $v_r$ is chosen to be 200 km/s and $9.51\times10^{-31}~{\rm s}^{-2}$ when $v_r$ is chosen to be $-200$ km/s. 
  
Additionally, we report the statistical summary of the absolute values of the combined and the fourth square bracket term for these real millisecond pulsars in table \ref{tb:simulrealvalsMSP} for $v_r=$ 50 km/s. We see that the median of the absolute values of the `combined term' is within three orders of magnitude of that of the values of the fourth square bracket term. We have also found that, for 19.35\% pulsars, the magnitude of the fourth square bracket term is larger than the magnitude of the combined term. Further, 66.67\% of this particular set of pulsars (or 6.45\% of the total population) have the magnitude of the fourth square bracket term larger than the magnitude of the combined term by more than one order of magnitude. These facts again support our conclusion that the first three terms should not be ignored, even when we are working with the orbital frequency and its derivatives.

\subsection{Pulsar population near the Galactic centre}
\label{subsec:MCMCGC}

In our earlier work \citep[figure 4]{pb18}, we saw that at low values of $R$ and $z$ (vertical height of the pulsar from the Galactic disk), both $|\frac{\partial \Phi}{ \partial R}|$ and $|\frac{\partial \Phi}{ \partial z}|$ peak where $\Phi$ is the gravitational potential of the Galaxy. Moreover, the slopes of the $|\frac{\partial \Phi}{ \partial R}| ~ {\rm vs} ~ R$ and $|\frac{\partial \Phi}{ \partial R}| ~{\rm vs} ~ z$ curves are steepest near the peak. As $-\frac{\partial \Phi}{ \partial R}$ and $-\frac{\partial \Phi}{ \partial z}$ provide the acceleration of the pulsars parallel and perpendicular to the Galactic disk, we expect that both the acceleration and jerk would be large for pulsars located in this region, which corresponds to the region close to the Galactic centre. Unfortunately, no such pulsar is known at present in this region, so we decided to work on a synthetic set of pulsars.

Like section \ref{subsec:MCMCField}, here too, we separately studied the millisecond pulsar population and the normal pulsar population, and we took a uniform distribution between $-200$ and 200 km/s, for $v_r$. We constrained our simulations to the cases with $l$ varying uniformly between 0 and 5 degrees as well as between 355 and 360 degrees, $b$ varying uniformly between -5 and 5 degrees, and $d$ varying uniformly between 7.8 and 8.2 kpc. 

For these parameters, we generated the same number of cases (124115) corresponding to the normal pulsar population as those generated in the earlier subsection, so that we could use the same sample of $f_{\rm s}$ as obtained from PsrPopPy earlier. Similarly, we generated the same number of cases (2791) corresponding to the millisecond pulsar population and used the corresponding sample of $f_{\rm s}$ as obtained from PsrPopPy earlier.

For the rest of the parameters ($\mu_l$, $\mu_b$, $\dot{f}_{\rm s, obs}$, $\ddot{f}_{\rm s, obs}$, $f_{\rm b}$, $\dot{f}_{\rm b, obs}$, and $\ddot{f}_{\rm b, obs}$) we used the same distributions as generated in \ref{subsec:MCMCField} for the millisecond pulsars. Similarly, for the normal pulsars, we used the same distributions generated in \ref{subsec:MCMCField} for the remaining corresponding parameters ($\mu_l$, $\mu_b$, $\dot{f}_{\rm s, obs}$ and, $\ddot{f}_{\rm s, obs}$). The underlying assumption here is that the population of pulsars near Galactic centre is not way too different than the overall disk population. This assumption is not perfectly valid, however our only aim was to obtain a qualitative comparison between various terms in eq. (\ref{eq:fddotex4}).

\begin{figure*}
\begin{center}
\hskip-0.8cm\subfigure[]{\label{subfig:gc123SBmsp}\includegraphics[width=0.52\textwidth, angle=0]{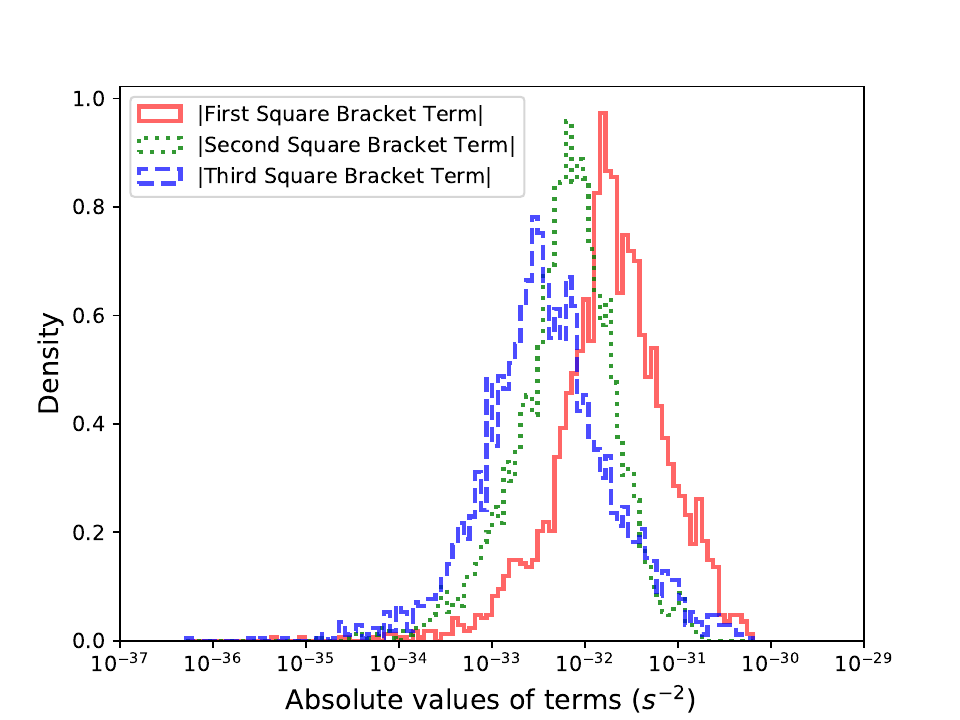}}
 \subfigure[]{\label{subfig:gc4SBspincombnmsp}\includegraphics[width=0.52\textwidth, angle=0]{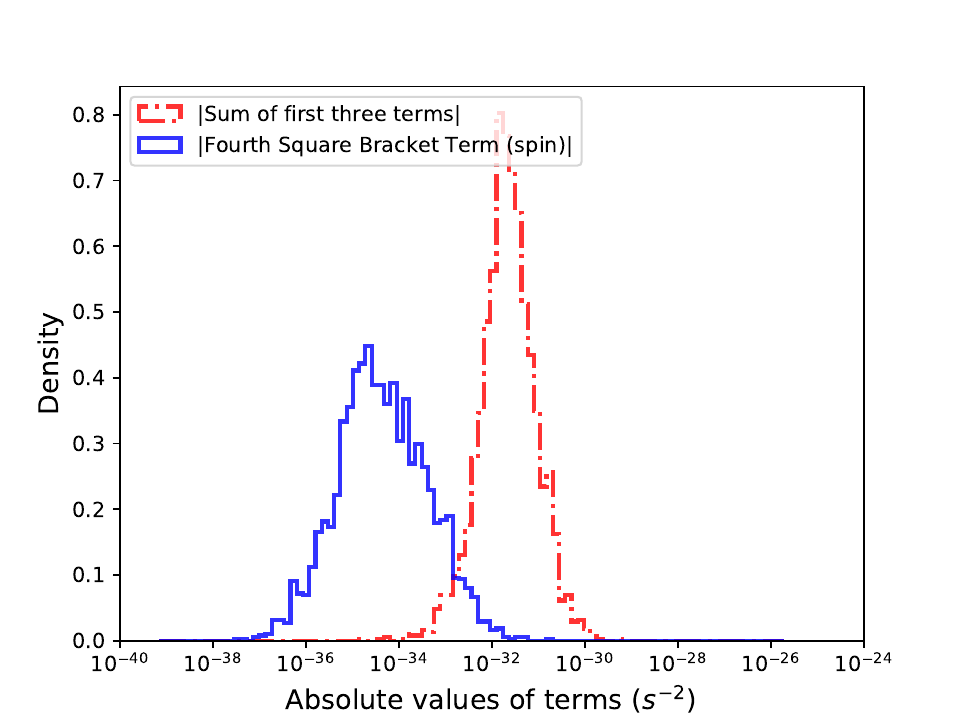}}\\
  \subfigure[]{\label{subfig:gc4SBorbcombmsp}\includegraphics[width=0.52\textwidth, angle=0]{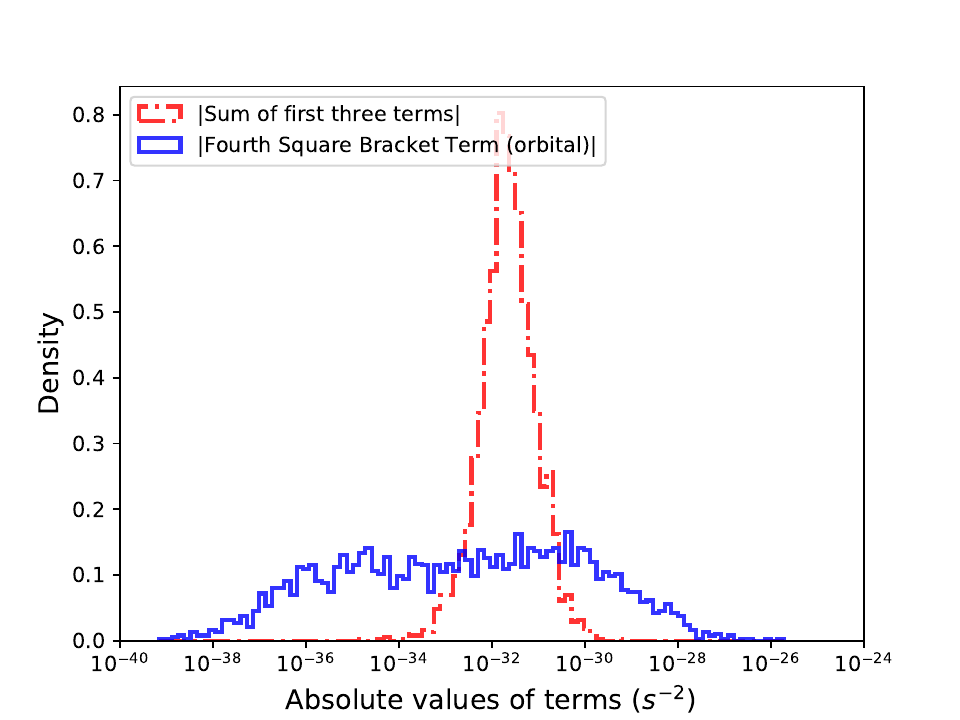}}\\
\end{center}
\caption{Density distribution of the absolute values of various square bracket terms in eq. (\ref{eq:fddotex4}) for simulated millisecond pulsars near the Galactic centre. The subplots are as follow: a) comparison of line histograms for the first square bracket term, the second square bracket term, and the third square bracket term, b) comparison of line histograms for the sum of the first three terms and the fourth square bracket term using the spin frequency and its derivatives, and c) comparison of line histograms for the sum of the first three terms and the fourth square bracket term using the orbital frequency and its derivatives.}
\label{fig:histMSPGC}
\end{figure*}

\begin{figure*}
\begin{center}
\hskip-0.8cm\subfigure[]{\label{subfig:gc123SBnorm}\includegraphics[width=0.52\textwidth, angle=0]{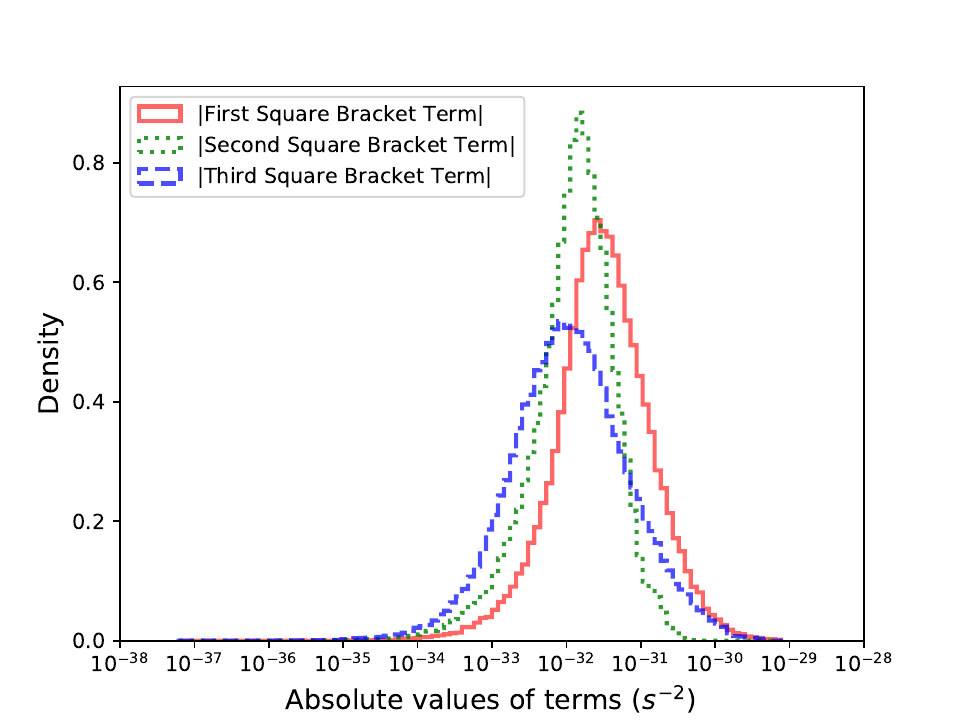}}
 \subfigure[]{\label{subfig:gc4SBspincombnorm}\includegraphics[width=0.52\textwidth, angle=0]{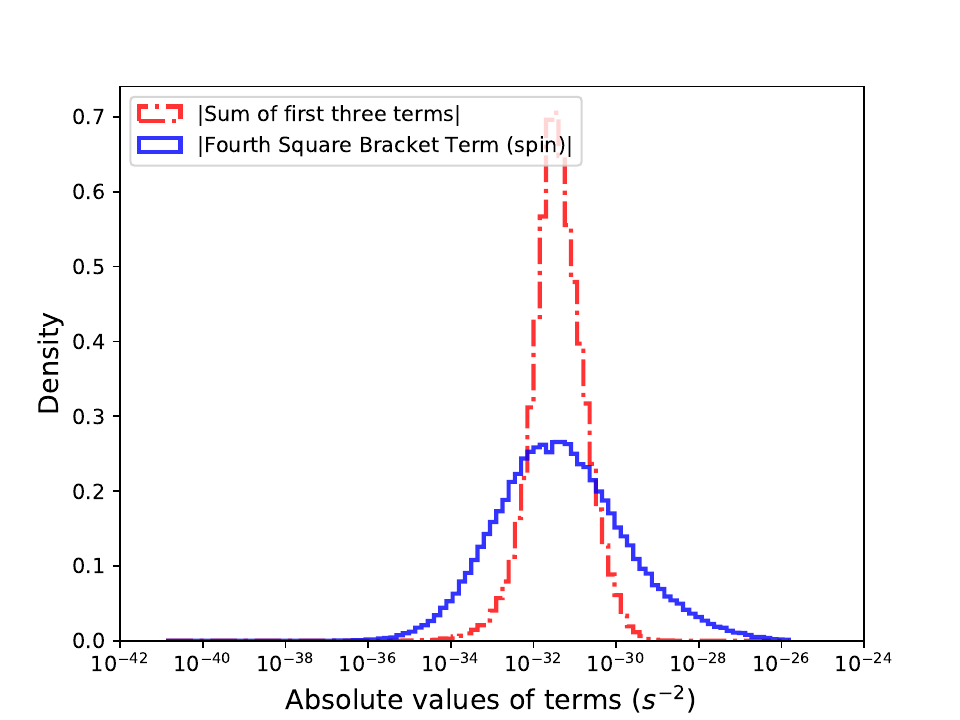}}\\
\end{center}
\caption{Density distribution of the absolute values of various square bracket terms in eq. (\ref{eq:fddotex4}) for the simulated normal pulsars near the Galactic centre. The subplots are as follow: a) comparison of line histograms for the first square bracket term, the second square bracket term, and the third square bracket term, and b) comparison of line histograms for the sum of the first three terms and the fourth square bracket term using the spin frequency and its derivatives.}

\label{fig:histNormGC}
\end{figure*}

Like in the previous subsection, here also, we calculate all square bracket terms of eq. (\ref{eq:fddotex4}) separately. We summarize these results in the histograms of figure \ref{fig:histMSPGC} for millisecond pulsars and figure \ref{fig:histNormGC} for normal pulsars. Additionally, table \ref{tb:simulvalsMSPGC} contains the statistical summary for the synthetic millisecond pulsars while table \ref{tb:simulvalsNormGC} contains the statistical summary for the synthetic normal pulsars.

\begin{table*}
\begin{center}
\caption{Statistical summary of the of all square bracket terms appearing in of eq. (\ref{eq:fddotex4}) for the simulated millisecond pulsars near the Galactic centre.  We compare the absolute values of all the terms here. }
\begin{tabular}{|l|l|l|l|l|}
\hline
Term & Minimum & Mean & Median & Maximum \\ 
  & (${\rm s}^{-2}$) & (${\rm s}^{-2}$) & (${\rm s}^{-2}$)  & (${\rm s}^{-2}$) \\ \hline
  First Square Bracket Term & $4.49\times10^{-36}$ & $5.23\times10^{-32}$ & $1.96\times10^{-32}$ & $4.69\times10^{-30}$ \\ \hline
 Second Square Bracket Term & $1.23\times10^{-35}$ & $1.17\times10^{-32}$ & $6.65\times10^{-33}$ & $1.58\times10^{-31}$ \\ \hline
 Third Square Bracket Term & $5.12\times10^{-37}$ & $1.50\times10^{-32}$ & $3.67\times10^{-33}$ & $6.39\times10^{-31}$ \\ \hline
 Combined Term & $1.69\times10^{-35}$ & $5.79\times10^{-32}$ & $2.15\times10^{-32}$ & $4.69\times10^{-30}$ \\ \hline
 Fourth Square Bracket Term (spin) & $5.43\times10^{-40}$ & $5.16\times10^{-34}$ & $4.00\times10^{-35}$ & $1.65\times10^{-31}$ \\ \hline
 Fourth Square Bracket Term (orbital) & $7.00\times10^{-40}$ & $2.02\times10^{-29}$ & $3.17\times10^{-33}$ & $1.89\times10^{-26}$ \\ \hline    
\end{tabular}
\label{tb:simulvalsMSPGC}
\end{center}
\end{table*}

\begin{table*}
\begin{center}
\caption{Statistical summary of the of all square bracket terms appearing in of eq. (\ref{eq:fddotex4}) for the simulated normal pulsars near the Galactic centre.  We compare the absolute values of all the terms here. }
\begin{tabular}{|l|l|l|l|l|}
\hline
Term & Minimum & Mean & Median & Maximum \\ 
  & (${\rm s}^{-2}$) & (${\rm s}^{-2}$) & (${\rm s}^{-2}$)  & (${\rm s}^{-2}$) \\ \hline
  First Square Bracket Term & $1.11\times10^{-36}$ & $9.97\times10^{-32}$ & $3.06\times10^{-32}$ & $2.29\times10^{-28}$ \\ \hline
 Second Square Bracket Term & $3.53\times10^{-38}$ & $2.38\times10^{-32}$ & $1.39\times10^{-32}$ & $8.33\times10^{-31}$ \\ \hline
 Third Square Bracket Term & $6.17\times10^{-38}$ & $5.72\times10^{-32}$ & $1.09\times10^{-32}$ & $7.70\times10^{-30}$ \\ \hline
 Combined Term & $2.74\times10^{-37}$ & $1.29\times10^{-31}$ & $3.75\times10^{-32}$ & $2.28\times10^{-28}$ \\ \hline
 Fourth Square Bracket Term (spin) & $1.28\times10^{-39}$ & $4.10\times10^{-29}$ & $3.94\times10^{-32}$ & $1.78\times10^{-25}$ \\ \hline    
\end{tabular}
\label{tb:simulvalsNormGC}
\end{center}
\end{table*}

Comparing the values reported in tables \ref{tb:simulrealvalsMSP} and \ref{tb:simulvalsMSPGC} for the millisecond pulsars, and the values reported in tables \ref{tb:simulrealvalsNorm} and \ref{tb:simulvalsNormGC} for the normal pulsars, we see that the median values of the absolute value of the first three square bracket terms are about one order of magnitude larger for the pulsar population near the Galactic centre than the general population. For the fourth square bracket term, we get almost the same median values for the two populations as expected, as in the fourth square bracket term, the dynamics dependent $\left( \frac{\dot{f}}{f} \right)_{\rm ex}$ is multiplied by $\left( \frac{\dot{f}}{f} \right)_{\rm obs}$ which have been simulated with the help of the observed values of the field pulsars.

As all four terms contribute to the observed value of the second derivative of the frequency (see eq. (\ref{eq:fddotex4})) separately, we can conclude that the measured value (through timing analysis) of the second derivative of the spin or orbital frequency of a pulsar near the Galactic centre would likely be contaminated by the dynamical terms by larger amount than the pulsars away from the Galactic centre, especially if the distributions of frequencies and its derivatives are not drastically different.

Note that, although our formalism can be used for pulsars close to the Galactic centre, it is not very accurate there, as gravitational pull of near-by stars on the pulsars likely to be significant. Nevertheless, if any pulsar close to the Galactic centre is discovered in the near future, our expressions can be used to obtain a first order correction.

\section{Applications}
\label{se:App}

In this section, we discuss some practical applications of our formalism of correcting dynamical effects from the second derivative of frequencies, and explore whether parameters that depend on the second derivative can be affected significantly by the dynamics.

\subsection{Properties of PSR J1024$-$0719}

As mentioned previously, \cite{kap16} and \cite{bas16} showed that PSR J1024$-$0719 is a wide-orbit binary pulsar. Its orbit is so wide that a good timing solution could be obtained even without fitting for orbital parameters. However, for this pulsar, after eliminating the contributions of the velocity, acceleration and jerk due to the Galactic potential from the measured first and second time derivative of the spin frequency, additional dynamical effects due to the orbital motion would remain and hence after using eqs. (\ref{eq:fdotint}) and (\ref{eq:fddotint}), what we obtain should be better called `residual' instead of `intrinsic' values of frequency derivatives. \cite{kap16} reported derivatives of spin periods instead of spin frequencies and used the values of $\dot{P}_{\rm s, res}$ and $\ddot{P}_{\rm s, obs}$ to put some constraints on the orbit of the pulsar. They did not correct for the dynamical effects in the second derivative as they correctly guessed that it would be very small. Indeed we see that $\ddot{P}_{\rm s, res} = 1.10\times10^{-31}~{\rm s^{-1}}$ while $\ddot{P}_{\rm s, obs} = 1.1 \times10^{-31}~{\rm s^{-1}}$ as reported by \cite{kap16}. Thus, for this particular pulsar, the contribution of the acceleration and jerk due to the gravitational potential of the Galaxy to the measured second derivative of the spin frequency is negligible. However this might not be the case always, as we will explore in the next subsection. Note that we obtain $\dot{P}_{\rm s, res} = -3.89\times10^{-20}~{\rm ss^{-1}}$ while \cite{kap16} had $\dot{P}_{\rm s, res}= -3.96\times10^{-20}~{\rm ss^{-1}}$.

\subsection{Intrinsic spin frequency second derivative and the braking index}
\label{subsec:brakind}

Instead of exploring the effect of the dynamical terms in the second derivative of the spin frequency alone, we decided to concentrate on the braking index $n$ which is associated with the basic emission model of rotation powered pulsars. The well known expression of $n$ for such a pulsar is found by equating the rate of loss of rotational kinetic energy to the rate of the electromagnetic energy emitted by a rotating magnetic dipole \citep{lk05}:

\begin{equation}
n = \frac{\ddot{f}_{\rm s, int}\,f_{\rm s}}{\dot{f}_{\rm s, int}^{2}}  ~.
\label{eq:braking2}
\end{equation} In the ideal scenario where the pulsar is a perfect magnetic dipole, and the mis-alignment of the spin axis and the magnetic axis is solely responsible for the emission of radio waves, the braking index turns out to be three \citep{lk05}. However, in reality, this may not be the case. In fact, people have reported values of $n$ differing from three and tried to find physics behind this \citep{hob04, ha12, da20}.

We first investigated whether those anomalous values of $n$ are affected by dynamical terms or there are other physical reasons intrinsic to the pulsars. \cite{hob04} reported measurements of $n$ for 374 pulsars. They found that the reported values of $n$ for a number of pulsars lie in the range $-2.6\times10^{8}$ to $2.5\times10^{8}$. These large deviations from the expected value 3 cannot be due to the timing noise as they had taken care of it by whitening the timing residuals. Although they calculated the values of $n$ using the measured values of the spin frequency and its derivatives instead of their intrinsic values, it is unlikely that the dynamical effects would change the value of $n$ from $ 3$ to $10^{8}$, so for pulsars having such large values of $n$, the conclusion of \cite{hob04} that the measured spin frequency second derivative values of the pulsars do not represent only the braking due to magnetic dipole radiation seems valid. However, we explored whether for the pulsars with measured $|n|<10$, dynamical terms play a significant role to shift the values from 3. Using the values of $l$, $b$, $f_{\rm s}$, $\dot{f}_{\rm s, obs}$, and $\ddot{f}_{\rm s, obs}$ directly from \cite{hob04}, $d$, calculated using NE2001 model \citep{cl02,cl03} based on the DM values given in \cite{hob04}, and $\mu_l$ and $\mu_b$, calculated using Right Ascension and Declination, and corresponding proper motion values given in \cite{hob04}, we evaluated $\dot{f}_{\rm int}$ using model-Lb of GalDynPsr, and $\ddot{f}_{\rm int}$ using the method described in section \ref{se:der1}. We took a nominal $v_{r}=50$ km/s, like \cite{liu19}. For all these pulsars, we didn't see any significant change by correcting for the dynamical terms. We repeated the above calculations for $v_{r}=0$ km/s and $v_{r}=-100$ km/s without any significant difference. Hence, we conclude that even for these pulsars, there are physical reasons for $n$ to differ from 3. Recently, \cite{da20} reported measurements of $n$ for 73 pulsars. Out of these, they report braking index values to be less than 10 for only two pulsars, namely, PSR J0157+6212 ($n=4.8$) and PSR J1743-3150 ($n=6.5$). Using the values of $d$, $f_{\rm s}$, $\dot{f}_{\rm s, obs}$, and $\ddot{f}_{\rm s, obs}$ directly from \cite{da20}, $l$, $b$, $\mu_l$, and $\mu_b$, from the Right Ascension and Declination values, and corresponding proper motion values given in \cite{da20}, and our formalism to correct for the dynamical terms, we found only a small change in the braking index value of PSR J0157+6212 ($n=5.0$) and a little higher change in PSR J1743-3150 ($n=8.3$), but still, in no case $n$ gets closer to 3.

Now the question arises, can there be a case where dynamical effects are large enough to alter the value of $n$ significantly? To answer this question, we performed calculations on simulated millisecond pulsars. We concentrated only on millisecond pulsars, as these are more stable than normal pulsars, so less likely to have glitch, red timing noise, etc., that might make the observed value of $n$ to deviate from 3 \citep{da20}. We aimed whether we could get a sufficient number of millisecond pulsars for which $n$ is close to three (theoretically expected), i.e., $2.5<n<3.5$, and $n_{\rm obs}<0$ or $n_{\rm obs}>6$, i.e., $n_{\rm obs}$ is significantly different than $n$ due to dynamical effects. Here $n_{\rm obs}$ means the value of the braking index we get by using the observed values of the spin frequency and its derivatives, and $n$ is the true braking index that we get after eliminating the dynamical contributions from the frequency derivatives. We used the same simulation approach as described in section \ref{subsec:MCMCField}. We did not find any cases where $2.5<n<3.5$, but $n_{\rm obs}<0$ or $n_{\rm obs}>6$. We even performed the investigation over a larger number of synthetic pulsars (10000 instead of 2791 as in section \ref{subsec:MCMCField}), still did not find any favorable case. We then generated 10000 pulsars near the Galactic centre where the distributions of various parameters are the same as section \ref{subsec:MCMCGC}. But still no favourable case could be found.

However, it is possible that there might exist unique pulsars whose parameters are missed by the standard representation of the population (generated by `PsrPopPy'), and have $2.5<n<3.5$, and $n_{\rm obs}<0$ or $n_{\rm obs}>6$. To take care of this possibility, we adopted a different approach to simulate parameters for a synthetic set of millisecond pulsars. We explored two populations of millisecond pulsars, one in the Galactic field and the other in the near-Galactic centre region, by using two approaches to simulate $l$, $b$, and $d$. For the first population, we just fitted the distribution from ATNF catalogue for the millisecond pulsars (excluding the ones in globular clusters, the Large Magellanic Cloud, and the Small Magellanic Cloud), i.e., we fitted the 142 values of a parameter, say, $l$ reported in the ATNF catalogue with a Cumulative Distribution Function (CDF) and then used the inverse CDF technique to generate 10000 synthetic values following the same CDF \footnote{We followed the same procedure whenever we fitted the parameters from the ATNF catalogue.}. For the second population we used the approach as described in section \ref{subsec:MCMCGC}.  For both the populations we generated 10000 values of these parameters. For $\mu_l$ and $\mu_b$ too, we simulated 10000 values based on the distribution of values in the ATNF catalogue for the millisecond pulsars. We also generated 10000 values of $v_r$ distributed uniformly between -200 to 200 km/s. However, for $f_{\rm s}$, $\dot{f}_{\rm s, obs}$, and $\ddot{f}_{\rm s, obs}$, we needed to take a closer look at the ATNF catalogue values, again excluding the ones in globular clusters, the Large Magellanic Cloud, and the Small Magellanic Cloud.

We found that the range of $f_{\rm s}$ spans from 34.657 Hz to 641.928 Hz for millisecond pulsars. We divided this range into three equal parts and generated 10000 uniformly distributed values for each range. For $\dot{f}_{\rm s, obs}$, the values span between the orders of $-1.0\times10^{-13}~{\rm s}^{-2}$ and $-1.0\times10^{-17}~{\rm s}^{-2}$ for millisecond pulsars. We divided this range into four parts with each range spanning over one order of magnitude and generated uniform distribution of 10000 values for each subrange. Since the ATNF catalogue only gives eight measurements of the $\ddot{f}_{\rm s, obs}$ for millisecond pulsars, we decided to extend the ranges. More specifically, the values of $\ddot{f}_{\rm s, obs}$ for these 8 millisecond pulsars lie in the ranges of $-1.0\times10^{-24}~{\rm s}^{-3}$ to $-1.0\times10^{-28}~{\rm s}^{-3}$ and $1.0\times10^{-27}~{\rm s}^{-3}$ to $1.0\times10^{-25}~{\rm s}^{-3}$, whereas, by fitting these values with a distribution function and generating a synthetic values in section \ref{subsec:MCMCField}, we found that $\ddot{f}_{\rm s, obs}$ vary between $-1.0\times10^{-24}~{\rm s}^{-3}$ to $-1.0\times10^{-30}~{\rm s}^{-3}$ and $1.0\times10^{-30}~{\rm s}^{-3}$ to $1.0\times10^{-24}~{\rm s}^{-3}$. We used these two sets and divided each set into 6 subranges which span over one order of magnitude and generated uniform distribution of 10000 values for each subrange.

In this way, we got 3 subranges of $f_{\rm s}$, 4 subranges of $\dot{f}_{\rm s, obs}$, and 12 subranges of $\ddot{f}_{\rm s, obs}$, each having 10000 uniformly generated values. For braking index calculations, we chose one subrange corresponding to each of these three parameters, in addition to the 10000 values generated for the parameters $l$, $b$, $d$, $\mu_{l}$, $\mu_{b}$, and $v_r$ each. These 10000 values for each parameter constitute the 10000 synthetic millisecond pulsars, concentrated in a specific sub-range of the multi-dimensional parameter space, out of the total 144 ($3\times4\times12$) of such sub-ranges. We computed the values of $n$ and $n_{\rm obs}$ for all 10000 synthetic pulsars in each sub-range and selected the favorable cases. In this approach, we implicitly assumed that the total number of millisecond pulsars over the full parameter space is much larger than 10000, which might not be very realistic. However, our quest here is to find unique combinations of parameters that can give large differences between the observed and intrinsic values of the braking index.  

The number of favourable cases for different combinations of $f_{\rm s}$, $\dot{f}_{\rm s, obs}$, and $\ddot{f}_{\rm s, obs}$ subranges are summarized in table \ref{tb:braknum}. We display six examples from each set (pulsars in the field and pulsars near the Galactic centre) in table \ref{tb:brakcompnew}, such that three of the Galactic field pulsars represent largest differences in $n$ and $n_{\rm obs}$ when $n_{\rm obs}>6$ (Pulsar1$_{\rm GF}$, Pulsar2$_{\rm GF}$, and Pulsar3$_{\rm GF}$) and the other three when $n_{\rm obs}<0$ (Pulsar4$_{\rm GF}$, Pulsar5$_{\rm GF}$, and Pulsar6$_{\rm GF}$). We also display six examples of the near-Galactic centre pulsars in table \ref{tb:brakcompnew}, such that three of them represent largest differences in $n$ and $n_{\rm obs}$ when $n_{\rm obs}>6$ (Pulsar1$_{\rm GC}$, Pulsar2$_{\rm GC}$, and Pulsar3$_{\rm GC}$) and the other three when $n_{\rm obs}<0$ (Pulsar4$_{\rm GC}$, Pulsar5$_{\rm GC}$, and Pulsar6$_{\rm GC}$). Note that even if we get large values of $n_{\rm obs}$ for pulsars near the Galactic field, those values are never as large as $10^{8}$ as reported for a number of field pulsars by \cite{hob04}.

We also see that the difference between the observed and intrinsic values of $\ddot{f}_{\rm s}$ are much larger than that for $\dot{f}_{\rm s}$ and the ratio of the observed and intrinsic values of braking index depends on both of these ratios as $n_{\rm obs} / n  = \left( {\dot{f}_{\rm s, int} / \dot{f}_{\rm s, obs} } \right)^2 \,\ddot{f}_{\rm s, obs} \, / \ddot{f}_{\rm s, int} $. This emphasizes the fact that if one wants to use the value of the second derivative of the spin frequency to get better insight of the properties of the pulsar, it is better to correct for the dynamical terms as accurately as possible, as this term is affected more by the dynamics. 

Moreover, the difference in the results of the simulated millisecond pulsars in the Galactic field and those near the Galactic centre in tables \ref{tb:braknum} and \ref{tb:brakcompnew} are noticeable. For the pulsars near the Galactic centre, considerably more number of favourable cases are generated and there is also a larger change in the values of braking index. In summary, these simulations establish the fact that it is possible that the dynamical terms change the value of $n$ significantly from three, especially near the Galactic centre. However, possibility of having such a system is not very likely as it needs very unique combination of various parameters.

\LTcapwidth=1.\linewidth
\begin{longtable}{|@{\hskip1.5pt}l@{\hskip1.5pt}|@{\hskip1.5pt}l@{\hskip1.5pt}|@{\hskip1.5pt}l@{\hskip1.5pt}|@{\hskip1.5pt}l@{\hskip1.5pt}|@{\hskip1.5pt}l@{\hskip1.5pt}|}
\caption{\footnotesize Number of favourable cases where $2.5<n<3.5$, and $n_{\rm obs}<0$ or $n_{\rm obs}>6$, for various ranges of $f_{\rm s}$, $\dot{f}_{\rm s, obs}$, and $\ddot{f}_{\rm s, obs}$ values. The fourth column displays favourable cases for Galactic field pulsars, whereas, the fifth column displays favourable cases for pulsars near Galactic centre. We display $f_{\rm s}$ rounded off to 2 decimal places. Each row represents one simulation run of 10000 simulated millisecond pulsars.}
\label{tb:braknum}\\
\hline
$f_{\rm s}$ range & \addstackgap{$\dot{f}_{\rm s, obs}$} range & \addstackgap{$\ddot{f}_{\rm s, obs}$} range & Favourable Cases & Favourable Cases\\
(Hz)& (${\rm s}^{-2}$) & (${\rm s}^{-3}$) & (Galactic Field) & (Near Galactic Centre)\\
\hline 
\multirow{10}{*} {34.66 to 273.08} & \multirow{4}{*} {$-10^{-13}$ to $-10^{-14}$}& $10^{-29}$ to $10^{-28}$& 1 & 3\\
\cline{3-5} 
 &  & $10^{-30}$ to $10^{-29}$& 3 & 6\\
 \cline{3-5}  
 &  & $-10^{-28}$ to $-10^{-29}$& 1 & 3\\
  \cline{3-5} 
 &  & $-10^{-29}$ to $-10^{-30}$& 8 & 63\\
 \cline{2-5} 
 & \multirow{4}{*} {$-10^{-14}$ to $-10^{-15}$} & $10^{-29}$ to $10^{-28}$& 0 & 1\\
\cline{3-5} 
 &  & $10^{-30}$ to $10^{-29}$& 18 & 62\\
 \cline{3-5}  
 &  & $-10^{-28}$ to $-10^{-29}$& 0 & 4\\
  \cline{3-5} 
 &  & $-10^{-29}$ to $-10^{-30}$& 18 & 53\\
 \cline{2-5}
  & \multirow{2}{*} {$-10^{-15}$ to $-10^{-16}$} & $10^{-30}$ to $10^{-29}$& 1 & 2\\
 \cline{3-5}  
 &  & $-10^{-29}$ to $-10^{-30}$& 0 & 1\\
 \cline{2-5}
 \hline
 \multirow{10}{*} {273.08 to 439.50} & \multirow{5}{*} {$-10^{-13}$ to $-10^{-14}$}& {$10^{-28}$ to $10^{-27}$} & 0 & 2\\
\cline{3-5} 
 &  & $10^{-29}$ to $10^{-28}$& 4 & 28\\
\cline{3-5} 
 &  & $10^{-30}$ to $10^{-29}$& 4 & 18\\
  \cline{3-5} 
 &  & $-10^{-28}$ to $-10^{-29}$& 10 & 32\\
  \cline{3-5} 
 &  & $-10^{-29}$ to $-10^{-30}$& 51 & 179\\
 \cline{2-5} 
 & \multirow{4}{*} {$-10^{-14}$ to $-10^{-15}$} & $10^{-29}$ to $10^{-28}$& 0 & 2\\
\cline{3-5} 
 &  & $10^{-30}$ to $10^{-29}$& 17 & 34\\
 \cline{3-5} 
 &  & $-10^{-28}$ to $-10^{-29}$& 0 & 1\\
  \cline{3-5} 
 &  & $-10^{-29}$ to $-10^{-30}$& 9 & 22\\
 \cline{2-5}
   & \multirow{1}{*} {$-10^{-15}$ to $-10^{-16}$} & $-10^{-29}$ to $-10^{-30}$& 0 & 1\\
 \cline{2-5}
 \hline
 \multirow{9}{*} {439.50 to 641.93} & \multirow{5}{*} {$-10^{-13}$ to $-10^{-14}$}& {$10^{-28}$ to $10^{-27}$} & 2 & 1\\
\cline{3-5} 
 &  & $10^{-29}$ to $10^{-28}$& 12 & 50\\
\cline{3-5} 
 &  & $10^{-30}$ to $10^{-29}$& 17 & 11\\
 \cline{3-5}  
 &  & $-10^{-28}$ to $-10^{-29}$& 18 & 55\\
  \cline{3-5} 
 &  & $-10^{-29}$ to $-10^{-30}$& 67 & 245\\
 \cline{2-5} 
 & \multirow{4}{*} {$-10^{-14}$ to $-10^{-15}$} & $10^{-29}$ to $10^{-28}$& 1 & 1\\
\cline{3-5} 
 &  & $10^{-30}$ to $10^{-29}$& 12 & 12\\
 \cline{3-5} 
 &  & $-10^{-28}$ to $-10^{-29}$& 0 & 1\\
  \cline{3-5} 
 &  & $-10^{-29}$ to $-10^{-30}$& 9 & 17\\
\hline
\end{longtable}

\LTcapwidth=1.3\linewidth
\begin{longtable}{|l|l|l|l|l|l|l|}
\caption{\footnotesize Parameters for simulated millisecond pulsars with $2.5<n<3.5$, and $n_{\rm obs}<0$ or $n_{\rm obs}>6$. Here, $l$ is the Galactic longitude, $b$ is the Galactic latitude, $d$ is the distance between the pulsar and the Solar system barycentre, $\mu_{l}$ is proper motion in $l$, $\mu_{b}$ is proper motion in $b$, $f_{\rm s}$ is the spin frequency, $v_{r}$ is the radial velocity, $\dot{f}_{\rm s, obs}$ is observed spin frequency derivative, $\dot{f}_{\rm s, int}$ is intrinsic spin frequency derivative, $\ddot{f}_{\rm s, obs}$ is observed spin frequency second derivative, $\ddot{f}_{\rm s, int}$ is intrinsic spin frequency second derivative, $n_{\rm obs}$ is the braking index based on observed spin frequency derivatives, and $n$ is the braking index based on intrinsic spin frequency derivatives. Top half displays the parameters of simulated millisecond pulsars in the Galactic field (represented by subscript `GF'). Pulsar1$_{\rm GF}$, Pulsar2$_{\rm GF}$, and Pulsar3$_{\rm GF}$ represent largest differences in $n$ and $n_{\rm obs}$ when $n_{\rm obs}>6$ whereas Pulsar4$_{\rm GF}$, Pulsar5$_{\rm GF}$, and Pulsar6$_{\rm GF}$ represent largest differences in $n$ and $n_{\rm obs}$ when $n_{\rm obs}<0$. Bottom half displays the parameters of simulated millisecond pulsars near the Galactic centre (represented by subscript `GC'). Pulsar1$_{\rm GC}$, Pulsar2$_{\rm GC}$, and Pulsar3$_{\rm GC}$ represent largest differences in $n$ and $n_{\rm obs}$ when $n_{\rm obs}>6$ whereas Pulsar4$_{\rm GC}$, Pulsar5$_{\rm GC}$, and Pulsar6$_{\rm GC}$ represent largest differences in $n$ and $n_{\rm obs}$ when $n_{\rm obs}<0$. We display the results till second decimal place.}
\label{tb:brakcompnew}\\
 \hline 
  \multicolumn{7}{|c|}{\addstackgap{Galactic Field pulsars}}  \\
  \hline 
 Parameters & Pulsar1$_{\rm GF}$ & Pulsar2$_{\rm GF}$ & Pulsar3$_{\rm GF}$ & Pulsar4$_{G\rm F}$ & Pulsar5$_{\rm GF}$ & Pulsar6$_{\rm GF}$ \\ 
[1em]\hline 
$l$ (deg) & 306.41 & 231.33 & 287.52  & 349.02 & 318.66 & 349.33  \\ 
 &  &  &   &  &  &   \\ 
$b$ (deg) & 48.06 & 21.16 & 6.58 & 6.65 & -8.35 & 27.86  \\
 &  &  &   &  &  &   \\ 
$d$ (kpc) & 1.67 & 0.51 & 0.63 & 0.62 & 0.18 & 0.31  \\   
 &  &  &   &  &  &   \\ 
$\mu_{l}$ (mas/yr)  & -9.00 & -18.76 & -24.62 & -20.79 & 18.22 & 19.48  \\   
 &  &  &   &  &  &   \\ 
$\mu_{b}$ (mas/yr)  & -2.37 & -5.68 & 3.16 & -2.43 & -1.46 & -6.27  \\   
 &  &  &   &  &  &   \\ 
$v_{r}$ (km/s) & 157.09 & 176.56 & 172.92 & -153.22 & -135.60 & -187.75 \\ 
 &  &  &   &  &  &   \\ 
$f_{\rm s}$ (Hz)  & 588.59 & 132.95 & 598.36 & 575.70 & 262.89 & 479.26  \\   
 &  &  &   &  &  &   \\ 
\addstackgap{$\dot{f}_{\rm s, obs}$ ($\times10^{-15}~\rm s^{-2}$)} & -1.01 & -0.98 & -8.61 & -6.30 & -2.50 & -8.22 \\   
 &  &  &   &  &  &   \\ 
\addstackgap{$\ddot{f}_{\rm s, obs}$ ($\times10^{-30}~\rm s^{-3}$)} & 1.88 & 2.22 & 14.94 & -8.73 & -2.68 & -8.76  \\   
  &  &  &   &  &  &   \\ 
\addstackgap{$\dot{f}_{\rm s, int}$ ($\times10^{-15}~\rm s^{-2}$)} & -0.94 &-0.92 & -8.07 & -5.89 & -2.46 & -8.07 \\
 &  &  &   &  &  &   \\ 
\addstackgap{$\frac{(\dot{f}_{\rm s,int}-\dot{f}_{\rm s,obs}) \times 100}{\dot{f}_{\rm s,obs}}$} & -6.53 & -5.96 & -6.18 & -6.53 & -1.51 & -1.75 \\   
 &  &  &   &  &  &   \\ 
\addstackgap{$\ddot{f}_{\rm s, int}$ ($\times10^{-31}~\rm s^{-3}$)} & 0.04 & 0.18 & 3.53 & 1.96 & 0.64 & 3.50  \\ 
 &  &  &   &  &  &   \\ 
\addstackgap{$\frac{(\ddot{f}_{\rm s,int}-\ddot{f}_{\rm s,obs}) \times 100}{\ddot{f}_{\rm s,obs}}$} & -99.78 & -99.20 & -97.64 & -102.25 & -102.39 & -104.00  \\  
 &  &  &   &  &  &   \\ 
$n_{\rm obs}$  & 1081.41 & 306.73 & 120.70 & -126.69 & -112.67 & -62.22 \\   
 &  &  &   &  &  &   \\ 
$n$ &  2.69 & 2.78 & 3.24 & 3.26 & 2.78 & 2.58 \\   
\hline 
  \multicolumn{7}{|c|}{\addstackgap{Near Galactic Centre pulsars}}  \\ 
  \hline
   & Pulsar1$_{\rm GC}$ & Pulsar2$_{\rm GC}$ & Pulsar3$_{\rm GC}$ & Pulsar4$_{\rm GC}$ & Pulsar5$_{\rm GC}$ & Pulsar6$_{\rm GC}$ \\ 
[1em]\hline 
$l$ (deg) & 356.09 & 1.20 & 0.06 & 3.32 & 357.26 & 4.35  \\
 &  &  &   &  &  &   \\ 
$b$ (deg) & 2.35 & -1.40 & -1.64 & 3.67 & 1.78 & -0.69  \\
 &  &  &   &  &  &   \\ 
$d$ (kpc) & 7.92 & 7.89 & 7.92 & 8.12 & 8.13 & 8.17  \\   
 &  &  &   &  &  &   \\ 
$\mu_{l}$ (mas/yr)  & -0.24 & -4.53 & -5.18 & -3.44 & 4.65 & 1.41  \\   
 &  &  &   &  &  &   \\ 
$\mu_{b}$ (mas/yr)  & -2.56 & 7.08 & -0.82 & -0.41 & -4.14 & -10.76  \\   
 &  &  &   &  &  &   \\ 
$v_{r}$ (km/s) & -168.11 & -80.87 & 156.31 & 79.36 & 196.72 & 69.24 \\ 
 &  &  &   &  &  &   \\ 
$f_{\rm s}$ (Hz)  & 196.77 & 153.40 & 492.92 & 274.49 & 203.81 & 509.90  \\   
 &  &  &   &  &  &   \\ 
\addstackgap{$\dot{f}_{\rm s, obs}$ ($\times10^{-15}~\rm s^{-2}$)} & -0.88 & -1.41 & -9.65 & -0.25 & -0.84 & -3.04  \\   
  &  &  &   &  &  &   \\ 
\addstackgap{$\ddot{f}_{\rm s, obs}$ ($\times10^{-30}~\rm s^{-3}$)} & 4.01 & 3.85 & 46.16 & -2.11 & -4.92 & -6.85  \\   
   &  &  &   &  &  &   \\ 
\addstackgap{$\dot{f}_{\rm s, int}$ ($\times10^{-15}~\rm s^{-2}$)} & -0.96 &-1.10 & -9.09 & -0.53 & -0.99 & -2.55 \\ 
 &  &  &   &  &  &   \\ 
\addstackgap{$\frac{(\dot{f}_{\rm s,int}-\dot{f}_{\rm s,obs}) \times 100}{\dot{f}_{\rm s,obs}}$} & 8.45 & -21.96 & -5.83 & 109.40 & 17.60 & -15.98 \\  
 &  &  &   &  &  &   \\ 
\addstackgap{$\ddot{f}_{\rm s, int}$ ($\times10^{-32}~\rm s^{-3}$)} & 1.56 & 2.34 & 46.18 & 0.27 & 1.66 & 4.37  \\ 
 &  &  &   &  &  &   \\ 
\addstackgap{$\frac{(\ddot{f}_{\rm s,int}-\ddot{f}_{\rm s,obs}) \times 100}{\ddot{f}_{\rm s,obs}}$} & -99.61 & -99.39 & -99.0 & -100.13 & -100.34 & -100.64  \\  
  &  &  &   &  &  &   \\ 
$n_{\rm obs}$  & 1013.69 & 296.35 & 244.37 & -9082.72 & -1421.58 & -378.78 \\   
  &  &  &   &  &  &   \\ 
$n$ &  3.39 & 2.95 & 2.76 & 2.64 & 3.47 & 3.42 \\
\hline  
\end{longtable}

\subsection{Exploring the cases with measured second derivative of the orbital frequency}

In this section, we study dynamical contributions in the observed values of the second derivative of the orbital frequency, $\ddot{f}_{\rm b,obs}$. We find eight pulsars with reported values of $\ddot{f}_{\rm b,obs}$ the ATNF catalogue - PSRs J0023$+$0923, J1048$+$2339, J1731$-$1847, J2339$-$0533, J0024$-$7204J, J0024$-$7204V, J0024$-$7204O, and J0024$-$7204W, and two additional pulsars whose $\ddot{f}_{\rm b,obs}$ value is not reported in ATNF catalogue - PSRs J1723$-$2837 \citep{cr13}, and J2051$-$0827 \citep{go16}. All these ten pulsars are millisecond pulsars.

 Among these pulsars, five are black-widows, which are PSRs J0023$+$0923 \citep{abb18}, J0024$-$7204J \citep{fr17}, J0024$-$7204O \citep{fr17}, J1731$-$1847 \citep{nbb14}, and J2051$-$0827 \citep{go16} and four are Red-Backs, which are PSRs J0024$-$7204W \citep{ru16}, J1048$+$2339 \citep{drc16}, J1723$-$2837 \citep{cr13}, and J2339$-$0533 \citep{pc15}. Additionally, pulsars PSRs J0024$-$7204J, J0024$-$7204V, J0024$-$7204O, and J0024$-$7204W, belong to the Globular Cluster 47 Tucane \citep{fr17,ru16}. 

We calculated the intrinsic values of the first and second derivatives of the orbital frequency, $\dot{f}_{\rm b, int}$ and $\ddot{f}_{\rm b, int}$ respectively, for these pulsars using eqs. (\ref{eq:doppler2b}), (\ref{eq:fddotex1}), and GalDynPsr. We took the parameters as given in the ATNF catalogue except a few cases as mentioned below. The $f_{\rm b}$, $\dot{f}_{\rm b,obs}$, and $\ddot{f}_{\rm b,obs}$ values for PSR J0023$+$0923 were taken from \cite{abb18}. The $\ddot{f}_{\rm b,obs}$ value for PSR J1723$-$2837 was taken from \cite{cr13}, and for PSR J2051$-$0827 was taken from \cite{go16}. The proper motion values for PSR J0024$-$7204V were taken from \cite{ru16}. For all of the pulsars, we used independent distance estimates if available, otherwise NE2001 based distance \citep{cl02,cl03}. We took a nominal value for $v_{r}$ as 50 km/s for all cases. We see that for all the cases, there is not any perceptible difference between $\dot{f}_{\rm b,obs}$ and $\dot{f}_{\rm b,int}$ values, and $\ddot{f}_{\rm b,obs}$ and $\ddot{f}_{\rm b,int}$ values. Note that, for pulsars in globular clusters, additional correction is needed to account for the cluster potential. 

However, it is not wise to make any strong conclusion based on such a small number of pulsars with measurements of the second derivatives of the orbital frequency. We again used the simulated millisecond pulsar population generated in section \ref{subsec:MCMCField}, to study the dynamical contributions in $\ddot{f}_{\rm b,obs}$. Here, we aimed whether we could get a sufficient number of millisecond pulsars for which the values of $\ddot{f}_{\rm b,obs}$ and $\ddot{f}_{\rm b,int}$ differ at least in the first decimal place.

Here too, even when we performed the investigation over a larger number of synthetic pulsars (10000 instead of 2791 as in section \ref{subsec:MCMCField}), we did not find any favorable case. We also generated 10000 pulsars near the Galactic centre where the distributions of various parameters are the same as section \ref{subsec:MCMCGC}. But we still could not find any favourable case. 

Consequently, we adopted a different approach to simulate parameters for a synthetic set of millisecond pulsars. Similar to the approach used in section \ref{subsec:brakind}, in order to explore two populations of millisecond pulsars, one in the Galactic field and the other in the near Galactic centre region, we used two approaches to simulate $l$ , $b$, and $d$. For the first population, we just fitted the distribution from ATNF catalogue for the millisecond pulsars (excluding the ones in globular clusters, the Large Magellanic Cloud, and the Small Magellanic Cloud), and for the second population we used the approach as described in section \ref{subsec:MCMCGC}. For both the populations we generated 10000 values of these parameters. For $\mu_l$ and $\mu_b$ too, we simulated 10000 values based on the distribution of values in the ATNF catalogue for the millisecond pulsars. We also generated 10000 values of $v_r$ distributed uniformly between -200 to 200 km/s. For $\dot{f}_{\rm b, obs}$ and $\ddot{f}_{\rm b, obs}$, we generated uniform distributions in sub-ranges, and the logic of selection of the sub-ranges is described below.

From the ATNF catalogue, we see that the observed values of $\dot{f}_{\rm b, obs}$ lie in the ranges of $-1.0\times10^{-17}~{\rm s}^{-2}$ to $-1.0\times10^{-25}~{\rm s}^{-2}$ and $1.0\times10^{-25}~{\rm s}^{-2}$ to $1.0\times10^{-17}~{\rm s}^{-2}$. We divided each set into ranges which span over one order of magnitude and generated 10000 uniformly distributed values for each range. Since the ATNF catalogue only gives four measurements of the $\ddot{f}_{\rm b, obs}$ for millisecond pulsars, we decided to extend its range. More specifically, the values of $\ddot{f}_{\rm s, obs}$ for these four millisecond pulsars lie in the ranges of $-1.0\times10^{-26}~{\rm s}^{-3}$ to $-1.0\times10^{-27}~{\rm s}^{-3}$ and $1.0\times10^{-28}~{\rm s}^{-3}$ to $1.0\times10^{-26}~{\rm s}^{-3}$, while by fitting these values with a distribution function and generating a synthetic values in section \ref{subsec:MCMCField}, we found that $\ddot{f}_{\rm b, obs}$ vary between $-1.0\times10^{-25}~{\rm s}^{-3}$ to $-1.0\times10^{-30}~{\rm s}^{-3}$ and $1.0\times10^{-30}~{\rm s}^{-3}$ to $1.0\times10^{-25}~{\rm s}^{-3}$. We used these two sets and divided each set into 5 subranges each of which span over one order of magnitude and generated uniform distribution of 10000 values for each subrange.

In this way, we got 16 subranges of $\dot{f}_{\rm b, obs}$, and 10 subranges of $\ddot{f}_{\rm b, obs}$, each having 10000 uniformly generated values. Similar to the technique used for simulation and calculation in section \ref{subsec:brakind}, we chose one subrange corresponding to $\dot{f}_{\rm b, obs}$ and $\ddot{f}_{\rm b, obs}$ each, in addition to the 10000 values generated for the parameters $l$, $b$, $d$, $\mu_{l}$, $\mu_{b}$, and $v_r$ each. These 10000 values for each parameter constitute the 10000 synthetic millisecond pulsars, concentrated in the specific sub-range in the multi-dimensional parameter space, out of the total 160 ($16\times10$) such sub-ranges. We computed the values of $\dot{f}_{\rm b, int}$ and $\ddot{f}_{\rm b, int}$ for all 10000 synthetic pulsars in each sub-range and selected the favorable cases. In this approach, we implicitly assumed that the total number of millisecond pulsars over the full parameter space is much larger than 10000, which is a bit over-estimation. However, our quest here is to find unique combinations of parameters that can give large differences between the observed and intrinsic values of the second derivative of the orbital period.

Since we did not get any favourable result, we then used uniform distributions for $\mu_{l}$ and $\mu_{b}$ too. As per the ATNF catalogue, for millisecond pulsars, $\mu_{l}$ varies between -52.8 mas/yr and 74.485 mas/yr, and $\mu_{b}$ varies between -103 mas/yr and 120.820 mas/yr. We generated 10000 uniformly distributed values between these maximum and minimum values of $\mu_{l}$ and $\mu_{b}$, respectively. However, we did not get any favourable case, not even for near-Galactic centre pulsars. We then checked the maximum and minimum values of $\mu_{l}$ and $\mu_{b}$ from ATNF catalogue for the set of all the pulsars and found that $\mu_{l}$ varies between -336.73 mas/yr and 193.8 mas/yr, and $\mu_{b}$ varies between -314.1 mas/yr and 176 mas/yr. We then generated 10000 uniformly distributed values between these maximum and minimum values of $\mu_{l}$ and $\mu_{b}$, respectively. We found only one favourable case among the simulated Galactic field pulsars but multiple for the simulated near-Galactic centre pulsars.

The number of favourable cases (where $\ddot{f}_{\rm b,obs}$ and $\ddot{f}_{\rm b,int}$ values differ atleast in the first decimal place) for different ranges of $\dot{f}_{\rm b,obs}$ and $\ddot{f}_{\rm b,obs}$ are summarized in table \ref{tb:orbitlist} for both sets, i.e., the simulated Galactic field pulsars, as well as, the simulated near-Galactic centre pulsars. We display the one favourable case from the simulated Galactic field pulsar and five examples from the set of favourable simulated near-Galactic centre pulsars in table \ref{tb:orbitpsrs} with maximum percentage change in $\ddot{f}_{\rm b,obs}$. From table \ref{tb:orbitlist}, it is evident that the number of favourable cases significantly increase when we consider the simulated pulsars near the Galactic centre.

\begin{table*}
\begin{center}
\caption{Number of favourable simulated millisecond pulsars in the Galactic field, as well as, near the Galactic centre where $\ddot{f}_{\rm b,obs}$ and $\ddot{f}_{\rm b,int}$ values differ atleast in first decimal place, for various ranges of $\dot{f}_{\rm b, obs}$, and $\ddot{f}_{\rm b, obs}$ values. One row represents single simulation run of 10000 millisecond pulsars. These entries represent the simulation runs when -336.73 $<$ $\mu_{l}$ $<$ 193.8 mas/yr and -314.1 $<$ $\mu_{b}$ $<$ 176 mas/yr. The third column represents the favourable cases for the Galactic field pulsars, and the fourth column represents the favourable cases for the pulsars near the Galactic centre.}
\begin{tabular}{|l|l|l|l|}
\hline
\addstackgap{$\dot{f}_{\rm b, obs}$} range  & \addstackgap{$\ddot{f}_{\rm b, obs}$} range  & Favourable Cases &  Favourable Cases \\
(${\rm s}^{-2}$) & (${\rm s}^{-3}$) & (Galactic Field) & (Near Galactic Centre) \\
\hline

 \addstackgap{$-10^{-17}$ to $-10^{-18}$ }& $-10^{-29}$ to $-10^{-30}$ & 1 & 5\\

 $-10^{-17}$ to $-10^{-18}$ & $10^{-30}$ to $10^{-29}$ & 0 & 3\\

 $10^{-18}$ to $10^{-17}$ & $-10^{-29}$ to $-10^{-30}$ & 0 & 5\\

 $10^{-18}$ to $10^{-17}$ & $10^{-30}$ to $10^{-29}$ & 0 & 1\\
\hline
\end{tabular}
\label{tb:orbitlist}
\end{center}
\end{table*}

\begin{table*}
\begin{center}
\caption{Parameters for simulated millisecond pulsars with $\ddot{f}_{\rm int}$ being different than $\ddot{f}_{\rm obs}$ atleast in the leading digit or in the first decimal place at the most. Meanings of $l$, $b$, $d$, $\mu_{l}$, $\mu_{b}$, and $v_{r}$ are explained in the caption of table \ref{tb:brakcompnew}. Here, $d_{\rm GC}$ is the distance between the pulsar and the Galactic centre, $\mu_{\rm Tot}$ is the total transverse proper motion, $f_{\rm b}$ is the orbital frequency, $\dot{f}_{\rm b,obs}$ is observed orbital frequency derivative, $\dot{f}_{\rm b,int}$ is intrinsic orbital frequency derivative, $\ddot{f}_{\rm b,obs}$ is observed orbital frequency second derivative, and $\ddot{f}_{\rm b,int}$ is intrinsic orbital frequency second derivative. The second column displays the parameters of the one favourable simulated Galactic field pulsar (represented by Pulsar$_{\rm GF}$), whereas, the columns 3, 4, 5, 6, and 7, display the parameters of the favourable simulated pulsars near the Galactic centre (represented by subscript `GC'). Displaying the results till second decimal place.}
\begin{tabular}{|l|l|l|l|l|l|l|}
 \hline 
 Parameters & Pulsar$_{\rm GF}$ & Pulsar1$_{\rm GC}$ & Pulsar2$_{\rm GC}$ & Pulsar3$_{\rm GC}$ & Pulsar4$_{\rm GC}$ & Pulsar5$_{\rm GC}$ \\ 
[1em]\hline 
$l$ (deg)  & 317.85 & 356.78 & 358.10 & 356.15 & 359.97 & 0.69 \\ 

$b$ (deg) & -43.66 & -3.83 & -4.97 & -3.69 & -1.22 & -4.35  \\

$d$ (kpc) &  7.61 & 8.15 & 8.01 & 7.82 & 8.02 & 8.10  \\ 

$d_{\rm GC}$ (kpc) & 7.52 & 0.72 & 0.74 & 0.76 & 0.17 & 0.62\\

$\mu_{l}$ (mas/yr)  & -329.61 & -284.68 & -336.14 & -313.63 & -334.68 & -214.96  \\  

$\mu_{b}$ (mas/yr)  & -270.21 & -290.50 & -197.44 & -250.02 & -304.15 & -308.93 \\      

$\mu_{\rm Tot}$ (mas/yr)  & 426.21 & 406.74 & 389.83 & 401.09 & 452.24 & 376.36 \\      

$v_{r}$ (km/s) & -157.14 & -30.54 & 27.67 & -40.97 & 12.78 & 199.42 \\ 

$f_{\rm b}$ (($\times10^{-6}$) Hz)  & 36.82 & 141.85 & 1.01 & 0.45 & 171.66 & 1.80 \\   

\addstackgap{$\dot{f}_{\rm b,obs}$ ($\times10^{-18}~\rm s^{-2}$)} & -9.87 & 9.50 & -8.70 & 9.68 & -9.19 & -9.55  \\   

\addstackgap{$\ddot{f}_{\rm b,obs}$ ($\times10^{-30}~\rm s^{-3}$)} & -2.39 & -1.16 & 1.85 & -2.40 & 3.25 & -2.70 \\   
 
\addstackgap{$\dot{f}_{\rm b,int}$ ($\times10^{-18}~\rm s^{-2}$)} & -9.75 & 9.97 & -8.70 & 9.68 & -8.50 & -9.54 \\   
 
\addstackgap{$\frac{(\dot{f}_{\rm b,int}-\dot{f}_{\rm b,obs}) \times 100}{\dot{f}_{\rm b,obs}}$} & -1.25 & 4.89 & -0.03 & 0.01 & -7.45 & -0.05 \\   

\addstackgap{$\ddot{f}_{\rm b,int}$ ($\times10^{-30}~\rm s^{-3}$)} & -2.46 & -1.10  & 1.80 & -2.34 & 3.18 & -2.75 \\

\addstackgap{$\frac{(\ddot{f}_{\rm b,int}-\ddot{f}_{\rm b,obs}) \times 100}{\ddot{f}_{\rm b,obs}}$} & 2.76 & -5.37 & -2.78 & -2.45 & -2.26 & 1.97  \\
 \hline   
\end{tabular}
\label{tb:orbitpsrs}
\end{center}
\end{table*}

\section{Conclusion}
\label{se:con}

The measured values of the first and the second derivatives of the frequency ($\dot{f}_{\rm obs}$ and $\ddot{f}_{\rm obs}$ respectively) of a pulsar differ from the intrinsic values ($\dot{f}_{\rm int}$ and $\ddot{f}_{\rm int}$ respectively) due to its velocity, acceleration, and jerk. These derivatives can be either of the spin frequency or of the orbital frequency. In the present work, we provide expressions for $\dot{f}_{\rm int}$ and $\ddot{f}_{\rm int}$ in terms of other measurable parameters with the assumptions that the gravitational potential of the Galaxy is the only cause of the acceleration and jerk of the pulsar. For additional sources of acceleration and jerk, e.g., unmodelled orbital motion, local potential, etc., one will need to work on a case-by-case basis. In our earlier work \citep{pb18}, we explored the first derivative (although of the period instead of the frequency), hence in the present work, we concentrate on the second derivative. The main result of this paper, i.e., eq. (\ref{eq:fddotex4}) will be useful to eliminate the effect of the gravitational potential of the Galaxy and then to model additional dynamical terms if present.

Some of the intermediate equations will be useful for other purposes, e.g., our eq. (\ref{eq:fddotex1}) can be used to model the gravitational potential of a globular cluster if the second derivative of the spin frequency of some pulsars in that cluster is measured. Even if we neglect the terms involving the time derivative of the unit vector, there are extra terms in eq. (\ref{eq:fddotex1}) in comparison to commonly used expression (eq. (6) of \cite{prf17}; eq. (7) of \cite{fr17}). 

As the expressions needed to be evaluated to estimate $\ddot{f}_{\rm int}$, i.e. eqs. (\ref{eq:fddotex4}) and (\ref{eq:fddotint}) are moderately large, first we explored whether any one or more of the terms can be ignored. Our simulations established the fact that all terms are of nearly equal importance, and should be kept. We also found that the total dynamical contribution would be much larger for pulsars located near the Galactic centre, so when such pulsars will be discovered and timed, one should not forget to correct for dynamical contributions from the first and second derivatives of the frequencies. However, we emphasize that the values reported in this paper for Galactic centre are not very accurate, as the gravitational potential of the Galaxy in that region is not very well modelled. If such a model is available in the future, one can easily implement that in a code based on our analytical expressions.

We then investigated potential cases where the dynamical contributions might lead to confusing results if not corrected for. As our expression is valid for the second derivatives of the spin frequency as well as the second derivatives of the orbital frequency frequency, we studied both. We paid special attention to the pulsars with reported values of breaking index being different from 3 \citep{hob04, da20}. Although from these real pulsars, we didn't see any significant contribution of dynamical terms to the value of the braking index, our simulations resulted in a few such cases. We also saw that it is very rare to have the second derivative of the orbital frequency contaminated by the dynamical terms. 

Although our paper is not the first one studying dynamical contributions in the second derivative of the frequency of a pulsar, this is the first time an accurate analytical expression is given, see \cite{liu18} and references there in for earlier approximated approaches to this issue. We have also presented detailed mathematical derivations in two appendices those might be useful for further exploration of dynamics of pulsars in the Galaxy.

In the present work, we reported dynamical effect corrected values of the second derivative of frequencies without any errors, as we calculated the accelerations and jerks using a model of the Galactic potential that does not report errors. The lack of error estimation does not alter the conclusions of this work, because, first, most of the results we report are for simulated pulsars, second, dynamical effects are not that large for the second derivative of the frequencies for the real pulsars we have. However, as we have seen, it will be large for pulsars near the Galactic centre, one should have a much improved model of the Galactic potential including uncertainties on various parameters. It will not be difficult to adopt the standard error propagation technique on our analytical expressions which are the main results of the present work.

\section*{Acknowledgements}

The authors thank the anonymous reviewer for insightful suggestions that have improved the paper.

\appendix

\section{Unit vectors and their derivatives}
\label{appendix:unitvector}

\begin{figure*}
\begin{center}
\subfigure[]{\label{subfig:3Dplot}\includegraphics[width=0.40\textwidth]{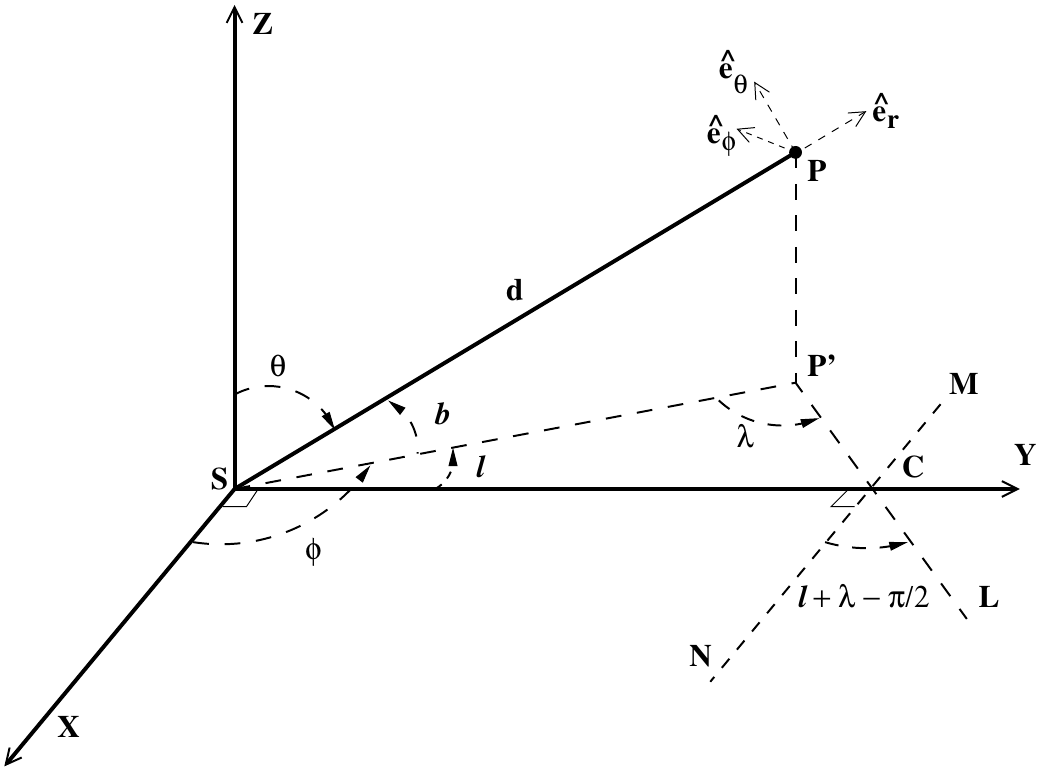}}
\hskip 1.cm \subfigure[]{\label{subfig:topview}\includegraphics[width=0.40\textwidth]{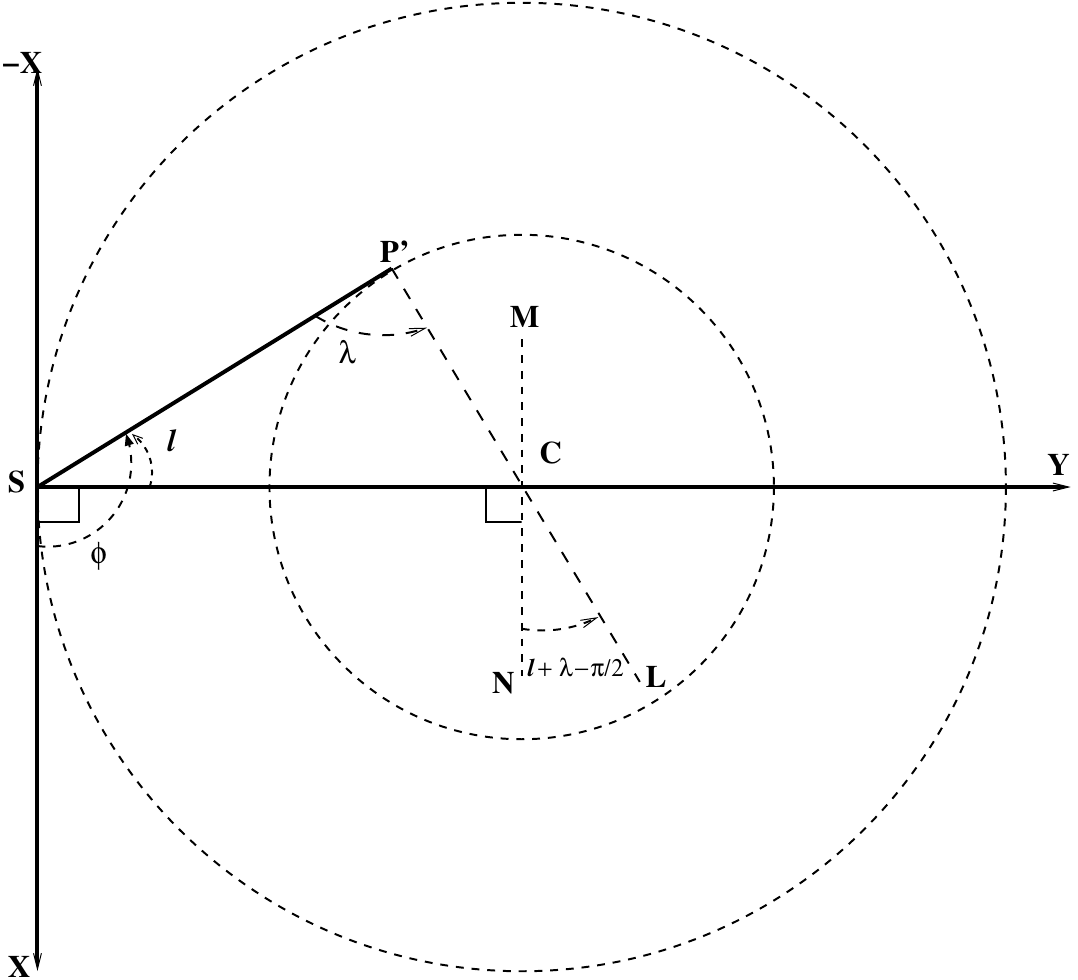}}
\end{center}
\caption{a) Schematic 3D-diagram describing Pulsar position. b) Top view of the Galactic plane shown in the left panel.}
\label{fig:posppr3}
\end{figure*}

As the equations in section \ref{se:der1} involve various unit vector and their derivatives, in this appendix we explain the derivatives of unit vectors used in our calculations. Here we use a Sun centered spherical coordinate system as well as a Sun centered cartesian coordinate system shown in figure \ref{fig:posppr3}. In this figure, the Sun is denoted by `S' and the pulsar as `P'. Additionally, $\theta$ is the angle the radius vector ($\overrightarrow{SP}$) makes with the positive Z-axis, $\phi$ is the angle that the projection of the radius vector on the XY-plane makes with the positive X-axis, and the distance of the pulsar from the Sun $SP=d$ is equivalent to its radial coordinate $r$, $b$ is the Galactic latitude, $l$ is the Galactic longitude, i.e., the spherical coordinates ($r$, $\theta$, $\phi$) are equivalent to ($d$, $90^{\circ} - b$, $90^{\circ} + l$). Note that the Sun centered cartesian coordinate system is chosen in such a way that the Galactic centre `C' is along the Y-axis having the ($x$, $y$, $z$) coordinates as ($0, R_{\rm s}$, 0), where $R_{\rm s}$ is the Galactocentric distance of the Sun. SP is projected on the Galactic plane as SP$^{\prime}$, i.e., P$^{\prime}$ is the projected location of the pulsar on the Galactic plane. The value of P$^{\prime}$C is denoted by $R_{\rm p'}$, which is the Galactocentric distance of the pulsar projection on the Galactic plane.  As the Sun is orbiting around the Galactic centre, the stable configuration demands that at any instant, the velocity of the Sun has two components, along the X-axis and along the Z-axis while the acceleration of the Sun has components along the Y-axis and Z-axis. 

We can, however, drop the terms containing the component of the velocity and the acceleration of the Sun perpendicular to the Galactic plane (${v}_{\rm s,z}$ and ${a}_{\rm s,z}$ respectively) as the motion of the Sun in that direction is negligible. We substantiate this claim below.

The vertical velocity of stars in the Galaxy with respect to the local standard of rest is defined as $W=\dot{d} \sin b + \mu_{b} \, d \, \cos b$ \citep{bovy10}. The first term contains the radial velocity $v_r =\dot{d}$ which is difficult to measure. For five binary pulsars, optical spectroscopy of their binary companions made it possible to measure $v_r$ \citep{liu18}) whose absolute value lie in the range of $42 - 185 ~{\rm km ~s^{-1}}$, and in their simulation, they vary $v_r$ from $-200 ~{\rm km ~s^{-1}}$  to $200 ~{\rm km ~s^{-1}}$. The second term in the expression of $W$ is usually smaller, for the 106 pulsars in the Galactic field with reported values of $\mu_b$ and dispersion measure independent measurements of distances, we find that the values of this term can be fitted with a Gaussian of mean $-12.49  ~{\rm km \, s^{-1}}$ and standard deviation  $172  ~{\rm km \, s^{-1}}$. These two terms can add up if they have the same sign. On the other hand, for the sun, the latest measurement of $W$ is only $7.25^{+0.37}_{-0.36}~ {\rm km ~s^{-1}}$ \citep{sbd10}. Note that, \citep{sbd10} also reported other two components of the motion of the Sun with respect to the local standard of rest. These are the radial ($U$) and along the Galactic rotation ($V$) velocities, values being $11.1^{+0.69}_{-0.75}~ {\rm km ~s^{-1}}$  and $12.24^{+0.47}_{-0.47}~ {\rm km ~s^{-1}}$ respectively. These values would contribute in the net velocity of the Sun parallel to the disk ${v}_{\rm s,pl} = \sqrt{(\Theta_S + V)^2 + U^2}$ where $\Theta_S$ is the Galactic rotation at the location of the Sun, whose value ($240 \pm 8~ {\rm km ~s^{-1}}$ according to \citet{rmb14}) is much larger than the values of $U$ and $V$.

As the Sun is located in the Galactic plane with a vertical height zero, the value of the z-gradient of the gravitational potential of the Galaxy, which causes the vertical acceleration of the stars, is zero at this location \citep{pb18}. Although the measured value of the vertical acceleration of the sun is non-zero, $3.95 \pm 0.47 ~{\rm mm \, s^{-1} \, yr^{-1}}$ \citep{xwz12} which translates (dividing by $c$) into $4.17 \times 10^{-19}~{\rm s^{-1}}$. The cause of this acceleration is not yet understood, but surely something in addtion to the acceleration due to the standard Galactic gravitational potential. However, this value is much smaller to than the z-component of Galactic acceleration at most of the places in the Galaxy, as shown in figure 4b of \citep{pb18}, and hence can be ignored. Hence, everywhere we will consider ${v}_{\rm s,z}=0$ and ${a}_{\rm s,z}=0$.

Now using the properties of spherical coordinates, we can write:

\begin{equation}
\widehat{e}_{r} = \sin \theta\,\cos \phi\,\widehat{e}_{x} + \sin \theta\,\sin \phi\,\widehat{e}_{y} + \cos \theta\,\widehat{e}_{z},
\label{eq:er}
\end{equation}

\begin{equation}
\widehat{e}_{\theta} = \cos \theta\,\cos \phi\,\widehat{e}_{x} + \cos \theta\,\sin \phi\,\widehat{e}_{y} - \sin \theta\,\widehat{e}_{z},
\label{eq:etheta}
\end{equation}

\begin{equation}
\widehat{e}_{\phi} = -\sin \phi\,\widehat{e}_{x} + \cos \phi\,\widehat{e}_{y} ~, 
\label{eq:ephi}
\end{equation} where $\widehat{e}_{x}$, $\widehat{e}_{x}$, $\widehat{e}_{x}$ are the cartesian unit vectors and $\widehat{e}_{r}$, $\widehat{e}_{\theta}$, $\widehat{e}_{\phi}$  are the unit vectors of the spherical coordinate system as mentioned above. Also, $\theta = 90^{\circ} - b$ gives $\sin \theta = \cos b$,  $\cos \theta= \sin b$, $\dot{\theta} = - \dot{b}$, and $\phi = 90^{\circ} + l $ gives $\sin \phi = \cos l$, $\cos \phi = - \sin l$, $\dot{\phi} = \dot{l}$.

Consequently, the time-derivatives of $\widehat{e}_{r}$, $\widehat{e}_{\theta}$, and $\widehat{e}_{\phi}$ are: 
\begin{equation}
\dot{\widehat{e}}_{r} =  \dot{\theta}\,\widehat{e}_{\theta} + \dot{\phi}\,\sin \theta\,\widehat{e}_{\phi},
\label{eq:erdot}
\end{equation}

\begin{equation}
\dot{\widehat{e}}_{\theta} =  -\dot{\theta}\,\widehat{e}_{r} + \dot{\phi}\,\cos \theta\,\widehat{e}_{\phi},
\label{eq:ethetadot}
\end{equation}

\begin{equation}
\dot{\widehat{e}}_{\phi} = -\dot{\phi}\,\left(\sin \theta\,\widehat{e}_{r} + \cos \theta\,\widehat{e}_{\theta}\right)  = -\dot{\phi}\,\left(\cos \phi\,\widehat{e}_{x} + \sin \phi\,\widehat{e}_{y}\right)~.
\label{eq:ephidot}
\end{equation}

From figure \ref{subfig:3Dplot}, we can see that The unit vector from the Sun to the pulsar $\widehat{n}_{\rm sp}$ is the radial unit vector $\widehat{e}_{r}$. Hence, $\widehat{n}_{\rm sp} = \widehat{e}_{r}$ gives $\dot{\widehat{n}}_{\rm sp} = \dot{\widehat{e}}_{r}$ and $\ddot{\widehat{n}}_{\rm sp} = \ddot{\widehat{e}}_{r}$, which can be written as:

\begin{align}
\ddot{\widehat{n}}_{\rm sp} &= \ddot{\widehat{e}}_{r} \nonumber \\
 &= \ddot{\theta}\,\widehat{e}_{\theta} + \dot{\theta}\,\dot{\widehat{e}}_{\theta} +  \ddot{\phi}\,\sin \theta\,\widehat{e}_{\phi} + \dot{\phi}\,\dot{\theta}\,\cos \theta\,\widehat{e}_{\phi} + \dot{\phi}\,\sin \theta\,\dot{\widehat{e}}_{\phi}  ~~~~ {\rm [Used ~eq. (\ref{eq:erdot})]} \nonumber \\
 &= \ddot{\theta}\,\widehat{e}_{\theta} + \dot{\theta}\,\left(-\dot{\theta}\,\widehat{e}_{r} + \dot{\phi}\,\cos \theta\,\widehat{e}_{\phi}\right) + \ddot{\phi}\,\sin \theta\,\widehat{e}_{\phi} + \dot{\phi}\,\dot{\theta}\,\cos \theta\,\widehat{e}_{\phi} + \dot{\phi}\,\sin \theta\,\left(-\dot{\phi}\,\left(\sin \theta\,\widehat{e}_{r} + \cos \theta\,\widehat{e}_{\theta}\right)\right) \nonumber \\
 &= \left(\ddot{\theta} - {\dot{\phi}}^{2}\,\sin \theta\,\cos \theta\right)\,\widehat{e}_{\theta}  - \left({\dot{\theta}}^{2} + {\dot{\phi}}^{2}\,\sin^{2} \theta\right)\,\widehat{e}_{r} + \left(2 \, \dot{\theta}\,\dot{\phi}\,\cos \theta + \ddot{\phi}\,\sin \theta\right)\,\widehat{e}_{\phi}~.
\label{eq:nspddot}
\end{align}

\section{Simplifying terms in eq. (\ref{eq:pddot4})}
\label{appendix:B}

The proper motion comes into various terms in eq. (\ref{eq:pddot4}, so let us first define those. $\mu_{l} = \dot{l} \, \cos b $ is the proper motion in the Galactic longitude and $\mu_{b} = \dot{b} $ is the proper motion in the Galactic latitude. The total transverse proper motion is defined as, $\mu_{\rm T} = \sqrt{\mu_{l}^{2} + \mu_{b}^{2}} =  v_T/d$. 

\subsection{Simplifying the second term on the right hand side of eq. (\ref{eq:pddot4})}
\label{appendix:B1}

Let us now focus on the second term on the right hand side of eq. (\ref{eq:pddot4}):

\begin{align}\hskip -0.6cm
\frac{\left(\vec{a}_{\rm p} - \vec{a}_{\rm s}\right)\cdot\dot{\widehat{n}}_{\rm sp}}{c} = &\frac{\left(\vec{a}_{\rm p} - \vec{a}_{\rm s}\right)\cdot\dot{\widehat{e}}_{r}}{c} \nonumber \\
= &\frac{1}{c}\,{\left(\vec{a}_{\rm p,pl} + \vec{a}_{\rm p,z} - \vec{a}_{\rm s,pl} - \vec{a}_{\rm s,z}\right)\cdot\left(\dot{\theta}\,\widehat{e}_{\theta} + \dot{\phi}\,\sin \theta\,\widehat{e}_{\phi}\right)} \nonumber \\
= &\frac{1}{c}\,\left(\vec{a}_{\rm p,pl} - \vec{a}_{\rm s,pl} + \vec{a}_{\rm p,z} - \vec{a}_{\rm s,z}\right)\cdot\left(\dot{\theta}\,\cos \theta\,\cos \phi\,\widehat{e}_{x} + \dot{\theta}\,\cos \theta\,\sin \phi\,\widehat{e}_{y} - \dot{\theta}\,\sin \theta\,\widehat{e}_{z} \right. \nonumber \\ 
&\left.-\dot{\phi}\,\sin \theta\,\sin \phi\,\widehat{e}_{x} + \dot{\phi}\,\sin \theta\,\cos \phi\,\widehat{e}_{y} \right) ~,
\label{eq:term2.1}
\end{align} where $\vec{a}_{\rm p,pl}$ and $\vec{a}_{\rm p,z}$ are the components of the acceleration of the pulsar parallel and perpendicular to the Galactic plane, and $\vec{a}_{\rm s,pl}$ and $\vec{a}_{\rm s,z}$ are the components of the acceleration of the Sun parallel and perpendicular to the Galactic plane.

Considering the parallel component, we get,
\begin{align}
\frac{1}{c}\,&{\left(\vec{a}_{\rm p,pl} - \vec{a}_{\rm s,pl}\right)\cdot\left(\dot{\theta}\,\cos \theta\,\cos \phi\,\widehat{e}_{x} + \dot{\theta}\,\cos \theta\,\sin \phi\,\widehat{e}_{y} - \dot{\theta}\,\sin \theta\,\widehat{e}_{z} -\dot{\phi}\,\sin \theta\,\sin \phi\,\widehat{e}_{x} + \dot{\phi}\,\sin \theta\,\cos \phi\,\widehat{e}_{y} \right)}  \nonumber \\
= &\frac{1}{c}\,\left[\left({a}_{\rm p,pl}\,\cos (l + \lambda - \frac{\pi}{2})\,\widehat{e}_{x} + {a}_{\rm p,pl}\,\sin (l + \lambda - \frac{\pi}{2})\,\widehat{e}_{y} - {a}_{\rm s,pl}\,\widehat{e}_{y}\right) \cdot\left(\dot{\theta}\,\cos \theta\,\cos \phi\,\widehat{e}_{x} + \dot{\theta}\,\cos \theta\,\sin \phi\,\widehat{e}_{y} \right. \right. \nonumber \\ 
&\left. \left. - \dot{\theta}\,\sin \theta\,\widehat{e}_{z} -\dot{\phi}\,\sin \theta\,\sin \phi\,\widehat{e}_{x} + \dot{\phi}\,\sin \theta\,\cos \phi\,\widehat{e}_{y} \right)\right] \nonumber \\
= &\frac{1}{c}\,\left[-\dot{\theta}\,\sin b\,\{{a}_{\rm p,pl}\,\cos \lambda + {a}_{\rm s,pl}\,\cos l \} - \dot{\phi}\,\cos b\,\{{a}_{\rm p,pl}\,\sin \lambda - {a}_{\rm s,pl}\,\sin l\}\right]\nonumber \\
= &-\frac{\dot{\theta}\,\sin b}{c}\,\left[{a}_{\rm p,pl}\,\cos \lambda + {a}_{\rm s,pl}\,\cos l\right] - \frac{\dot{\phi}\,\cos b}{c}\,\left[{a}_{\rm p,pl}\,\frac{R_{\rm s}\,\sin l}{R_{\rm p'}} - {a}_{\rm s,pl}\,\sin l \right] \nonumber \\
&\hskip5cm [{\rm Using}~ \sin \lambda = \frac{R_{\rm s}\,\sin l}{R_{\rm p'}}, {\rm from ~sine ~law~in}~\triangle {\rm SP'C} ~{\rm of~figure~\ref{subfig:topview}}]  \nonumber \\ 
= &-\frac{\dot{\theta}\,\sin b}{c}\,\left[{a}_{\rm p,pl}\,\sqrt{\left(1-\frac{R_{\rm s}^{2}\,\sin^{2} l}{{R_{\rm p'}}^{2}}\right)} + {a}_{\rm s,pl}\,\cos l\right] - \frac{\dot{\phi}\,\cos b\,\sin l}{c}\,\left[{a}_{\rm p,pl}\,\frac{R_{\rm s}}{R_{\rm p'}} - {a}_{\rm s,pl} \right]~.
\label{eq:pl1}
\end{align}

Now, considering the perpendicular component, we get,
\begin{align}
& \frac{1}{c}\,\left(\vec{a}_{\rm p,z} - \vec{a}_{\rm s,z}\right)\cdot\left(\dot{\theta}\,\cos \theta\,\cos \phi\,\widehat{e}_{x} + \dot{\theta}\,\cos \theta\,\sin \phi\,\widehat{e}_{y} - \dot{\theta}\,\sin \theta\,\widehat{e}_{z} -\dot{\phi}\,\sin \theta\,\sin \phi\,\widehat{e}_{x} + \dot{\phi}\,\sin \theta\,\cos \phi\,\widehat{e}_{y} \right)  \nonumber \\
&= \frac{1}{c}\,\left(\vec{a}_{\rm p,z} - \vec{a}_{\rm s,z}\right)\cdot(-\dot{\theta}\,\sin \theta\,\widehat{e}_{z}) \nonumber \\
&= \frac{-\dot{\theta}\,\sin \theta}{c}\,\left(\vec{a}_{\rm p,z} - \vec{a}_{\rm s,z}\right)\cdot(\,\widehat{e}_{z}) \nonumber \\
&= \frac{-\dot{\theta}\,\sin \theta}{c}\,\left(\vec{a}_{\rm p,z} - \vec{a}_{\rm s,z}\right)\cdot \left(\frac{\widehat{e}_{r}}{\cos \theta} - \frac{(\sin \theta\,\cos \phi\,\widehat{e}_{x} + \sin \theta\,\sin \phi\,\widehat{e}_{y})}{\cos \theta} \right) \hskip2cm ~~~[{\rm Using~eq.}\,(\ref{eq:er})]\nonumber \\
&=  \frac{-\dot{\theta}\,\sin \theta}{c}\,\left(\vec{a}_{\rm p,z} - \vec{a}_{\rm s,z}\right)\cdot \left( \frac{\widehat{e}_{r}}{\cos \theta} \right)  \nonumber \\
&= -\dot{\theta}\,\tan \theta\,\frac{\left(\vec{a}_{\rm p,z} - \vec{a}_{\rm s,z}\right)}{c}\cdot(\,\widehat{n}_{\rm sp}) \nonumber \\
&= \dot{\theta}\,\cot b\,\frac{1}{c}(a_{\rm p,z}\,\sin b) \nonumber \\
&= \frac{\dot{\theta}\,\cos b}{c}\,a_{\rm p,z}~.
\label{eq:z1}
\end{align}

So, from eq. (\ref{eq:pl1}) and eq. (\ref{eq:z1}), we get: 

\begin{align}
\frac{\left(\vec{a}_{\rm p} - \vec{a}_{\rm s}\right)\cdot\dot{\widehat{n}}_{\rm sp}}{c} = &-\frac{\dot{\theta}\,\sin b}{c}\,\left[{a}_{\rm p,pl}\,\sqrt{\left(1-\frac{R_{\rm s}^{2}\,\sin^{2} l}{{R_{\rm p'}}^{2}}\right)} + {a}_{\rm s,pl}\,\cos l\right] - \frac{\dot{\phi}\,\cos b\,\sin l}{c}\,\left[{a}_{\rm p,pl}\,\frac{R_{\rm s}}{R_{\rm p'}} - {a}_{\rm s,pl} \right] \nonumber \\ 
&+ \frac{\dot{\theta}\,\cos b}{c}\,a_{\rm p,z} \nonumber \\
= &\mu_{b}\,\frac{\sin b}{c}\,\left[{a}_{\rm p,pl}\,\sqrt{\left(1-\frac{R_{\rm s}^{2}\,\sin^{2} l}{{R_{\rm p'}}^{2}}\right)} + {a}_{\rm s,pl}\,\cos l\right] - \mu_{b}\,\frac{\cos b}{c}\,a_{\rm p,z} - \mu_{l}\,\frac{\sin l}{c}\,\left[{a}_{\rm p,pl}\,\frac{R_{\rm s}}{R_{\rm p'}} - {a}_{\rm s,pl} \right] ~.
\label{eq:term2final}
\end{align}

We have used eq. (\ref{eq:term2final}) in eq. (\ref{eq:fddotex1}) of section \ref{se:der1}.

\subsection{Simplifying the third term on the right hand side of eq. (\ref{eq:pddot4})}
\label{appendix:B2}

Now, we consider the third term on the right hand side of eq. (\ref{eq:pddot4}), i.e., $\frac{(\vec{v}_{\rm p} - \vec{v}_{\rm s})\cdot\ddot{\widehat{n}}_{\rm sp}}{c}$. First, using eq. (\ref{eq:erdot}), we get:

\begin{subequations}
\begin{align}
\vec{v}_{\rm p} - \vec{v}_{\rm s} &= \vec{v}_{\rm sp} = \frac{d}{dt}\,(d\,\widehat{e}_{r}) = \dot{d}\,\widehat{e}_{r} + d\,\dot{\widehat{e}}_{r} =  \dot{d}\,\widehat{e}_{r} + d\,\dot{\theta}\,\widehat{e}_{\theta} + d\,\dot{\phi}\,\sin \theta\,\widehat{e}_{\phi}~.
\label{eq:vsp}
\end{align}
Also,
\begin{align}
\vec{v}_{\rm p} - \vec{v}_{\rm s} &= \vec{v}_{\rm sp} = \frac{d}{dt}\,(d\,\widehat{e}_{r}) = \dot{d}\,\widehat{e}_{r} + d\,\dot{\widehat{e}}_{r} =  v_{r}\,\,\widehat{e}_{r} + v_{\rm T}\,\,\widehat{e}_{\rm T,v}~.
\label{eq:vsp2}
\end{align}
\end{subequations}
Here, $v_r$ is the magnitude of the radial component and $v_T$ is the magnitude of the transverse component of the relative velocity $\vec{v}_{\rm sp}$ while $\widehat{e}_{r}$ and $\widehat{e}_{\rm T,v}$ are unit vectors in those directions, $d\,\dot{\widehat{e}}_{r} = d\,\mu_{\rm T}\,\widehat{e}_{\rm T,v} = v_{\rm T}\,\,\widehat{e}_{\rm T,v}$ and $\dot{d} = v_{r}$.

In order to obtain the expression of $\dot{\widehat{e}}_{\rm T,v}$, let us consider a 2-dimensional Cartesian plane (X$^{\prime} \, Y^{\prime}$) containing the unit vectors $\widehat{e}_{r}$ and $\widehat{e}_{\rm T,v}$, such that $\widehat{e}_{r}$ makes an angle $\gamma$ with $\widehat{e}_{x^{\prime}}$. In that case,

\begin{subequations}
\begin{align}
\widehat{e}_{r} = \cos \gamma \widehat{e}_{x^{\prime}} + \sin \gamma \widehat{e}_{y^{\prime}}~,
\label{eq:er1}
\end{align}
and
\begin{align}
\widehat{e}_{\rm T,v} = -\sin \gamma \widehat{e}_{x^{\prime}} + \cos \gamma \widehat{e}_{y^{\prime}}~.
\label{eq:eT1}
\end{align}
\end{subequations} 

Taking time-derivatives of above equations, we get,
\begin{subequations}
\begin{align}
\dot{\widehat{e}}_{r} &= \dot{\gamma} (-\sin \gamma \widehat{e}_{x^{\prime}} + \cos \gamma \widehat{e}_{y^{\prime}}) \nonumber \\
&= \mu_{\rm T}\,\widehat{e}_{\rm T,v}~,  \hskip1cm [\text{Using eq. (\ref{eq:eT1})}]
\label{eq:erdot1}
\end{align}
and
\begin{align}
\dot{\widehat{e}}_{\rm T,v} &= - \dot{\gamma} (\cos \gamma \widehat{e}_{x^{\prime}} + \sin \gamma \widehat{e}_{y^{\prime}}) \nonumber \\
&= -\mu_{\rm T}\,\widehat{e}_{r}~, \hskip1cm [\text{Using eq. (\ref{eq:er1})}]
\label{eq:eTdot1}
\end{align}
\end{subequations}  where $\dot{\gamma} = \mu_{\rm T}$ is the total transverse proper motion. Hence, $\dot{\widehat{e}}_{\rm T,v} = -\mu_{\rm T} \, \widehat{e}_{r}$ which we use in eq. (\ref{eq:asp1}).

Using the expressions of $ \vec{v}_{\rm p} - \vec{v}_{\rm s}$ from eq. (\ref{eq:vsp}) and $\ddot{\widehat{n}}_{\rm sp}$ from eq. (\ref{eq:nspddot}), we get,

\begin{align}\hskip -0.9cm
\frac{(\vec{v}_{\rm p} - \vec{v}_{\rm s})\cdot\ddot{\widehat{n}}_{\rm sp}}{c} = &\frac{1}{c}\,\left[\dot{d}\,\widehat{e}_{r} + d\,\dot{\theta}\,\widehat{e}_{\theta} + d\,\dot{\phi}\,\sin \theta\,\widehat{e}_{\phi}\right]\cdot\left[\left(\ddot{\theta} - {\dot{\phi}}^{2}\,\sin \theta\,\cos \theta\right)\,\widehat{e}_{\theta}  - \left({\dot{\theta}}^{2} + {\dot{\phi}}^{2}\,\sin^{2} \theta\right)\,\widehat{e}_{r} \right.\nonumber \\ 
 &\left. +\left(2 \, \dot{\theta}\,\dot{\phi}\,\cos \theta + \ddot{\phi}\,\sin \theta\right)\,\widehat{e}_{\phi}\right] \nonumber \\
= &\frac{1}{c}\,\left[d\,\dot{\theta}\,(\ddot{\theta} - {\dot{\phi}}^{2}\,\sin \theta\,\cos \theta) - \dot{d}\,({\dot{\theta}}^{2} + {\dot{\phi}}^{2}\,\sin^{2} \theta) + d\,\dot{\phi}\,\sin \theta\,(2\,\dot{\theta}\,\dot{\phi}\,\cos \theta + \ddot{\phi}\,\sin \theta)\right] \nonumber \\
= &\frac{1}{c}\,\left[ d\,\dot{b}\,\ddot{b} - \dot{d}\,{\dot{b}}^{2} - \dot{d}\,{\dot{l}}^{2}\,\cos^{2} b - d\,\dot{b}\,{\dot{l}}^{2}\,\cos b\,\sin b + d\,\dot{l}\,\cos b\,\ddot{l}\,\cos b \right] \nonumber \\      &\hskip9cm [{\rm As}~ \theta = 90-b,\,\dot{\theta} = -\dot{b}~{\rm and}~ \dot{\phi} = \dot{l} ]\nonumber \\
= &\frac{1}{c}\,\left[ d\,\dot{b}\,\ddot{b} - \dot{d}\,{\dot{b}}^{2} - \dot{d}\,{\dot{l}}^{2}\,\cos^{2} b - d\,\dot{b}\,{\dot{l}}^{2}\,\cos^{2} b\,\tan b + d\,\dot{l}\,\cos b\,\ddot{l}\,\cos b \right]  \nonumber \\
= &\frac{1}{c}\,\left[ d\,\mu_{b}\,\dot{\mu}_{b} - \dot{d}\,{\mu_{b}}^{2} - \dot{d}\,{\mu_{l}}^{2} - d\,\mu_{b}\,{\mu_{l}}^{2}\,\tan b + d\,\mu_{l}\,\ddot{l}\,\cos b \right] \nonumber \\
 &\hskip7.8cm [{\rm As}~\mu_{b} = \dot{b},\,\dot{\mu}_{b} = \ddot{b} ~{\rm and} ~\mu_{l} = \dot{\phi}\,\sin \theta = \dot{l}\,\cos b]  \nonumber \\
= &\frac{1}{c}\,\left[ d\,\mu_{b}\,\dot{\mu}_{b} - \dot{d}\,{\mu_{b}}^{2} - \dot{d}\,{\mu_{l}}^{2} - d\,\mu_{b}\,{\mu_{l}}^{2}\,\tan b + d\,\mu_{l}\,\{\dot{\mu}_{l} + \mu_{l}\,\mu_{b}\,\tan b\} \right] \nonumber \\
 &\hskip7.5cm [{\rm As} ~ \dot{\mu}_{l} = \ddot{l}\cos b - \dot{l}\sin b\,\dot{b} = \ddot{l}\cos b - \mu_{l}\tan b\,\mu_{b}] \nonumber \\
= &\frac{1}{c}\,\left[d\,(\mu_{b}\,\dot{\mu}_{b} + \mu_{l}\,\dot{\mu}_{l}) - \dot{d}\,({\mu_{b}}^{2} + {\mu_{l}}^{2})\right]~.
\label{eq:term3}
\end{align}

On further simplifying, we get
\begin{align}
\frac{(\vec{v}_{\rm p} - \vec{v}_{\rm s})\cdot\ddot{\widehat{n}}_{\rm sp}}{c} &= \frac{1}{c}\,\left[d\,(\mu_{b}\,\dot{\mu}_{b} + \mu_{l}\,\dot{\mu}_{l}) - \dot{d}\,({\mu_{b}}^{2} + {\mu_{l}}^{2})\right] \nonumber \\
&= \frac{1}{c}\,\left[d\,\mu_{\rm T}\,\dot{\mu}_{\rm T} - \dot{d}\,{\mu_{\rm T}}^{2}\right]  & \hskip2cm [{\rm As}~ \mu_{\rm T}^2 = \mu_{l}^{2} + \mu_{b}^{2}] ~~.
\label{eq:tsbterm}
\end{align}

Here, we see that $\mu_{\rm T}$ is an observable but we need an expression for $\dot{\mu}_{\rm T}$ in terms of other observables like $l$, $b$, $\mu_l$, $\mu_b$, $d$, and $v_r$. In order to do so, we need to examine the time derivative of the relative velocity, i.e., the relative acceleration.

Taking time derivative of eq. (\ref{eq:vsp2}) we get the expression of relative acceleration as,
\begin{align}
\vec{a}_{\rm p} - \vec{a}_{\rm s} &= \dot{\vec{v}}_{\rm p} - \dot{\vec{v}}_{\rm s} = \dot{\vec{v}}_{\rm sp}\nonumber \\
&=  \dot{v}_{r}\,\,\widehat{e}_{r} + v_{r}\,\,\dot{\widehat{e}}_{r} + \dot{v}_{\rm T}\,\,\widehat{e}_{\rm T,v} + v_{\rm T}\,\,\dot{\widehat{e}}_{\rm T,v} \nonumber \\
&= \dot{v}_{r}\,\,\widehat{e}_{r} + \frac{v_{r}\,v_{\rm T}}{d}\,\widehat{e}_{\rm T,v} + \dot{v}_{\rm T}\,\,\widehat{e}_{\rm T,v} - \frac{{v_{\rm T}}^{2}}{d}\,\widehat{e}_{r} \hskip1cm [{\rm As}~ \dot{\widehat{e}}_{r} = \frac{v_T}{d} \, \widehat{e}_{\rm T,v} ~~ , \dot{\widehat{e}}_{\rm T,v} = -\mu_{\rm T} \, \widehat{e}_{r} = - \frac{v_{\rm T}}{d} \, \widehat{e}_{r}]  \nonumber \\
&= \left(\dot{v}_{r} - \frac{{v_{\rm T}}^{2}}{d}\right)\,\widehat{e}_{r} + \left(\dot{v}_{\rm T} + \frac{v_{r}\,v_{\rm T}}{d}\right)\,\widehat{e}_{\rm T,v} ~.
\label{eq:asp1}
\end{align}

Since, in this work, we are considering only the effect of the Galactic potential on the relative acceleration and relative jerk terms, we write relative acceleration as,
\begin{align}
\vec{a}_{\rm p} - \vec{a}_{\rm s} = \vec{a}_{\rm sp} = a_{r}\,\widehat{e}_{r} + a_{\rm T}\,\widehat{e}_{\rm T,a}~, 
\label{eq:asp2}
\end{align}where $\widehat{e}_{\rm T,a}$ is the unit vector in the direction of the transverse relative acceleration based on the Galactic potential. The expression for $a_{\rm T}$ in terms of other measurable quantities is derived in Appendix B.4 (refer to eq. (\ref{eq:at1})).

\begin{figure*}
\begin{center}
\includegraphics[width=0.45\textwidth]{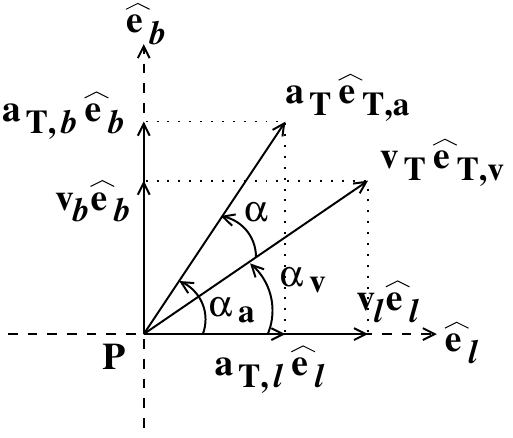}
\end{center}
\caption{Schematic diagram showing components of transverse acceleration and transverse velocity in the orthogonal coordinate frame ($\widehat{e}_{l}$, $\widehat{e}_{b}$), centred at the Pulsar position.}
\label{fig:aTvT}
\end{figure*}

An important thing to note here is the distinction between the directions of transverse acceleration ($\widehat{e}_{\rm T,a}$) and the transverse velocity ($\widehat{e}_{\rm T,v}$). Consider a plane at the pulsar position to which $\widehat{e}_{r}$  is a normal. We can compare $\widehat{e}_{\rm T,a}$ and $\widehat{e}_{\rm T,v}$ using a set of orthogonal pair of coordinates, ($\widehat{e}_{l}$, $\widehat{e}_{b}$), drawn on this plane. Positive $\widehat{e}_{l}$ is defined as the direction of positive $v_{l} = d\,\mu_{l}$, and positive $\widehat{e}_{b}$ is, similarly, defined as the direction of positive $v_{b} = d\,\mu_{b}$. From figure \ref{fig:aTvT} we can see that, the angle that $\widehat{e}_{\rm T,v}$ makes with the $\widehat{e}_{l}$ direction is given by $\alpha_{v} = \tan^{-1}\left(\frac{v_{b}}{v_{l}}\right) = \tan^{-1}\left(\frac{\mu_{b}}{\mu_{l}}\right)$, whereas, the angle that $\widehat{e}_{\rm T,a}$ makes with the $\widehat{e}_{l}$ direction is given by $\alpha_{a} = \tan^{-1}\left(\frac{a_{{\rm T},b}}{a_{{\rm T},l}}\right)$, where $a_{{\rm T},l}$ and $a_{{\rm T},b}$ are the components of $\vec{a}_{\rm T}$ along $\widehat{e}_{l}$ and $\widehat{e}_{b}$ respectively (refer eqs. (\ref{eq:aTl}) and (\ref{eq:aTb}) for their derivation). Hence, as shown in figure \ref{fig:aTvT}, the angle ($\alpha$) between $\widehat{e}_{\rm T,a}$ and $\widehat{e}_{\rm T,v}$ is given by the absolute difference of these angles, i.e., $\alpha = |\alpha_{a} - \alpha_{v}|$.

To understand the relationship between $\widehat{e}_{\rm T,a}$ and $\widehat{e}_{\rm T,v}$, we try to understand the origins of the transverse velocity and the acceleration. We have assumed that the gravitational potential of the Galaxy is the only cause of the transverse acceleration. The transverse velocity that we considered till now is based on the observed total proper motion of the pulsar. But this observed transverse velocity is a vector sum of the transverse velocity due the Galactic potential (in the direction of the transverse acceleration) and an extra transverse velocity term caused by other factors like the initial supernova kick, etc. Hence, we can write,

\begin{align}
v_{\rm T} \,\widehat{e}_{\rm T,v} = v_{\rm T,gal}\,\widehat{e}_{\rm T,a} + \vec{v}_{\rm T,ex}~, 
\label{eq:vtrans}
\end{align} where $v_{\rm T,gal}$ is the magnitude of the transverse velocity due the Galactic potential, $\widehat{e}_{\rm T,a}$ is its unit vector in its direction, and $\vec{v}_{\rm T,ex}$ is the extra transverse velocity term.

So, $\widehat{e}_{\rm T,a}$ can be written in terms of $\widehat{e}_{\rm T,v}$ as,
\begin{align}
 \widehat{e}_{\rm T,a} = \frac{v_{\rm T}\,\widehat{e}_{\rm T,v} - \vec{v}_{\rm T,ex}}{v_{\rm T, gal}}~. 
\label{eq:etransa}
\end{align}

Also, taking dot product with $\widehat{e}_{\rm T,v}$ on both sides of eq. (\ref{eq:vtrans}), we get,
\begin{align}
v_{\rm T} &= v_{\rm T,gal}\,\widehat{e}_{\rm T,a}\cdot\widehat{e}_{\rm T,v} + \vec{v}_{\rm T,ex} \cdot\widehat{e}_{\rm T,v} \nonumber \\
\implies\, v_{\rm T} &= v_{\rm T,gal}\,\cos \alpha + \vec{v}_{\rm T,ex} \cdot\widehat{e}_{\rm T,v} \hskip1cm [{\rm As}~ \alpha {\text{ is the angle between $\widehat{e}_{\rm T,a}$ and $\widehat{e}_{\rm T,v}$}} ] 
\label{eq:vtrans2}
\end{align} 

Substituting $\widehat{e}_{\rm T,a}$ in eq. (\ref{eq:asp2}) using eq. (\ref{eq:etransa}), we get,

\begin{align}
\vec{a}_{\rm p} - \vec{a}_{\rm s} &= a_{r}\,\widehat{e}_{r} + a_{\rm T}\,\frac{v_{\rm T}\,\widehat{e}_{\rm T,v} - \vec{v}_{\rm T,ex}}{v_{\rm T, gal}}  \nonumber \\
 &= a_{r}\,\widehat{e}_{r} + a_{\rm T}\,\frac{(v_{\rm T,gal}\,\cos \alpha + \vec{v}_{\rm T,ex} \cdot\widehat{e}_{\rm T,v})\,\widehat{e}_{\rm T,v} - \vec{v}_{\rm T,ex}}{v_{\rm T, gal}} \hskip1cm [{\text{Substituting $v_{\rm T}$ from eq. (\ref{eq:vtrans2})}}] \nonumber \\
 &= a_{r}\,\widehat{e}_{r} + a_{\rm T}\, \left[ \cos \alpha \,\widehat{e}_{\rm T,v} - \frac{\{\vec{v}_{\rm T,ex} - (\vec{v}_{\rm T,ex} \cdot\widehat{e}_{\rm T,v}) \,\widehat{e}_{\rm T,v}\}}{v_{\rm T, gal}}    \right] \nonumber \\
  &= a_{r}\,\widehat{e}_{r} + a_{\rm T}\, \cos \alpha \,\widehat{e}_{\rm T,v} - a_{\rm T}\,\frac{\{\vec{v}_{\rm T,ex} - (\vec{v}_{\rm T,ex} \cdot\widehat{e}_{\rm T,v}) \,\widehat{e}_{\rm T,v}\}}{v_{\rm T, gal}}    ~, 
\label{eq:asp3}
\end{align}

Now, let us define $\vec{n} = (\vec{v}_{\rm T,ex} - (\vec{v}_{\rm T,ex} \cdot\widehat{e}_{\rm T,v}) \,\widehat{e}_{\rm T,v})$. We can see that $\vec{n}$ is perpendicular to $\widehat{e}_{\rm T,v}$ as $\vec{n} \cdot \widehat{e}_{\rm T,v} =  \vec{v}_{\rm T,ex}  \cdot \widehat{e}_{\rm T,v} - (\vec{v}_{\rm T,ex} \cdot\widehat{e}_{\rm T,v}) \, ( \widehat{e}_{\rm T,v} \cdot \widehat{e}_{\rm T,v}) = 0$. Hence, the coefficient of $\widehat{e}_{\rm T,v}$ in eq. (\ref{eq:asp3}) is just $a_{\rm T}\, \cos \alpha$. Furthermore, comparing with the coefficient of $\widehat{e}_{\rm T,v}$ in eq. (\ref{eq:asp1}), we get,     

\begin{align}
a_{\rm T}\, \cos \alpha \approx \left(\dot{v}_{\rm T} + \frac{v_{r}\,v_{\rm T}}{d}\right)
\label{eq:atcos}
\end{align}

Now we can find the expression of $\dot{\mu}_{\rm T}$ in terms of the other measurable parameters. We take the time derivative of both sides of the expression $\mu_{\rm T} = \frac{v_{\rm T}}{d}$, and we get,

\begin{align}
\dot{\mu}_{\rm T} &= \frac{\dot{v}_{\rm T}}{d} - \frac{v_{\rm T}\,\dot{d}}{d^{2}} \nonumber \\
&= \frac{1}{d}\left(a_{\rm T}\,\cos \alpha - \frac{v_{r}\,v_{\rm T}}{d}\right) - \frac{v_{\rm T}\,v_{r}}{d^{2}} \hskip1cm [{\rm Using\,eq.\,(\ref{eq:atcos})}]\nonumber \\
&= \frac{a_{\rm T}\,\cos \alpha}{d} - 2\,\frac{v_{\rm T}\,v_{r}}{d^{2}} \nonumber \\
&= \frac{a_{\rm T}\,\cos \alpha}{d} - 2\,\frac{\mu_{\rm T}\,v_{r}}{d} ~.
\label{eq:utdot}
\end{align}
Using expression of $\dot{\mu}_{\rm T}$ from eq. (\ref{eq:utdot}) in eq. (\ref{eq:tsbterm}), we obtain,
\begin{align}
\frac{(\vec{v}_{\rm p} - \vec{v}_{\rm s})\cdot\ddot{\widehat{n}}_{\rm sp}}{c} &= \frac{1}{c}\,\left[d\,\mu_{\rm T}\,\dot{\mu}_{\rm T} - \dot{d}\,{\mu_{\rm T}}^{2}\right] \nonumber \\
&= \frac{1}{c}\,\left[d\,\mu_{\rm T}\,\left(\frac{a_{\rm T}\,\cos \alpha}{d} - 2\,\frac{\mu_{\rm T}\,v_{r}}{d}\right) - v_{r}\,{\mu_{\rm T}}^{2}\right] \nonumber \\
&= \frac{1}{c}\,\left[\mu_{\rm T}\,a_{\rm T}\,\cos \alpha - 3\,v_{r}\,{\mu_{\rm T}}^{2}\right] ~. 
\label{eq:tsbterm2}
\end{align}

We have used eq. (\ref{eq:tsbterm2}) in eq. (\ref{eq:fddotex1}) of section \ref{se:der1}.

\begin{figure*}
\begin{center}
\subfigure[]{\label{subfig:2DaTl}\includegraphics[width=0.25\textwidth]{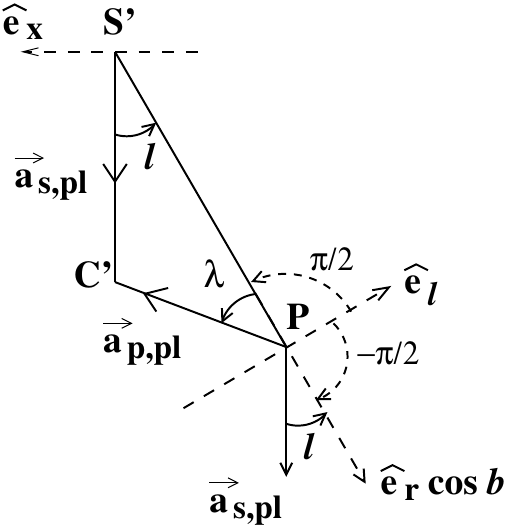}}
\hskip 1.cm \subfigure[]{\label{subfig:2DaTb}\includegraphics[width=0.45\textwidth]{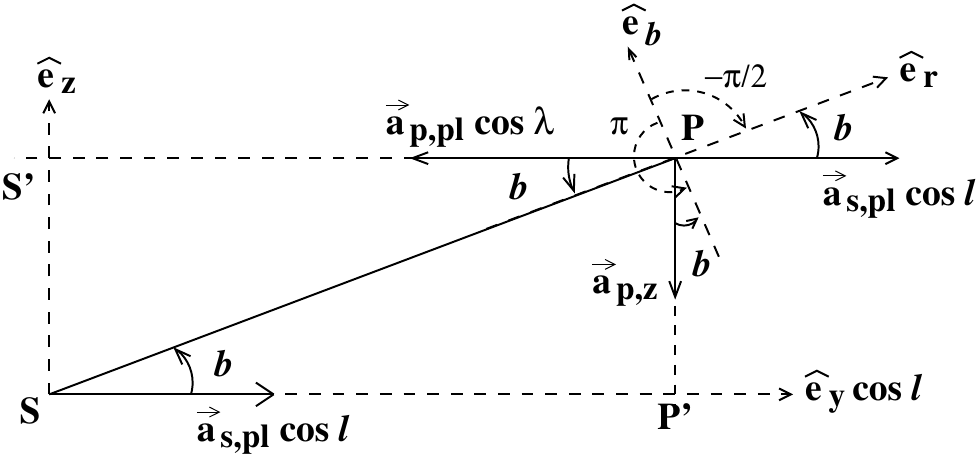}}
\end{center}
\caption{Schematic diagram showing components of Solar and pulsar accelerations, parallel and perpendicular to the Galactic plane. The subplots are:  a) Figure on the left shows the top view of the Galactic plane shown in the left panel. S' and C' are the projections of the Sun (S) and the Galactic centre (C) on the plane containing the Pulsar (P) and parallel to the Galactic plane. The components of accelerations along $\widehat{e}_{l}$ are examined at P. Angles are measured from $\widehat{e}_{l}$ in the anti-clockwise direction.  b) Figure on the right shows the plane containing the Sun (S), the Pulsar (P) and its projection on the Galactic plane (P') in the edge-on view of the Galactic plane. The components of accelerations along $\widehat{e}_{b}$ are examined at P. Angles are measured from $\widehat{e}_{b}$ in the anti-clockwise direction. }
\label{fig:aTb1}
\end{figure*}

Now, in order to calculate $\alpha = |\alpha_{a} - \alpha_{v}|$, we need to get $\alpha_{a} = \tan^{-1}\left(\frac{a_{{\rm T},b}}{a_{{\rm T},l}}\right)$, and  $\alpha_{v}= \tan^{-1}\left(\frac{\mu_{b}}{\mu_{l}}\right)$, which we can directly obtain using the observables $\mu_{l}$ and $\mu_{b}$. We can also obtain the expressions of $a_{{\rm T},l}$ and $a_{{\rm T},b}$ (as shown in figure \ref{fig:aTb1}), in terms of $a_{\rm p,pl}$, $a_{\rm s,pl}$, $a_{\rm p,z}$, and galactic coordinates as described below.

From figure \ref{subfig:2DaTl}, we can write expression for  $a_{{\rm T},l}$ as, 
\begin{align}
a_{{\rm T},l} &= a_{\rm T}\,\widehat{e}_{\rm T,a} \cdot \widehat{e}_{l} \nonumber \\
&= (\vec{a}_{\rm p} - \vec{a}_{\rm s} - a_{r}\,\widehat{e}_{r})\cdot \widehat{e}_{l} \hskip3.5cm {\rm [used~ eqn.~ (\ref{eq:asp2})]} \nonumber \\
&= (\vec{a}_{\rm p} - \vec{a}_{\rm s})\cdot \widehat{e}_{l}  \hskip4.5cm [{\rm As}~ \widehat{e}_{r}\perp \widehat{e}_{l}] \nonumber \\
&= (\vec{a}_{\rm p,pl} + \vec{a}_{\rm p,z} - \vec{a}_{\rm s,pl} - \vec{a}_{\rm s,z})\cdot \widehat{e}_{l}\nonumber \\
&= \vec{a}_{\rm p,pl}\cdot \widehat{e}_{l} + \vec{a}_{\rm p,z}\cdot \widehat{e}_{l} - \vec{a}_{\rm s,pl}\cdot \widehat{e}_{l} - \vec{a}_{\rm s,z}\cdot \widehat{e}_{l} \hskip1.5cm [{\rm As} ~ \widehat{e}_{z}\perp \widehat{e}_{l} {\text{ and }} \vec{a}_{\rm s,z}\approx 0]  \nonumber \\
&= a_{\rm p,pl}\,\cos (\pi/2 + \lambda)  - a_{\rm s,pl}\cos (-\pi/2-l)   \hskip1.3cm [\text{Refer figure \ref{subfig:2DaTl}} ]\nonumber \\
&= - (a_{\rm p,pl}\, \sin \lambda - a_{\rm s,pl}\, \sin l) .
\label{eq:aTl}
\end{align}

From figure \ref{subfig:2DaTb}, we can write expression for  $a_{{\rm T},b}$ as,

\begin{align}
a_{{\rm T},b} &= a_{\rm T}\,\widehat{e}_{\rm T,a} \cdot \widehat{e}_{b} \nonumber \\
&= (\vec{a}_{\rm p} - \vec{a}_{\rm s} - a_{r}\,\widehat{e}_{r})\cdot \widehat{e}_{b}\nonumber \\
&= (\vec{a}_{\rm p} - \vec{a}_{\rm s})\cdot \widehat{e}_{b}  \hskip1.5cm [{\rm As}~ \widehat{e}_{r}\perp \widehat{e}_{b}] \nonumber \\
&= (\vec{a}_{\rm p,pl} + \vec{a}_{\rm p,z} - \vec{a}_{\rm s,pl} - \vec{a}_{\rm s,z})\cdot \widehat{e}_{b}\nonumber \\
&= \vec{a}_{\rm p,pl}\cdot \widehat{e}_{b} + \vec{a}_{\rm p,z}\cdot \widehat{e}_{b} - \vec{a}_{\rm s,pl}\cdot \widehat{e}_{b} \hskip1.5cm [{\rm As}~ \vec{a}_{\rm s,z}\approx 0]  \nonumber \\
&= a_{\rm p,pl}\,\cos \lambda \cos (\pi/2 - b) + a_{\rm p,z}\,\cos (\pi - b) - a_{\rm s,pl}\,\cos l \cos (3\pi/2 -b) \hskip0.5cm [\text{Refer figure \ref{subfig:2DaTb}} ] \nonumber \\
&= - (a_{\rm p,z}\,\cos b - a_{\rm p,pl}\,\cos \lambda \, \sin b - a_{\rm s,pl}\,\cos l \, \sin b).
\label{eq:aTb}
\end{align}

Using the these expressions of $a_{{\rm T},l}$, and $a_{{\rm T},b}$, we calculate $\alpha_{a}$, and subsequently, $\alpha$. We then use $\cos \alpha$ in eq. (\ref{eq:fddotex4}).

\subsection{Simplifying the fourth term on the right hand side of eq. (\ref{eq:pddot4})}
\label{appendix:B3}

The fourth term on the right hand side of eq. (\ref{eq:pddot4}) is $2\left(\frac{\dot{f}}{f} \right)_{\rm ex}\left(\frac{\vec{a}_{\rm s}\cdot \widehat{n}_{\rm sp}}{c} + \frac{\vec{v}_{\rm s}\cdot\dot{\widehat{n}}_{\rm sp}}{c} \right)$. We focus on the part $\left(\vec{a}_{\rm s}\cdot \widehat{n}_{\rm sp} + \vec{v}_{\rm s}\cdot\dot{\widehat{n}}_{\rm sp} \right)$.

First, considering $\vec{a}_{\rm s}\cdot \widehat{n}_{\rm sp}$:

\begin{align}
\vec{a}_{\rm s}\cdot \widehat{n}_{\rm sp} &= \left(\vec{a}_{\rm s,pl} + \vec{a}_{\rm s,z}\right)\cdot \widehat{n}_{\rm sp} \nonumber \\
&= \left({a}_{\rm s,pl}\,\widehat{e}_{y} + {a}_{\rm s,z}\,\widehat{e}_{z}\right)\cdot \left(\cos b\,\widehat{n}_{sp'} + \sin b\,\widehat{e}_{z}\right)  \hskip1.3cm [\text{Refer figure \ref{subfig:3Dplot}} ] \nonumber\\
&= {a}_{\rm s,pl} \cos b\,\widehat{e}_{y}\cdot\widehat{n}_{sp'} +  {a}_{\rm s,z}\,\sin b \nonumber\\
&= \cos b\,\cos l\,{a}_{\rm s,pl} + {a}_{\rm s,z}\,\sin b ~. \hskip1.3cm [\text{As per figure \ref{subfig:3Dplot}, angle between $\widehat{n}_{sp'}$ and $\widehat{e}_{y}$ is $l$} ] 
\label{eq:asnsp}
\end{align} 
Where we have used the fact that at any instant, the acceleration of the Sun has components along the Y-axis and Z-axis.

Considering $\vec{v}_{\rm s}\cdot\dot{\widehat{n}}_{\rm sp}$:
\begin{align}
\vec{v}_{\rm s}\cdot\dot{\widehat{n}}_{\rm sp} &= (\vec{v}_{\rm s,pl} +\vec{v}_{\rm s,z})\cdot (\dot{\theta}\,\widehat{e}_{\theta} + \dot{\phi}\,\sin \theta\,\widehat{e}_{\phi}) \hskip1cm [{\rm Used ~eq. } ~ \ref{eq:erdot}; {\rm ~ and ~the ~fact ~that~} \widehat{n}_{\rm sp} = \widehat{e}_r; ` \dot{\widehat{n}}_{\rm sp} = \dot{\widehat{e}}_r ] \nonumber \\
&= ({v}_{\rm s,pl}\,\widehat{e}_{x} +{v}_{\rm s,z}\,\widehat{e}_{z})\cdot (\dot{\theta}\,\widehat{e}_{\theta} + \dot{\phi}\,\sin \theta\,\widehat{e}_{\phi})  \nonumber \\
&= ({v}_{\rm s,pl}\,\widehat{e}_{x} +{v}_{\rm s,z}\,\widehat{e}_{z})\cdot (-\mu_{b}\,\widehat{e}_{\theta} + \mu_{l}\,\widehat{e}_{\phi}) \hskip1cm [{\rm As}~ \dot{\theta} = -\dot{b}= -\mu_{b}, \dot{\phi}\,\sin \theta = \dot{l}\,\cos b = \mu_{l} ] \nonumber \\
&= ({v}_{\rm s,pl}\,\widehat{e}_{x} +{v}_{\rm s,z}\,\widehat{e}_{z})\cdot \{-\mu_{b}\,(\cos \theta\,\cos \phi\,\widehat{e}_{x} + \cos \theta\,\sin \phi\,\widehat{e}_{y} - \sin \theta\,\widehat{e}_{z}) + \mu_{l}\,(-\sin \phi\,\widehat{e}_{x} + \cos \phi\,\widehat{e}_{y})\}  \nonumber \\
&\hskip9cm [\text{Using eqs. (\ref{eq:etheta}) and (\ref{eq:ephi})}] \nonumber \\
&= {v}_{\rm s,pl}\,(-\mu_{b}\cos \theta\,\cos \phi\, -\mu_{l}\sin \phi) +{v}_{\rm s,z}\,(\mu_{b}\sin \theta)  \nonumber \\
&=  \mu_{b}\,{v}_{\rm s,pl}\,\sin b\,\sin l - \mu_{l}\,{v}_{\rm s,pl}\,\cos l + \mu_{b}\,{v}_{\rm s,z}\,\cos b ~. \hskip0.6cm [{\rm As}~ \sin \theta = \cos b,\cos \theta= \sin b,\sin \phi = \cos l, \cos \phi = - \sin l]
\label{eq:vsnsp}
\end{align} Where we have used the fact that at any instant, the velocity of the Sun can have components along the X-axis and Z-axis.

Therefore,

\begin{align}
\left(\frac{\vec{a}_{\rm s}\cdot \widehat{n}_{\rm sp}}{c} + \frac{\vec{v}_{\rm s}\cdot\dot{\widehat{n}}_{\rm sp}}{c} \right) &= \cos b\,\cos l\,\frac{{a}_{\rm s,pl}}{c} + \frac{{a}_{\rm s,z}}{c}\,\sin b + \mu_{b}\,\frac{{v}_{\rm s,pl}}{c}\,\sin b\,\sin l - \mu_{l}\,\frac{{v}_{\rm s,pl}}{c}\,\cos l + \mu_{b}\,\frac{{v}_{\rm s,z}}{c}\,\cos b ~. \nonumber \\
&= \cos b\,\cos l\,\frac{{a}_{\rm s,pl}}{c}  + \mu_{b}\,\frac{{v}_{\rm s,pl}}{c}\,\sin b\,\sin l - \mu_{l}\,\frac{{v}_{\rm s,pl}}{c}\,\cos l  ~.
\label{eq:fourthterm}
\end{align} 
Here, we have taken the Z component of the Solar velocity (${v}_{\rm s,z}$) and the Z component of the Solar acceleration (${a}_{\rm s,z}$) to be negligibly small (as discussed in \ref{appendix:unitvector}). We have used eq. (\ref{eq:fourthterm}) in eq. (\ref{eq:fddotex1}) of section \ref{se:der1}.

\subsection{Simplifying the first term on the right hand side of eq. (\ref{eq:pddot4})}
\label{appendix:B4}

Here we write $ \frac{\left(\dot{\vec{a}}_{\rm p}-\dot{\vec{a}}_{\rm s}\right)\cdot \widehat{n}_{\rm sp}}{c}$ of eq. (\ref{eq:pddot4}) in terms of measurable quantities. Defining $\vec{j}_{\rm p} = \dot{\vec{a}}_{\rm p}$ and $\vec{j}_{\rm s} = \dot{\vec{a}}_{\rm s}$, and assuming Galactic potential only as the source of these jerk terms, we write:

\begin{align}
\frac{\left(\vec{j}_{\rm p}-\vec{j}_{\rm s}\right)\cdot \widehat{n}_{\rm sp}}{c} &= \frac{1}{c}\,\left(\dot{\vec{a}}_{\rm p} -\dot{\vec{a}}_{\rm s}\right)\cdot \widehat{n}_{\rm sp}  \nonumber \\
&= \frac{1}{c}\,\left( \frac{d}{dt}(\vec{a}_{\rm p} -\vec{a}_{\rm s})\right)\cdot \widehat{e}_{r}\nonumber \\
&= \frac{1}{c}\,\left( \frac{d}{dt}(a_{r}\,\widehat{e}_{r} + a_{\rm T}\,\widehat{e}_{\rm T,a})\right)\cdot \widehat{e}_{r} \hskip1cm [{\text{Using eq. (\ref{eq:asp2})}}]\nonumber \\
%&= \frac{1}{c}\,\left( \frac{d}{dt}(a_{r}\,\widehat{e}_{r} + a_{\rm T}\,\cos \alpha\,\widehat{e}_{\rm T,v})\right)\cdot \widehat{e}_{r} \hskip1.5cm [\text{Note: Using eq. (\ref{eq:etaetv})}] \nonumber \\
&= \frac{1}{c}\,\left( \dot{a}_{r}\,\widehat{e}_{r} + a_{r}\,\dot{\widehat{e}}_{r} + \dot{a}_{\rm T}\,\widehat{e}_{\rm T,a} + a_{\rm T}\,\dot{\widehat{e}}_{\rm T,a}\right)\cdot \widehat{e}_{r} \nonumber \\
&= \frac{1}{c}\,\left( \dot{a}_{r}\,\widehat{e}_{r} + a_{r}\mu_T\,\widehat{e}_{\rm T,v} + \dot{a}_{\rm T}\,\widehat{e}_{\rm T,a}  + a_{\rm T}\,\dot{\widehat{e}}_{\rm T,a}\right)\cdot \widehat{e}_{r}  \hskip1cm [\because \dot{\widehat{e}}_{r} = \mu_T\,\widehat{e}_{\rm T,v} ]\nonumber \\
&= \frac{1}{c}\,\left( \dot{a}_{r} + a_{\rm T}\,\dot{\widehat{e}}_{\rm T,a}\cdot \widehat{e}_{r}\right) \hskip1cm [\because \widehat{e}_{\rm T,v} \cdot\widehat{e}_{r} = 0, \, \widehat{e}_{\rm T,a} \cdot\widehat{e}_{r} = 0 ] ~. 
\label{eq:jerk1}
\end{align} 

Let us focus on the term $\dot{\widehat{e}}_{\rm T,a}\cdot \widehat{e}_{r}$ in eq. (\ref{eq:jerk1}). Taking time derivative of eq. (\ref{eq:etransa}), we get,
\begin{align}
\dot{\widehat{e}}_{\rm T,a}\cdot \widehat{e}_{r} &= \left(\frac{\dot{v}_{\rm T}\,\widehat{e}_{\rm T,v} +  v_{\rm T}\,\dot{\widehat{e}}_{\rm T,v} - \dot{\vec{v}}_{\rm T,ex}}{v_{\rm T, gal}} - \frac{v_{\rm T}\,\widehat{e}_{\rm T,v} - \vec{v}_{\rm T,ex}}{{v_{\rm T, gal}}^{2}}\,\dot{v}_{\rm T, gal} \right)\cdot \widehat{e}_{r} \nonumber \\
&= \left(\frac{\dot{v}_{\rm T}\,\widehat{e}_{\rm T,v} +  v_{\rm T}\,\dot{\widehat{e}}_{\rm T,v} - \dot{\vec{v}}_{\rm T,ex}}{v_{\rm T, gal}} - \frac{\widehat{e}_{\rm T,a}}{v_{\rm T, gal}}\,\dot{v}_{\rm T, gal} \right)\cdot \widehat{e}_{r} \hskip1cm [{\text{Using eq. (\ref{eq:etransa})}}] \nonumber \\
&=\frac{(v_{\rm T}\,\dot{\widehat{e}}_{\rm T,v}\cdot \widehat{e}_{r}- \dot{\vec{v}}_{\rm T,ex}\cdot \widehat{e}_{r}) }{v_{\rm T, gal}} \hskip1cm [\because \widehat{e}_{\rm T,v} \cdot\widehat{e}_{r} = 0, \, \widehat{e}_{\rm T,a} \cdot\widehat{e}_{r} = 0] \nonumber \\
&=\frac{(-v_{\rm T}\,\mu_{\rm T}\widehat{e}_{r}\cdot \widehat{e}_{r}- \dot{\vec{v}}_{\rm T,ex}\cdot \widehat{e}_{r}) }{v_{\rm T, gal}} \hskip1cm [\because \dot{\widehat{e}}_{\rm T,v} = -\mu_{\rm T}\,\widehat{e}_{r}] \nonumber \\
&=\frac{(-v_{\rm T}\,\mu_{\rm T}- \dot{\vec{v}}_{\rm T,ex}\cdot \widehat{e}_{r}) }{v_{\rm T, gal}}  \nonumber \\
&=\frac{\{-(v_{\rm T,gal}\,\cos \alpha + \vec{v}_{\rm T,ex} \cdot\widehat{e}_{\rm T,v})\,\mu_{\rm T}- \dot{\vec{v}}_{\rm T,ex}\cdot \widehat{e}_{r}\} }{v_{\rm T, gal}}   \hskip1cm [{\text{Using eq. (\ref{eq:vtrans2})}}] \nonumber \\
&= -\cos \alpha\,\mu_{\rm T} - \frac{( \vec{v}_{\rm T,ex} \cdot (\mu_{\rm T}\widehat{e}_{\rm T,v}) + \dot{\vec{v}}_{\rm T,ex}\cdot \widehat{e}_{r}) }{v_{\rm T, gal}}   \nonumber \\
&= -\cos \alpha\,\mu_{\rm T} - \frac{( \vec{v}_{\rm T,ex} \cdot \dot{\widehat{e}}_{r} + \dot{\vec{v}}_{\rm T,ex}\cdot \widehat{e}_{r}) }{v_{\rm T, gal}}\hskip1cm [\because \dot{\widehat{e}}_{r} = \mu_{\rm T}\,\widehat{e}_{\rm T,v}]  \nonumber \\
&= -\cos \alpha\,\mu_{\rm T} - \frac{\frac{d}{dt}(\vec{v}_{\rm T,ex}\cdot\widehat{e}_{r}) }{v_{\rm T, gal}}  \nonumber \\
&= -\cos \alpha\,\mu_{\rm T} - \frac{\frac{d}{dt}((v_{\rm T} \,\widehat{e}_{\rm T,v} - v_{\rm T,gal}\,\widehat{e}_{\rm T,a})\cdot\widehat{e}_{r}) }{v_{\rm T, gal}}\hskip1cm [{\text{Using eq. (\ref{eq:vtrans})}}]  \nonumber \\
&= -\cos \alpha\,\mu_{\rm T} \hskip1cm [\because \widehat{e}_{\rm T,v} \cdot\widehat{e}_{r} = 0, \, \widehat{e}_{\rm T,a} \cdot\widehat{e}_{r} = 0] 
\label{eq:doteta}
\end{align}

Hence, on substituting expression for $\dot{\widehat{e}}_{\rm T,a}\cdot \widehat{e}_{r}$ in eq. (\ref{eq:jerk1}), we get
\begin{align}
\frac{\left(\vec{j}_{\rm p}-\vec{j}_{\rm s}\right)\cdot \widehat{n}_{\rm sp}}{c} = \frac{1}{c}\,\left( \dot{a}_{r} - a_{\rm T}\,\mu_{\rm T}\,\cos \alpha \right)
\label{eq:jerk2}
\end{align}

We need to now derive the expressions of $\dot{a}_{r}$ and $a_{\rm T}$ in terms of observable parameters like $l$, $b$, $\mu_l$, $\mu_b$, $d$, and $v_r$, as well as, calculable quantities like $a_{\rm p,pl}$, $a_{\rm s,pl}$, and $a_{\rm p,z}$. Method of estimating $\alpha$ is described in the Appendix B.2. 

For $a_{\rm T}$, using eq. (\ref{eq:asp2}), we can write,

\begin{align}
a_{\rm T} &=  (\vec{a}_{\rm sp}\cdot\vec{a}_{\rm sp} - a_{r}^{2})^{1/2} \nonumber \\
&= \left( (\vec{a}_{\rm p} - \vec{a}_{\rm s})\cdot(\vec{a}_{\rm p} - \vec{a}_{\rm s}) - ((\vec{a}_{\rm p} - \vec{a}_{\rm s})\cdot\widehat{e}_{r})^{2} \right)^{1/2}  \hskip2cm [{\rm As} ~ \widehat{e}_{\rm T,v} \cdot\widehat{e}_{r}= 0]\nonumber \\
&= \left((\vec{a}_{\rm p,pl} + \vec{a}_{\rm p,z} - \vec{a}_{\rm s,pl} - \vec{a}_{\rm s,z})\cdot(\vec{a}_{\rm p,pl} + \vec{a}_{\rm p,z} - \vec{a}_{\rm s,pl} - \vec{a}_{\rm s,z}) - ((\vec{a}_{\rm p} - \vec{a}_{\rm s})\cdot\widehat{n}_{\rm sp})^{2} \right)^{1/2}\nonumber \\
&= \left((\vec{a}_{\rm p,pl} - \vec{a}_{\rm s,pl} + \vec{a}_{\rm p,z} - \vec{a}_{\rm s,z})\cdot(\vec{a}_{\rm p,pl} - \vec{a}_{\rm s,pl} + \vec{a}_{\rm p,z} - \vec{a}_{\rm s,z}) - ((\vec{a}_{\rm p} - \vec{a}_{\rm s})\cdot\widehat{n}_{\rm sp})^{2} \right)^{1/2}\nonumber \\
&= \left((\vec{a}_{\rm p,pl} - \vec{a}_{\rm s,pl})\cdot(\vec{a}_{\rm p,pl} - \vec{a}_{\rm s,pl}) + (\vec{a}_{\rm p,z} - \vec{a}_{\rm s,z})\cdot(\vec{a}_{\rm p,z} - \vec{a}_{\rm s,z}) - c^2 \left[-\left( \frac{\dot{f}}{f} \right)_{\rm ex, Galpl} -\left( \frac{\dot{f}}{f} \right)_{\rm ex, Galz} \right]^{2} \right)^{1/2}\nonumber \\
&\hskip1cm [\text{From eqs. (3) and (4) of \cite{pb18} and using the fact $\left( \frac{\dot{f}}{f} \right) = -\left( \frac{\dot{P}}{P} \right)$, where P is period.}] \nonumber \\
&= \left( {a}_{\rm p,pl}^{2} + {a}_{\rm s,pl}^{2} - 2\,{a}_{\rm p,pl}\,{a}_{\rm s,pl}\,\cos(\pi-(l+\lambda)) + {a}_{\rm p,z}^{2} - c^2 \, \left[\left( \frac{\dot{f}}{f} \right)_{\rm ex, Galpl} + \left( \frac{\dot{f}}{f} \right)_{\rm ex, Galz} \right]^{2} \right)^{1/2} \nonumber \\ & \hskip1cm [{\rm As} ~ {a}_{\rm s,z} = 0  ~ {\rm and ~the ~ angle ~ between} ~ \vec{a}_{\rm p,pl} ~{\rm  and}~ \vec{a}_{\rm p,pl}{\rm ~ is} ~ \pi-(l+\lambda) ~{\rm as ~shown ~ in ~figure ~\ref{subfig:3Dplot}} ] \nonumber \\
&= \left( {a}_{\rm p,pl}^{2} + {a}_{\rm s,pl}^{2} + 2\,{a}_{\rm p,pl}\,{a}_{\rm s,pl}\,\cos(l+\lambda) + {a}_{\rm p,z}^{2} - c^2 \left[ \left( \frac{\dot{f}}{f} \right)_{\rm ex, Galpl} + \left( \frac{\dot{f}}{f} \right)_{\rm ex, Galz} \right]^{2} \right)^{1/2}~.
\label{eq:at1}
\end{align}

The non-observables in eq. (\ref{eq:at1}) like, $\left(\frac{\dot{f}}{f} \right)_{\rm ex, Galpl}$ and $\left( \frac{\dot{f}}{f} \right)_{\rm ex, Galz}$, and $\lambda$ can be calculated in terms of observable parameters. Our previously developed package GalDynPsr \citep{pb18} can be used to calculate $\left(\frac{\dot{f}}{f} \right)_{\rm ex, Galpl}$ and $\left( \frac{\dot{f}}{f} \right)_{\rm ex, Galz}$. In order to estimate $\lambda$, we can use sine law in $\triangle$ SP'C of figure \ref{fig:posppr3} to get,

\begin{equation}
\sin \lambda = \frac{R_{\rm s}\,\sin l}{R_{\rm p'}}~.
\label{eq:sinlaw}
\end{equation}

For $\dot{a}_{r}$, we consider the relation $a_{r} = \left(\vec{a}_{\rm p}-\vec{a}_{\rm s}\right)\cdot \widehat{n}_{\rm sp}$ and take its time-derivative, which gives,

\begin{align}
\dot{a}_{r} = &\frac{d}{dt} \left[\left(\vec{a}_{\rm p}-\vec{a}_{\rm s}\right)\cdot \widehat{n}_{\rm sp}\right] \nonumber \\
= & \frac{d}{dt} \left[\left(\vec{a}_{\rm p, pl} + \vec{a}_{\rm p, z} -\vec{a}_{\rm s, pl} - \vec{a}_{\rm s, z}\right)\cdot \widehat{n}_{\rm sp}\right] \nonumber \\
= & \frac{d}{dt} \left[\left(\vec{a}_{\rm p, pl} -\vec{a}_{\rm s, pl} \right) \cdot \widehat{n}_{\rm sp} + \vec{a}_{\rm p, z} \cdot \widehat{n}_{\rm sp}\right] \nonumber \\
=  &\frac{d}{dt} \left[ - \cos b \,\left({a}_{\rm p,pl}\,\cos \lambda + {a}_{\rm s,pl}\,\cos l  \right) - (a_{\rm p,z}\,\sin b)\right] \nonumber \\
& \hskip0.8cm \text{[{\rm From eqs. (5), (14) and (15) of~}\cite{pb18}]} \nonumber \\
= &- \frac{d}{dt} \left[ \cos b \left({a}_{\rm p,pl}\,\cos \lambda + {a}_{\rm s,pl}\,\cos l \right) + a_{\rm p,z}\,\sin b \right] \nonumber \\
=  &- \left[-\sin b\left({a}_{\rm p,pl}\,\cos \lambda + {a}_{\rm s,pl}\,\cos l \right)\,\dot{b} +  \cos b \left(\dot{a}_{\rm p,pl}\,\cos \lambda - a_{\rm p,pl}\,\sin \lambda\,\dot{\lambda} + \dot{a}_{\rm s,pl}\,\cos l - a_{\rm s,pl}\,\sin l\,\dot{l} \right) \right. \nonumber \\
 &\left. + \dot{a}_{\rm p,z}\,\sin b + a_{\rm p,z}\,\cos b\,\dot{b} \right]  \nonumber \\
= &\sin b\left({a}_{\rm p,pl}\,\cos \lambda + {a}_{\rm s,pl}\,\cos l \right)\,\mu_b -  \cos b \left(\dot{a}_{\rm p,pl}\,\cos \lambda + \dot{a}_{\rm s,pl}\,\cos l\right) - \dot{a}_{\rm p,z}\,\sin b + a_{\rm p,pl}\,\cos b \,\sin \lambda\,\dot{\lambda} \nonumber \\
& + a_{\rm s,pl}\,\sin l\,\mu_l - a_{\rm p,z}\,\cos b\,\mu_b~ \hskip1.4cm [{\rm As}~ \mu_{l} = \dot{l} \, \cos b; ~ \mu_b = \dot{b} ].  
\label{eq:ardot}
\end{align}

Here also, the non-observables in eq. (\ref{eq:ardot}) like, $\lambda$, $\dot{\lambda}$, $\dot{a}_{\rm p,pl}$, $\dot{a}_{\rm s,pl}$, and $\dot{a}_{\rm p,z}$ can be calculated in terms of observable parameters. For estimation of $\lambda$ refer eq. (\ref{eq:sinlaw}). For calculating $\dot{\lambda}$, we take time-derivative of eq. (\ref{eq:sinlaw}), to obtain,

\begin{align}
\cos \lambda \dot{\lambda} &= \frac{R_{\rm s} \cos l}{R_{\rm p'}}\,\dot{l} - \frac{R_{\rm s}\sin l}{R_{\rm p'}^{2}} \frac{dR_{\rm p'}}{dt} \nonumber \\
\implies \dot{\lambda} &= \frac{R_{\rm s} \cos l}{\cos \lambda \,R_{\rm p'}}\,\dot{l} - \frac{R_{\rm s}\sin l}{\cos \lambda \,R_{\rm p'}^{2}} \frac{dR_{\rm p'}}{dt} \nonumber \\
\implies \dot{\lambda} &= \frac{R_{\rm s} \cos l}{\cos \lambda \,R_{\rm p'}}\,\frac{\dot{l}\, \cos b}{\cos b} - \frac{R_{\rm s}\sin l}{\cos \lambda \,R_{\rm p'}^{2}} \frac{dR_{\rm p'}}{dt} \nonumber \\
\implies \dot{\lambda} &= \frac{R_{\rm s} \cos l}{\cos \lambda \,R_{\rm p'}}\,\frac{\mu_{l}}{\cos b} - \frac{R_{\rm s}\sin l}{\cos \lambda \,R_{\rm p'}^{2}} \frac{dR_{\rm p'}}{dt} ~.
\label{eq:lambdadot}
\end{align}

Here, we need to further obtain an expression of $\frac{dR_{\rm p'}}{dt}$ in terms of observable quantities. In order to do so, we first consider the expression of $R_{\rm p'}$, which can be obtained using the cosine law in $\triangle$ SP'C of figure \ref{fig:posppr3} as,

\begin{equation}
R_{\rm p'}^2 = R_{\rm s}^2 + (d \cos b)^2 - 2 R_{\rm s} (d \cos b) \cos l ~.
\label{eq:Rp1}
\end{equation}

Taking time derivative of eq. (\ref{eq:Rp1}), we get:
\begin{align}
\frac{dR_{\rm p'}}{dt} &= \frac{1}{R_{\rm p'}}\left(d\cos^{2}b \dot{d} - d^{2} \cos b \sin b \dot{b} - R_{\rm s}\cos b \cos l \dot{d} + R_{\rm s}d\sin b \cos l \dot{b} + R_{\rm s}d\cos b \sin l \dot{l}\right) \nonumber \\
&= \frac{1}{R_{\rm p'}}\left((d\cos^{2}b - R_{\rm s}\cos b \cos l)v_{r} + (R_{\rm s}d\sin b \cos l - d^{2} \cos b \sin b)\mu_{b} + R_{\rm s}d\sin l\, \mu_{l}\right) ~.
\label{eq:Rpdot}
\end{align}

The quantities that remain to be calculated in eq. (\ref{eq:ardot}) are $\dot{a}_{\rm p,pl}$, $\dot{a}_{\rm p,z}$, and $\dot{a}_{\rm s,pl}$. We write the gravitational potential of the Milky Way as $\Phi_{MW}(R,z)$ where $R$ is the Galactocentric radius and $z$ is the perpendicular distance from the Galactic plane. Then, $\vec{a}_{\rm p} = -\vec{\nabla}\Phi_{MW}(R_{\rm p'},z)$ and $\vec{a}_{\rm s} = -\vec{\nabla}\Phi_{MW}(R_{\rm s},z)$. So, 

\begin{align}
\dot{a}_{\rm p, pl} &= -\frac{d}{dt}\frac{\partial \Phi_{MW}(R,z)}{\partial R}\vert_{(R = R_{\rm p'})} \nonumber \\
&= -\left(\frac{\partial^{2}\Phi_{MW}(R,z)}{\partial R^{2}}\frac{dR}{dt} + \frac{\partial^{2}\Phi_{MW}(R,z)}{\partial z\partial R}\frac{dz}{dt} \right)\vert_{(R = R_{\rm p'})}~.
\label{eq:jppl}
\end{align}  
Similarly,
\begin{align}
\dot{a}_{\rm s,pl} &=  -\left(\frac{\partial^{2}\Phi_{MW}(R,z)}{\partial R^{2}} \frac{dR}{dt} + \frac{\partial^{2}\Phi_{MW}(R,z)}{\partial z\partial R} \frac{dz}{dt}\right)\vert_{(R = R_{\rm s})}~. 
\label{eq:jspl}
\end{align}

For the perpendicular component:
\begin{align}
\dot{a}_{\rm p,z}& = -\frac{d}{dt}\frac{\partial \Phi_{MW}(R,z)}{\partial z}\vert_{(R = R_{\rm p'})} \nonumber \\
&= -\left(\frac{\partial^{2}\Phi_{MW}(R,z)}{\partial R\partial z}\frac{dR}{dt} + \frac{\partial^{2}\Phi_{MW}(R,z)}{\partial z^{2}}\frac{dz}{dt} \right)\vert_{(R = R_{\rm p'})}~.
\label{eq:jpz}
\end{align}  

For above eqs. (\ref{eq:jppl}), (\ref{eq:jspl}), and (\ref{eq:jpz}), we can use eq. (\ref{eq:Rpdot}) for the expression of $\left(\frac{dR}{dt}\right)\vert_{(R = R_{\rm p'})}$, and for the expression of $\frac{dz}{dt}$, we use $z = d\sin b$, and take its time-derivative as follows,
\begin{align}
\frac{dz}{dt} &= \dot{d}\sin b + d\cos b \,\dot{b} = v_r \sin b + \mu_b d\cos b ~.
\label{eq:zdot}
\end{align}

One can evaluate the partial derivatives of $\Phi_{MW}(R,z)$ as appeared in eqs. (\ref{eq:jppl}), (\ref{eq:jspl}), and (\ref{eq:jpz}) in various way. One option is to use the publicly available package `galpy' \citep{bovy15} that provides models of the potential of the Galaxy as well as these derivatives. More specifically, functions `evaluateR2derivs', `evaluatez2derivs', and `evaluateRzderivs' evaluate $\frac{\partial^{2}\Phi_{MW}(R,z)}{\partial R^{2}}$, $\frac{\partial^{2}\Phi_{MW}(R,z)}{\partial z^{2}}$, and $\frac{\partial^{2}\Phi_{MW}(R,z)}{\partial R\partial z}$ respectively.

\section{Comparison with \citet{liu18}}
\label{appendix:compareLiu}
 
Here, we compare our results with that of \citet{liu18}. In the right hand side of eq. (1) of \citet{liu18}, they appear to have made the approximation, $\left(1- \frac{\vec{v}_{\rm p}\cdot\hat{ {r}}}{c}\right) \left(1- \frac{\vec{v}_{\rm s}\cdot\hat{ {r}}}{c}\right)^{-1} \approx \left(1- \frac{\vec{v}_{\rm p}\cdot\hat{ {r}}}{c}\right) \left(1+ \frac{\vec{v}_{\rm s}\cdot\hat{ {r}}}{c}\right) \approx \left(1- \frac{(\vec{v}_{\rm p}-\vec{v}_{\rm s})\cdot\hat{ {r}}}{c}\right) \approx \left(1-\frac{ \vec{\varv}\cdot\hat{ {r}}}{c}\right)$. Here they ignored higher power terms of $\frac{\vec{v}_{\rm s}\cdot\hat{ {r}}}{c}$ in the numerator and defined $\vec{\varv} = \vec{v}_{\rm p}-\vec{v}_{\rm s}$, where $\vec{v}_{\rm p}$ is the velocity of the pulsar and $\vec{v}_{\rm s}$ is the velocity of the Solar system barycentre ($\approx$ the Sun). Subsequently, \citet{liu18} gave the expression of $\dot{f}_{\rm obs}$ and $\ddot{f}_{\rm obs}$ in terms of dynamical parameters and intrinsic frequency derivatives. In our work however, we do not make such an approximation in the first step. We take first and second derivatives and express intrinsic frequency derivatives in terms of the measured frequency derivatives and then in the denominator we approximate $1 + \frac{\vec{v}_{\rm s} . \widehat{n}_{\rm sp}}{c} \approx 1$ and $1 + \frac{\vec{v}_{\rm p} . \widehat{n}_{\rm sp}}{c} \approx 1$.

We re-write our eq. (\ref{eq:fddotex1}) such that $\ddot{f}_{\rm obs}$ is expressed in terms of the dynamical terms and the intrinsic frequency derivatives to get,

\begin{align}
\left( \frac{\ddot{f}}{f}\right)_{\rm ex} = &-\left[ \frac{\left(\dot{\vec{a}}_{\rm p}-\dot{\vec{a}}_{\rm s}\right)\cdot \widehat{n}_{\rm sp}}{c} + 2\frac{\left(\vec{a}_{\rm p} - \vec{a}_{\rm s}\right)\cdot\dot{\widehat{n}}_{\rm sp}}{c} + \frac{(\vec{v}_{\rm p} - \vec{v}_{\rm s})\cdot\ddot{\widehat{n}}_{\rm sp}}{c} +2\left(\frac{\dot{f}}{f} \right)_{\rm ex}\left(\frac{\vec{a}_{\rm s}\cdot \widehat{n}_{\rm sp}}{c} + \frac{\vec{v}_{\rm s}\cdot\dot{\widehat{n}}_{\rm sp}}{c} \right)\right. \nonumber \\ 
&\left.-2\left(\frac{\dot{f}}{f}\right)_{\rm ex}\left(\frac{\dot{f}}{f}\right)_{\rm obs}\right] \nonumber \\
=&-\left[ \frac{\left(\dot{\vec{a}}_{\rm p}-\dot{\vec{a}}_{\rm s}\right)\cdot \widehat{n}_{\rm sp}}{c} + 2\frac{\left(\vec{a}_{\rm p} - \vec{a}_{\rm s}\right)\cdot\dot{\widehat{n}}_{\rm sp}}{c} + \frac{(\vec{v}_{\rm p} - \vec{v}_{\rm s})\cdot\ddot{\widehat{n}}_{\rm sp}}{c} +2\left(\frac{\dot{f}}{f} \right)_{\rm ex}\left(\frac{\vec{a}_{\rm s}\cdot \widehat{n}_{\rm sp}}{c} + \frac{\vec{v}_{\rm s}\cdot\dot{\widehat{n}}_{\rm sp}}{c} \right)\right. \nonumber \\ 
&\left.-2\left(\frac{\dot{f}}{f}\right)_{\rm ex}\left(\left(\frac{\dot{f}}{f}\right)_{\rm int} + \left(\frac{\dot{f}}{f}\right)_{\rm ex}\right)\right]  \hskip1cm [\text{Used eq. (\ref{eq:doppler2bn})}]\nonumber \\
= &-\frac{\left(\dot{\vec{a}}_{\rm p}-\dot{\vec{a}}_{\rm s}\right)\cdot \widehat{n}_{\rm sp}}{c} - 2\frac{\left(\vec{a}_{\rm p} - \vec{a}_{\rm s}\right)\cdot\dot{\widehat{n}}_{\rm sp}}{c} - \frac{(\vec{v}_{\rm p} - \vec{v}_{\rm s})\cdot\ddot{\widehat{n}}_{\rm sp}}{c} - 2\left(\frac{\dot{f}}{f} \right)_{\rm ex}\left(\frac{\vec{a}_{\rm s}\cdot \widehat{n}_{\rm sp}}{c} + \frac{\vec{v}_{\rm s}\cdot\dot{\widehat{n}}_{\rm sp}}{c} \right) \nonumber \\ 
&+2\left(-\frac{ (\vec{a}_{\rm p} - \vec{a}_{\rm s}) \cdot \widehat{n}_{\rm sp}}{c} - \frac{1}{c} (\vec{v}_{\rm p} - \vec{v}_{\rm s}) \cdot \dot{\widehat{n}}_{\rm sp}\right) \left(\frac{\dot{f}}{f}\right)_{\rm int} + 2\left(\frac{\dot{f}}{f}\right)_{\rm ex}^2  \hskip1cm [\text{Used eq. (\ref{eq:doppler2bn})}]\nonumber \\
= &-\frac{\left(\dot{\vec{a}}_{\rm p}-\dot{\vec{a}}_{\rm s}\right)\cdot \widehat{e}_{r}}{c} - 2\frac{\left(a_{r}\,\widehat{e}_{r} + a_{\rm T}\,\widehat{e}_{\rm T,a}\right)\cdot\dot{\widehat{e}}_{r}}{c} - \frac{(\vec{v}_{\rm p} - \vec{v}_{\rm s})\cdot\ddot{\widehat{e}}_{r}}{c} - 2\left(\frac{\dot{f}}{f} \right)_{\rm ex}\left(\frac{\vec{a}_{\rm s}\cdot \widehat{e}_{r}}{c} + \frac{\vec{v}_{\rm s}\cdot\dot{\widehat{e}}_{r}}{c} \right) \nonumber \\ 
&+2\left(-\frac{ (a_{r}\,\widehat{e}_{r} + a_{\rm T}\,\widehat{e}_{\rm T,a})\cdot \widehat{e}_{r}}{c} - \frac{1}{c} (v_{r}\,\widehat{e}_{r} + v_{\rm T}\,\widehat{e}_{\rm T,v}) \cdot \dot{\widehat{e}}_{r}\right) \left(\frac{\dot{f}}{f}\right)_{\rm int} + 2\left(\frac{\dot{f}}{f}\right)_{\rm ex}^2  \hskip0.4cm [\text{Used eqs. (\ref{eq:vsp2}), (\ref{eq:asp2}) and $ \widehat{n}_{\rm sp}= \widehat{e}_{r}$}]\nonumber \\
= &-\frac{j_{r}}{c} - 2\frac{\left(a_{r}\,\widehat{e}_{r} + a_{\rm T}\,\widehat{e}_{\rm T,a}\right)\cdot \mu_{\rm T}\widehat{e}_{\rm T,v}}{c} - \frac{(\vec{v}_{\rm p} - \vec{v}_{\rm s})\cdot\ddot{\widehat{e}}_{r}}{c} - 2\left(\frac{\dot{f}}{f} \right)_{\rm ex}\left(\frac{\vec{a}_{\rm s}\cdot \widehat{e}_{r}}{c} + \frac{\vec{v}_{\rm s}\cdot\dot{\widehat{e}}_{r}}{c} \right) \nonumber \\ 
&+2\left(-\frac{a_{r}}{c} - \frac{1}{c} (v_{r}\,\widehat{e}_{r} + v_{\rm T}\,\widehat{e}_{\rm T,v}) \cdot \mu_{\rm T}\widehat{e}_{\rm T,v}\right) \left(\frac{\dot{f}}{f}\right)_{\rm int} + 2\left(\frac{\dot{f}}{f}\right)_{\rm ex}^2  \hskip1cm [{\rm As} ~  \dot{\widehat{e}}_{r} = \mu_{\rm T}\widehat{e}_{\rm T,v}, ~ \widehat{e}_{r}\cdot \widehat{e}_{\rm T,a}=0 ] \nonumber 
\end{align}

Thus,
\begin{align}
\left( \frac{\ddot{f}}{f}\right)_{\rm ex} = &-\frac{j_{r}}{c} - 2\frac{a_{\rm T}\,\mu_{\rm T} \widehat{e}_{\rm T,a}\cdot \widehat{e}_{\rm T,v}}{c} - \frac{(\vec{v}_{\rm p} - \vec{v}_{\rm s})\cdot\ddot{\widehat{e}}_{r}}{c} - 2\left(\frac{\dot{f}}{f} \right)_{\rm ex}\left(\frac{\vec{a}_{\rm s}\cdot \widehat{e}_{r}}{c} + \frac{\vec{v}_{\rm s}\cdot\dot{\widehat{e}}_{r}}{c} \right) \nonumber \\ 
&+2\left(-\frac{a_{r}}{c} - \frac{1}{c} (v_{\rm T}\,\mu_{\rm T})\right) \left(\frac{\dot{f}}{f}\right)_{\rm int} + 2\left(\frac{\dot{f}}{f}\right)_{\rm ex}^2  \hskip1cm [{\rm As} ~ \widehat{e}_{r}\cdot \widehat{e}_{\rm T,a} =0, ~\widehat{e}_{r}\cdot \widehat{e}_{\rm T,v} =0 ] \nonumber \\
= &-\frac{j_{r}}{c} - 2\frac{a_{\rm T}\,v_{\rm T} \cos \alpha}{dc} - \frac{(\vec{v}_{\rm p} - \vec{v}_{\rm s})\cdot\ddot{\widehat{e}}_{r}}{c} - 2\left(\frac{\dot{f}}{f} \right)_{\rm ex}\left(\frac{\vec{a}_{\rm s}\cdot \widehat{e}_{r}}{c} + \frac{\vec{v}_{\rm s}\cdot\dot{\widehat{e}}_{r}}{c} \right) \nonumber \\ 
&+2\left(-\frac{a_{r}}{c} - \frac{v_{\rm T}^2}{dc}\right) \left(\frac{\dot{f}}{f}\right)_{\rm int} + 2\left(\frac{\dot{f}}{f}\right)_{\rm ex}^2  \hskip1cm [{\rm As} ~ \widehat{e}_{r}\cdot \widehat{e}_{\rm T,a} =0, ~ \widehat{e}_{r}\cdot \widehat{e}_{\rm T,v} = 0, ~ \widehat{e}_{\rm T,a}\cdot \widehat{e}_{\rm T,v}=\cos \alpha,  ~ \mu_{\rm T} = v_{\rm T}/d ] \nonumber \\
= &-\frac{j_{r}}{c} - 2\frac{a_{\rm T}\,v_{\rm T} \cos \alpha}{dc} - \frac{\left(\mu_{\rm T}\,a_{\rm T}\,\cos \alpha - 3\,v_{r}\,{\mu_{\rm T}}^{2}\right)}{c} - 2\left(\frac{\dot{f}}{f} \right)_{\rm ex}\left(\frac{\vec{a}_{\rm s}\cdot \widehat{e}_{r}}{c} + \frac{\vec{v}_{\rm s}\cdot\dot{\widehat{e}}_{r}}{c} \right) \nonumber \\ 
&+2\left(-\frac{a_{r}}{c} - \frac{v_{\rm T}^2}{dc}\right) \left(\frac{\dot{f}}{f}\right)_{\rm int} + 2\left(\frac{\dot{f}}{f}\right)_{\rm ex}^2  \hskip1cm [\text{Used eq. (\ref{eq:tsbterm2})}] \nonumber \\
= &-\frac{j_{r}}{c} - 3\frac{\vec{v}_{\rm T}\cdot\vec{a}_{\rm T}}{dc} + \frac{3\,v_{r}\,v_{\rm T}^{2}}{d^{2}c} - 2\frac{a_{r}}{c}\left(\frac{\dot{f}}{f}\right)_{\rm int} -2\frac{v_{\rm T}^2}{dc}\left(\frac{\dot{f}}{f}\right)_{\rm int} + 2\left(\frac{\dot{f}}{f}\right)_{\rm ex}^2  - 2\left(\frac{\dot{f}}{f} \right)_{\rm ex}\left(\frac{\vec{a}_{\rm s}\cdot \widehat{e}_{r}}{c} + \frac{\vec{v}_{\rm s}\cdot\dot{\widehat{e}}_{r}}{c} \right) ~,
\label{eq:fddotex1comp}
\end{align}  where $\frac{j_{r}}{c}$ refers to $\frac{\left(\vec{j}_{\rm p}-\vec{j}_{\rm s}\right)\cdot \widehat{n}_{\rm sp}}{c}=\frac{\left(\dot{\vec{a}}_{\rm p}-\dot{\vec{a}}_{\rm s}\right)\cdot \widehat{e}_{r}}{c}$.

Now, the excess term from eq. (6) of \citet{liu18}, in terms of notations used by us, can be defined as,
\begin{equation}
\begin{split}
\label{eq:f2exliu}
\left(\frac{\ddot{f}}{f}\right)_{\rm ex, Liu2018} &= -2\frac{v^2_{T}}{dc} \left(\frac{\dot{f}}{f} \right)_{\rm int} -2 \left( \frac{a_{r}}{c}\frac{\dot{f}}{f} \right)_{\rm int}+\frac{3v_{r} v^{2}_{T}}{d^2 c}-\frac{3 \vec{v}_{T}\cdot \vec{a}_{T}}{dc} -\frac{j_{r}}{c} \\
 &= \left(\frac{\ddot{f}}{f}\right)_{\rm shk} + \left(\frac{\ddot{f}}{f}\right)_{\rm acc}+\left(\frac{\ddot{f}}{f}\right)_{r} + \left(\frac{\ddot{f}}{f}\right)_{T} + \left(\frac{\ddot{f}}{f}\right)_{\rm jerk} ~,
 \end{split}
 \end{equation}
 On comparing eqs. (\ref{eq:fddotex1comp}) and (\ref{eq:f2exliu}), we see that the terms in in our excess term exactly match with those obtained by \citet{liu18}, except that we also obtain an additional term, $2\left(\frac{\dot{f}}{f}\right)_{\rm ex}^2  - 2\left(\frac{\dot{f}}{f} \right)_{\rm ex}\left(\frac{\vec{a}_{\rm s}\cdot \widehat{e}_{r}}{c} + \frac{\vec{v}_{\rm s}\cdot\dot{\widehat{e}}_{r}}{c} \right)$. This is a consequence of the fact that we did not assume $1 + \frac{\vec{v}_{\rm s} . \widehat{n}_{\rm sp}}{c} \approx 1$ in the denominator in the first step before taking the time-derivatives.

Now, table 1 of \citet{liu18} provides various measured as well as calculated parameters for 62 pulsars. For the observable parameters like Galactic coordinates, proper motion, spin frequency, and distance, \citet{liu18} used measured values. They also used measured values of $v_{r}$ for 5 pulsars (table 2 of \citet{liu18}), while for the rest they used a nominal value of 50 km/s. However, for the parameters like position, velocity, acceleration, and jerk vectors, they used galpy orbit integrator along with a polynomial fitting. We compare the values of the calculated parameters given in table 1 of \citet{liu18} with our calculations based on our formalism. The parameters like $\frac{v^2_{T}}{dc}$ and $\frac{3v_{r} v^{2}_{T}}{d^2 c}$ are directly calculated using observed parameters, and those match with our values as expected.

\begin{figure*}
\begin{center}
\hskip-0.5cm\subfigure[]{\label{subfig:fdotint}\includegraphics[width=0.344\textwidth, angle=-90]{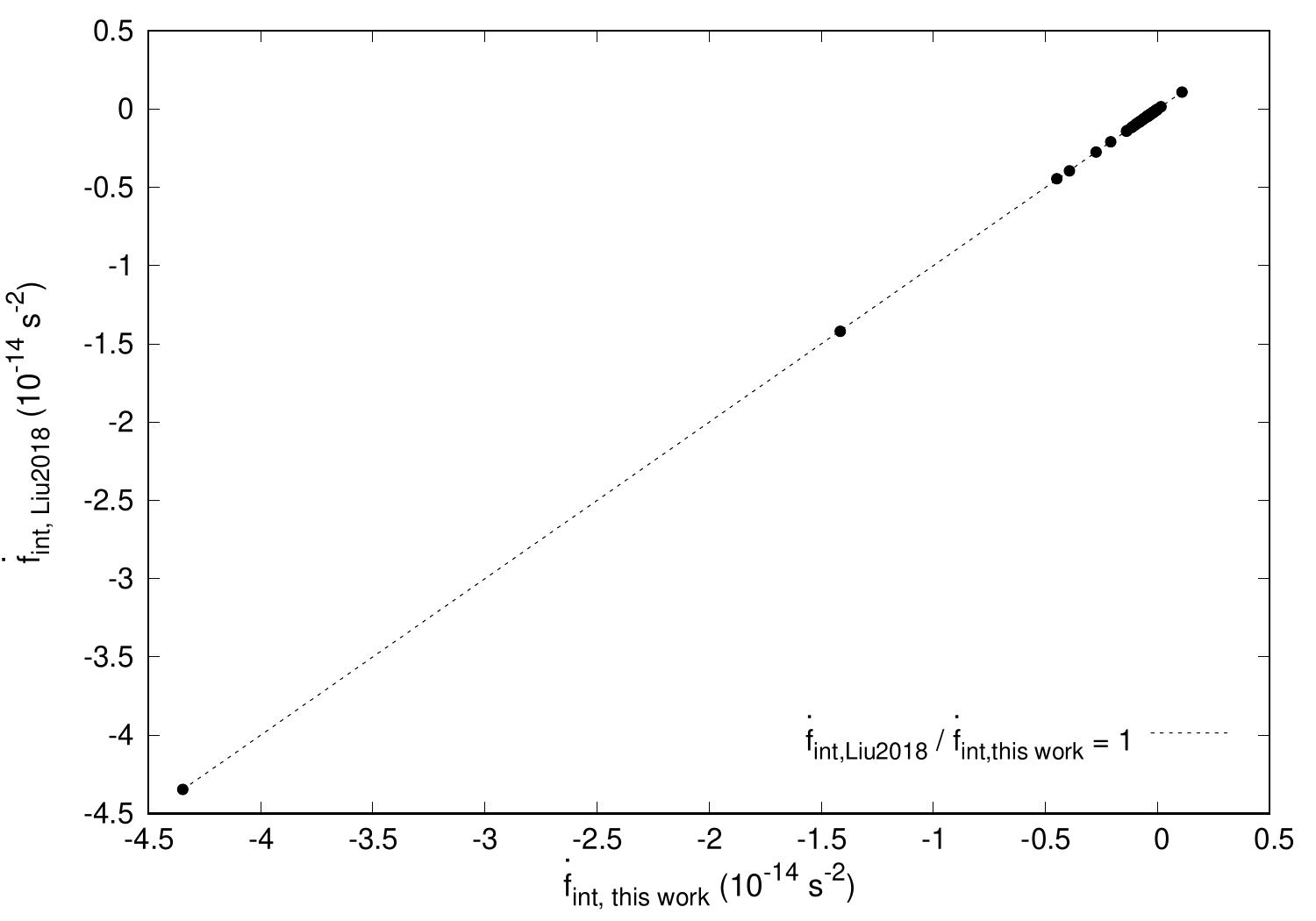}}
 \hskip0.5cm\subfigure[]{\label{subfig:arbyc}\includegraphics[width=0.344\textwidth, angle=-90]{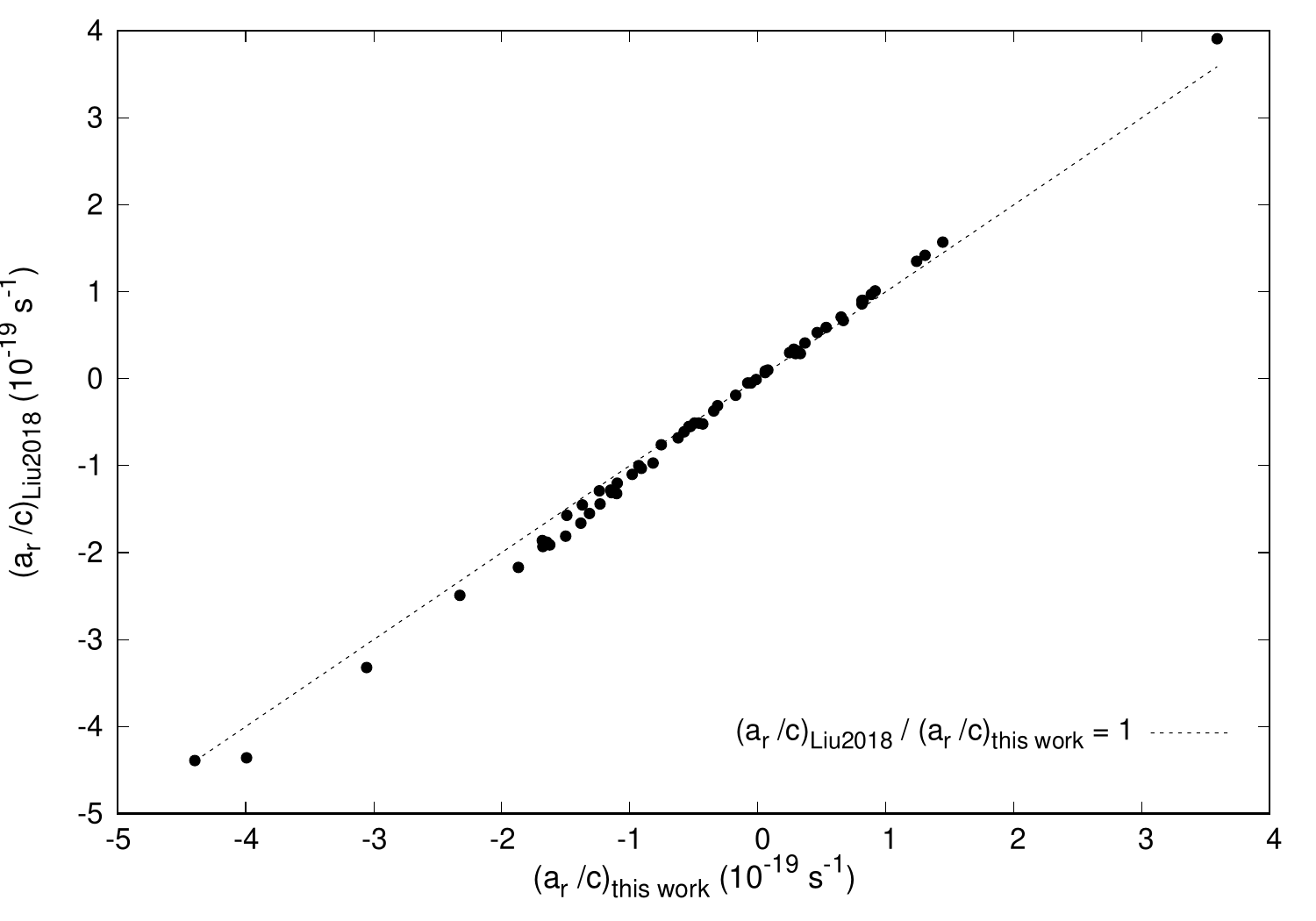}}\\
\hskip-0.5cm\subfigure[]{\label{subfig:vperaperx3byrc}\includegraphics[width=0.344\textwidth, angle=-90]{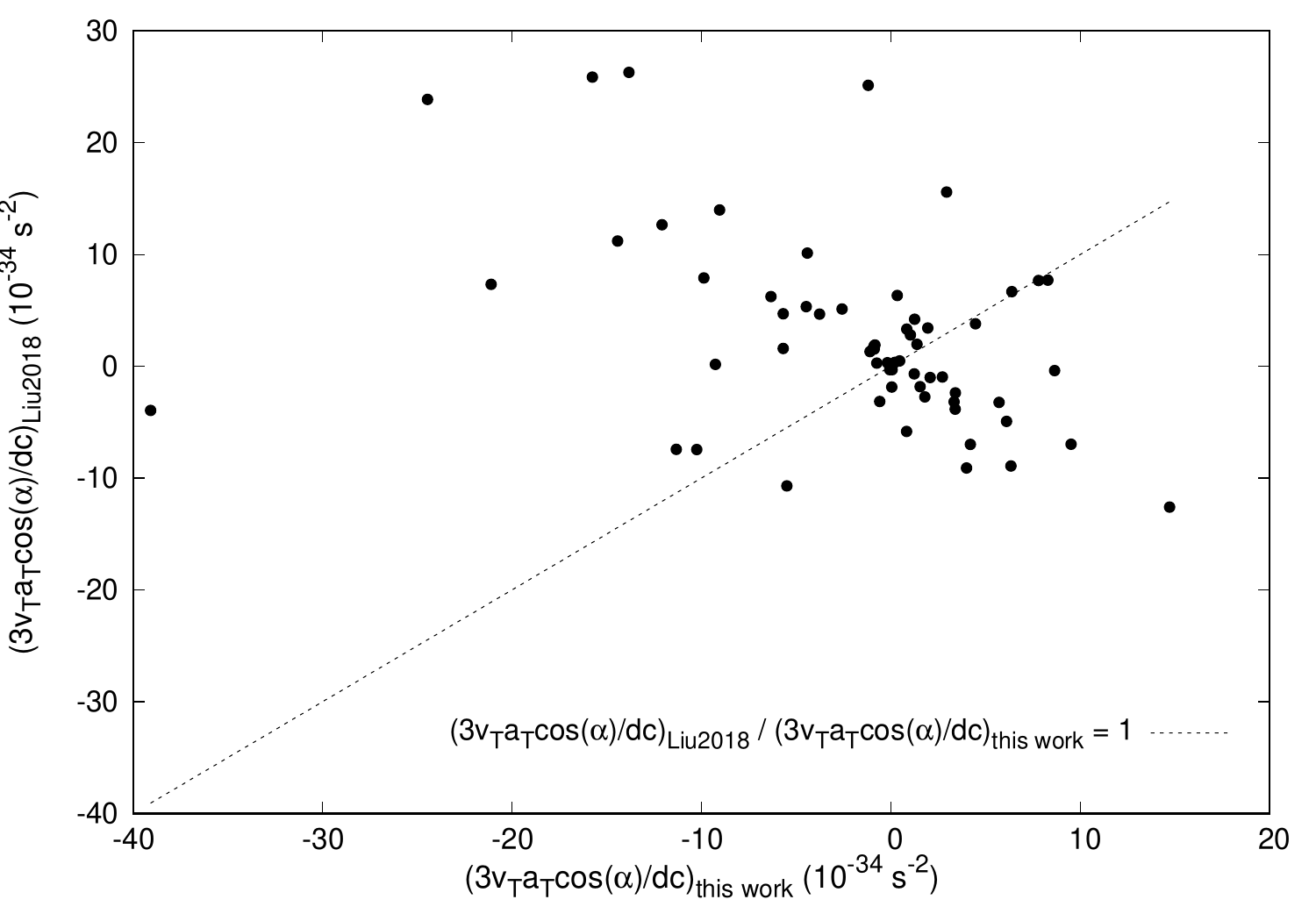}}
\hskip0.5cm\subfigure[]{\label{subfig:fddotfexNOJERK}\includegraphics[width=0.344\textwidth, angle=-90]{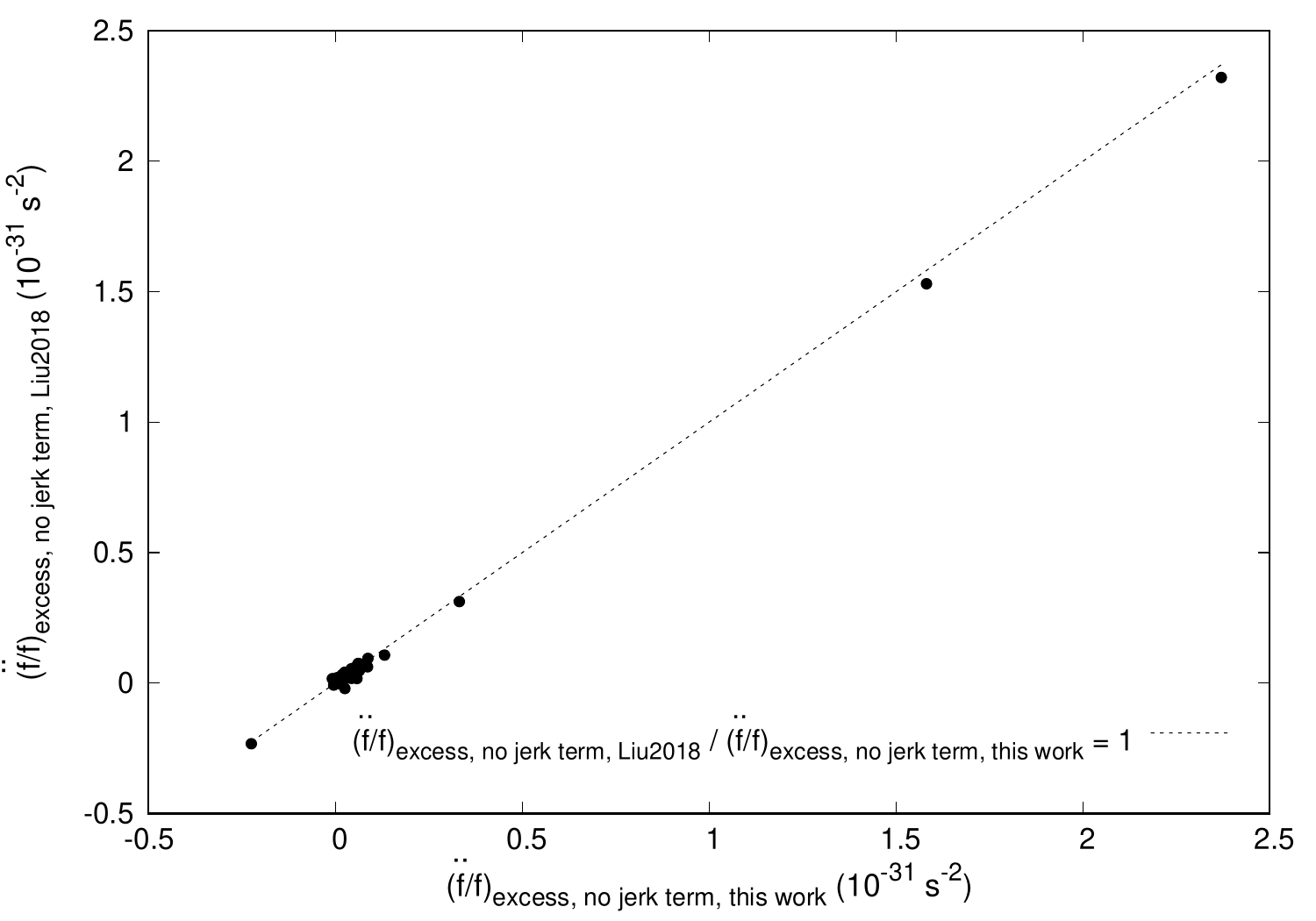}}\\
\hskip-0.5cm\subfigure[]{\label{subfig:jerk}\includegraphics[width=0.344\textwidth, angle=-90]{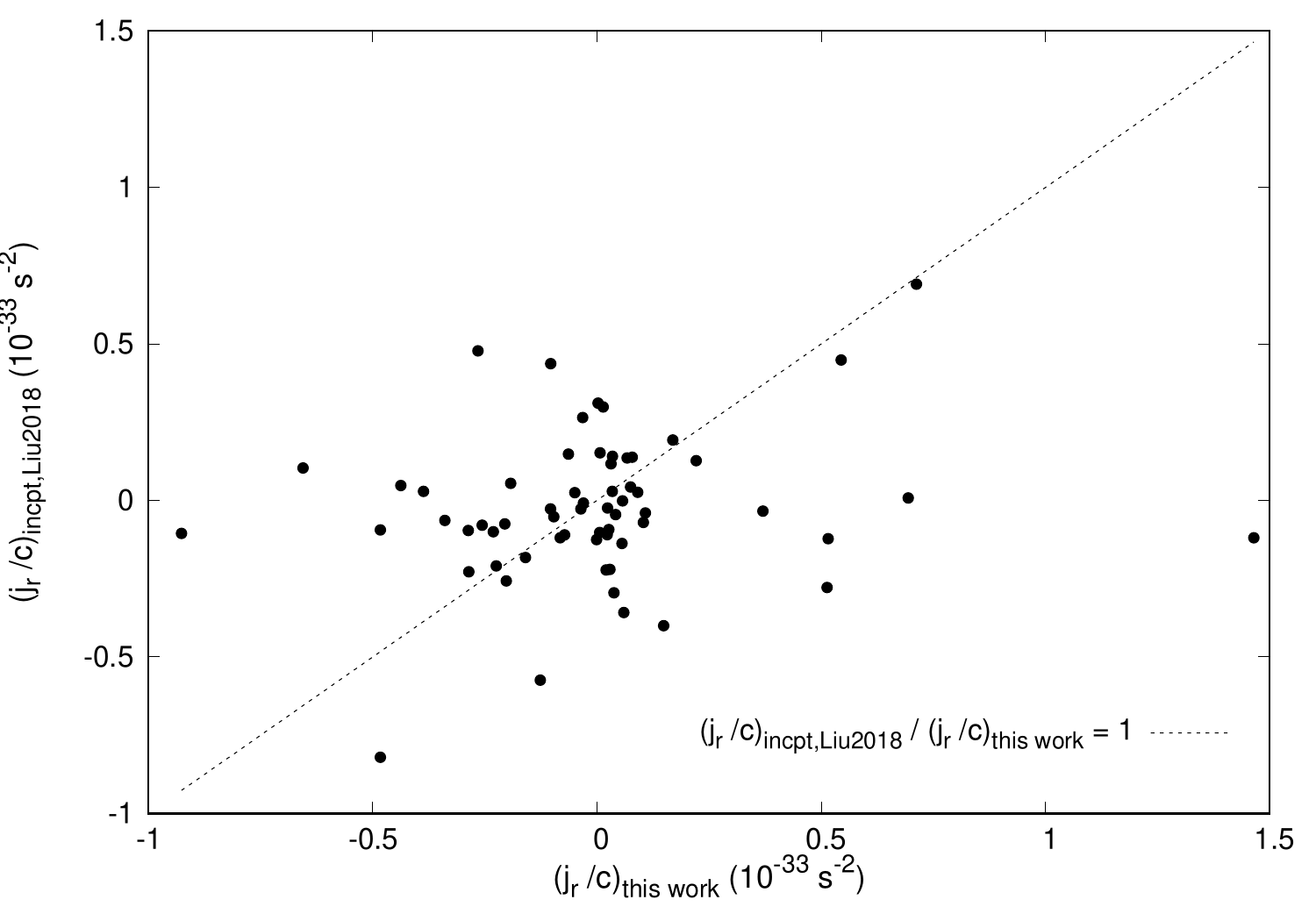}}
\hskip0.5cm\subfigure[]{\label{subfig:fddotfex}\includegraphics[width=0.344\textwidth, angle=-90]{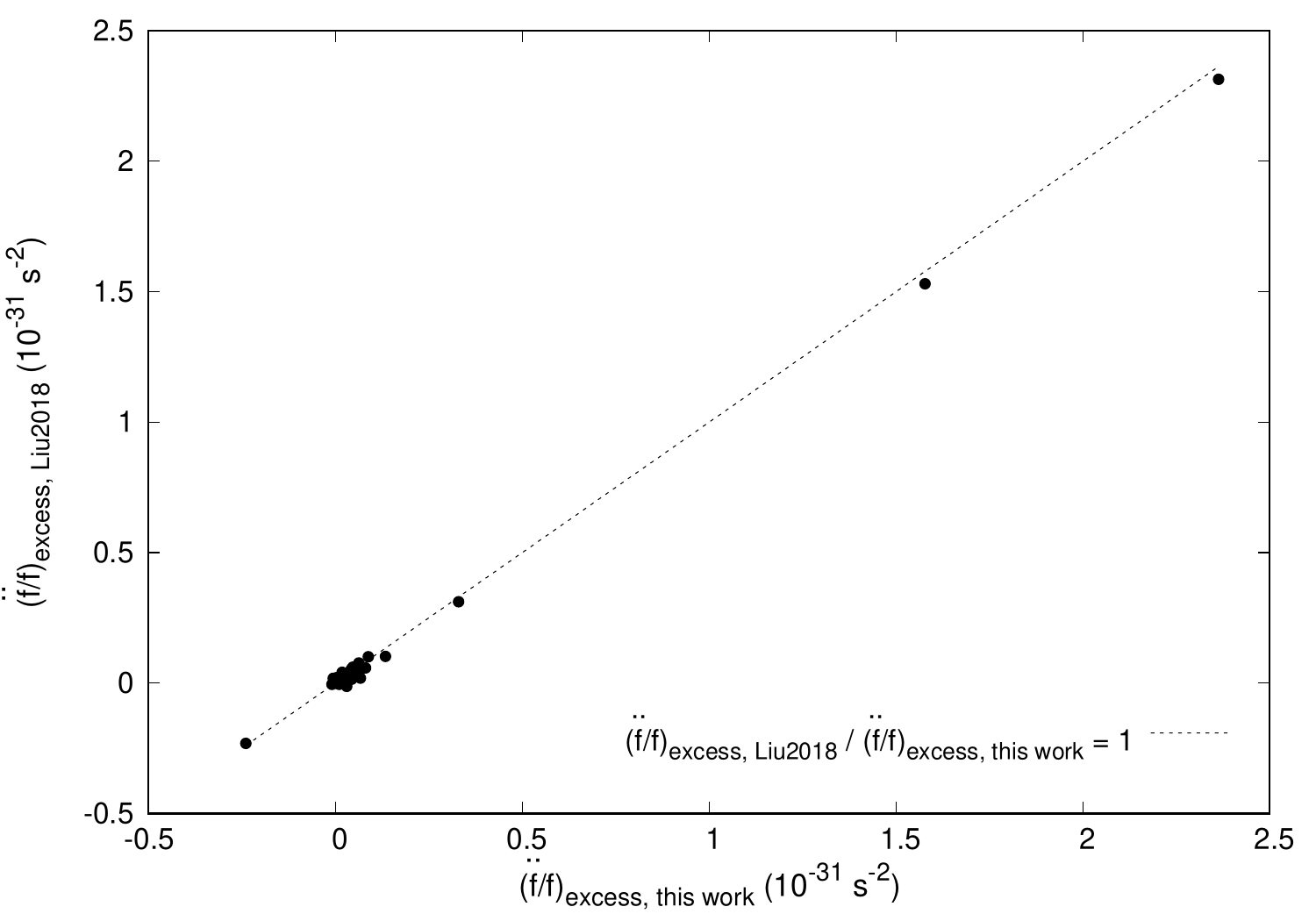}}\\
\end{center}
\caption{Scatter-plots comparing various terms calculated using our formalism and those calculated in \citet{liu18} for all the 62 pulsars given in table 1 of \citet{liu18}. The subplots are as follow: (a) $\dot{f}_{\rm int}$, (b) $a_{r}/c$, (c) $\frac{3 \vec{v}_{T}\cdot \vec{a}_{T}}{dc}$, d) excess term without the jerk term: $\left(\frac{\ddot{f}}{f}\right)_{\rm excess,\, no\, jerk\, term}$, (e) $\frac{j_{r}}{c}$ (using their values under the column ``$\frac{j_\parallel}{c}$ incpt'' of table 1 of \citet{liu18}), and (f) the complete excess term: $\left(\frac{\ddot{f}}{f}\right)_{\rm excess}$.}
\label{fig:liucomp}
\end{figure*}

We compare the values of $\dot{f}_{\rm int}$ and $\frac{a_{r}}{c}$ provided under columns ``$\dot{f}_{0}$'' and ``$\frac{a_\parallel}{c}$'' of table 1 of \citet{liu18}, respectively, with the values calculated by us using GalDynPsr. For the $\frac{j_{r}}{c}$ values, we compare the values provided under the column ``$\frac{j_\parallel}{c}$ incpt'' of table 1 of \citet{liu18} with our first square bracket term values. For $\frac{3 \vec{v}_{T}\cdot \vec{a}_{T}}{dc}$, we compare their values with $\frac{3 {v}_{T}{a}_{T}\cos \alpha}{dc}$ calculated by us. We also compare our excess term values ($\left(\frac{\ddot{f}}{f}\right)_{\rm ex}$ from eq. (\ref{eq:fddotex4})) with their excess term values ($\left(\frac{\ddot{f}}{f}\right)_{\rm ex, Liu2018}$ from eq. (\ref{eq:f2exliu})). We also compare the excess terms without the jerk term, i.e. $\left(\frac{\ddot{f}}{f}\right)_{\rm ex, Liu2018}$ without $\frac{j_{r}}{c}$ versus $\left(\frac{\ddot{f}}{f}\right)_{\rm ex}$ without the first square bracket term. We provide these comparisons as scatter-plots in figure \ref{fig:liucomp}. The idea is that, if the values match, they should lie along $y=x$ line.

We see that, from figure \ref{fig:liucomp}, for $\dot{f}_{\rm int}$ and $\frac{a_{r}}{c}$, our values match fairly closely with those calculated by \citet{liu18}. We also see that there is an absence of such a close match as far as the scatter plots for the values of $\frac{3 \vec{v}_{T}\cdot \vec{a}_{T}}{dc}$ and $\frac{j_{r}}{c}$ are concerned. This might be the result of the difference in method of calculating these terms, as we calculate using analytical expressions whereas \citet{liu18} resort to numerical fitting. However, in-spite of such discrepancies, we do get a usually get close match for the excess terms (both, excess terms without the jerk term, as well as, the complete excess term). We find that the minimum, median, and maximum of the absolute value of the term$ \left. \left[ \left(\frac{\ddot{f}}{f}\right)_{\rm ex} - \left(\frac{\ddot{f}}{f}\right)_{\rm ex, Liu2018} \right] \, \middle/ \, \left(\frac{\ddot{f}}{f}\right)_{\rm ex} \right. $  are 0.014, 0.507, and 43.059 respectively.

\newpage
\section{Details of simulating pulsar parameters based on observed values}
\label{appendix:DistriECDF}

For millisecond pulsars (MSPs), we needed to generate synthetic distributions of parameters from the observed distribution (as obtained from the reported values of the ATNF catalogue). The number of pulsars with the measured values of a particular parameter is much less than the number of synthetic values we wanted. To ensure that the parameter for the synthetic and the real pulsars follow the same distribution, we first fitted an Empirical Cumulative Distribution Function (ECDF) to the real distribution and then used the inverse CDF technique to generate the intended number of synthetic values following the same ECDF. The relevant parameters are Galactic longitude ($l$), Galactic lattitude ($b$), distance ($d$), proper motion in $l$ ($\mu_l$), proper motion in $b$ ($\mu_b$), spin frequency ($f_s$) and its first ($\dot{f}_s$) and second ($\ddot{f}_s$) time derivatives, as well as orbital frequency ($f_b$) and its first ($\dot{f}_b$) and second ($\ddot{f}_b$) time derivatives. Note that, since binary parameters are not reported for all MSPs, we did the above procedures for two sets of real MSPs. The first set consisted of the MSPs for which all relevant spin parameters are known and the second set consisted of the MSPs for which all relevant orbital parameters are known. We performed this task using the in-built functions in the `R' statistical package. 

\begin{figure}
\begin{center}
\subfigure[distribution of $l$ (real)]{\label{subfig:nogcatnflmspa}\includegraphics[width=0.17\textwidth, angle=-90]{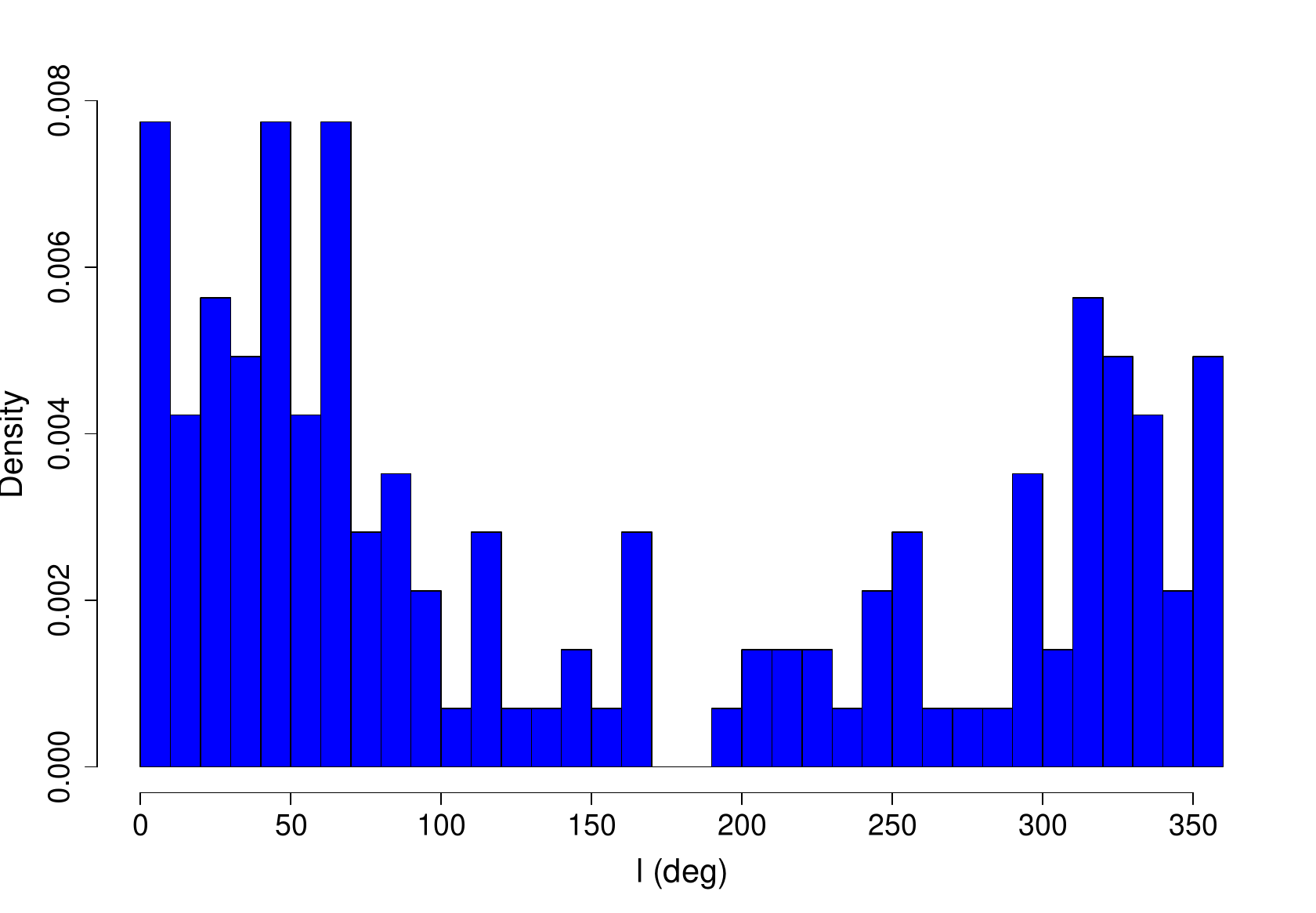}}
\subfigure[distribution of $l$ (synthetic)]{\label{subfig:nogcranlmspb}\includegraphics[width=0.17\textwidth, angle=-90]{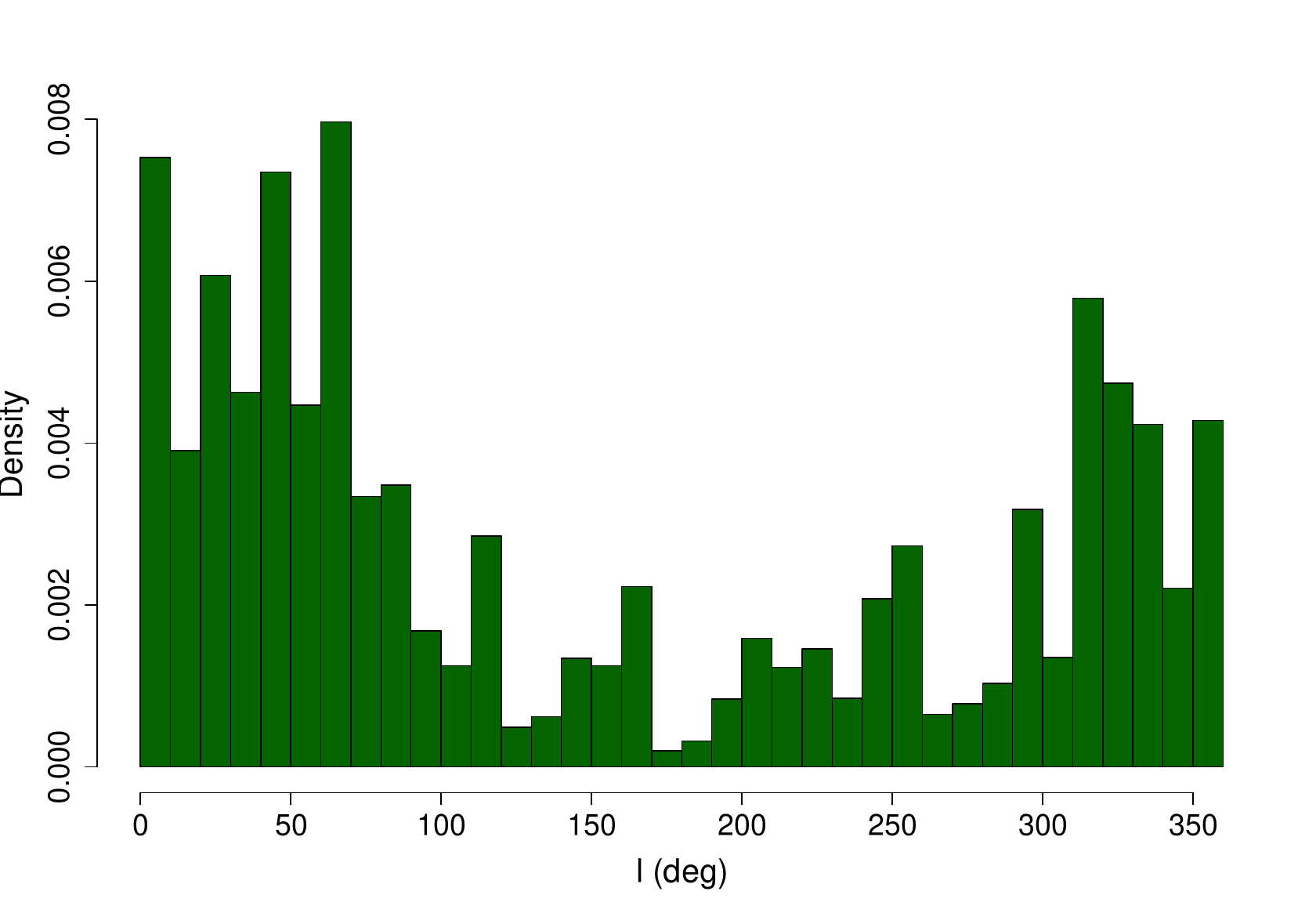}}
 \subfigure[distribution of $b$ (real)]{\label{subfig:nogcatnfbmspc}\includegraphics[width=0.17\textwidth, angle=-90]{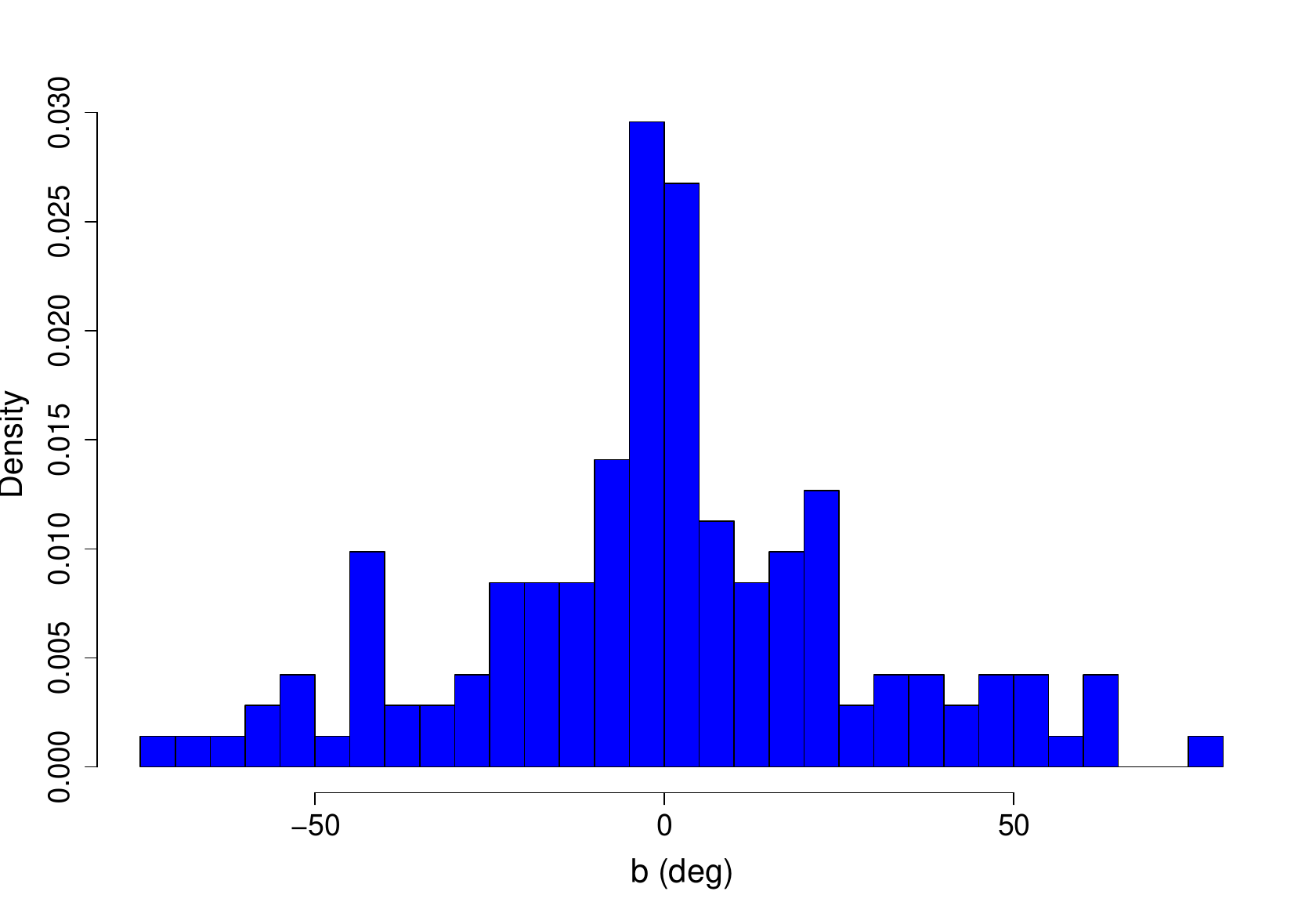}}
  \subfigure[distribution of $b$ (synthetic)]{\label{subfig:nogcranbmspd}\includegraphics[width=0.17\textwidth, angle=-90]{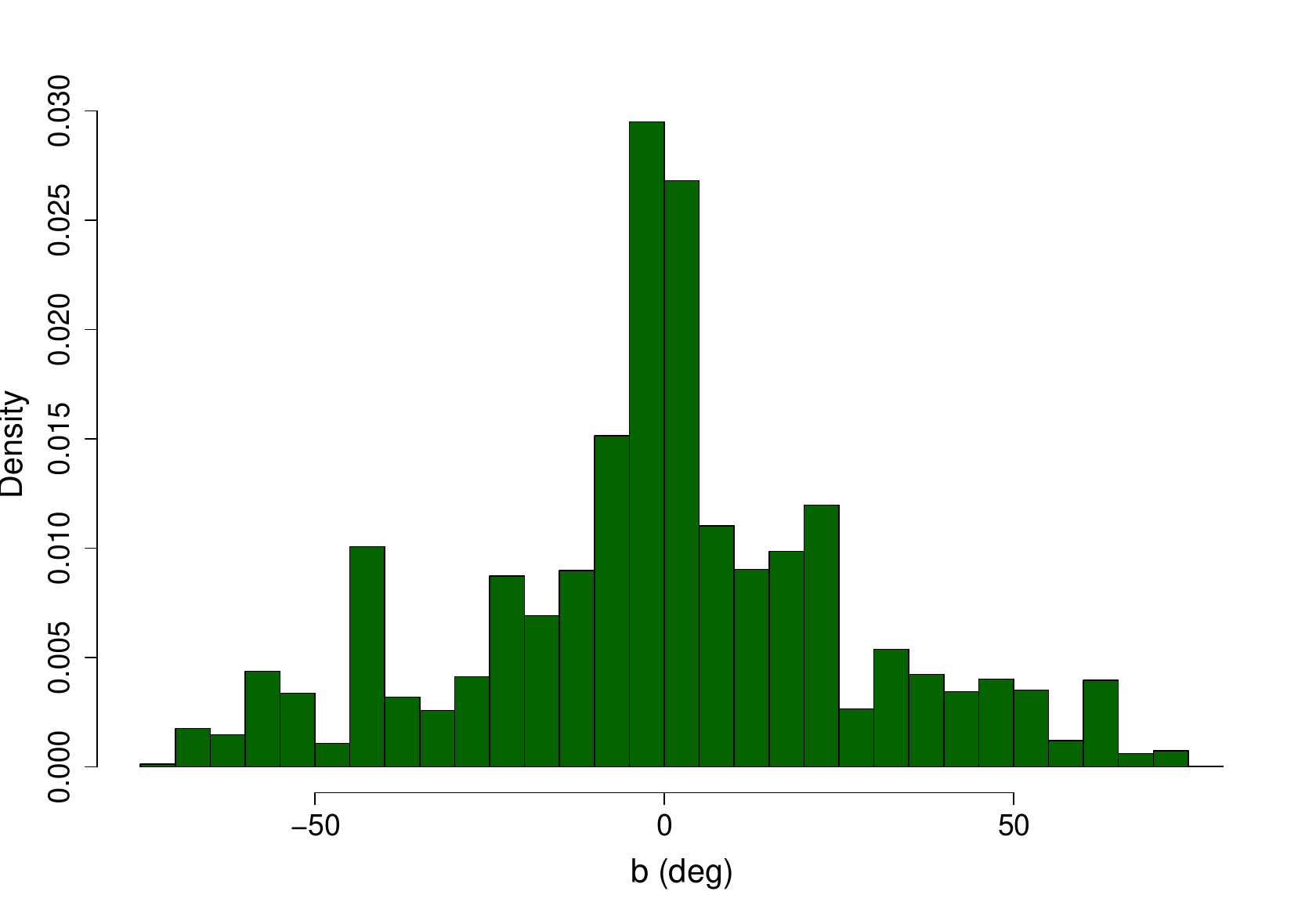}}\\
\subfigure[distribution of $d$ (real)]{\label{subfig:nogcatnfdmspe}\includegraphics[width=0.17\textwidth, angle=-90]{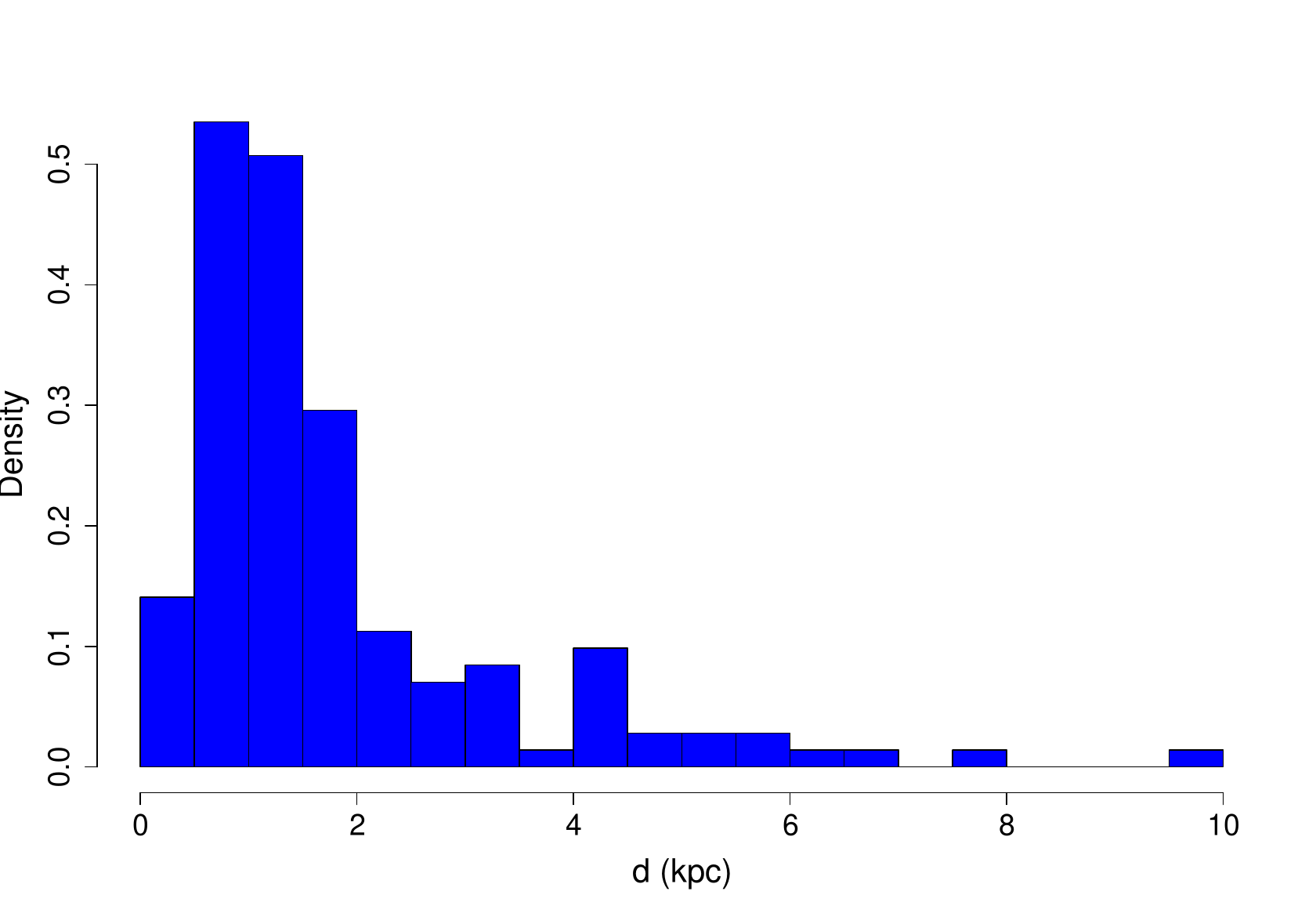}}
\subfigure[distribution of $d$ (synthetic)]{\label{subfig:nogcrandmspf}\includegraphics[width=0.17\textwidth, angle=-90]{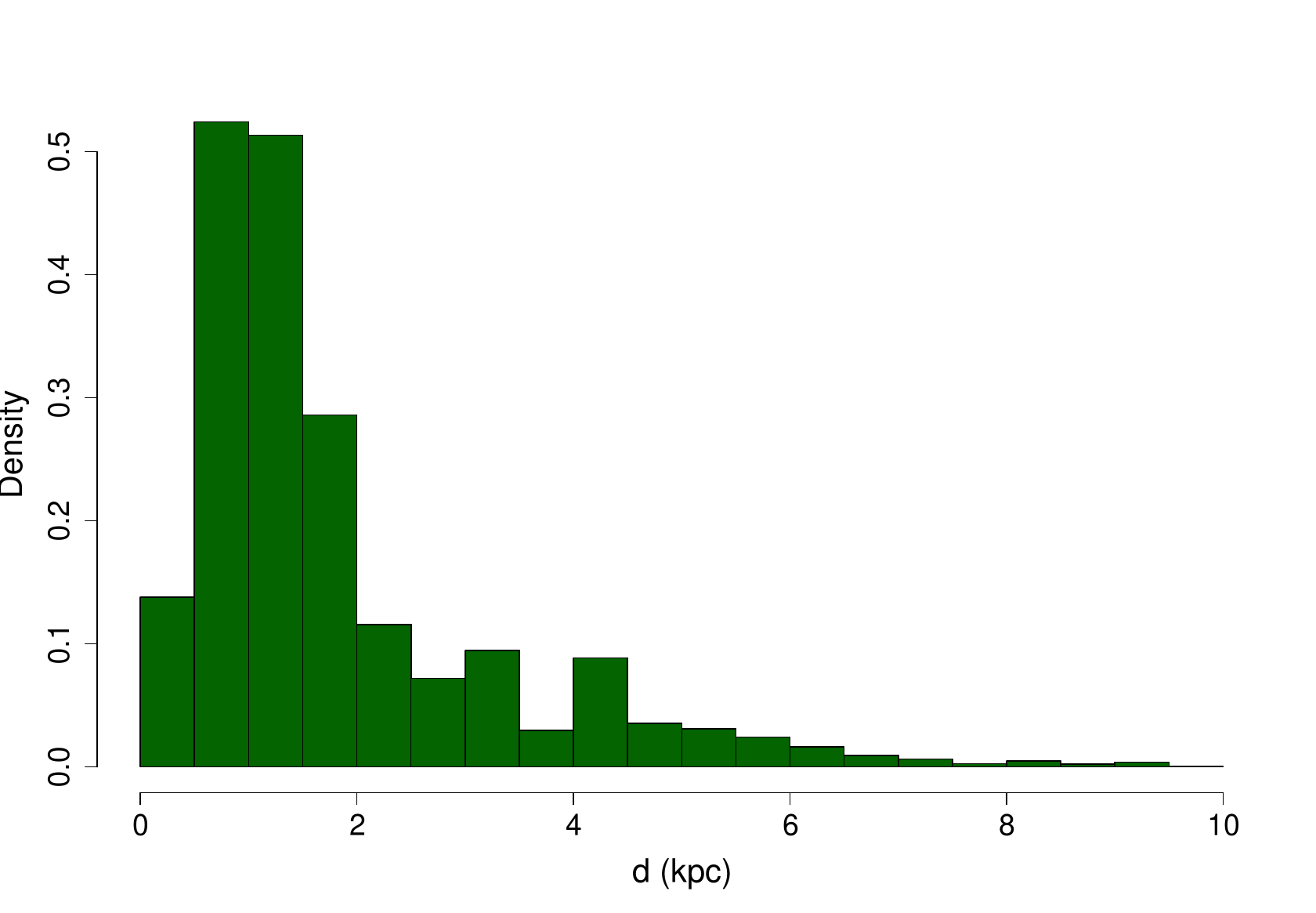}}
 \subfigure[distribution of $\mu_{l}$ (real)]{\label{subfig:nogcatnfmulmspg}\includegraphics[width=0.17\textwidth, angle=-90]{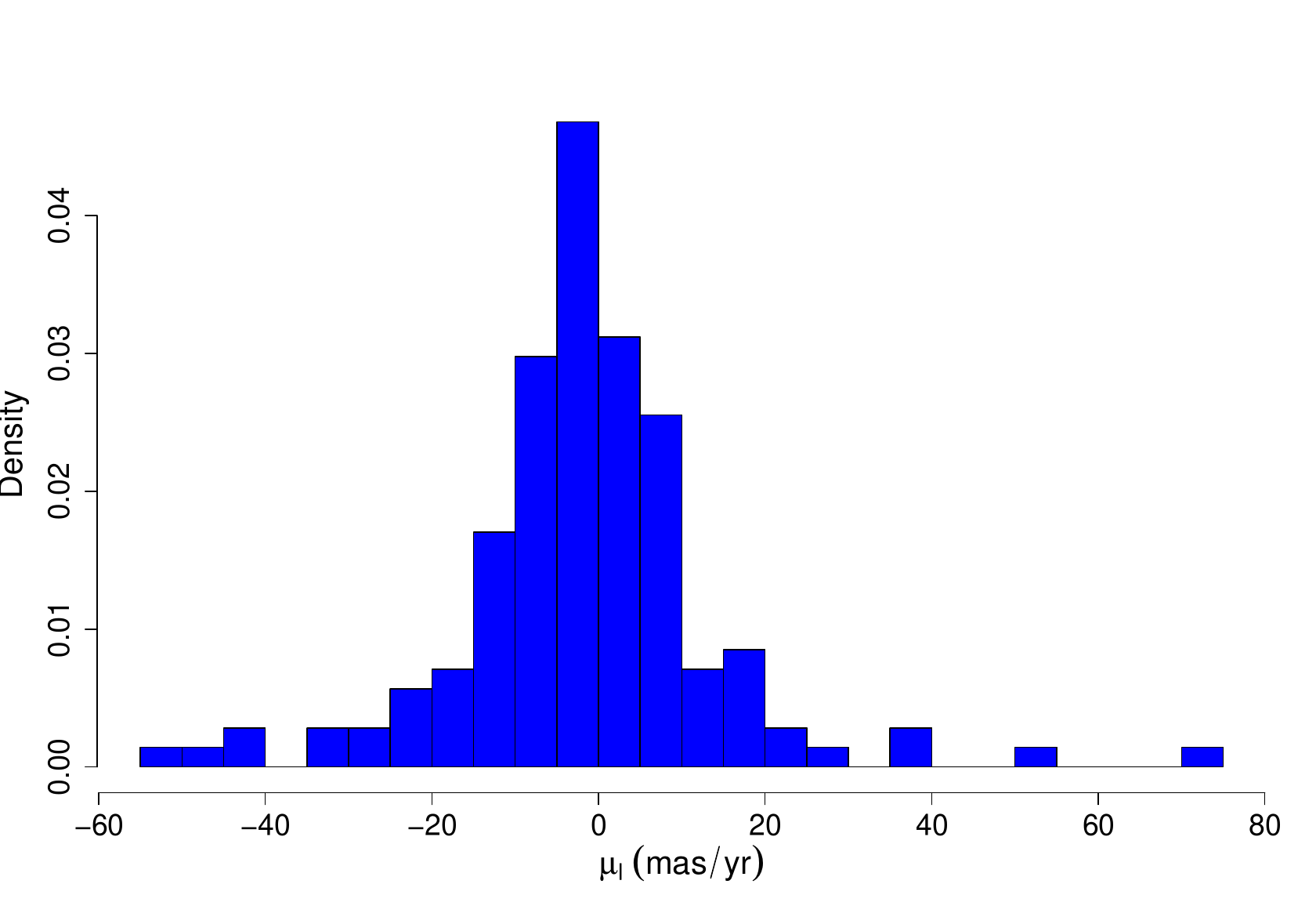}}
  \subfigure[distribution of $\mu_{l}$ (synthetic)]{\label{subfig:nogcranmulmsph}\includegraphics[width=0.17\textwidth, angle=-90]{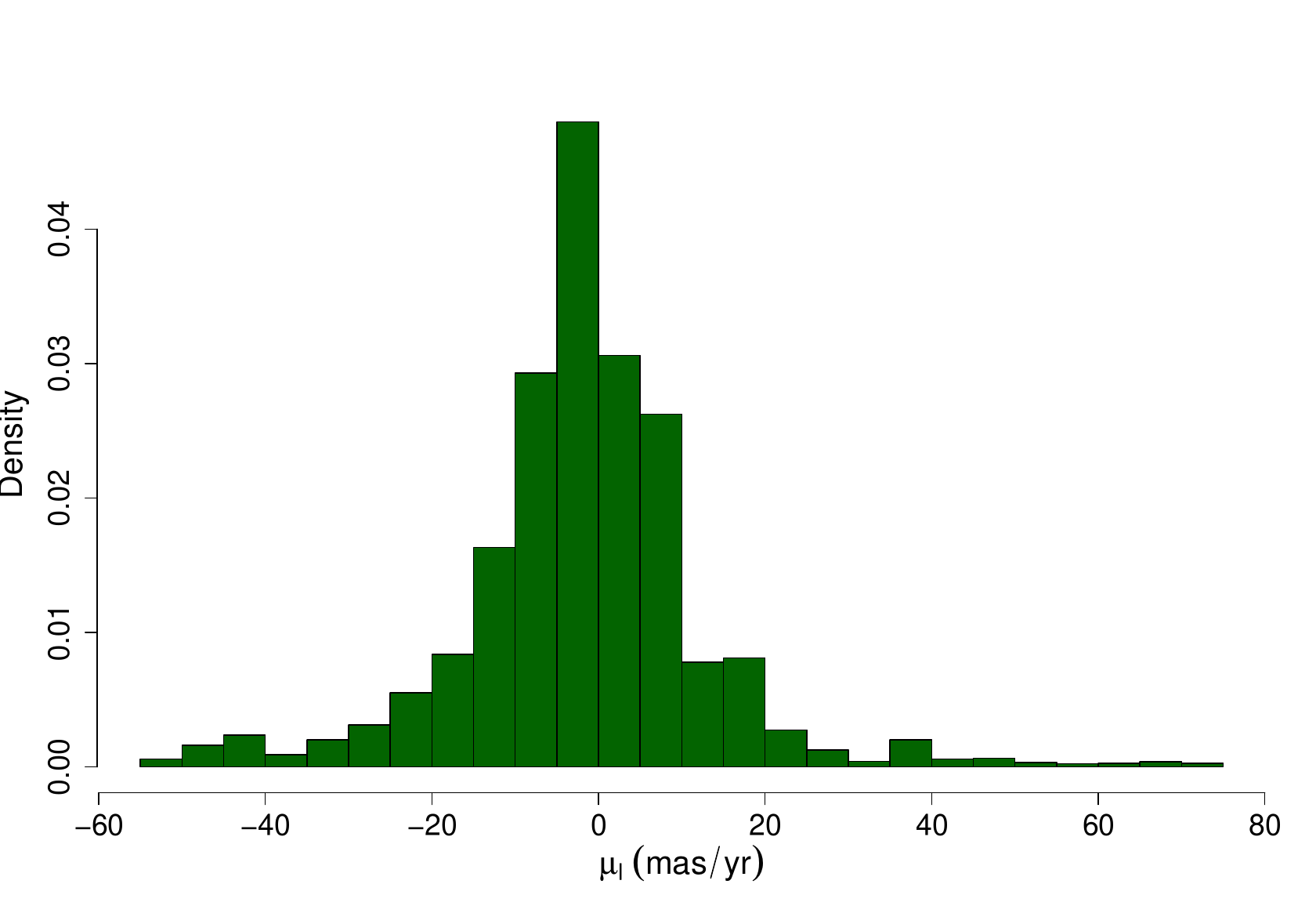}}\\
\subfigure[distribution of $\mu_{b}$ (real)]{\label{subfig:nogcatnfmubmspi}\includegraphics[width=0.17\textwidth, angle=-90]{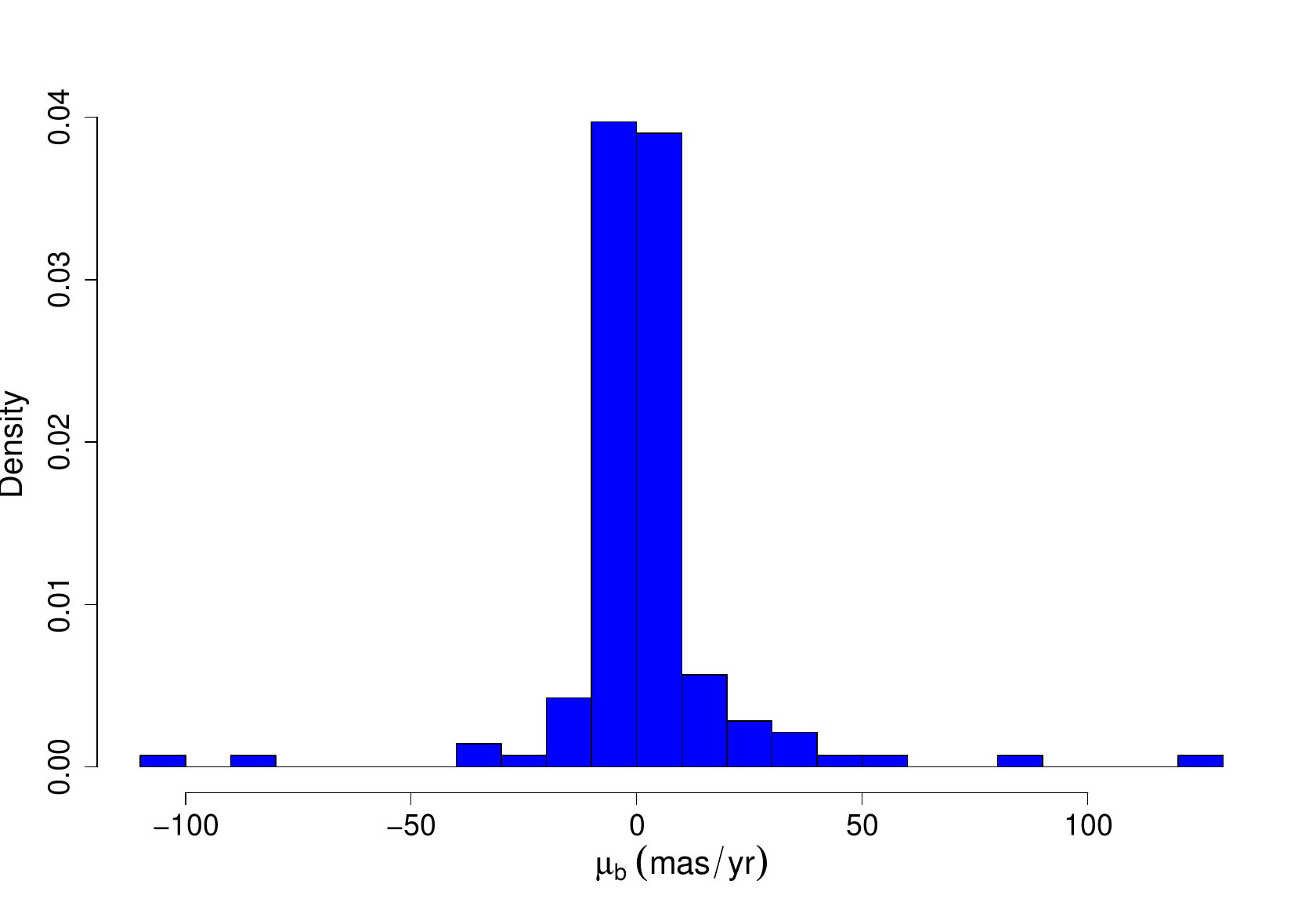}}
\subfigure[distribution of $\mu_{b}$ (synthetic)]{\label{subfig:nogcranmubmspj}\includegraphics[width=0.17\textwidth, angle=-90]{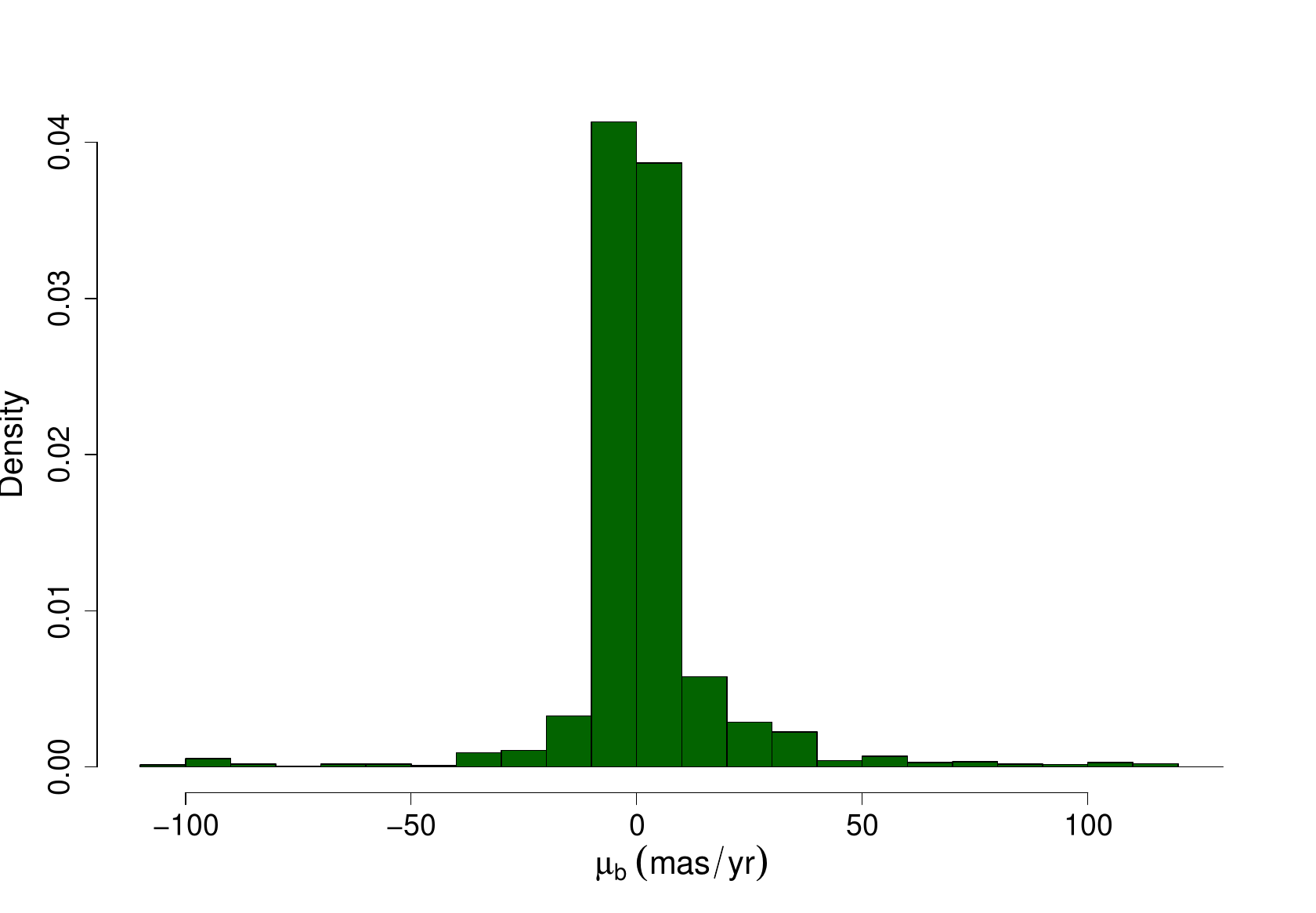}}
\subfigure[distribution of $f_{s}$ (real)]{\label{subfig:nogcatnff0mspk}\includegraphics[width=0.17\textwidth, angle=-90]{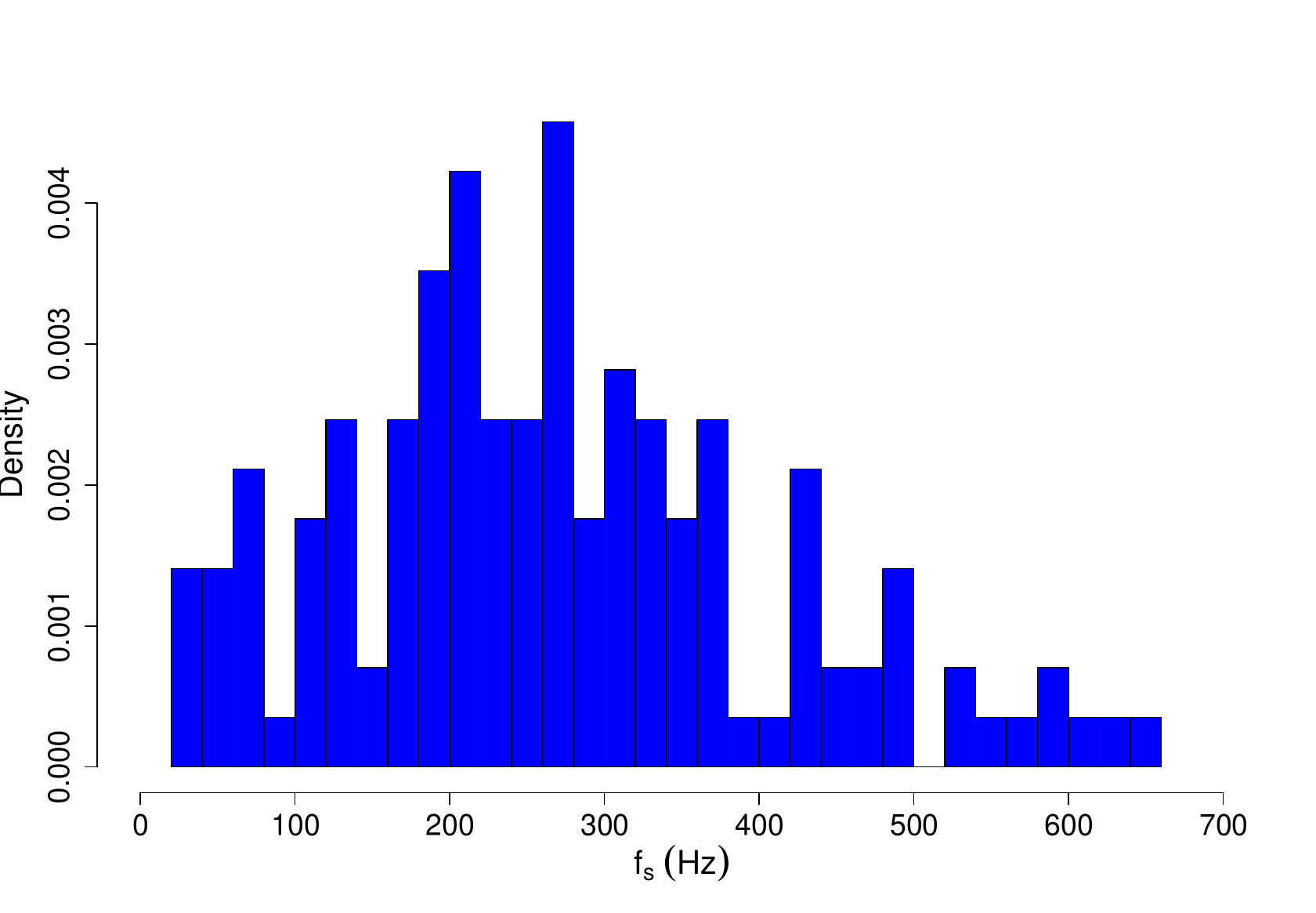}}
\subfigure[distribution of $f_{s}$ (synthetic)]{\label{subfig:nogcranf0mspl}\includegraphics[width=0.17\textwidth, angle=-90]{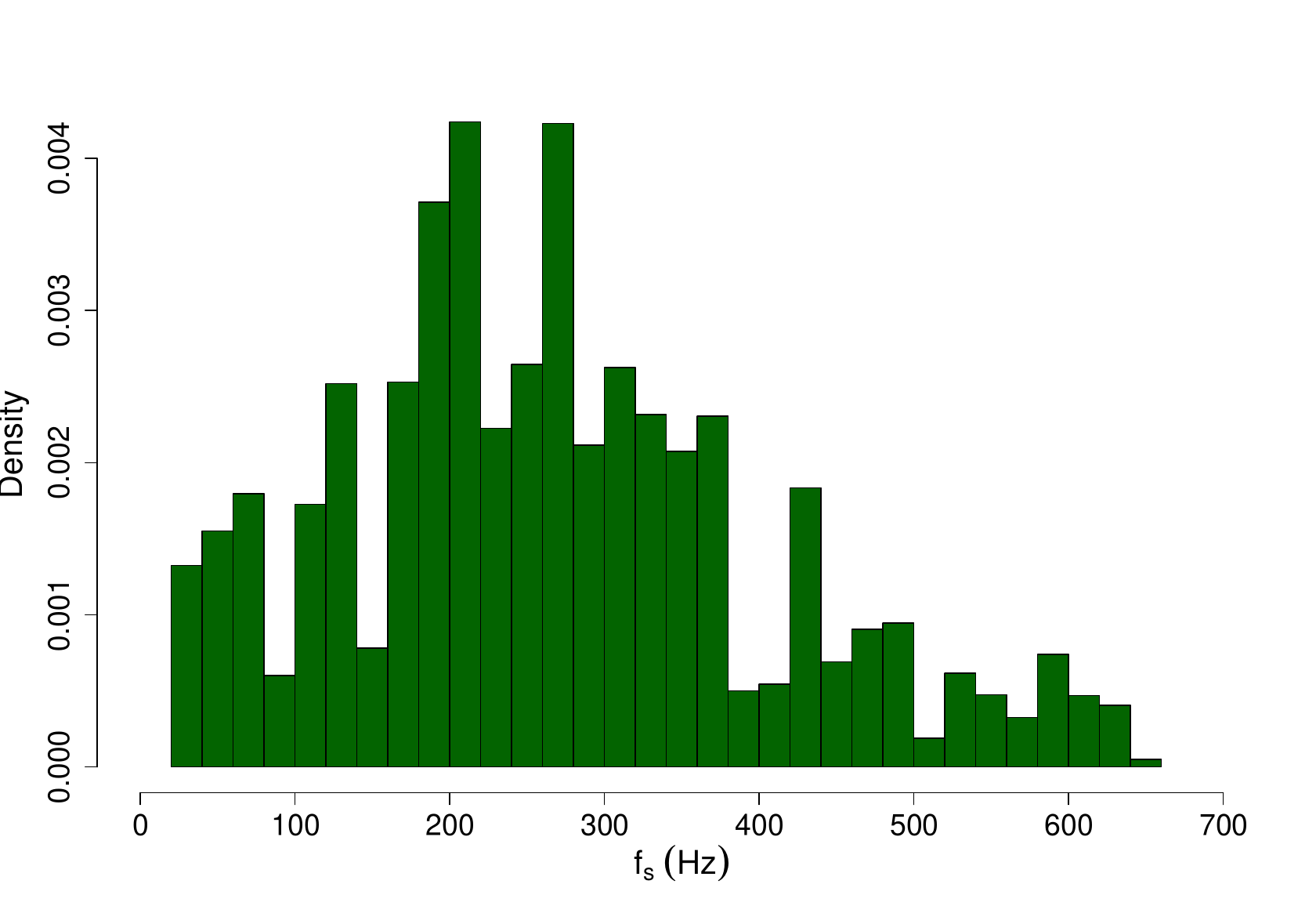}}\\
\subfigure[distribution of $\dot{f}_{s}$ (real)]{\label{subfig:nogcatnff1mspm}\includegraphics[width=0.17\textwidth, angle=-90]{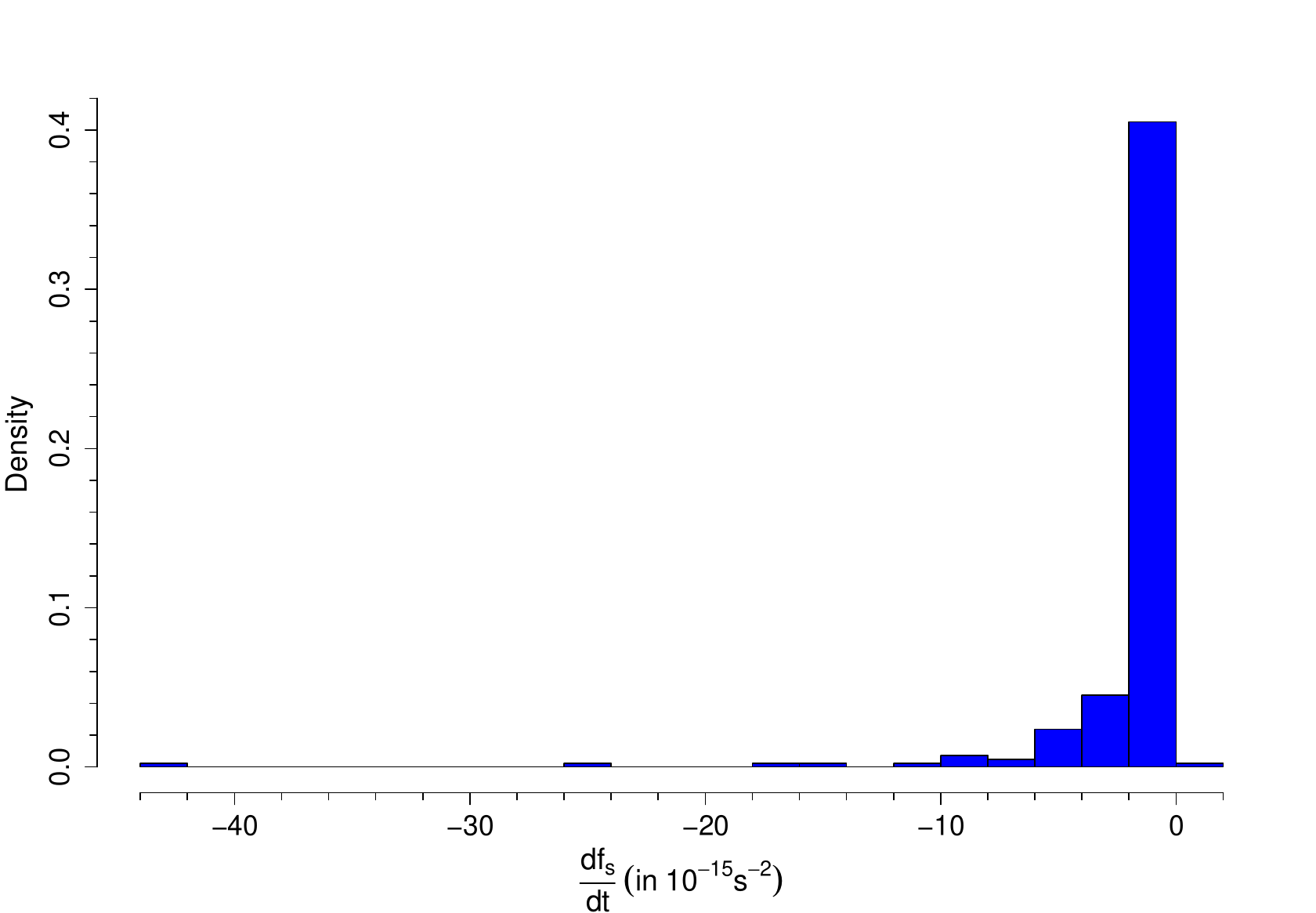}}
\subfigure[distribution of $\dot{f}_{s}$ (synthetic)]{\label{subfig:nogcranf1mspn}\includegraphics[width=0.17\textwidth, angle=-90]{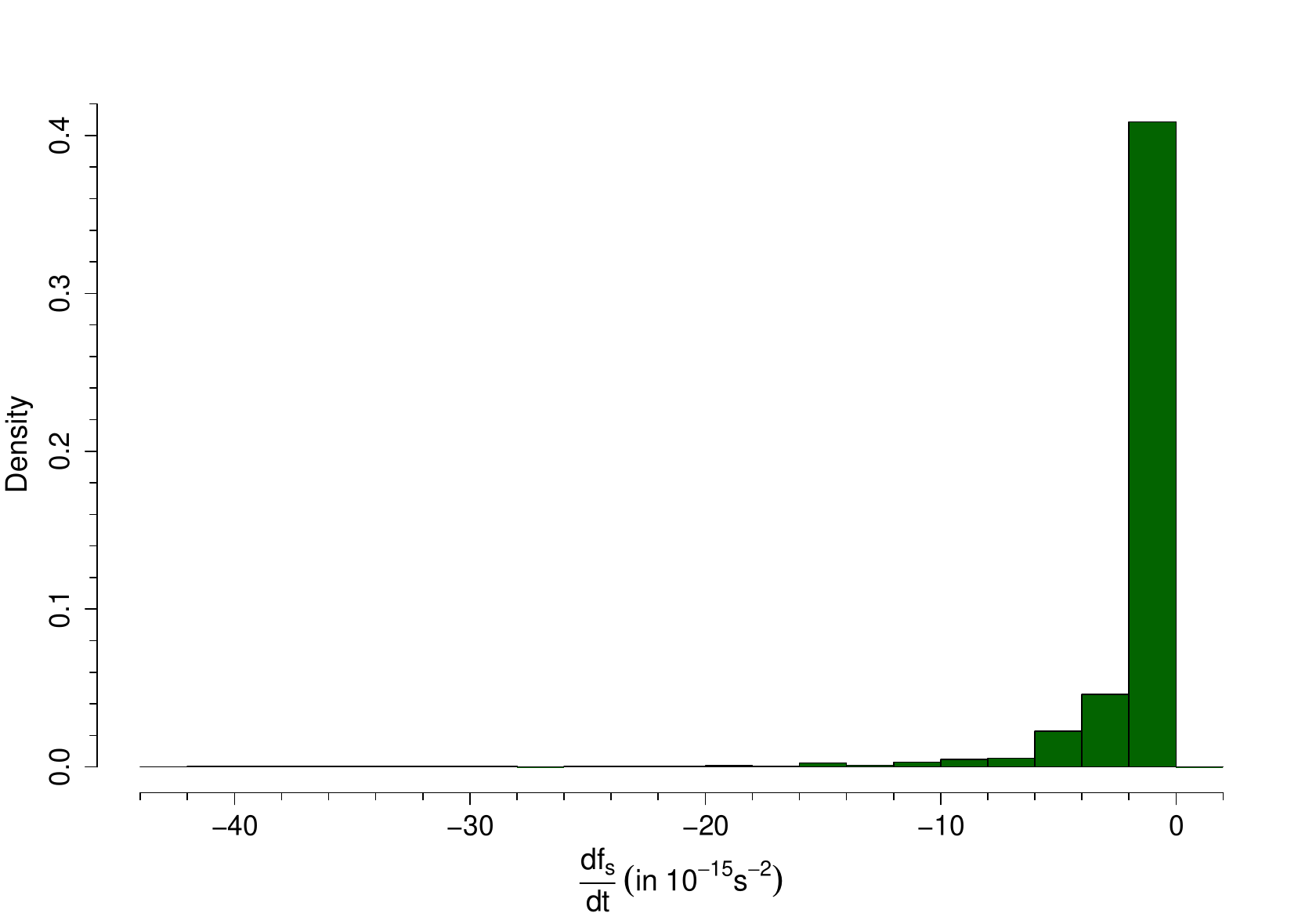}}
\subfigure[distribution of $\ddot{f}_{s}$ (real)]{\label{subfig:nogcatnff2mspo}\includegraphics[width=0.17\textwidth, angle=-90]{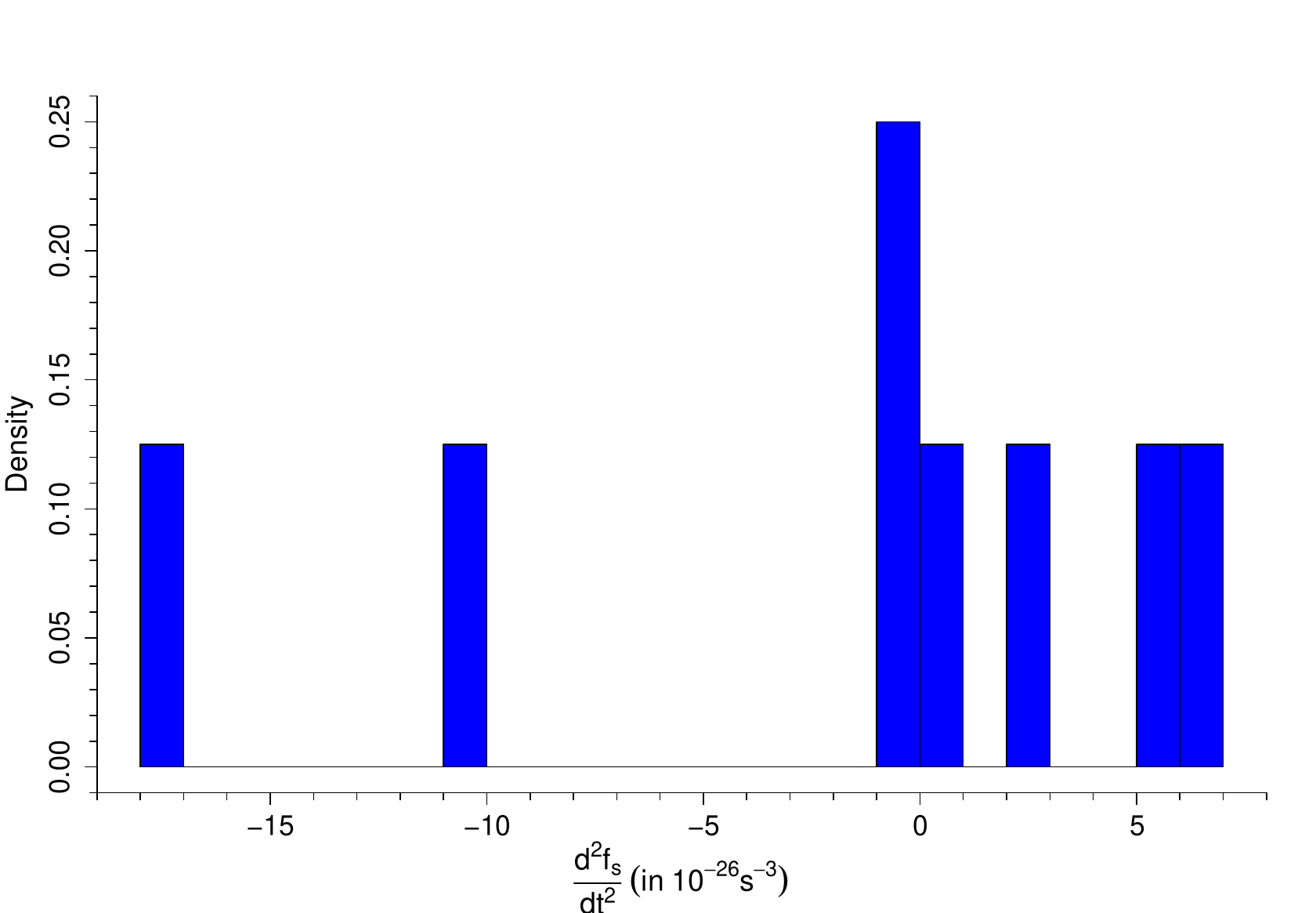}}
\subfigure[distribution of $\ddot{f}_{s}$ (synthetic)]{\label{subfig:nogcranf2mspp}\includegraphics[width=0.17\textwidth, angle=-90]{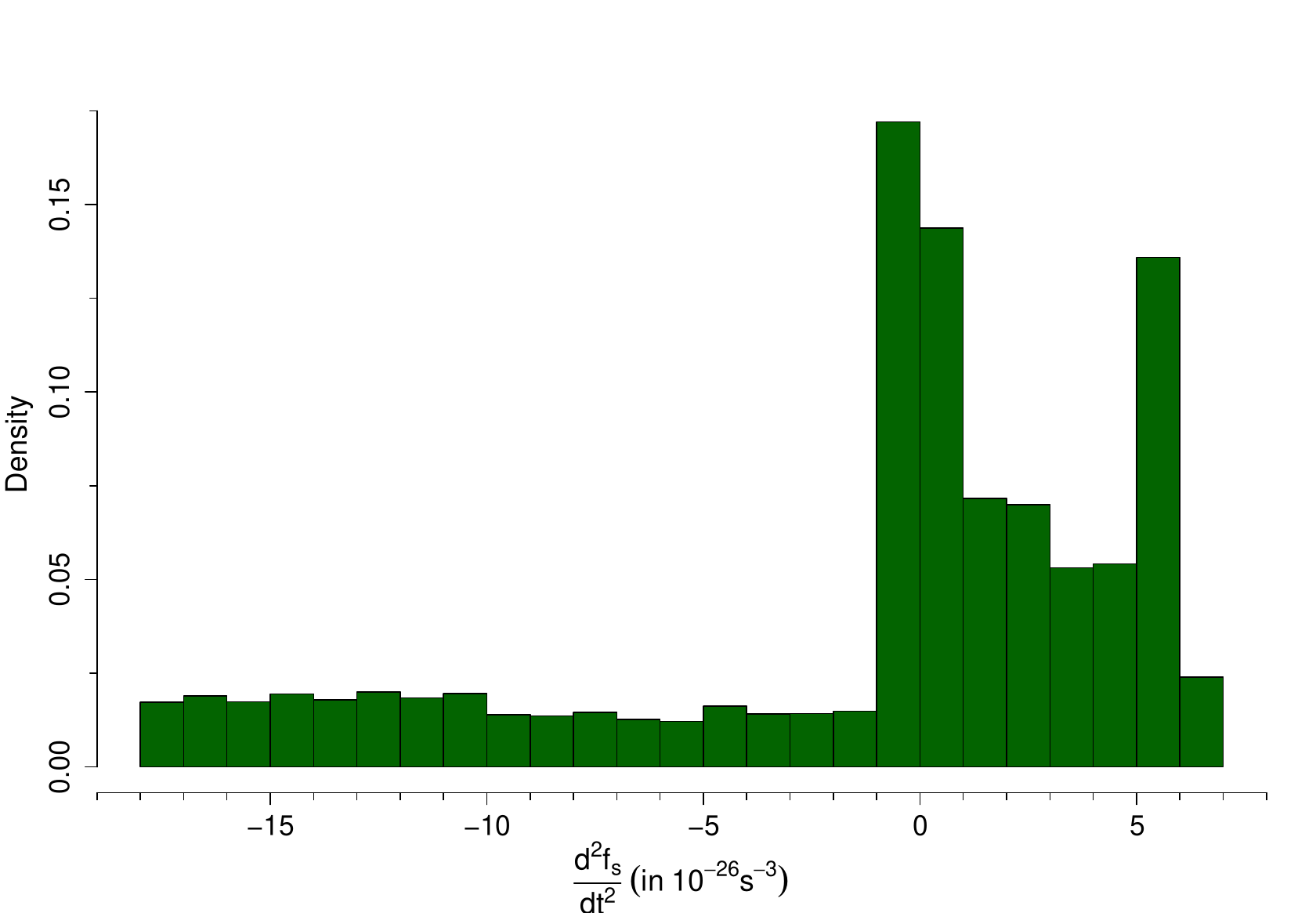}}
\end{center}
\caption[]{Comparison of observed and simulated density distribution of parameters of millisecond pulsars in the Galactic field for which we study spin frequency and its derivatives.    }
\label{fig:histMSPspin}
\end{figure}

\begin{figure}
\begin{center}
\subfigure[distribution of $l$ (real)]{\label{subfig:nogcatnflmsp}\includegraphics[width=0.17\textwidth, angle=-90]{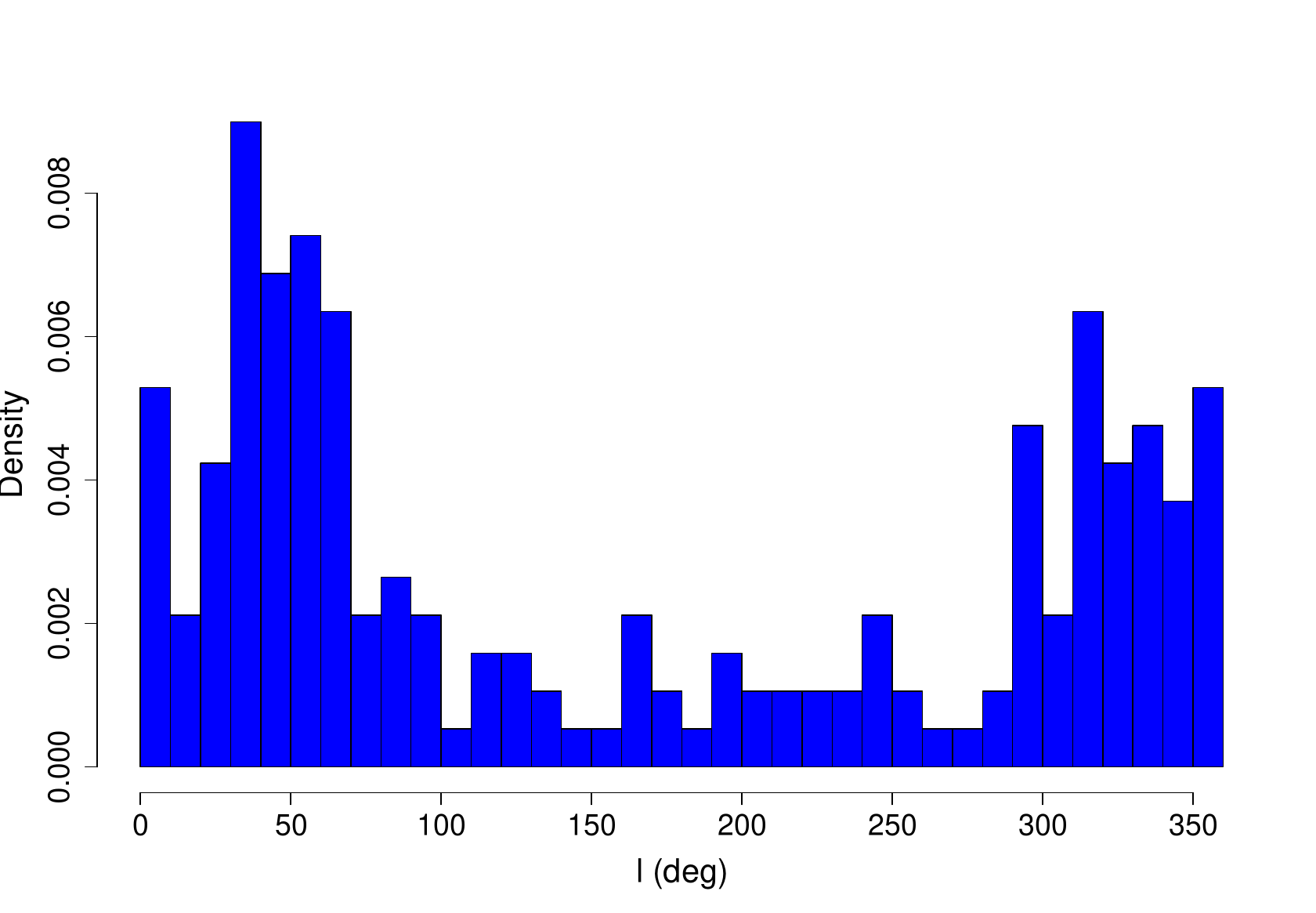}}
\subfigure[distribution of $l$ (synthetic)]{\label{subfig:nogcranlmsp}\includegraphics[width=0.17\textwidth, angle=-90]{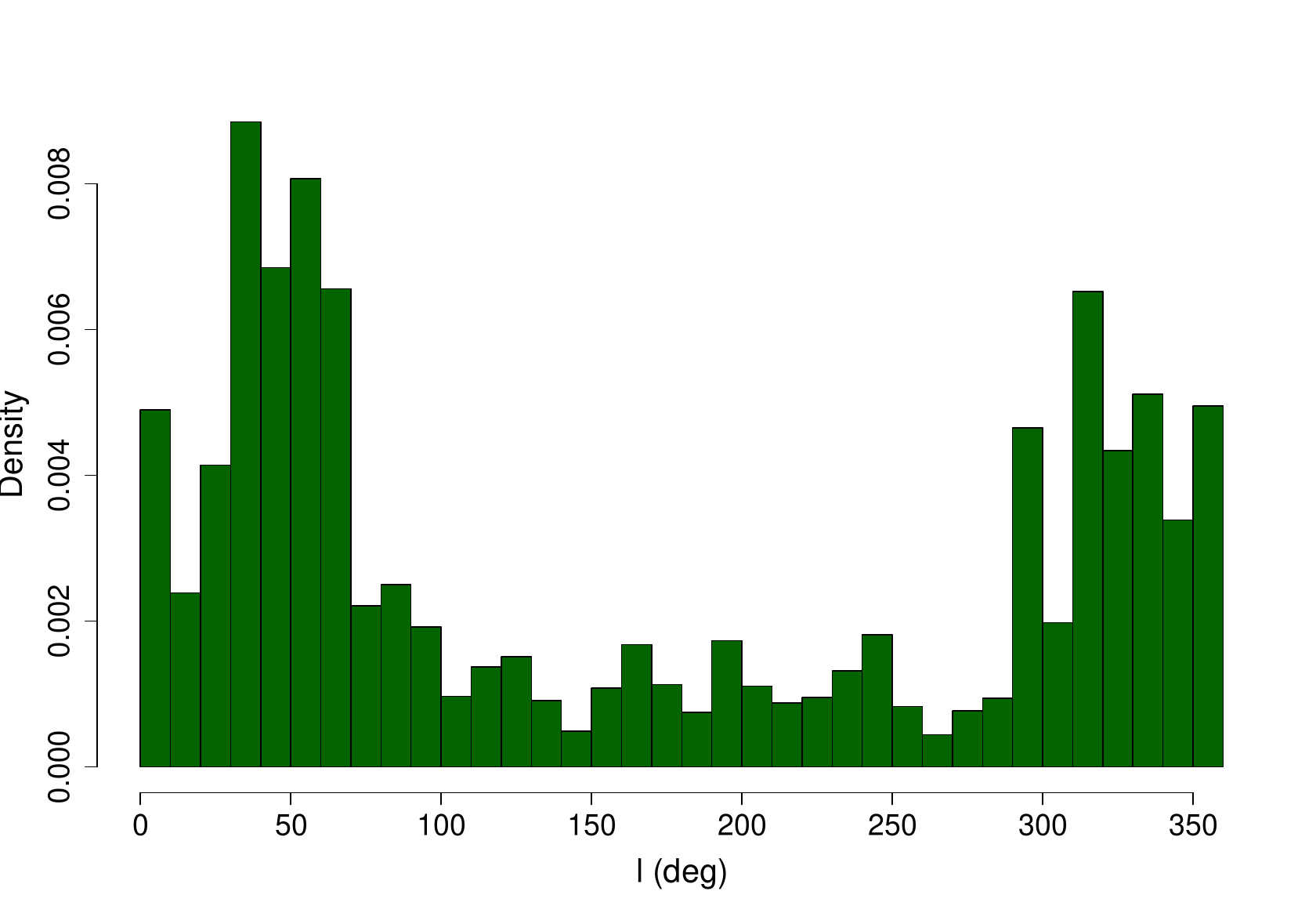}}
 \subfigure[distribution of $b$ (real)]{\label{subfig:nogcatnfbmsp}\includegraphics[width=0.17\textwidth, angle=-90]{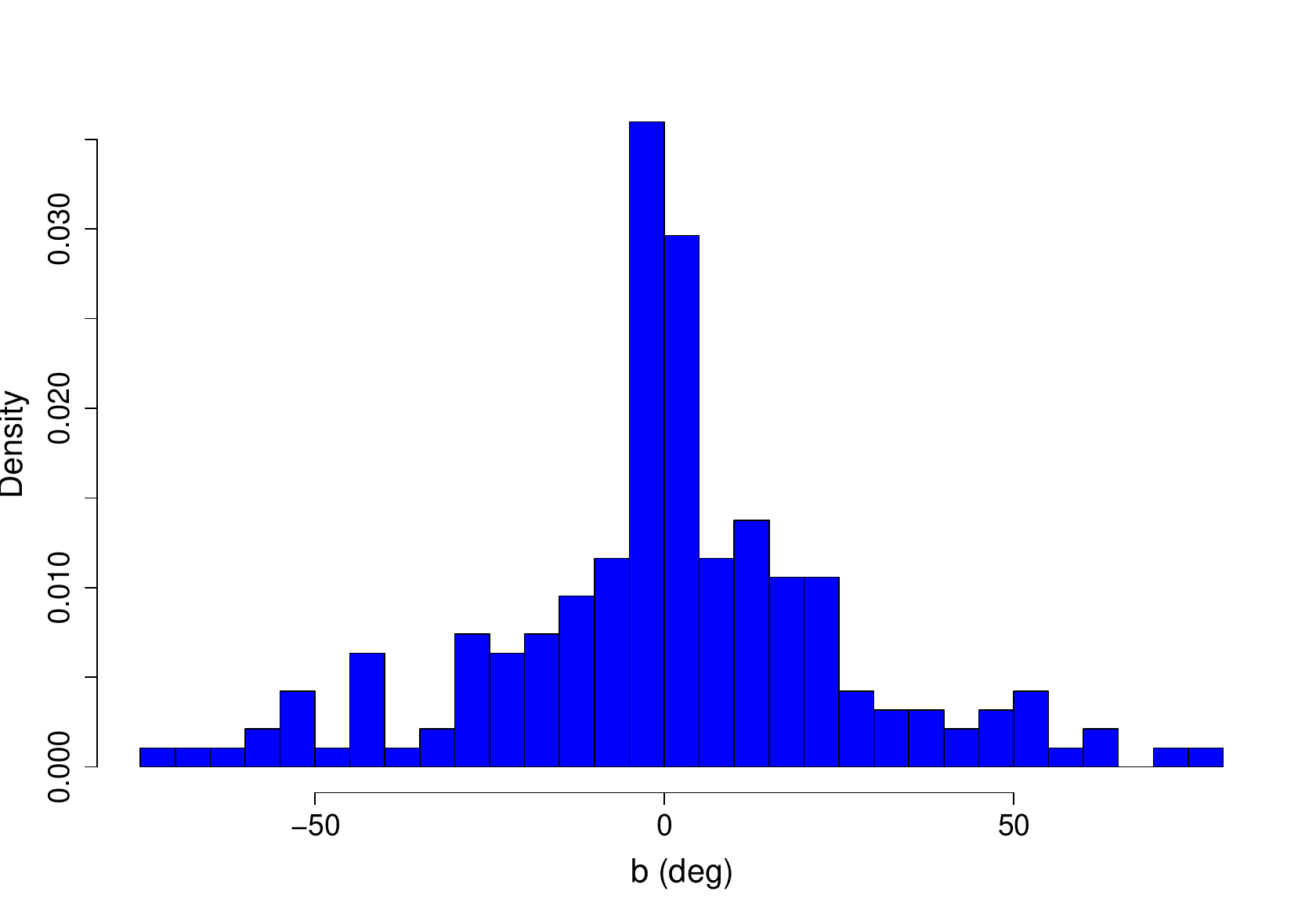}}
  \subfigure[distribution of $b$ (synthetic)]{\label{subfig:nogcranbmsp}\includegraphics[width=0.17\textwidth, angle=-90]{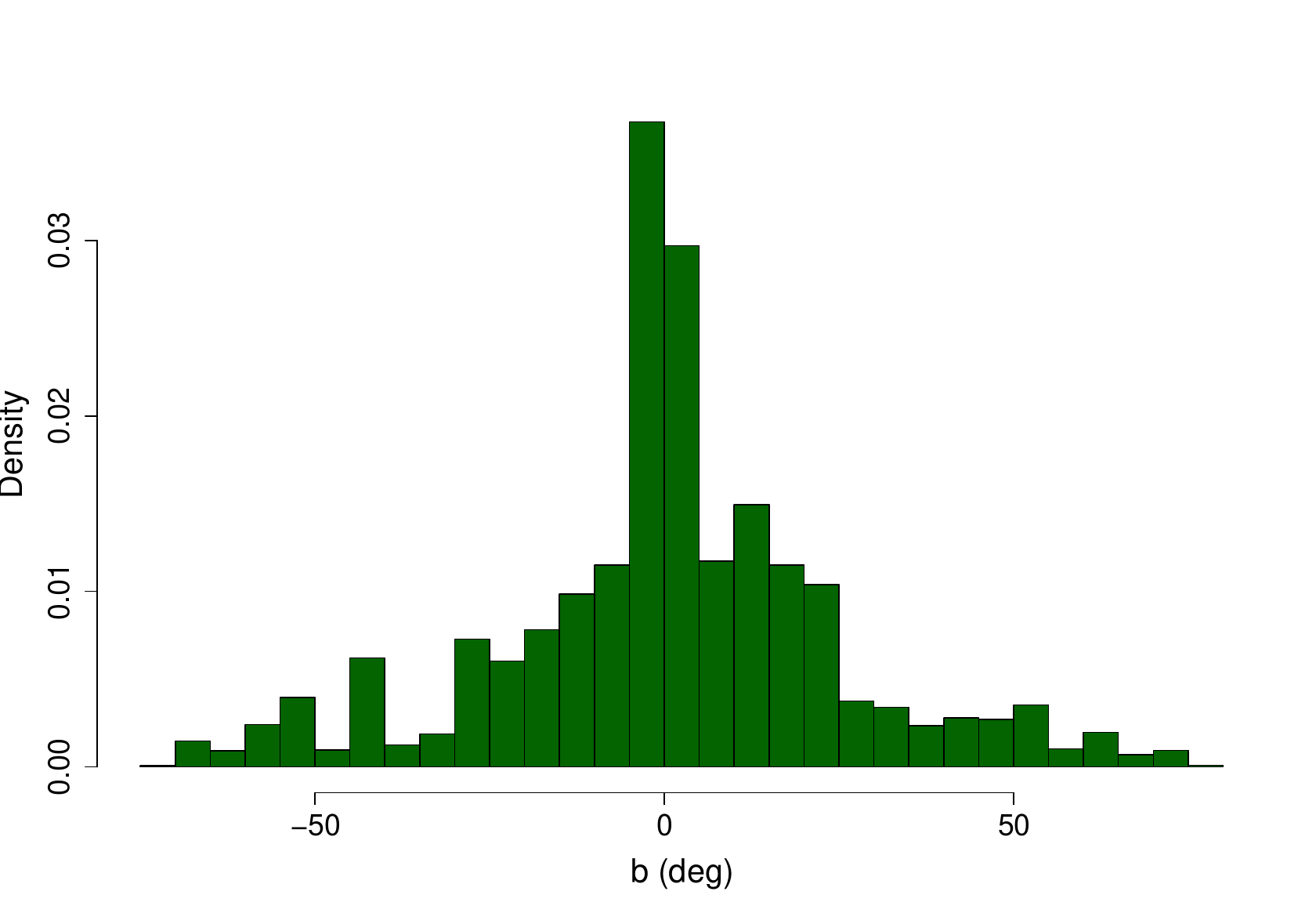}}\\
\subfigure[distribution of $d$ (real)]{\label{subfig:nogcatnfdmsp}\includegraphics[width=0.17\textwidth, angle=-90]{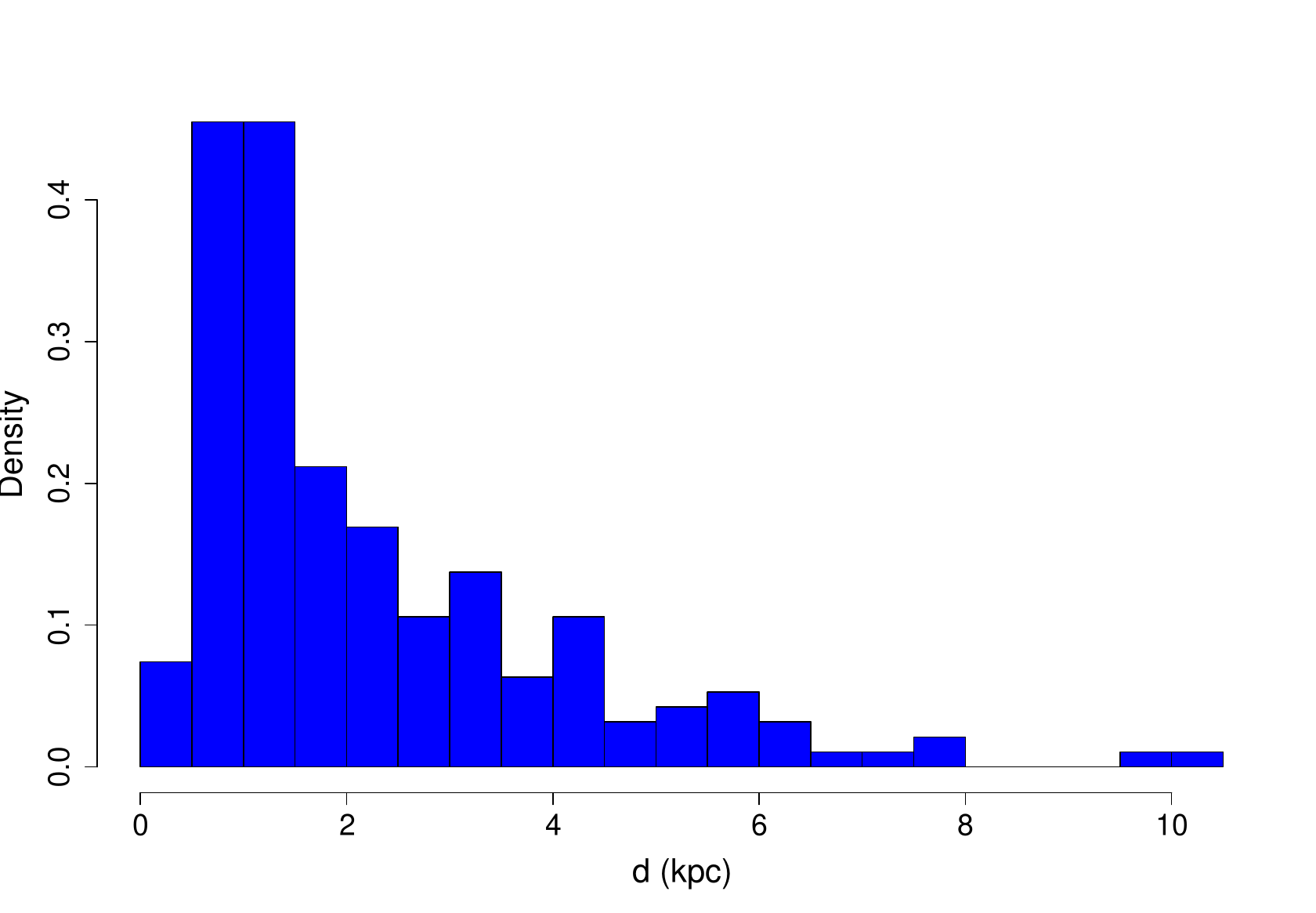}}
\subfigure[distribution of $d$ (synthetic)]{\label{subfig:nogcrandmsp}\includegraphics[width=0.17\textwidth, angle=-90]{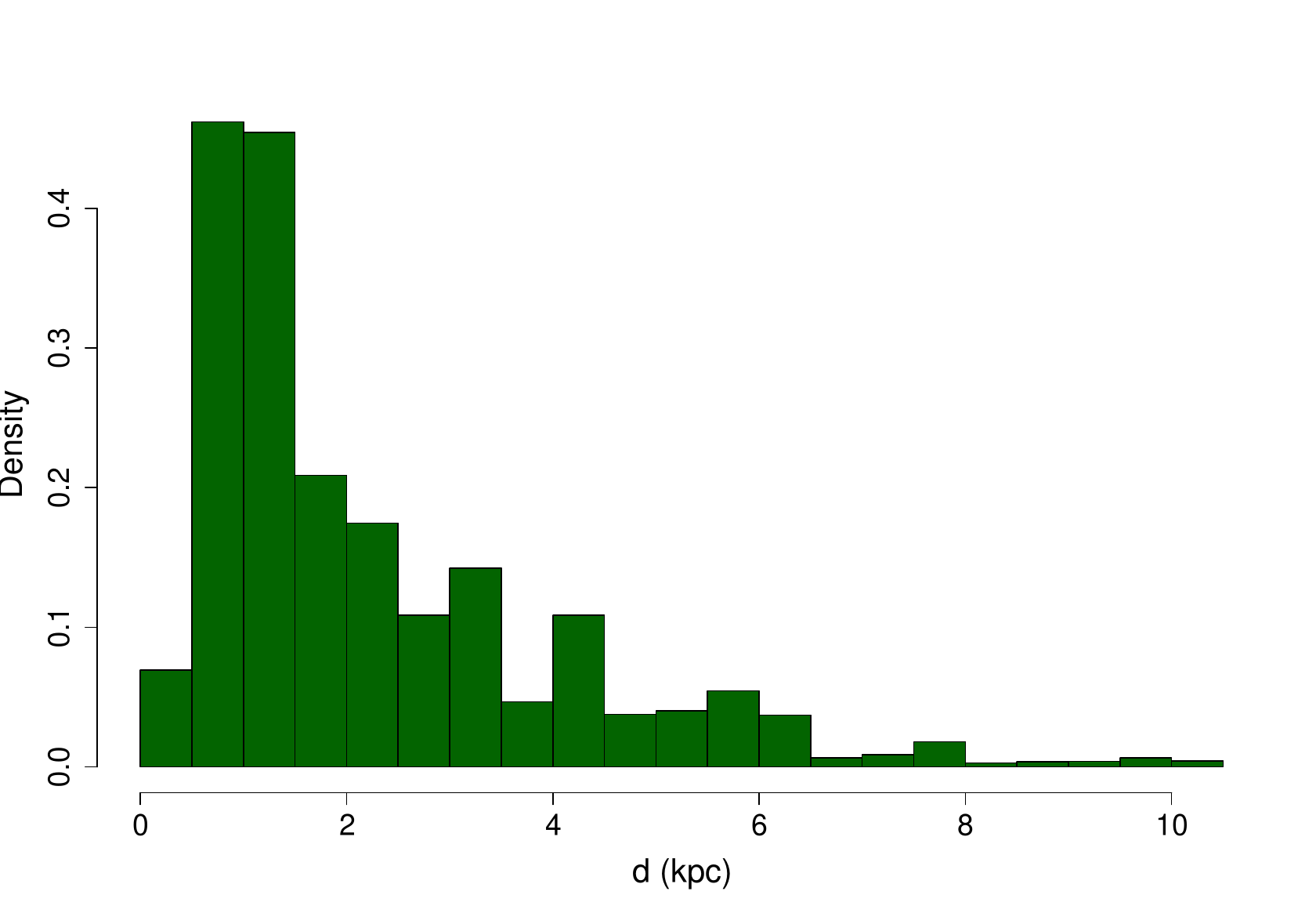}}
 \subfigure[distribution of $\mu_{l}$ (real)]{\label{subfig:nogcatnfmulmsp}\includegraphics[width=0.17\textwidth, angle=-90]{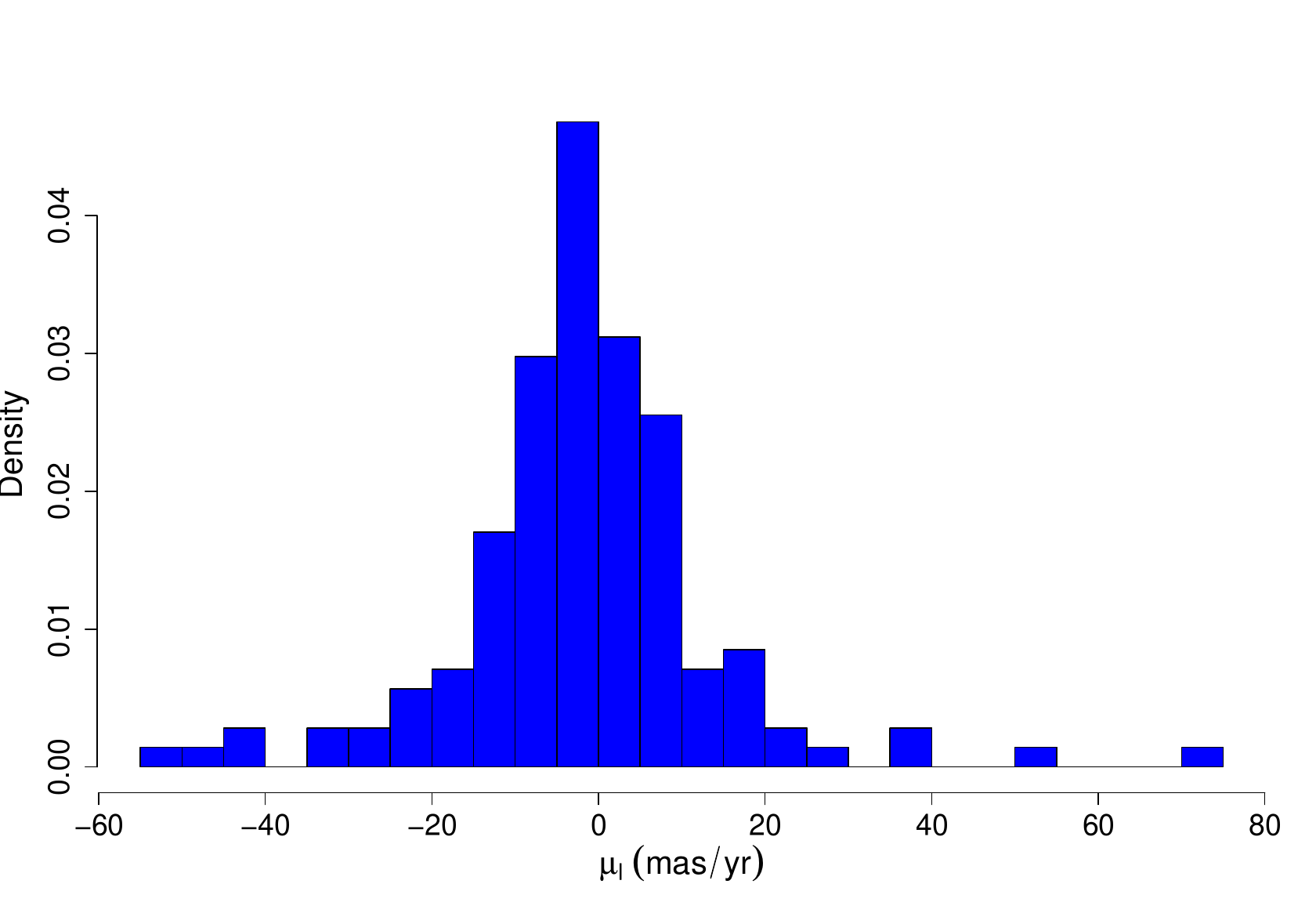}}
  \subfigure[distribution of $\mu_{l}$ (synthetic)]{\label{subfig:nogcranmulmsp}\includegraphics[width=0.17\textwidth, angle=-90]{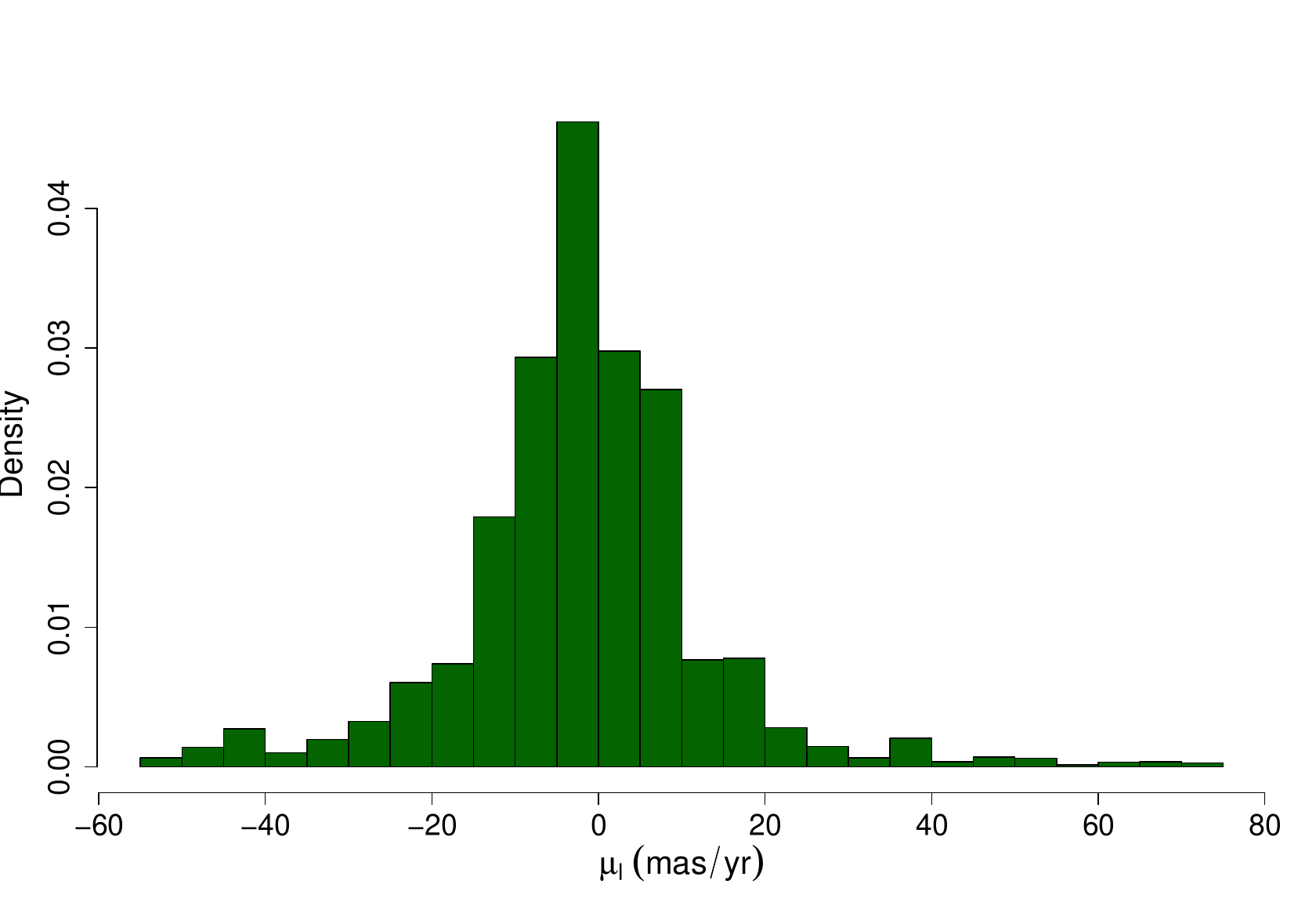}}\\
\subfigure[distribution of $\mu_{b}$ (real)]{\label{subfig:nogcatnfmubmsp}\includegraphics[width=0.17\textwidth, angle=-90]{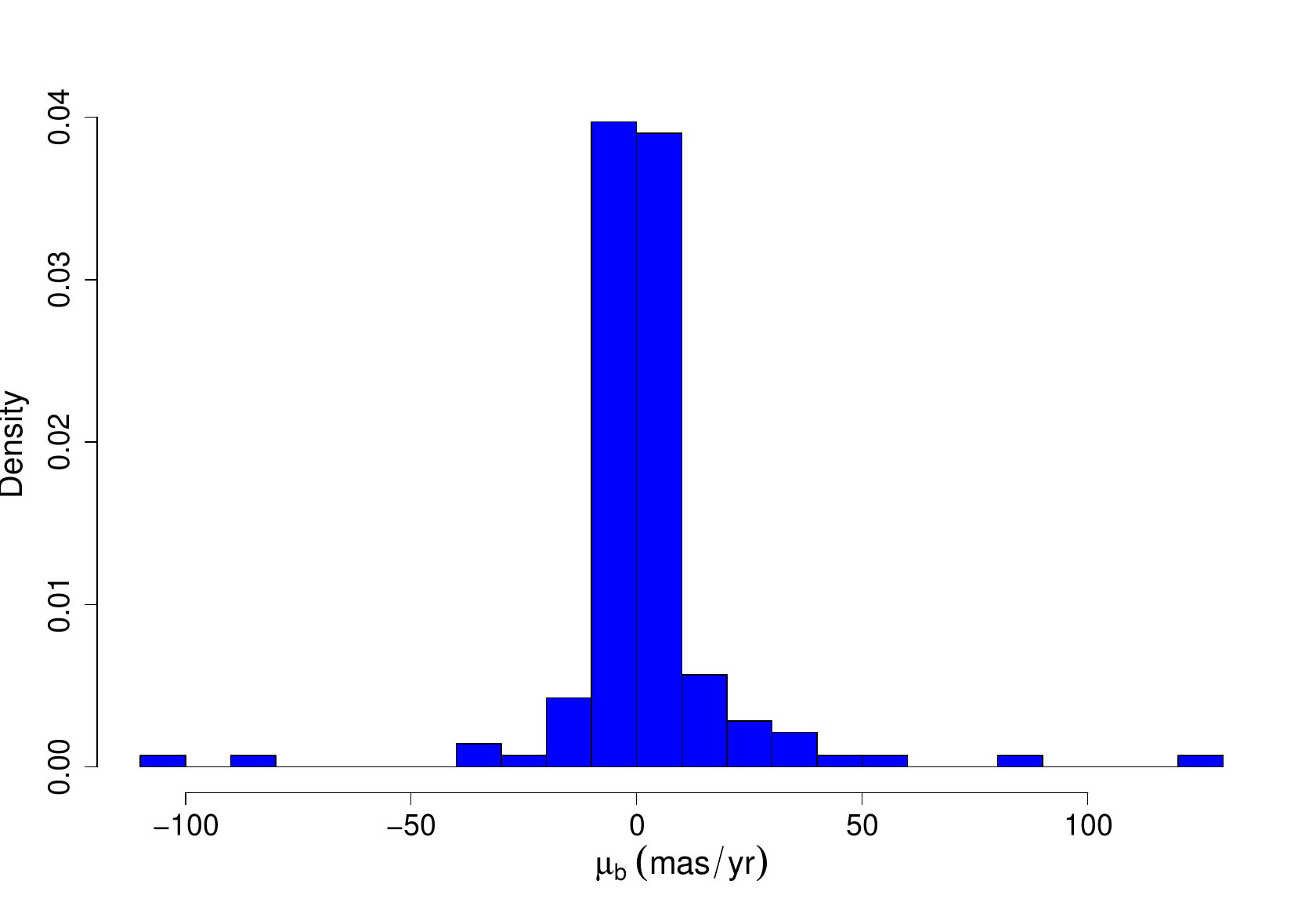}}
\subfigure[distribution of $\mu_{b}$ (synthetic)]{\label{subfig:nogcranmubmsp}\includegraphics[width=0.17\textwidth, angle=-90]{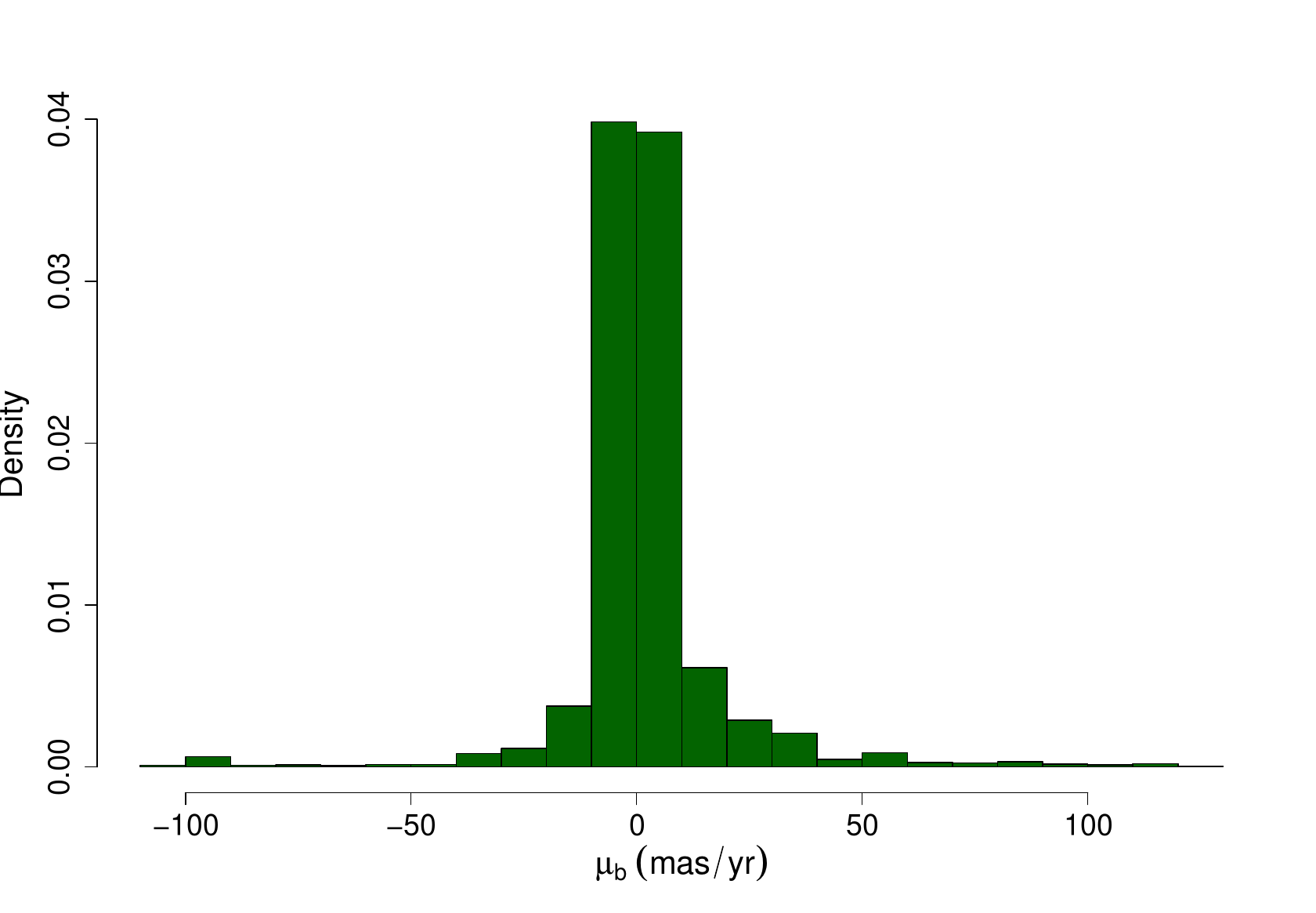}}
\subfigure[distribution of $f_{b}$ (real)]{\label{subfig:nogcatnff0msp}\includegraphics[width=0.17\textwidth, angle=-90]{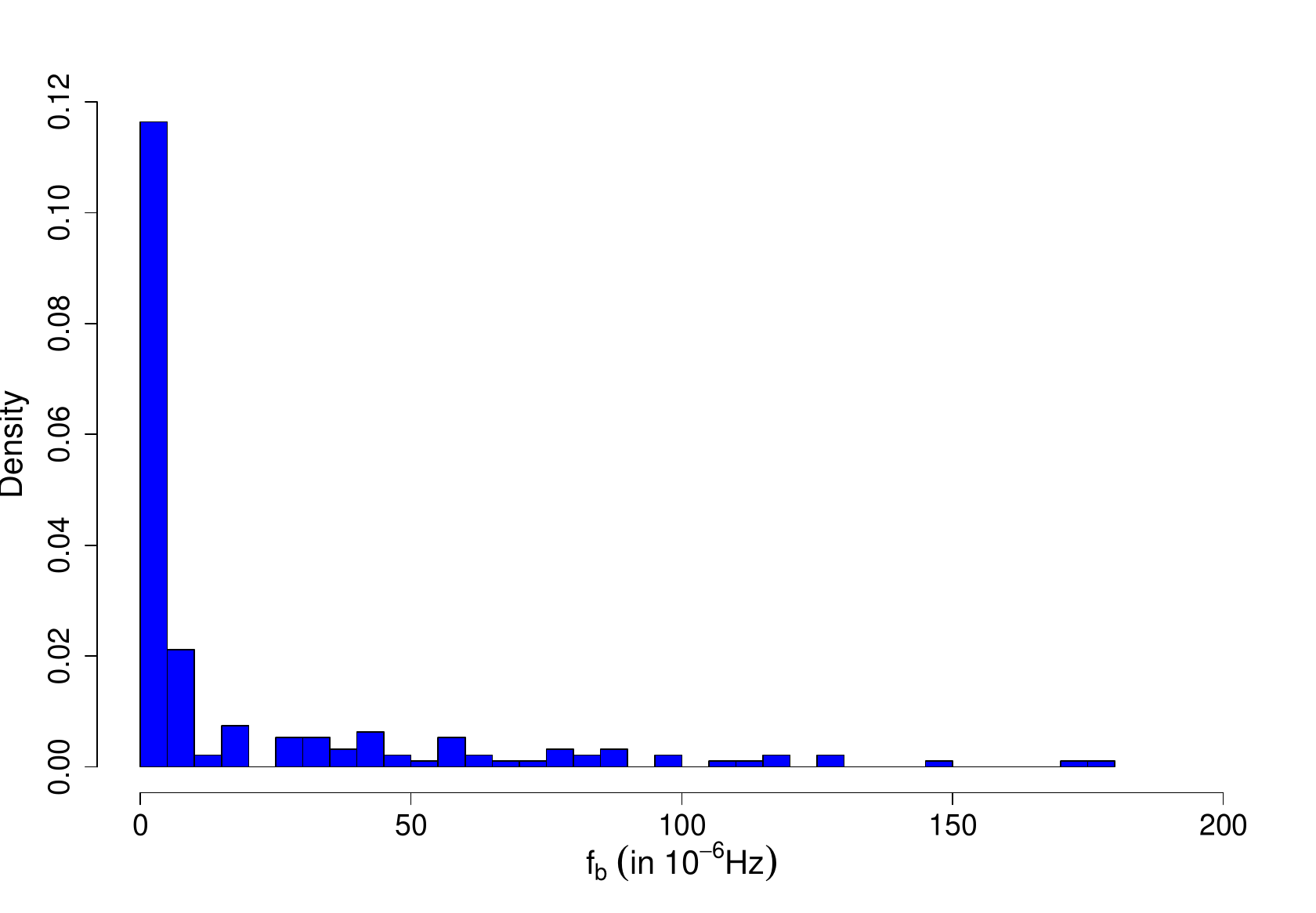}}
\subfigure[distribution of $f_{b}$ (synthetic)]{\label{subfig:nogcranf0msp}\includegraphics[width=0.17\textwidth, angle=-90]{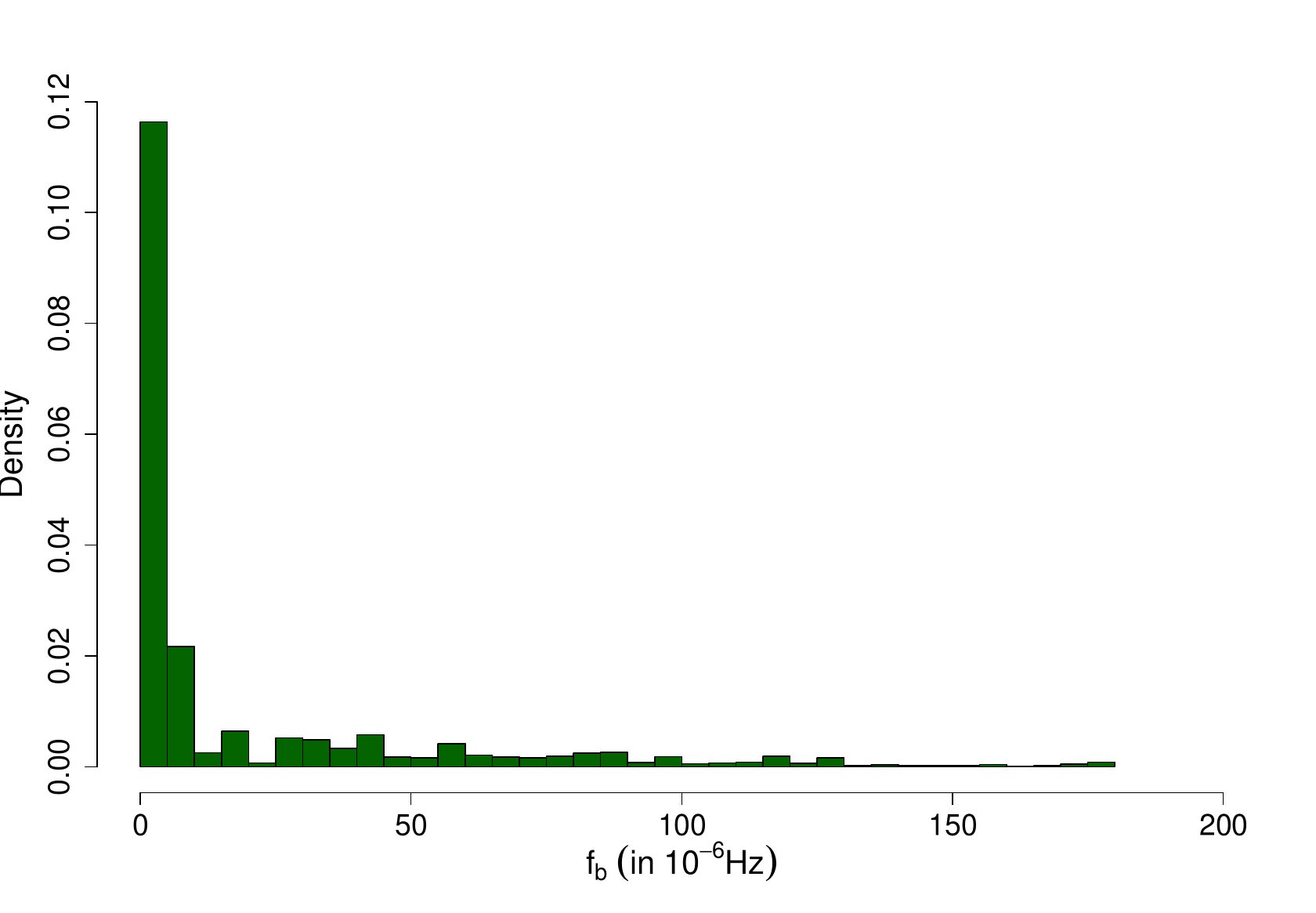}}\\
\subfigure[distribution of $\dot{f}_{b}$ (real)]{\label{subfig:nogcatnff1msp}\includegraphics[width=0.17\textwidth, angle=-90]{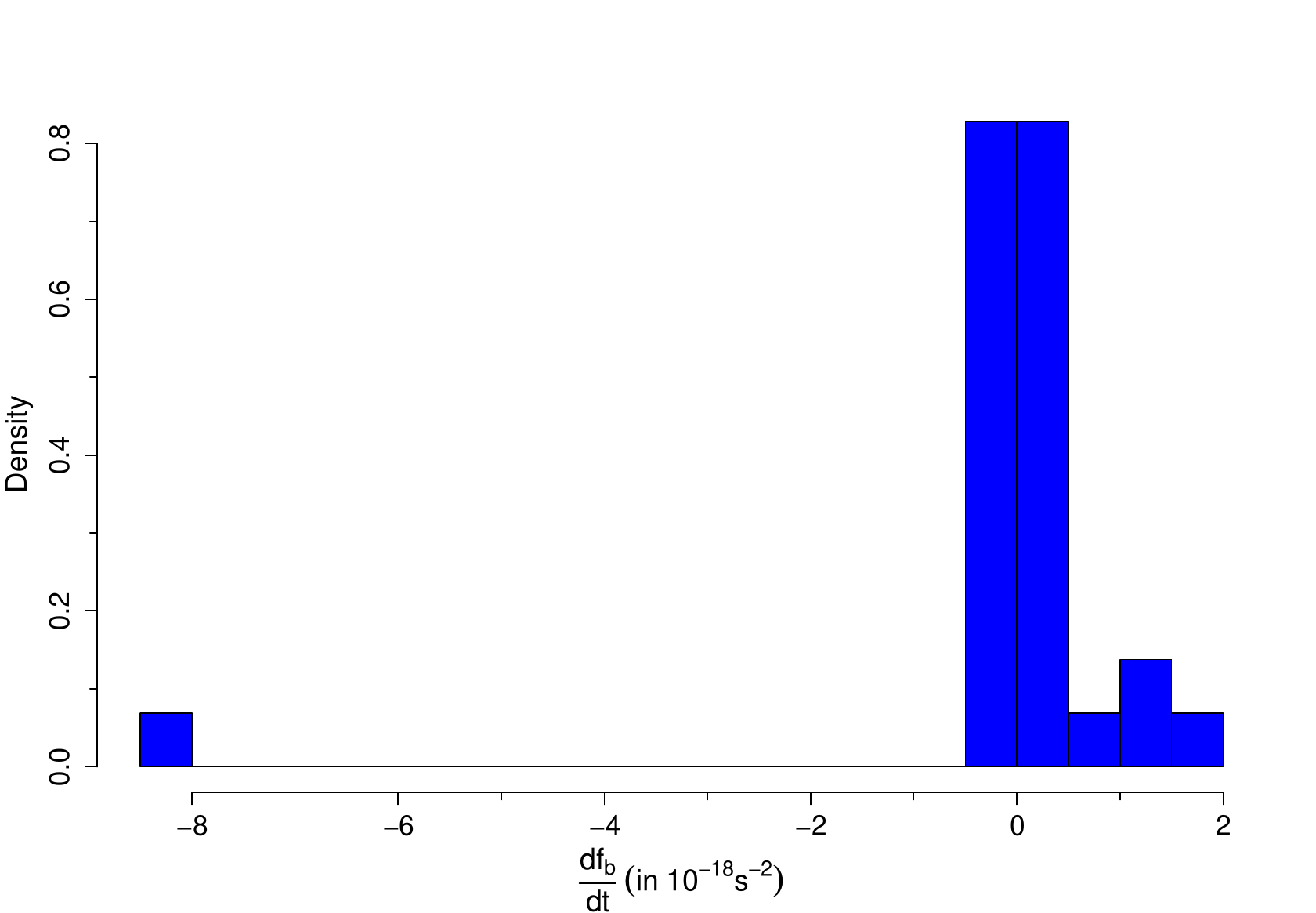}}
\subfigure[distribution of $\dot{f}_{b}$ (synthetic)]{\label{subfig:nogcranf1msp}\includegraphics[width=0.17\textwidth, angle=-90]{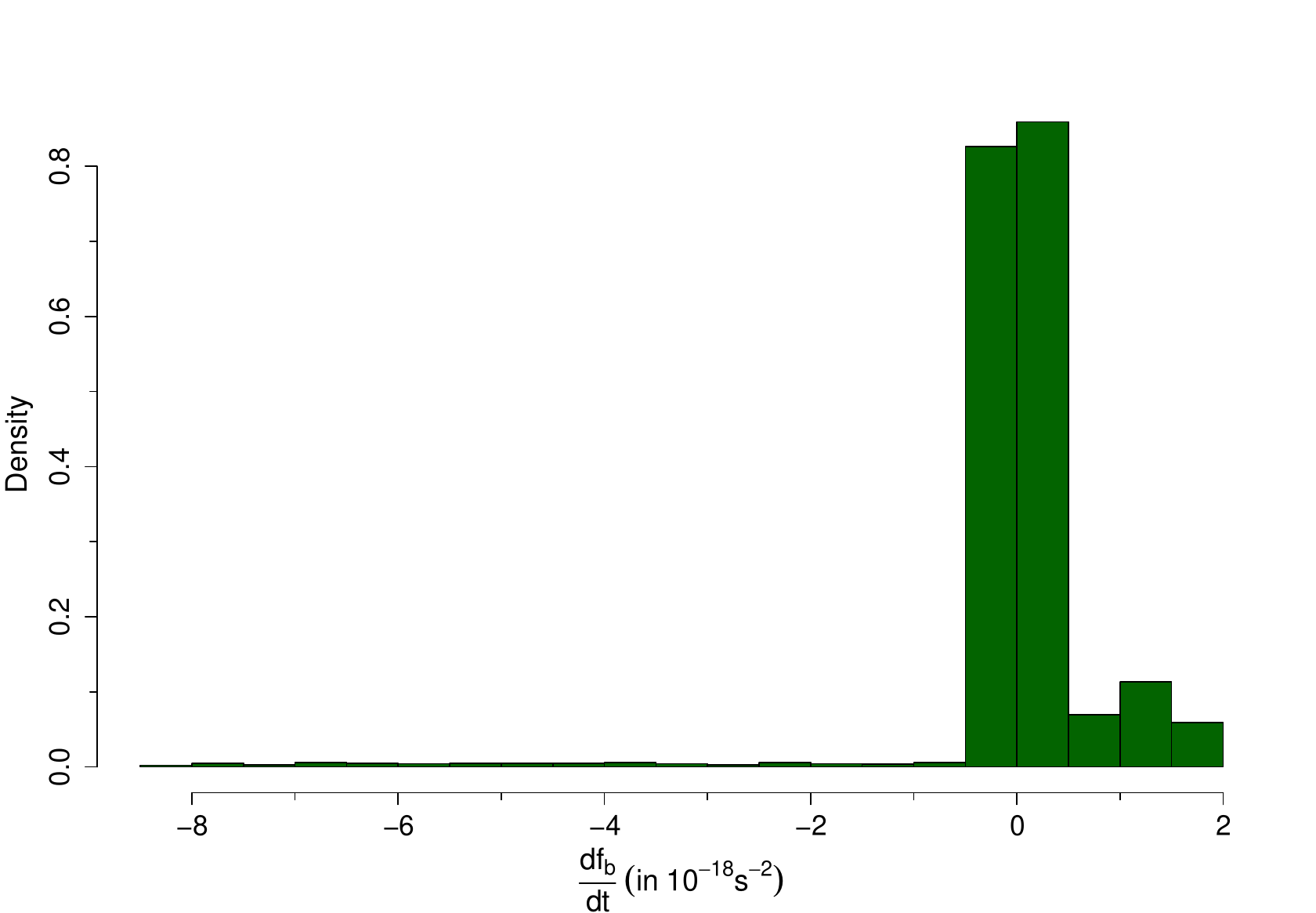}}
\subfigure[distribution of $\ddot{f}_{b}$ (real)]{\label{subfig:nogcatnff2msp}\includegraphics[width=0.17\textwidth, angle=-90]{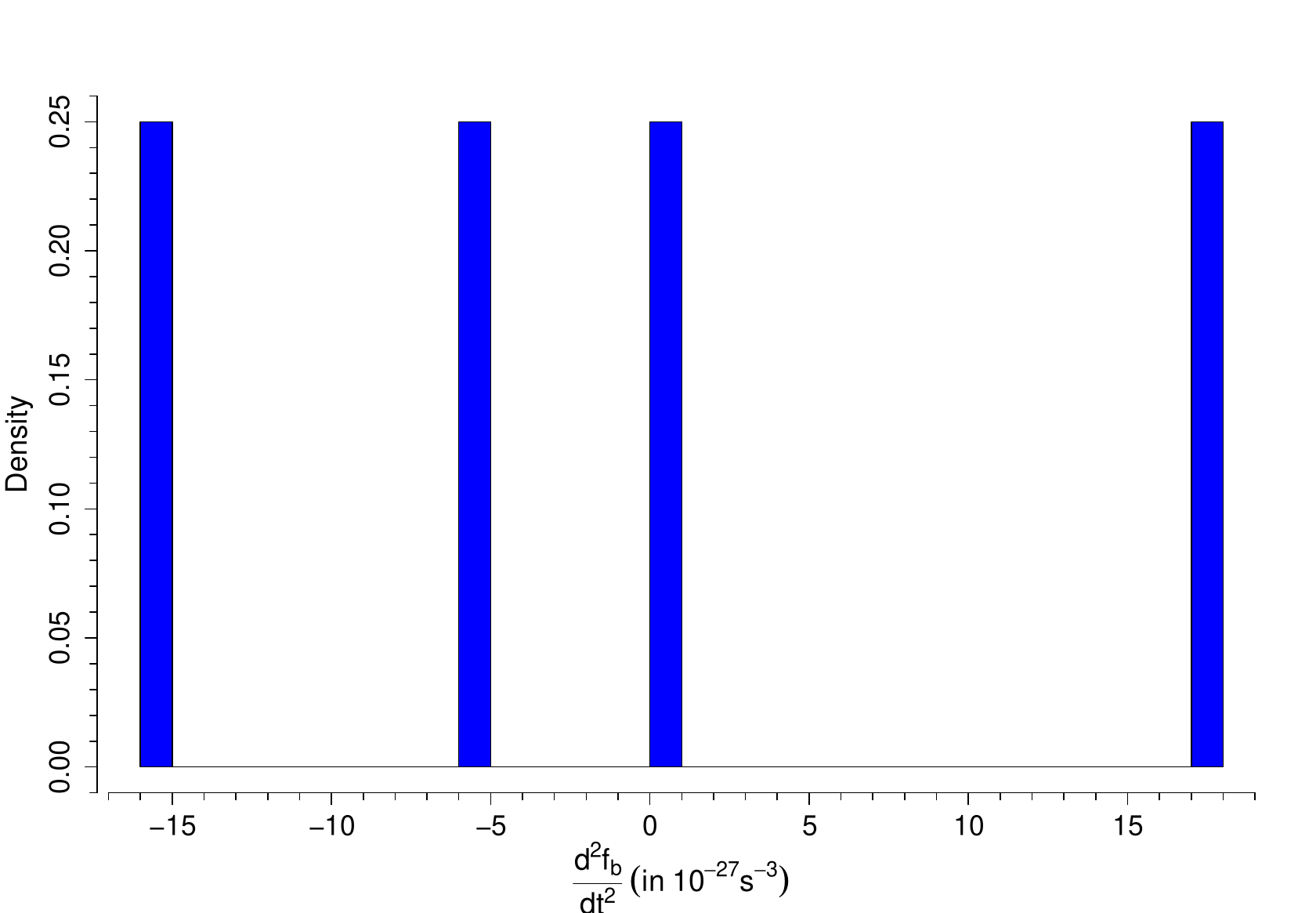}}
\subfigure[distribution of $\ddot{f}_{b}$ (synthetic)]{\label{subfig:nogcranf2msp}\includegraphics[width=0.17\textwidth, angle=-90]{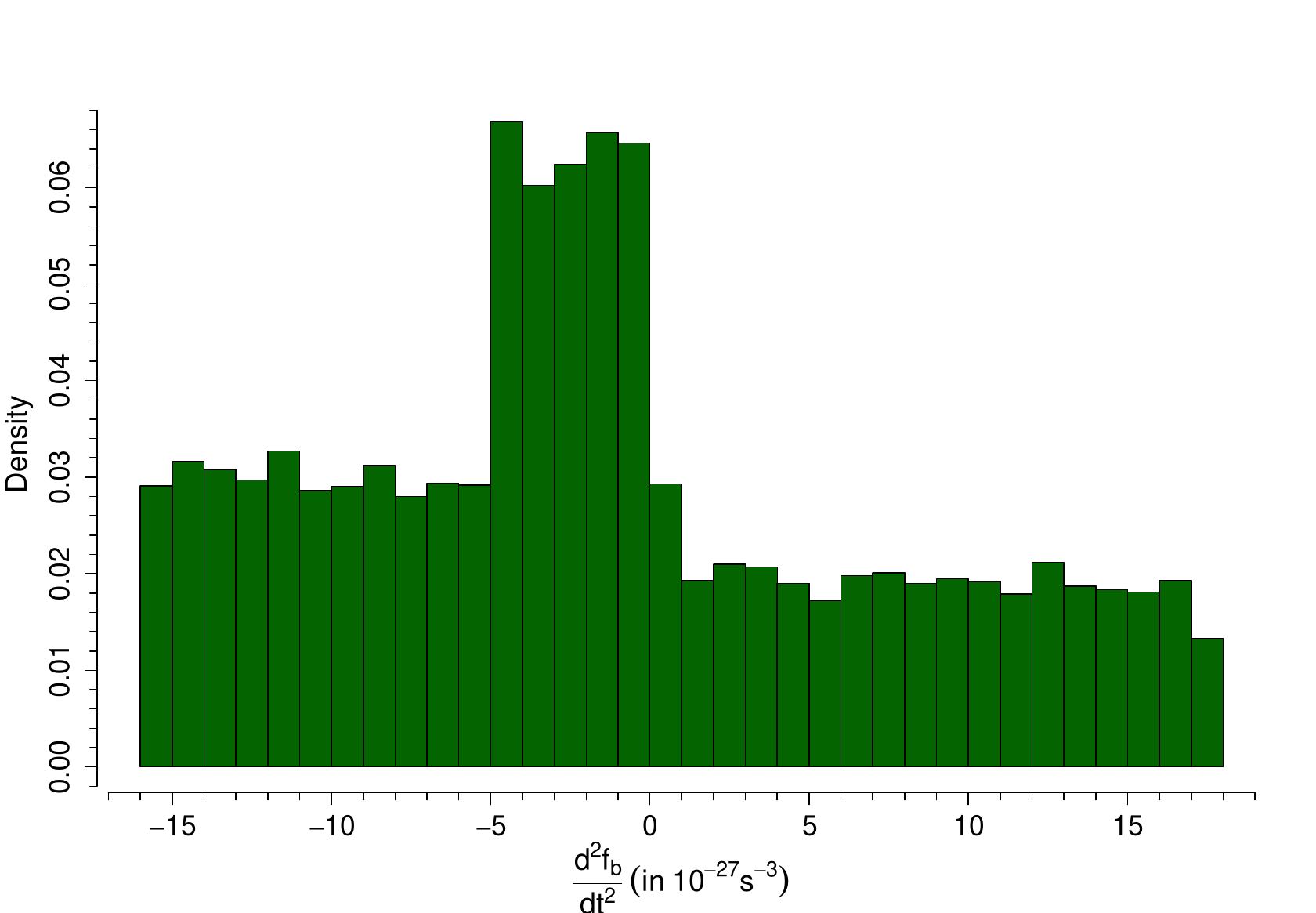}}
\end{center}
\caption[]{Comparison of observed and simulated density distribution of parameters of millisecond pulsars in the Galactic field for which we study orbital frequency and its derivatives.    }
\label{fig:histMSPorb}
\end{figure}

Comparison between the histograms of the parameter values for real pulsars with those of synthetic pulsars are demonstrated in Figs. \ref{fig:histMSPspin} and \ref{fig:histMSPorb}.

\begin{figure}
\begin{center}
\subfigure[]{\label{subfig:nogcatnflmspcdfa}\includegraphics[width=0.3\textwidth, angle=-90]{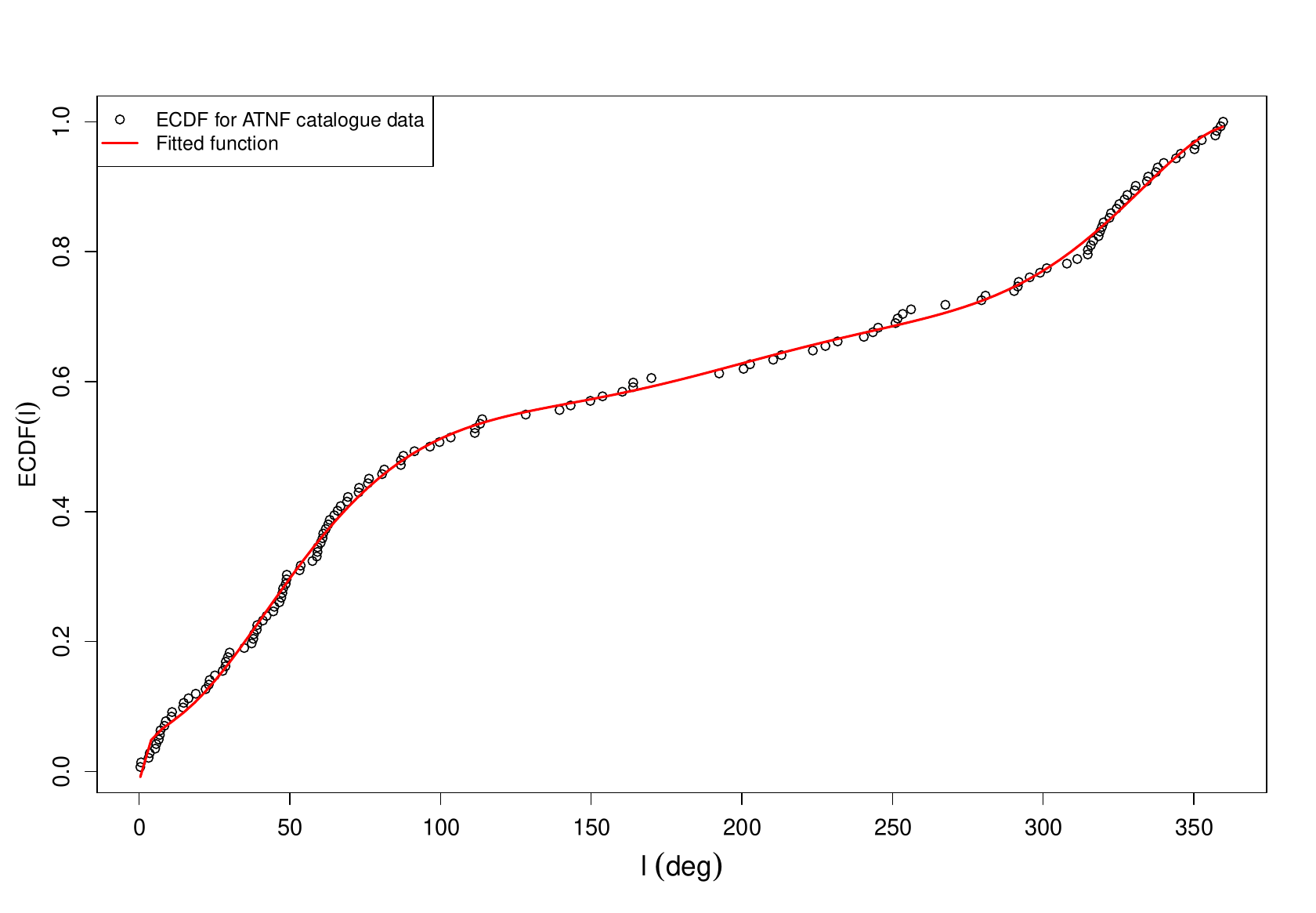}}
 \subfigure[]{\label{subfig:nogcatnfbmspcdfb}\includegraphics[width=0.3\textwidth, angle=-90]{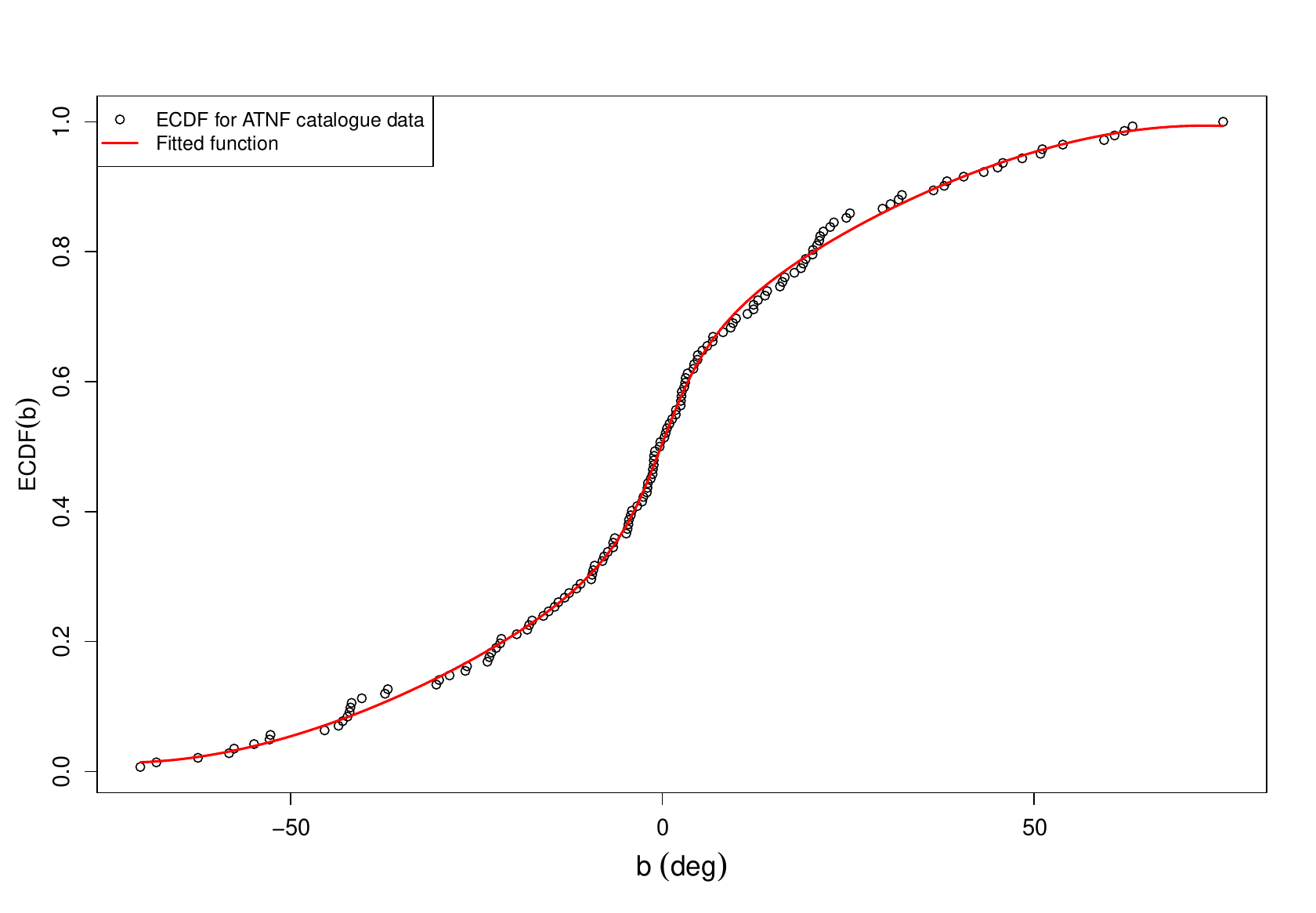}}\\
\subfigure[]{\label{subfig:nogcatnfdmspcdfc}\includegraphics[width=0.3\textwidth, angle=-90]{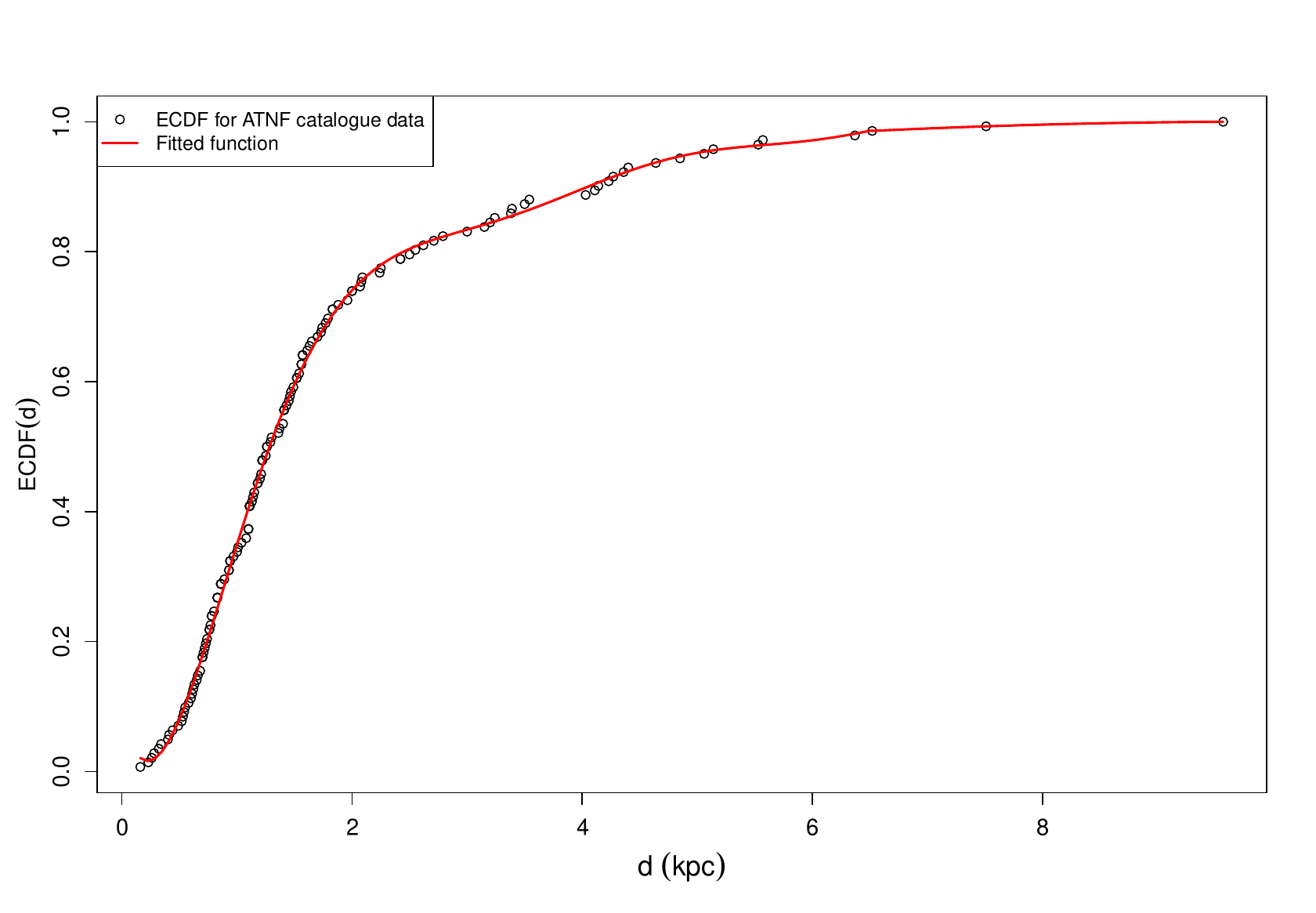}}
 \subfigure[]{\label{subfig:nogcatnfmulmspcdfd}\includegraphics[width=0.3\textwidth, angle=-90]{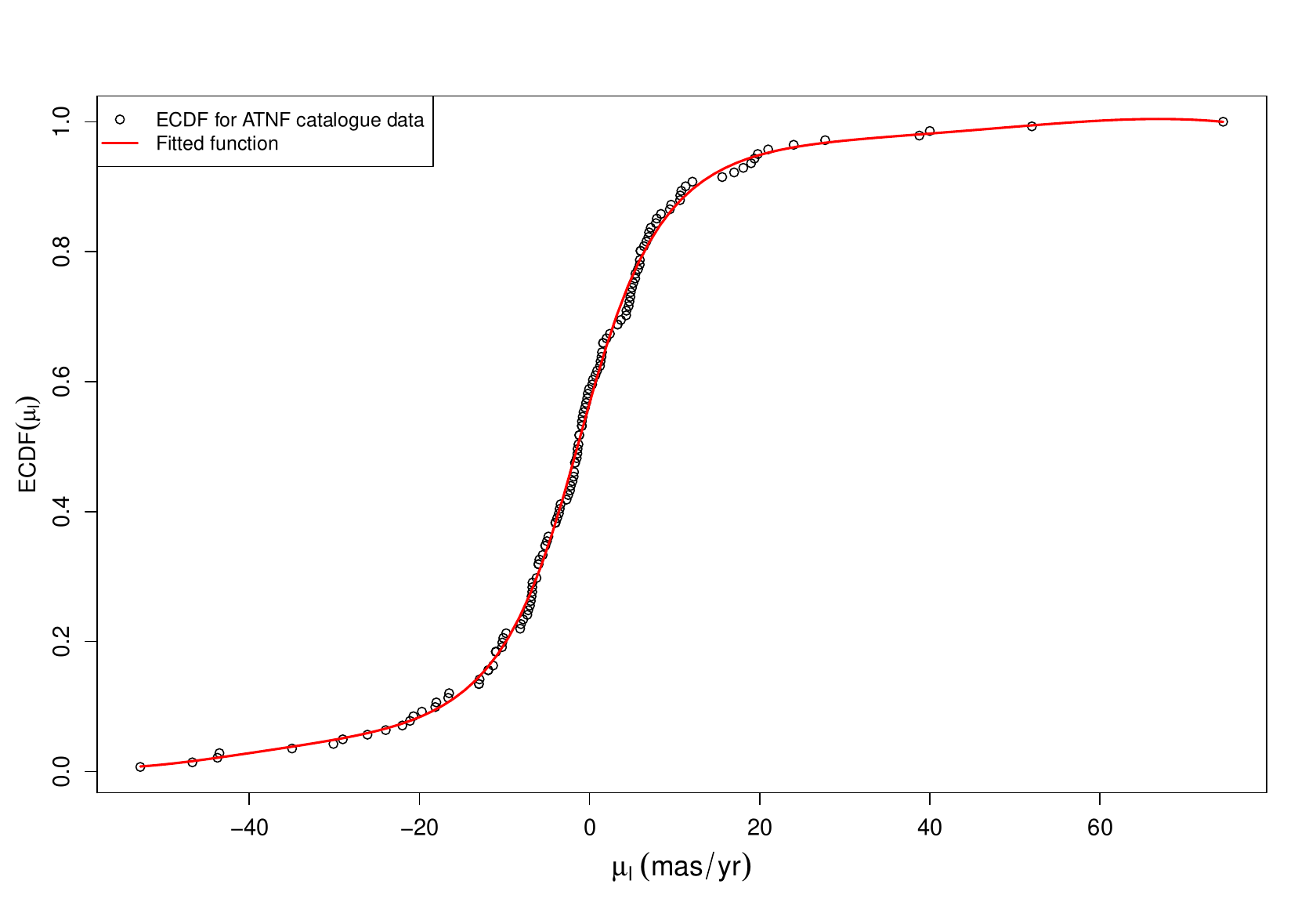}}\\
\subfigure[]{\label{subfig:nogcatnfmubmspcdfe}\includegraphics[width=0.3\textwidth, angle=-90]{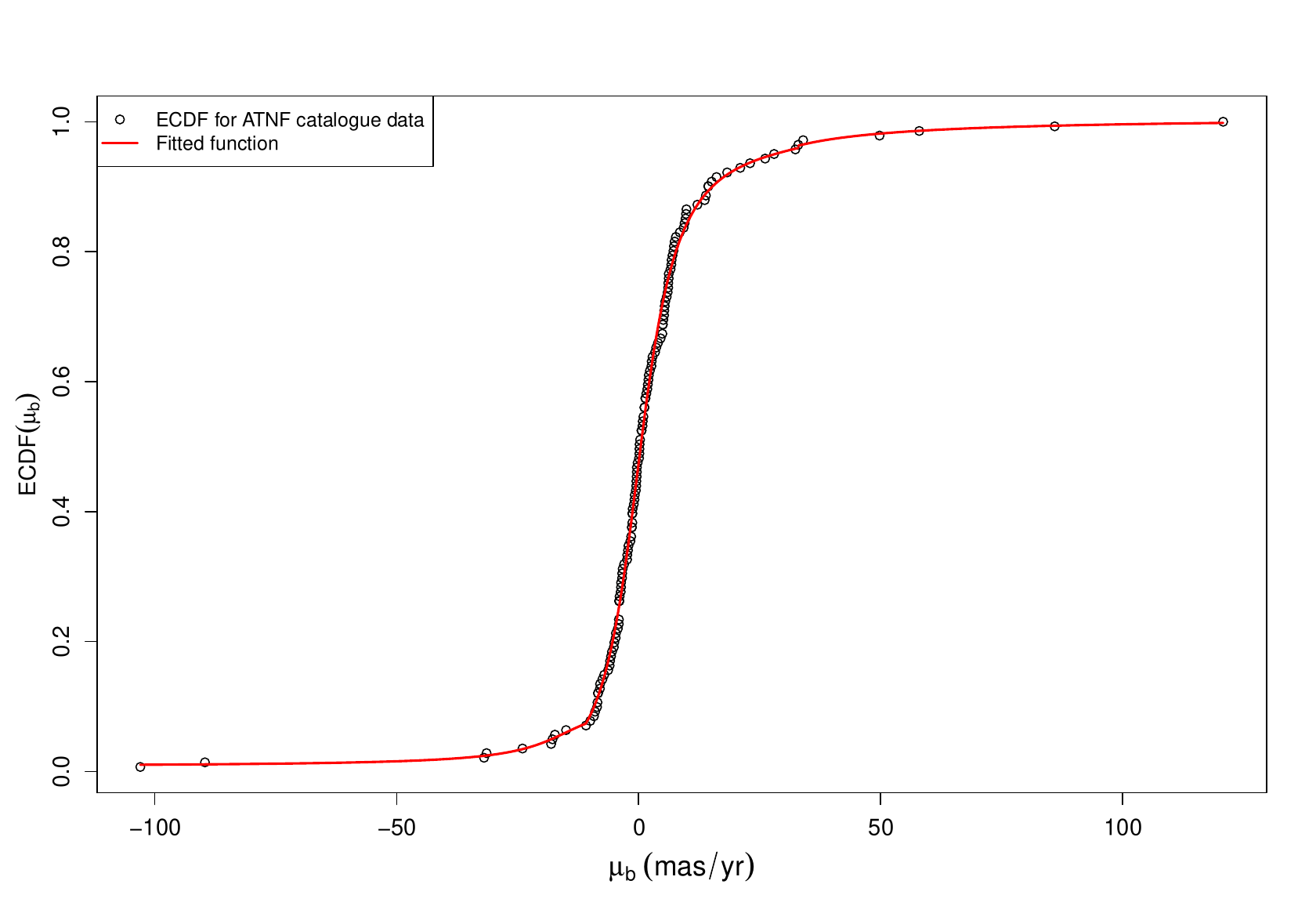}}
\subfigure[]{\label{subfig:nogcatnff0mspcdff}\includegraphics[width=0.3\textwidth, angle=-90]{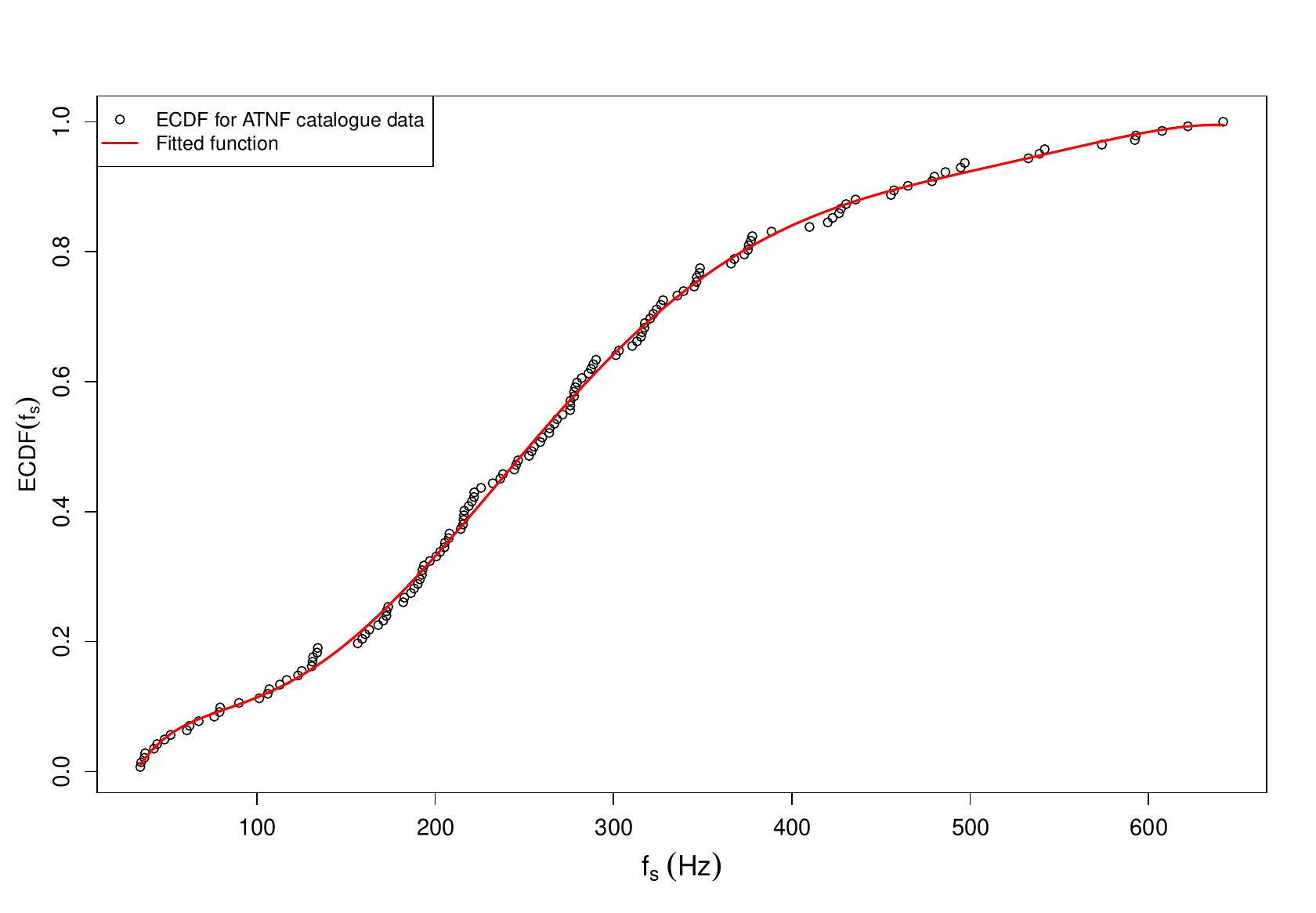}} \\
\subfigure[]{\label{subfig:nogcatnff1mspcdfg}\includegraphics[width=0.3\textwidth, angle=-90]{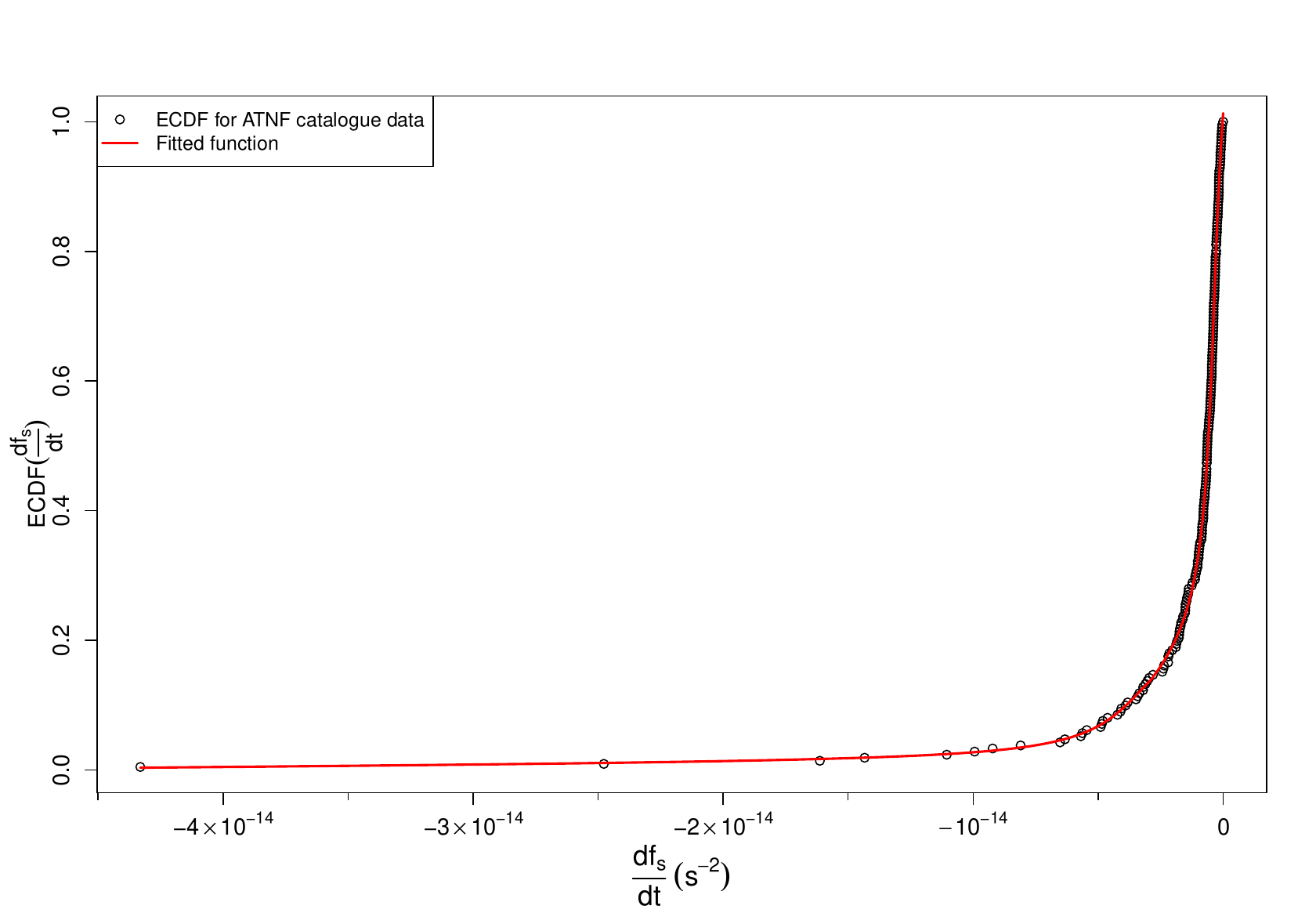}}
\subfigure[]{\label{subfig:nogcatnff2mspcdfh}\includegraphics[width=0.3\textwidth, angle=-90]{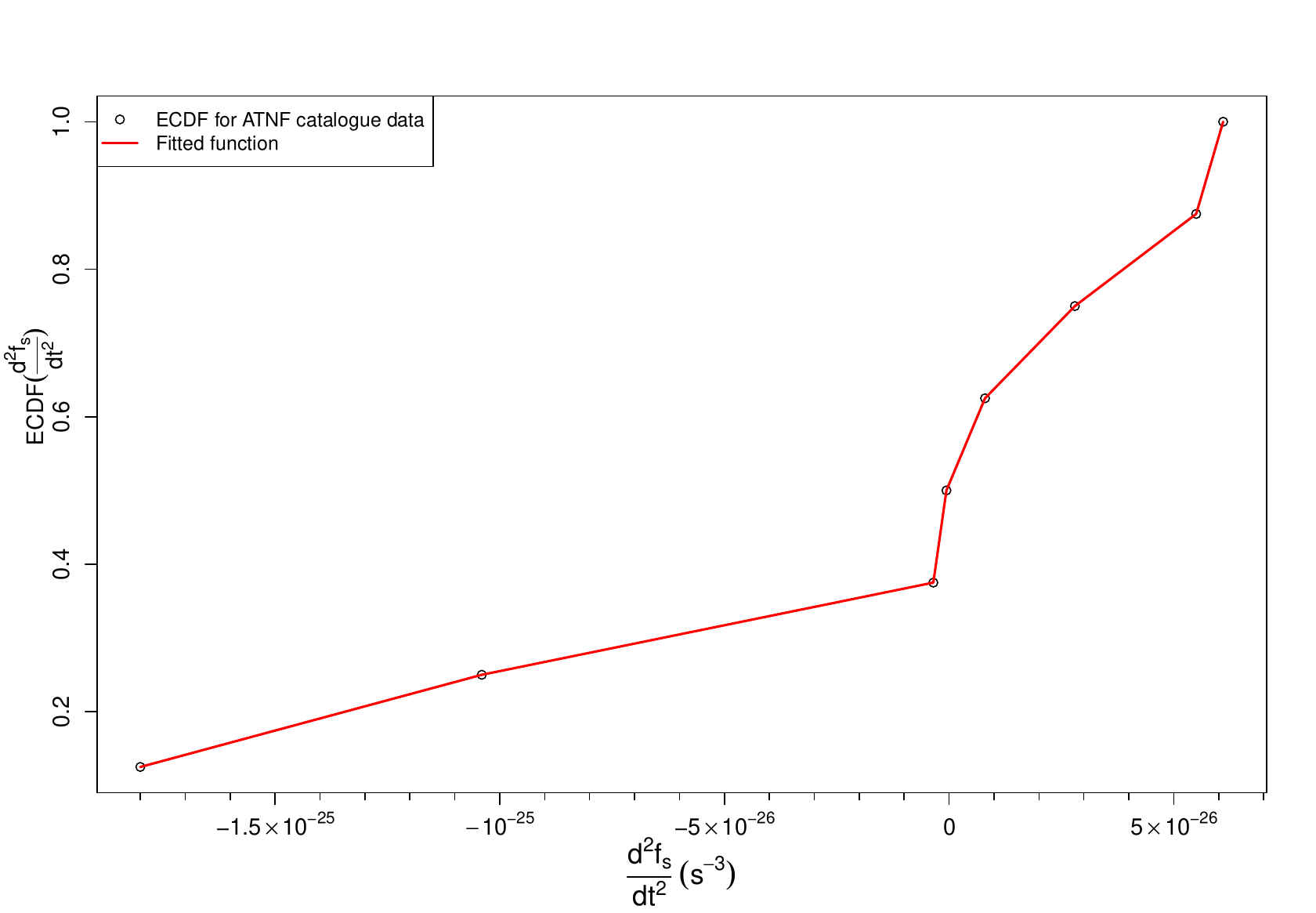}}
\end{center}
\caption[]{ Plots showing the Empirical Cumulative Distribution Function values for the ATNF catalogue parameters along with corresponding fitted functions for- (a) $l$,( b) $b$, (c) $d$, (d) $\mu_{l}$, (e) $\mu_{b}$, (f) $f_{s}$, (g) $\dot{f}_{s}$, and (h) $\ddot{f}_{s}$. This is for the cases when we study spin frequency and its derivatives.    }
\label{fig:CDFMSPspin}
\end{figure}

\begin{figure}
\begin{center}
\subfigure[]{\label{subfig:nogcatnflmsporbcdf}\includegraphics[width=0.3\textwidth, angle=-90]{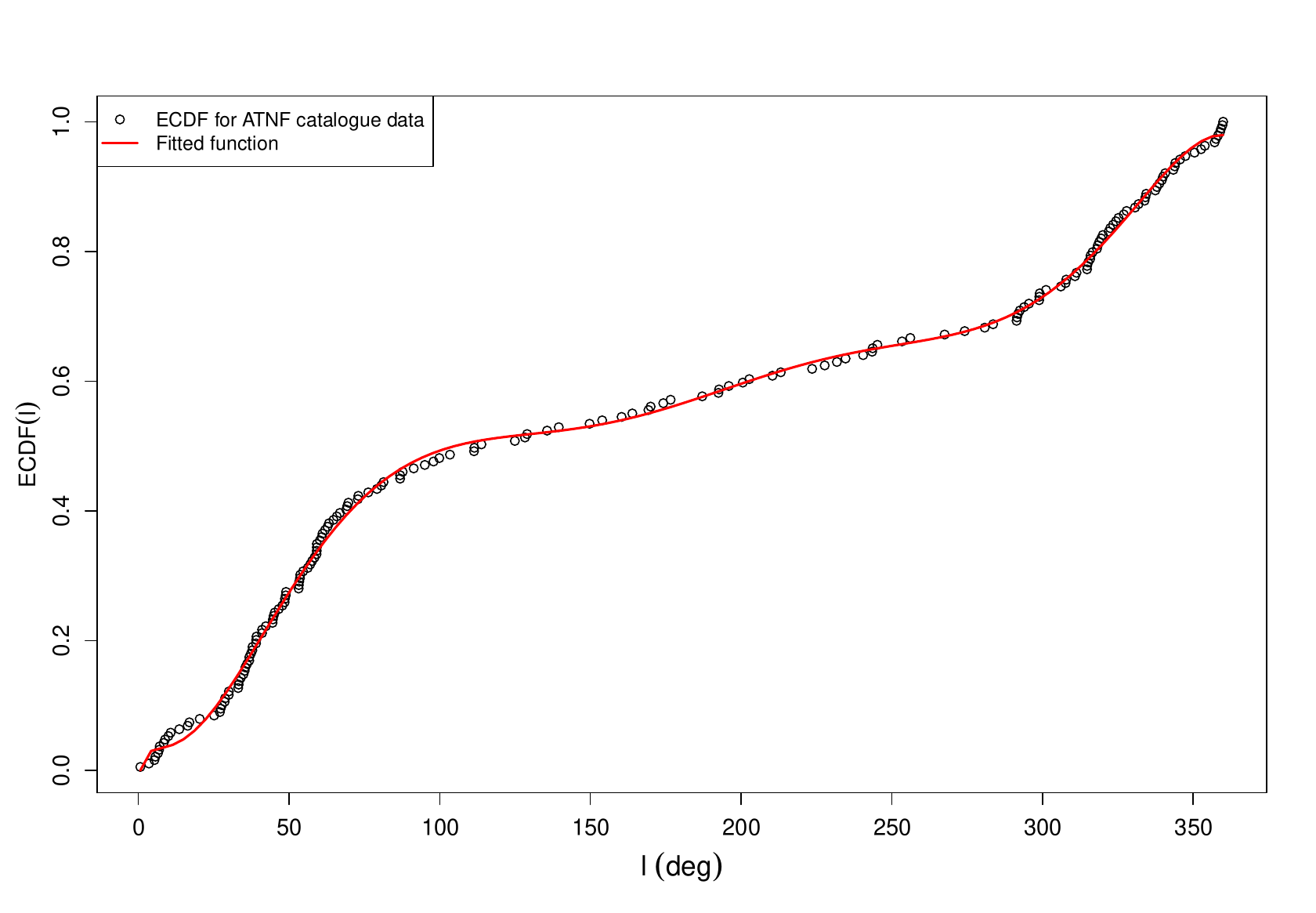}}
 \subfigure[]{\label{subfig:nogcatnfbmsporbcdf}\includegraphics[width=0.3\textwidth, angle=-90]{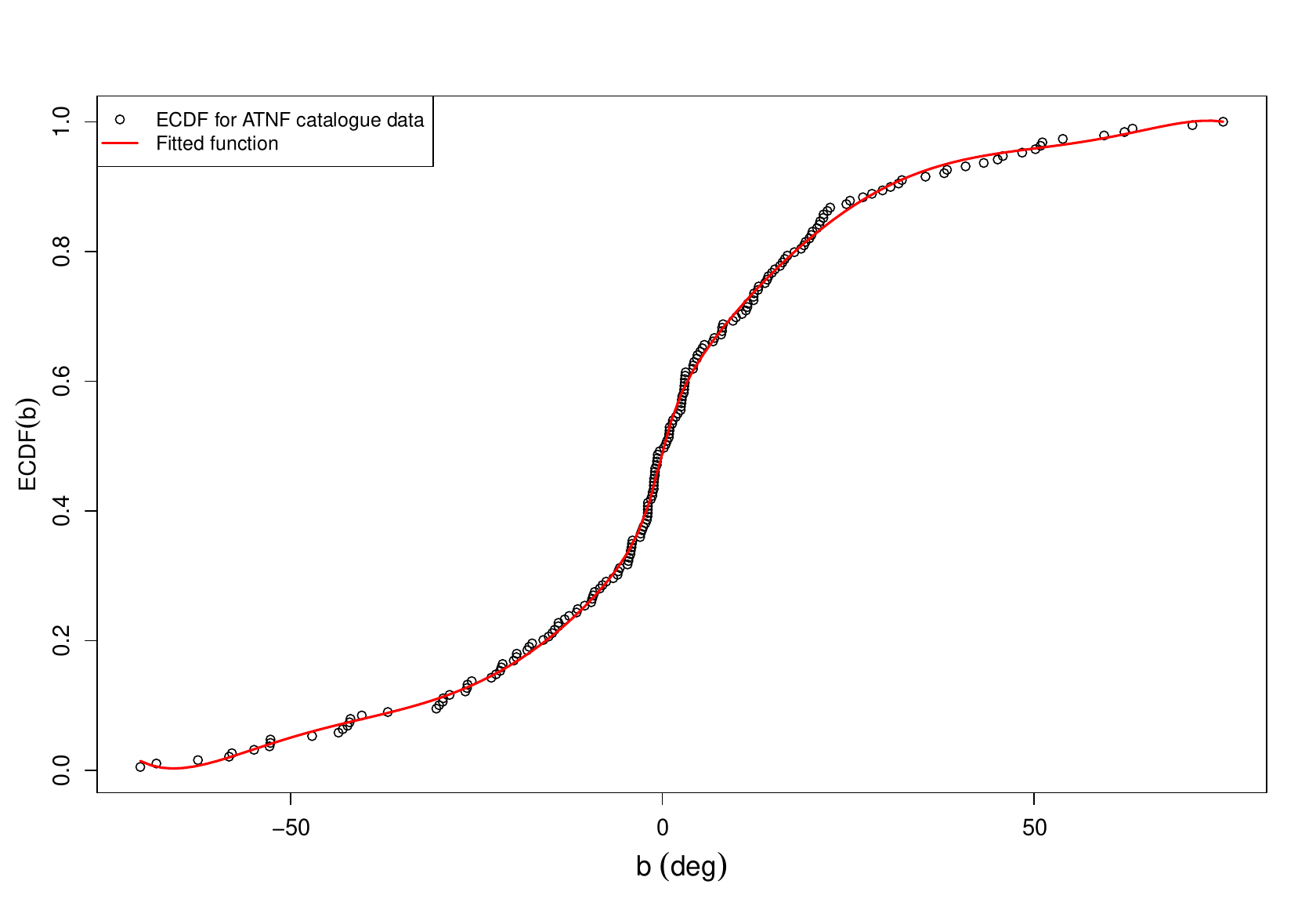}}\\
\subfigure[]{\label{subfig:nogcatnfdmsporbcdf}\includegraphics[width=0.3\textwidth, angle=-90]{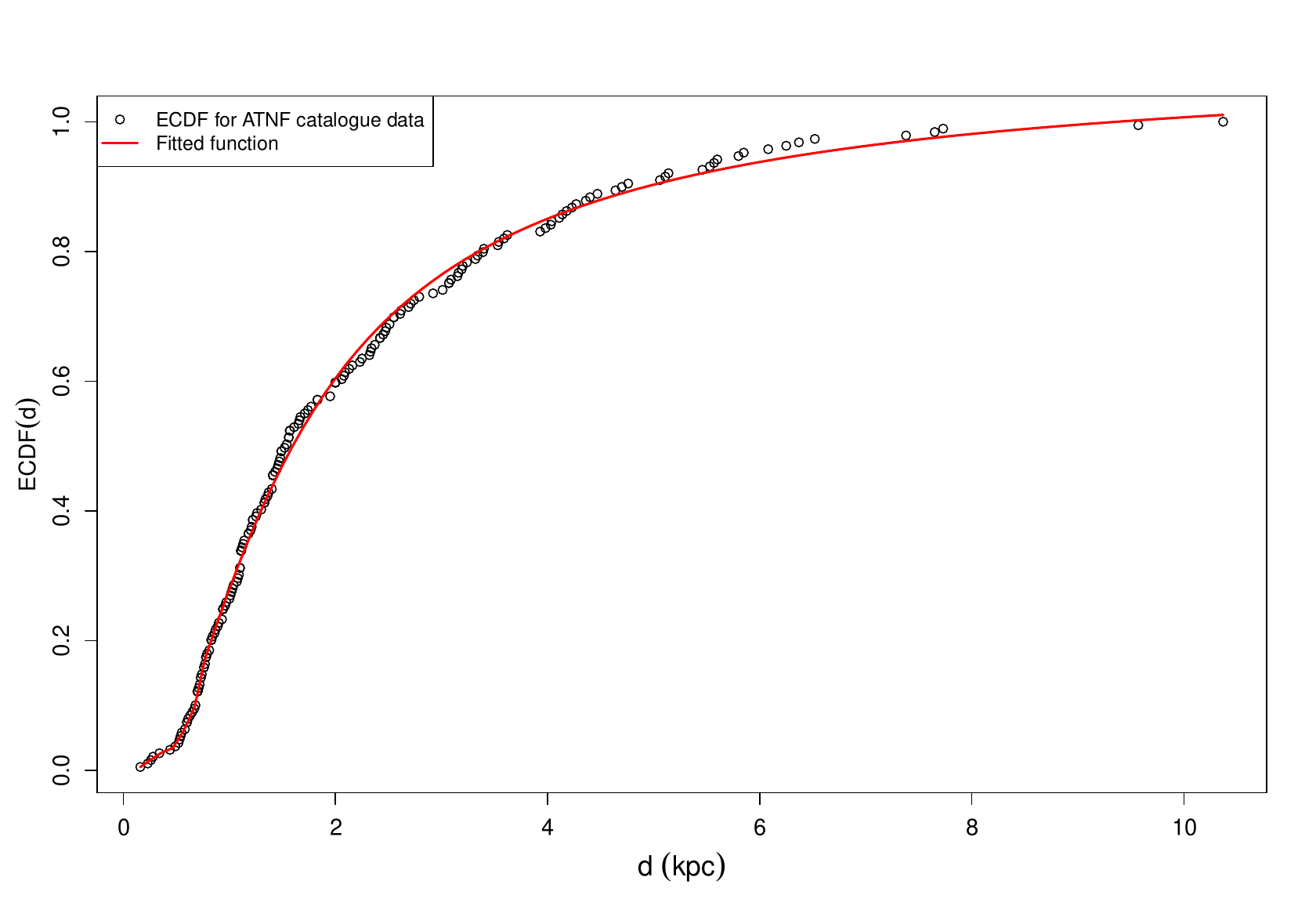}}
 \subfigure[]{\label{subfig:nogcatnfmulmsporbcdf}\includegraphics[width=0.3\textwidth, angle=-90]{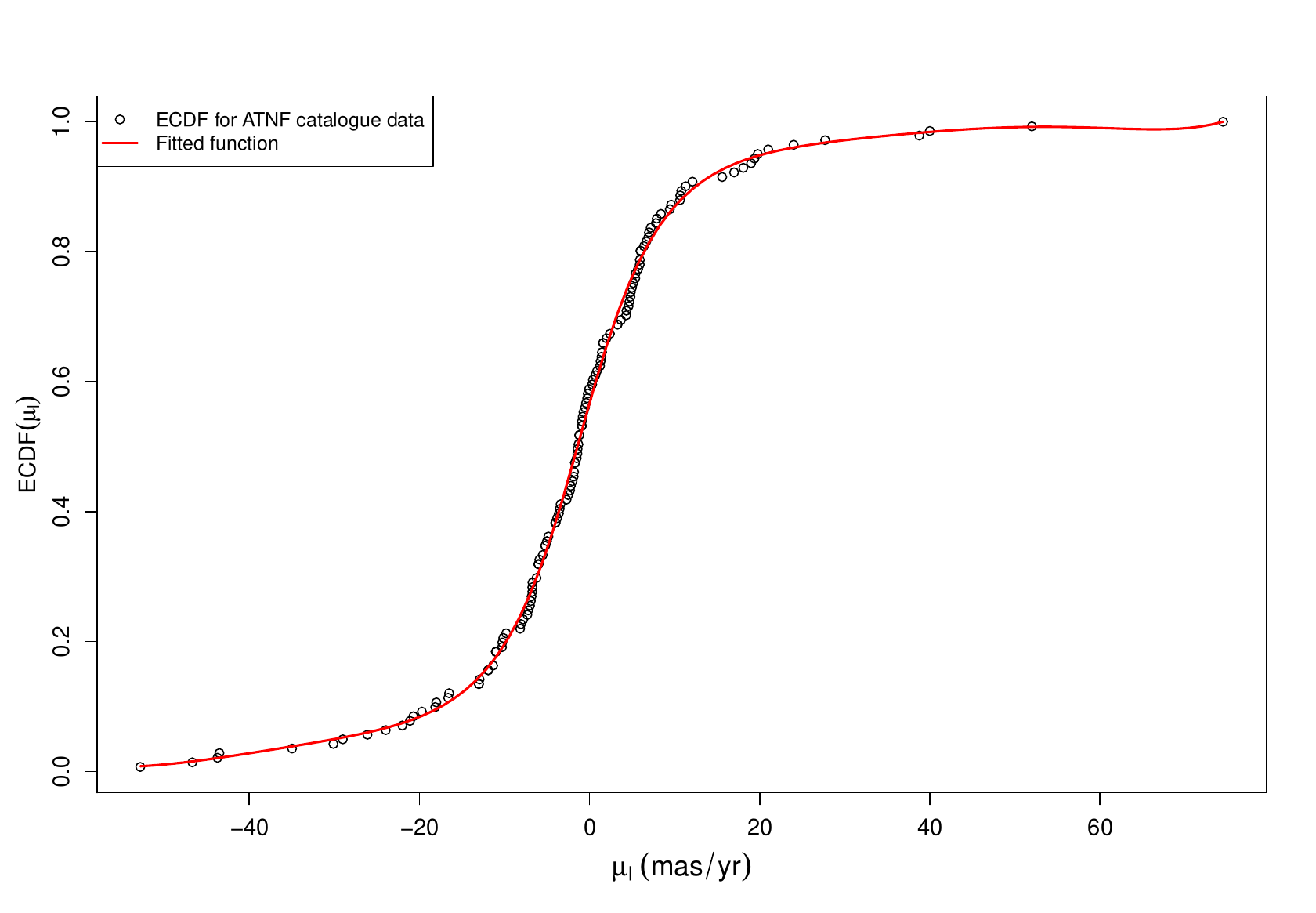}}\\
\subfigure[]{\label{subfig:nogcatnfmubmsporbcdf}\includegraphics[width=0.3\textwidth, angle=-90]{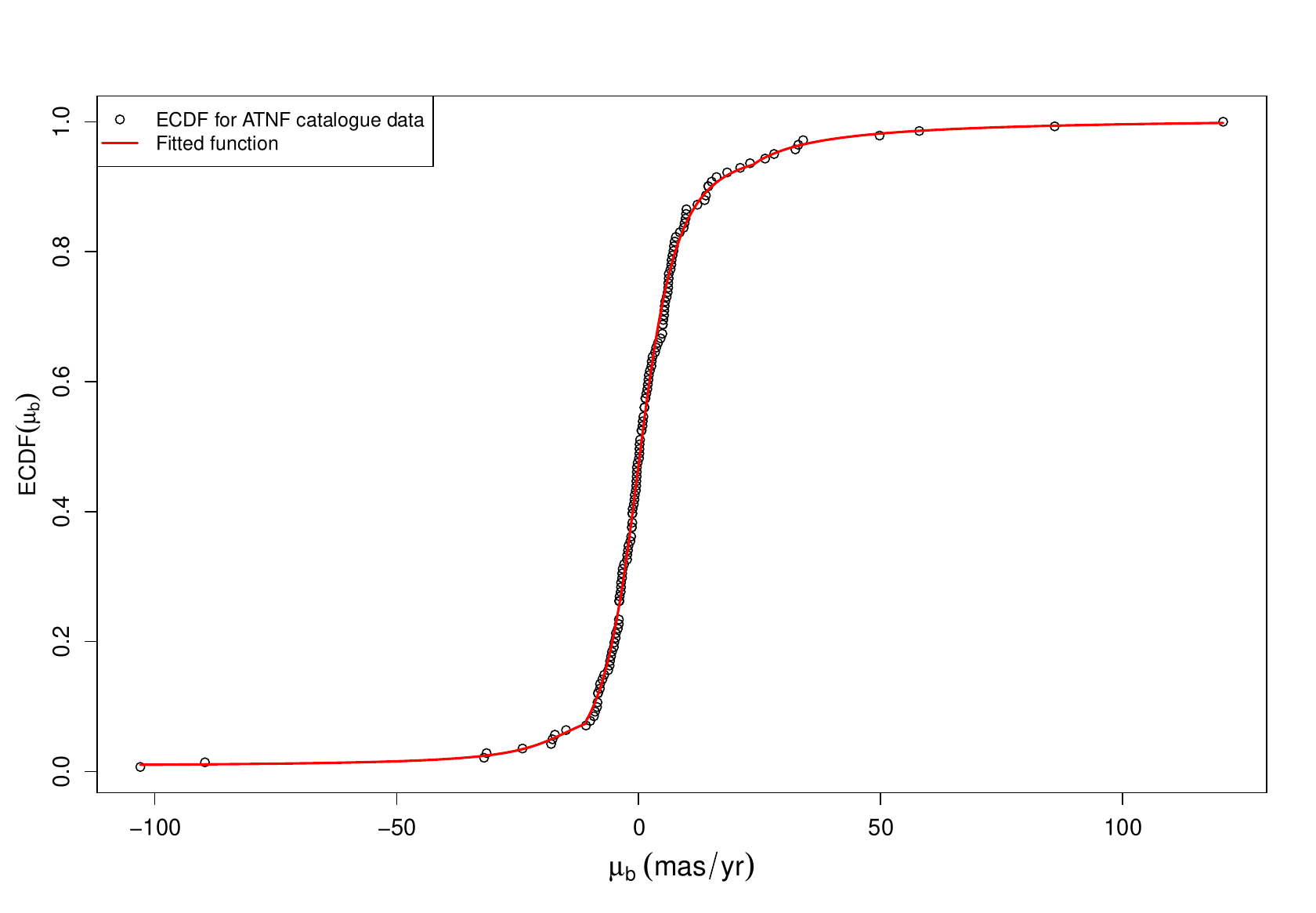}}
\subfigure[]{\label{subfig:nogcatnff0msporbcdf}\includegraphics[width=0.3\textwidth, angle=-90]{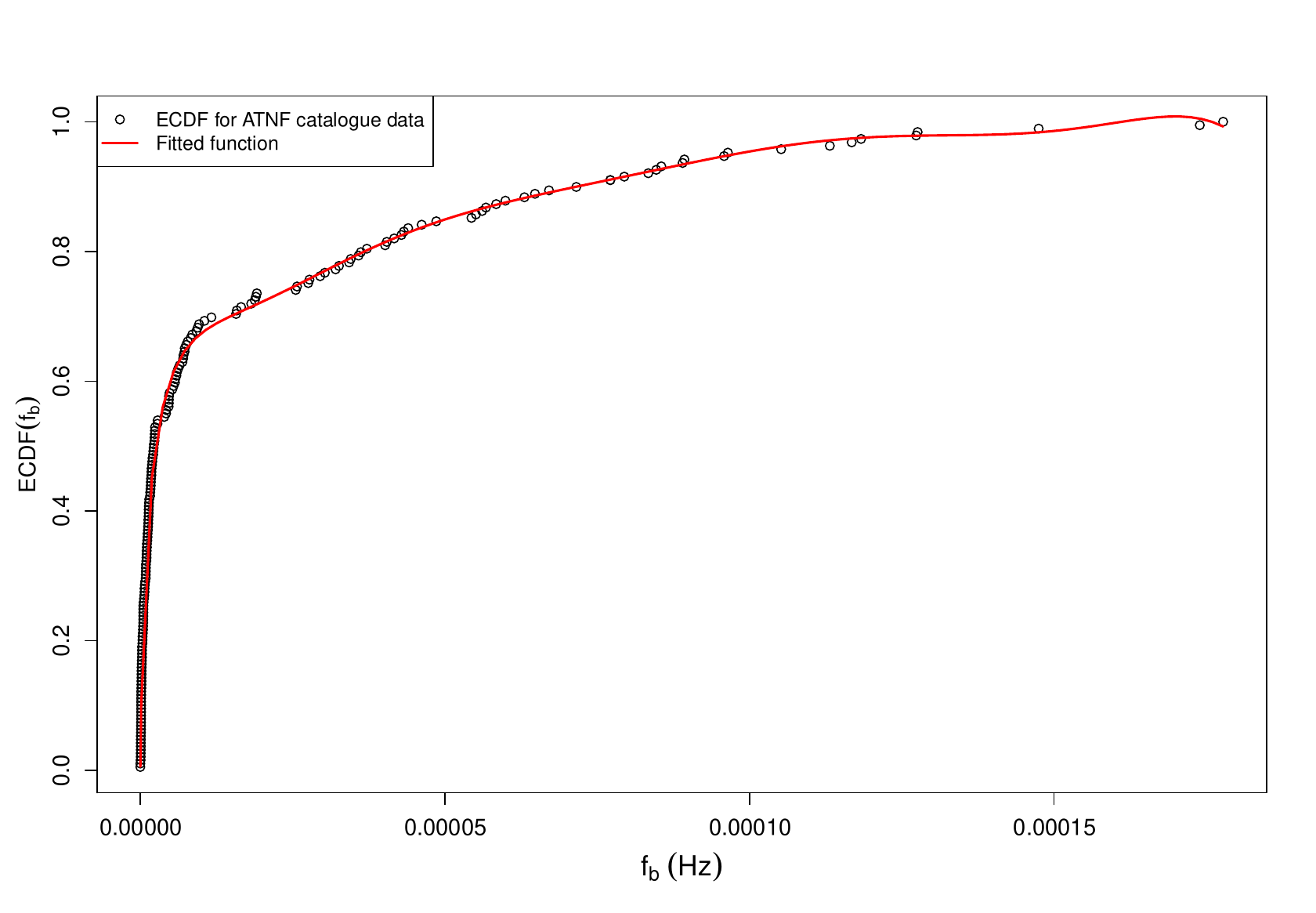}} \\
\subfigure[]{\label{subfig:nogcatnff1msporbcdf}\includegraphics[width=0.3\textwidth, angle=-90]{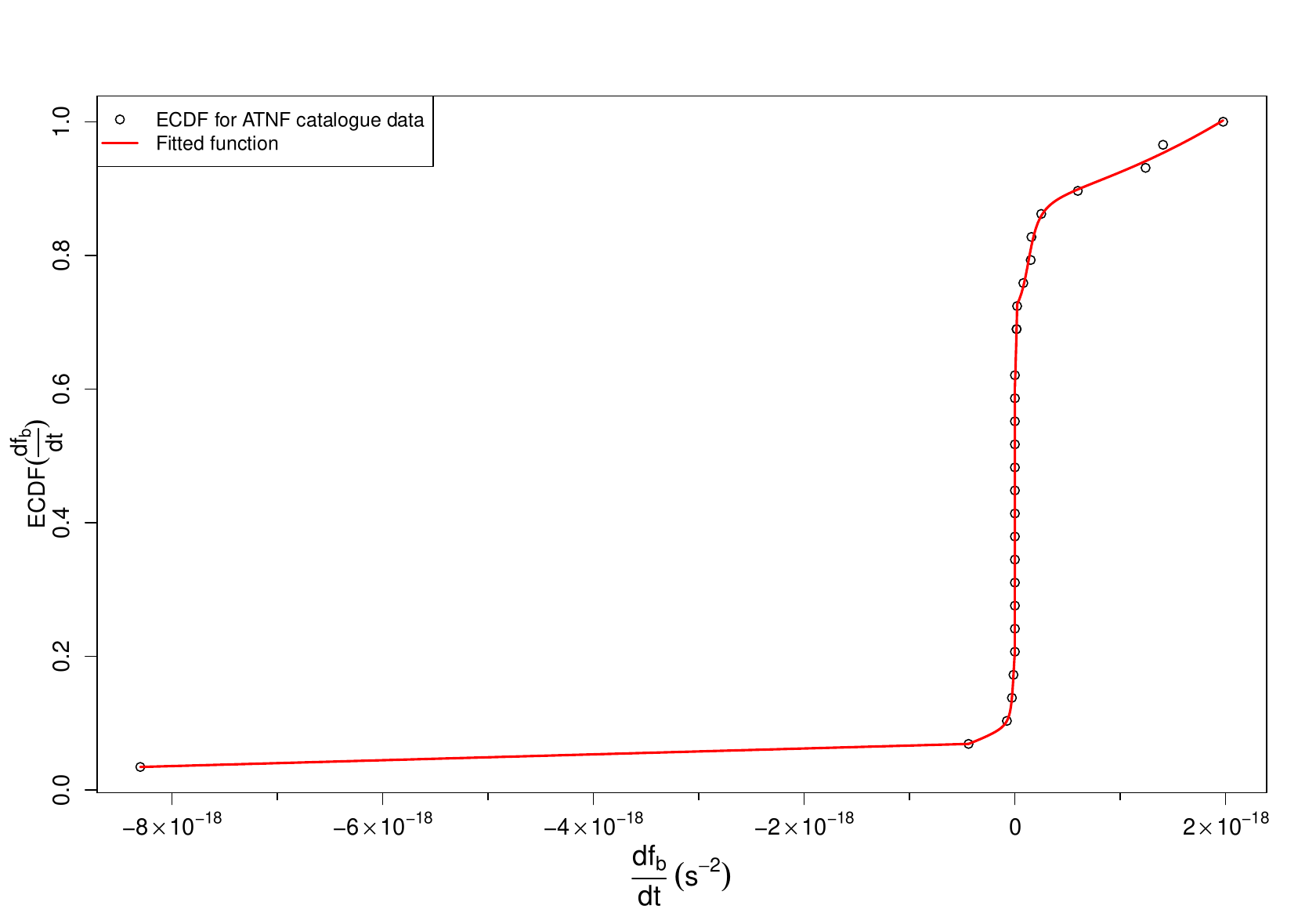}}
\subfigure[]{\label{subfig:nogcatnff2msporbcdf}\includegraphics[width=0.3\textwidth, angle=-90]{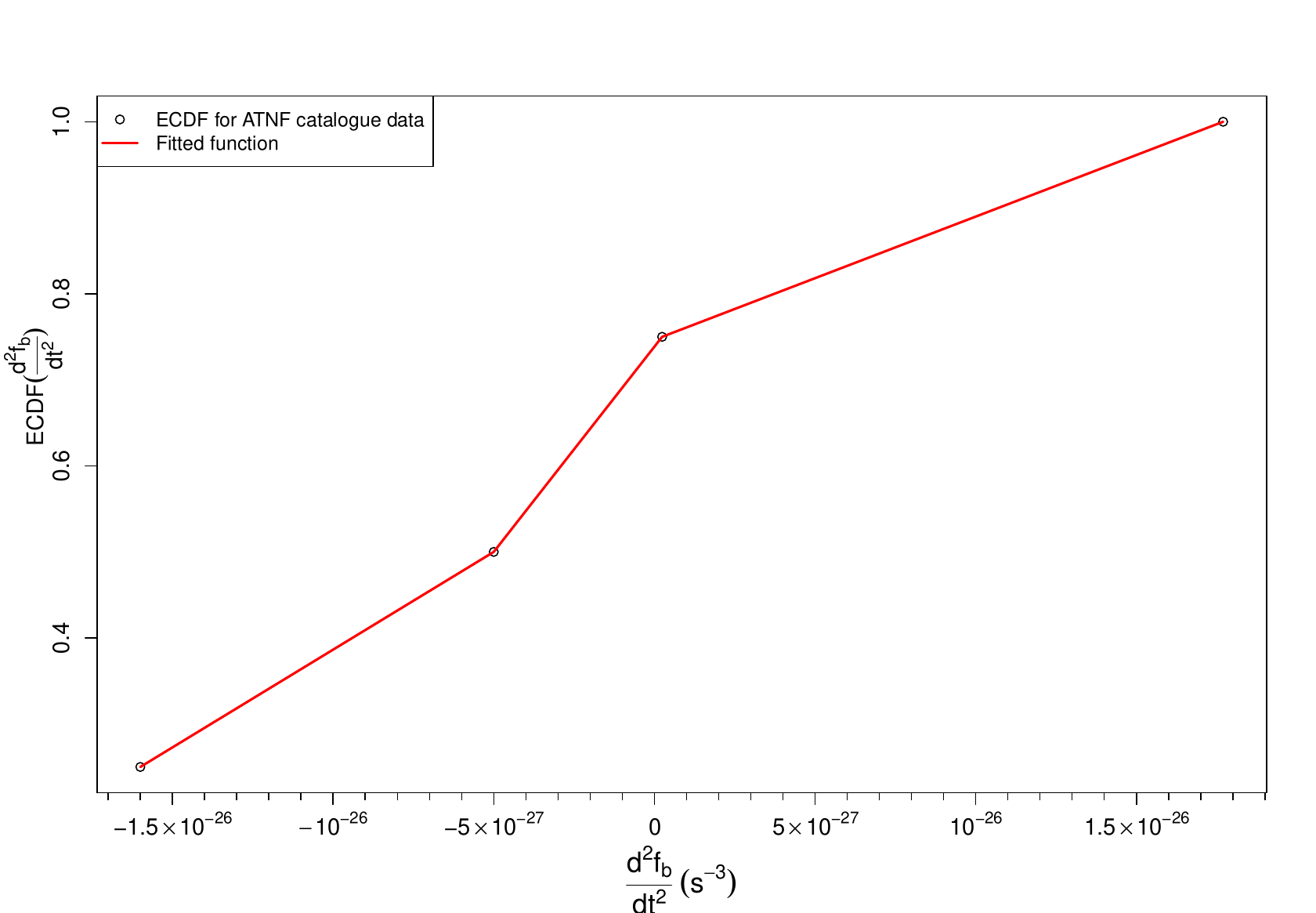}}
\end{center}
\caption{Plots showing the Empirical Cumulative Distribution Function values for the ATNF catalogue parameters along with corresponding fitted functions for- (a) $l$, (b) $b$, (c) $d$, (d) $\mu_{l}$, (e) $\mu_{b}$, (f) $f_{b}$, (g) $\dot{f}_{b}$, and (h) $\ddot{f}_{b}$. This is for the cases when we study orbital frequency and its derivatives. }
\label{fig:CDFMSPorb}
\end{figure}

Moreover, for each case, we also fit analytical functions to the ECDFs that can be used for future simulations. In Figs. \ref{fig:CDFMSPspin} and \ref{fig:CDFMSPorb}, we show the ECDFs of the catalogued parameters (black circles) and the corresponding fitted analytical functions (red line).  The analytical functions fitted for the ECDFs are given as follows:

\vskip 0.4cm

A) MSPs where we only study spin period and its derivatives

\vskip 0.2cm

For $l$,

\vskip 0.1cm

\begin{align}
f(x) = a + b x + c x^2 + d x^3 + e x^4 + f x^5 + g x^6 + h x^7  + i x^{0.5}
\label{eq:CDFlspin}
\end{align} where x is $l$ in degrees, a = -0.05297, b = -0.01771, c = 4.909$\times10^{-4}$, d = -6.360$\times10^{-6}$, e = 4.213$\times10^{-8}$, f = -1.506$\times10^{-10}$, g = 2.768$\times10^{-13}$, h = -2.046$\times10^{-16}$, and i = 0.08246.\\

For $b$,
\begin{align}
f(x) = a + b x + c x^3 + d \tan^{-1} (e x + f),
\label{eq:CDFbspin}
\end{align} where x is $b$ in degrees, a = 0.5038, b = 5.950$\times10^{-3}$, c = - 3.828$\times10^{-7}$, d = 0.1371, e = 0.1779, and f = 0.01238.\\

For $d$,
\begin{align}
f(x) = &a_1 + b_1 x + c_1 x^2  + d_1 x^3  + e_1 x^4 + f_1 x^5 + g_1 x^6 + h_1 \cos ( i_1 x + j_1 ) \,\,\,\,\text{for }0.16\leq x<6.52 \nonumber \\
       &a_2 + b_2 x + c_2 x^2 \,\,\,\,\hskip7.3cm\text{for } 6.52 \leq  x\leq9.57,
\label{eq:CDFdspin}
\end{align} where x is $d$ in kpc, $a_1$ = - 0.1793860, $b_1$ =  0.7355924, $c_1$ = 1.8799532, $d_1$ = - 1.3885579, $e_1$ = 0.3305595, $f_1$ = - 0.0325702, $g_1$ = 0.0011461, $h_1$ = 1.3421374, $i_1$ = 1.0323664, $j_1$ = 1.3759470, $a_2$ =  0.880219, $b_2$ = 0.024110, and $c_2$ = -0.001211.\\

For $\mu_{l}$,
\begin{align}
f(x) = a + b x + c x^3 + d x^4 + e x^5 + f x^6 + g \tan^{-1}( h x + i ),
\label{eq:CDFmulspin}
\end{align} where x is $\mu_{l}$ in mas/yr, a = 0.5076, b = - 4.955$\times10^{-3}$, c = 1.290$\times10^{-6}$, d = - 3.63$\times10^{-9}$, e = - 1.603$\times10^{-10}$, f = -8.583$\times10^{-13}$, g = 0.4475, h = 0.1171, and i = 0.1330.\\

For $\mu_{b}$,
\begin{align}
f(x) = &a_1 + b_1 \tan^{-1}( c_1 x + d_1 ) \,\,\,\,\hskip6.5cm\text{for }-103.0  \leq x< -10.88 \nonumber \\
       &a_2 + b_2 x + c_2 x^2  + d_2 x^3  + e_2 x^4 + f_2 x^5 + g_2 x^6 + h_2 \tan^{-1}( i_2 x + j_2 ) \,\,\,\,\text{for } -10.88  \leq x< 33.0 \nonumber \\
       &a_3 + b_3 \tan^{-1}( c_3 x + d_3 ) \,\,\,\,\hskip6.5cm\text{for } 33. \leq x \leq 120.82,
\label{eq:CDFmubspin}
\end{align} where x is $\mu_{b}$ in mas/yr, $a_1$ = 0.059922, $b_1$ = -0.033813, $c_1$ = -0.104716, $d_1$ = -1.579770, $a_2$ = 0.4534, $b_2$ = -4.695$\times10^{-3}$, $c_2$ = 5.246$\times10^{-5}$, $d_2$ = 5.883$\times10^{-7}$, $e_2$ = -9.476$\times10^{-9}$, $f_2$ = -2.872$\times10^{-11}$, $g_2$ = 4.542$\times10^{-13}$, $h_2$ = 0.4332, $i_2$ = 0.1494, $j_2$ = 0.04927, $a_3$ = -1610., $b_3$ = 1026., $c_3$ = 981., and $d_3$ = -9258.\\

For $f_{s}$,
\begin{align}
f(x) = a + b x + c x^3 + d x^4 + e x^5 + f x^6 + g x^7 + h\, {\rm log}(i x + j ),
\label{eq:CDFfspin}
\end{align} where x is $f_{s}$ in Hz, a = - 1.536, b = - 7.118$\times10^{-3}$, c = 2.528$\times10^{-7}$, d = - 1.110$\times10^{-9}$, e = 2.036$\times10^{-12}$, f = - 1.731$\times10^{-15}$, g = 5.566$\times10^{-19}$, h = 0.4535, i = 1.181, and j = 10.19.\\

For $\dot{f}_{s}$,
\begin{align}
f(x) = &a_1 + b_1 x  + c_1 \tan^{-1} ( d_1 x + e_1 ) \,\,\,\,\hskip4.5cm \text{for } -4.331143\times10^{-14}  \leq x<-2.805\times10^{-15} \nonumber \\
       &a_2 + b_2 x  + c_2 x^3 + d_2 x^4 + e_2 x^5 + f_2 x^6 + g_2\, \tan^{-1}(h_2 x + i_2 )  \,\,\,\,\text{for x} -2.805\times10^{-15} \leq x \leq 8.\times10^{-19},
\label{eq:CDFfdotspin}
\end{align} where x is $\dot{f}_{s}$ in s$^{-2}$,  $a_1$ = 0.1255, $b_1$ = 2.440$\times10^{11}$ $c_1$ = -0.07292, $d_1$ = -5.593$\times10^{14}$, $e_1$ = -1.817, $a_2$ = 0.7395, $b_2$ = 4.241$\times10^{13}$, $c_2$ = -6.675$\times10^{41}$, $d_2$ = -6.064$\times10^{55}$, $e_2$ = -1.981$\times10^{69}$, $f_2$ = -2.139$\times10^{82}$, $g_2$ =  0.3414, $h_2$ = 2.892$\times10^{15}$,and $i_2$ = 1.031. \\

For $\ddot{f}_{s}$,
\begin{align}
f(x) = &a_1 + b_1  x \,\,\,\,\text{for } -1.8\times10^{-25} \leq x<-1.04\times10^{-25} \nonumber \\
       &a_2 + b_2  x \,\,\,\,\text{for } -1.04\times10^{-25} \leq x<-3.5\times10^{-27}\nonumber \\
       &a_3 + b_3  x \,\,\,\,\text{for } -3.5\times10^{-27} \leq x<-5.9\times10^{-28}\nonumber \\
       &a_4 + b_4  x \,\,\,\,\text{for } -5.9\times10^{-28} \leq x<8.\times10^{-27}\nonumber \\
       &a_5 + b_5  x \,\,\,\,\text{for } \,\,\,\,\,\,\,8.\times10^{-27} \leq x<2.8\times10^{-26}\nonumber \\
       &a_6 + b_6  x \,\,\,\,\text{for } \,\,\,\,\,\,\,2.8\times10^{-26} \leq x<5.5\times10^{-26}\nonumber \\
       &a_7 + b_7  x \,\,\,\,\text{for } \,\,\,\,\,\,\,5.5\times10^{-26} \leq x \leq 6.1\times10^{-26},       
\label{eq:CDFfddotspin}
\end{align} where x is $\ddot{f}_{s}$ in s$^{-3}$, $a_1$ = 0.4211, $b_1$ = 1.645$\times10^{24}$, $a_2$ = 0.3794, $b_2$ = 1.244$\times10^{24}$, $a_3$ = 0.5253, $b_3$ = 4.296$\times10^{25}$, $a_4$ = 0.5086, $b_4$ = 1.455$\times10^{25}$, $a_5$ = 0.575, $b_5$ = 6.25$\times10^{24}$, $a_6$ = 0.6204, $b_6$ = 4.63$\times10^{24}$, $a_7$ = -0.2708, and $b_7$ = 2.083$\times10^{25}$.   \\

\vskip 0.3cm

B) MSPs where we only study orbital period and its derivatives

\vskip 0.2cm

For $l$,

\begin{align}
f(x) = a + b x + c x^2  + d x^3  + e x^4 + f x^5 + g x^6 + h x^7 + i x^{0.5},
\label{eq:CDFlorb}
\end{align} where x is $l$ in degrees, a = -0.04859, b = -0.02189, c = 6.774$\times10^{-4}$, d = -9.253$\times10^{-6}$, e = 6.367$\times10^{-8}$, f = -2.339$\times10^{-10}$, g = 4.380$\times10^{-13}$, h = -3.281$\times10^{-16}$, and i = 0.07764.\\

For $b$,
\begin{align}
f(x) = a + b x + c x^2  + d x^3  + e x^4 + f x^5 + g x^6 + h x^7 + i \tan^{-1} ( j x + k ),
\label{eq:CDFborb}
\end{align} where x is $b$ in degrees, a = 0.4778, b = 0.01177, c = 4.720$\times10^{-5}$, d = - 3.833$\times10^{-6}$, e = - 2.097$\times10^{-8}$, f = 8.940$\times10^{-10}$, g = 2.552$\times10^{-12}$, h = - 7.854$\times10^{-14}$, i = 0.08332, j = 0.4128, and k = 0.1518.\\

For $d$,
\begin{align}
f(x) = &a_1 + b_1 x + c_1 x^2 + d_1 x^3 + e_1 x^4 + f_1 x^5 \,\,\,\,\text{for } 0.16 \leq x<0.70 \nonumber \\
        &a_2 \tan^{-1} ( b_2 x + c_2 ) \,\,\,\,\,\hskip2.9cm\text{for } 0.70 \leq x \leq 10.37 ,
\label{eq:CDFdorb}
\end{align} where $d$ is in kpc, $a_1$ = 0.24179, $b_1$ = -4.01779, $c_1$ = 24.86681, $d_1$ = -69.71583, $e_1$ = 91.19579, $f_1$ = -44.62367, $a_2$ = 0.703997, $b_2$ = 0.738591, and $c_2$ = -0.319079. \\

For $\mu_{l}$,
\begin{align}
f(x) = a + b x + c x^2  + d x^3  + e x^4 + f x^5 + g x^6 + h x^7 + i \tan^{-1} ( j x + k ),
\label{eq:CDFmulorb}
\end{align} where x is $\mu_{l}$ in mas/yr, a = 0.5050, b = - 6.613$\times10^{-3}$, c = 4.728$\times10^{-6}$, d = 2.221$\times10^{-6}$, e = - 5.018$\times10^{-9}$, f = - 4.642$\times10^{-10}$, g = 8.919$\times10^{-13}$, h = 3.297$\times10^{-14}$, i = 0.4746, j = 0.1135, and k = 0.1306.\\

For $\mu_{b}$,
\begin{align}
f(x) = &a_1+ b_1 \tan^{-1} ( c_1 x + d_1 ) \,\,\,\,\hskip6.5cm\text{for }-103.0  \leq x< -10.88 \nonumber \\
       & a_2 + b_2 x + c_2 x^2  + d_2 x^3  + e_2 x^4 + f_2 x^5 + h_2 x^7 + i_2 \tan^{-1} ( j_2 x + k_2 ) \,\,\,\,\text{for }-10.88  \leq x< 24.0 \nonumber \\
       &a_3 + b_3 \tan^{-1} ( c_3 x + d_3 ) \,\,\,\,\hskip6.5cm\text{for } 24.0  \leq x \leq  120.82,
\label{eq:CDFmuborb}
\end{align} where x is $\mu_{b}$ in mas/yr, $a_1$ = 0.059922, $b_1$ = -0.033813, $c_1$ = -0.104716, $d_1$ = -1.579770, $a_2$ = 0.4630, $b_2$ = -5.730$\times10^{3}$, $c_2$ = 2.286$\times10^{5}$, $d_2$ = 1.348$\times10^{-6}$, $e_2$ = -2.167$\times10^{-9}$, $f_2$ = -1.692$\times10^{-10}$, $h_2$ = 7.069$\times10^{-15}$, $i_2$ = 0.4476, $j_2$ = 0.1461, $k_2$ = 0.02180, $a_3$ = -1.610$\times10^{3}$, $b_3$ = 1.026$\times10^{3}$, $c_3$ = 981, and $d_3$ = -9.258$\times10^{3}$.\\

For $f_{b}$,
\begin{align}
f(x) = &a_1 + b_1 x + c_1 x^3  \,\,\,\,\hskip7.cm\text{for }  1.7299\times10^{-28} \leq x<1.5127\times10^{-7} \nonumber \\ 
       &a_2 + b_2 x + c_2 x^2  + d_2 x^3  + e_2 x^4 + f_2 x^5 + g_2 x^6 + h_2 x^7 + i_2 x^{0.5} \,\,\,\,\text{for }1.5127\times10^{-7} \leq x \leq 1.7775\times10^{-4},
\label{eq:CDFforb}
\end{align} where x is $f_{b}$ in Hz, $a_1$ = 1.957$\times10^{-3}$, $b_1$ = 2.099$\times10^{5}$, $c_1$ = 1.815$\times10^{19}$, $a_2$ = -0.1036, $b_2$ = -1.246$\times10^{5}$, $c_2$ = 2.795$\times10^{9}$, $d_2$ = -5.180$\times10^{13}$, $e_2$ = 5.909$\times10^{17}$, $f_2$ = -3.924$\times10^{21}$, $g_2$ = 1.387$\times10^{25}$, $h_2$ = -2.011$\times10^{28}$, and $i_2$ = 566.2. \\

For $\dot{f}_{b}$,
\begin{align}
f(x) = &a_1 + b_1 x \,\,\,\,\hskip4.4cm\text{for } -8.3\times10^{-18} \leq x< -4.4\times10^{-19} \nonumber \\ 
&a_2 +  b_2 x +  c_2 \tan^{-1} ( d_2 x + e_2 )\,\,\,\,\hskip1.2cm\text{for } -4.4\times10^{-19} \leq x<-2.96\times10^{-23} \nonumber \\ 
&a_3 +  b_3 x \,\,\,\,\hskip4.4cm\text{for } -2.96\times10^{-23} \leq x<-5.0\times10^{-24} \nonumber \\ 
&a_4 +  b_4 \tan^{-1} ( c_4 x + d_4 )\,\,\,\,\hskip2.1cm\text{for } -5.0\times10^{-24} \leq x< 1.13\times10^{-22} \nonumber \\ 
&a_5 +  b_5 x \,\,\,\,\hskip4.4cm\text{for } 1.13\times10^{-22} \leq x< 1.82\times10^{-22} \nonumber \\ 
&a_6 +  b_6 x + c_6 x^2 \,\,\,\,\hskip3.3cm\text{for } 1.82\times10^{-22} \leq x<2.15\times10^{-20} \nonumber \\ 
&a_7 +  b_7 x + c_7 x^2 + d_7 \tan^{-1} ( e_7 x + f_7 )\,\,\,\,\text{for } 2.15\times10^{-20} \leq x \leq  1.9763\times10^{-18} ,
\label{eq:CDFfdotorb}
\end{align} where x is $\dot{f}_{b}$ in s$^{-2}$, $a_1$ = 0.07090, $b_1$ = 4.387$\times10^{15}$, $a_2$ = 0.1769, $b_2$ = 5.700$\times10^{16}$, $c_2$ = 0.05431, $d_2$ = 5.086$\times10^{19}$, $e_2$ = 0.6187, $a_3$ = 0.3361, $b_3$ = 4.353$\times10^{21}$, $a_4$ = 0.4179, $b_4$ = 0.1131, $c_4$ = 3.717$\times10^{23}$, $d_4$ = -0.4339, $a_5$ = 0.5761, $b_5$ = 1.486$\times10^{20}$, $a_6$ = 0.6024, $b_6$ = 4.680$\times10^{18}$, $c_6$ = 5.040$\times10^{37}$, $a_7$ = 0.7872, $b_7$ = 2.436$\times10^{16}$, $c_7$ = 1.742$\times10^{34}$, $d_7$ = 0.06457, $e_7$ = 1.287$\times10^{19}$, and $f_7$ = -1.645.\\

For $\ddot{f}_{b}$,
\begin{align}
f(x) = &a_1 + b_1  x \,\,\,\,\text{for } -1.6\times10^{-26} \leq x<-5.\times10^{-27} \nonumber \\
       &a_2 + b_2  x \,\,\,\,\text{for } -5.\times10^{-27} \leq x<2.4\times10^{-28}\nonumber \\
       &a_3 + b_3  x \,\,\,\,\text{for } \,\,\,\,\,\,\,2.4\times10^{-28} \leq x \leq 1.77\times10^{-26} ,
\label{eq:CDFfddotorb}
\end{align} where x is $\ddot{f}_{b}$ in s$^{-3}$, $a_1$ = 0.6136, $b_1$ = 2.273$\times10^{25}$, $a_2$ = 0.7385, $b_2$ = 4.771$\times10^{25}$, $a_3$ = 0.7466, and $b_3$ = 1.432$\times10^{25}$. \\

\end{document}